\documentclass[11pt, oneside]{Thesis} 

\hypersetup{linkcolor=blue,urlcolor=darkgreen,citecolor=darkgreen, colorlinks=true} 
\title{\ttitle} 

\usepackage[mathscr]{eucal}
\usepackage[margin=0.8cm]{caption}
\usepackage{comment}
\usepackage{graphicx}
\usepackage{subfigure}
\usepackage{amsthm}
\DeclareMathOperator*{\uint}{\scalerel*{\rotatebox{8}{$\!\scriptstyle\int\!$}}{\int}}
\usepackage{scalerel}
\usepackage{cprotect}
\usepackage{float}
\usepackage{nameref}
\usepackage{verbatimbox}

\usepackage{pgfornament}

\usepackage{tikz}
\usetikzlibrary{decorations.pathmorphing}
\usetikzlibrary{decorations.markings}
\usetikzlibrary{decorations.pathreplacing}
\usepackage{pgf}
\usepackage{pgffor}
\usepgfmodule{shapes}
\usepgfmodule{plot}
\usetikzlibrary{shapes.geometric}
\graphicspath{{Images/}}

\tikzstyle{solid}=                   [dash pattern=]
\tikzstyle{dotted}=                  [dash pattern=on \pgflinewidth off 2pt]
\tikzstyle{densely dotted}=          [dash pattern=on \pgflinewidth off 1pt]
\tikzstyle{loosely dotted}=          [dash pattern=on \pgflinewidth off 4pt]
\tikzstyle{dashed}=                  [dash pattern=on 3pt off 3pt]
\tikzstyle{densely dashed}=          [dash pattern=on 3pt off 2pt]
\tikzstyle{loosely dashed}=          [dash pattern=on 3pt off 6pt]
\tikzstyle{dashdotted}=              [dash pattern=on 3pt off 2pt on \the\pgflinewidth off 2pt]
\tikzstyle{densely dashdotted}=      [dash pattern=on 3pt off 1pt on \the\pgflinewidth off 1pt]
\tikzstyle{loosely dashdotted}=      [dash pattern=on 3pt off 4pt on \the\pgflinewidth off 4pt]

\definecolor{darkgreen}{rgb}{0,0.45,0}
\definecolor{darkblue}{rgb}{0,0,0.3}
\definecolor{darkred}{rgb}{0.7,0,0}
\definecolor{light gray}{RGB}{230,230,230}
\definecolor{dgreen}{RGB}{16,163,34}
\newcommand{\eg}{\textit{e.g.}\,\,}
\newcommand{\ie}{\textit{i.e.}\,}
\def\deltabar{\slash\hspace{-6pt} \delta}
\newcommand{\Example}[1]{\begin{example}\small{\emph{#1}}\end{example}}
\newcommand{\Lemma}[1]{\begin{lemma}\small{\emph{#1}}\end{lemma}}
\newcommand{\mrd}{\mathrm{d}}
\newcommand{\mt}{\mathrm{t}}
\newcommand{\mr}{\mathrm{r}}
\newcommand{\mrn}{\mathrm{n}}
\newcommand{\mrm}{\mathrm{m}}
\newcommand{\mrH}{\mathrm{H}}

\newcommand{\dotfillb}{
\begin{center}
\vspace*{0.7cm}
\textcolor{light gray}{\hrule height 1pt width 1\textwidth}

\vspace*{-0.4cm}
\textcolor{light gray}{\rule{0.5\textwidth}{1.5pt}}

\vspace*{-1.22cm}
\pgfornament[scale=1,width=6cm,
color = light gray,symmetry=h]{85}

\end{center}
\vspace*{0cm}}

\newcommand{\dotfille}{
\vspace*{-0.5cm}
\begin{center}
\pgfornament[scale=1,width=1.3cm,
color = light gray,symmetry=h]{41}
\hspace*{12.6cm}
\pgfornament[scale=1,width=1.3cm,
color = light gray,symmetry=c]{41}

\vspace*{-0.7cm}
\pgfornament[scale=1,width=6cm,
color = light gray]{85}
\end{center}}

\begin{document}

\frontmatter 

\setstretch{1.3} 

\fancyhead{} 
\rhead{\thepage} 
\lhead{} 

\pagestyle{fancy} 

\newcommand{\HRule}{\rule{\linewidth}{0.5mm}} 
\hypersetup{pdftitle={\ttitle}}
\hypersetup{pdfsubject=\subjectname}
\hypersetup{pdfauthor=\authornames}
\hypersetup{pdfkeywords=\keywordnames}


\begin{titlepage}

\begin{center}

\hspace*{-0.5cm}
\begin{minipage}{0.3\textwidth}
\begin{center}
\includegraphics[width=0.3\textwidth]{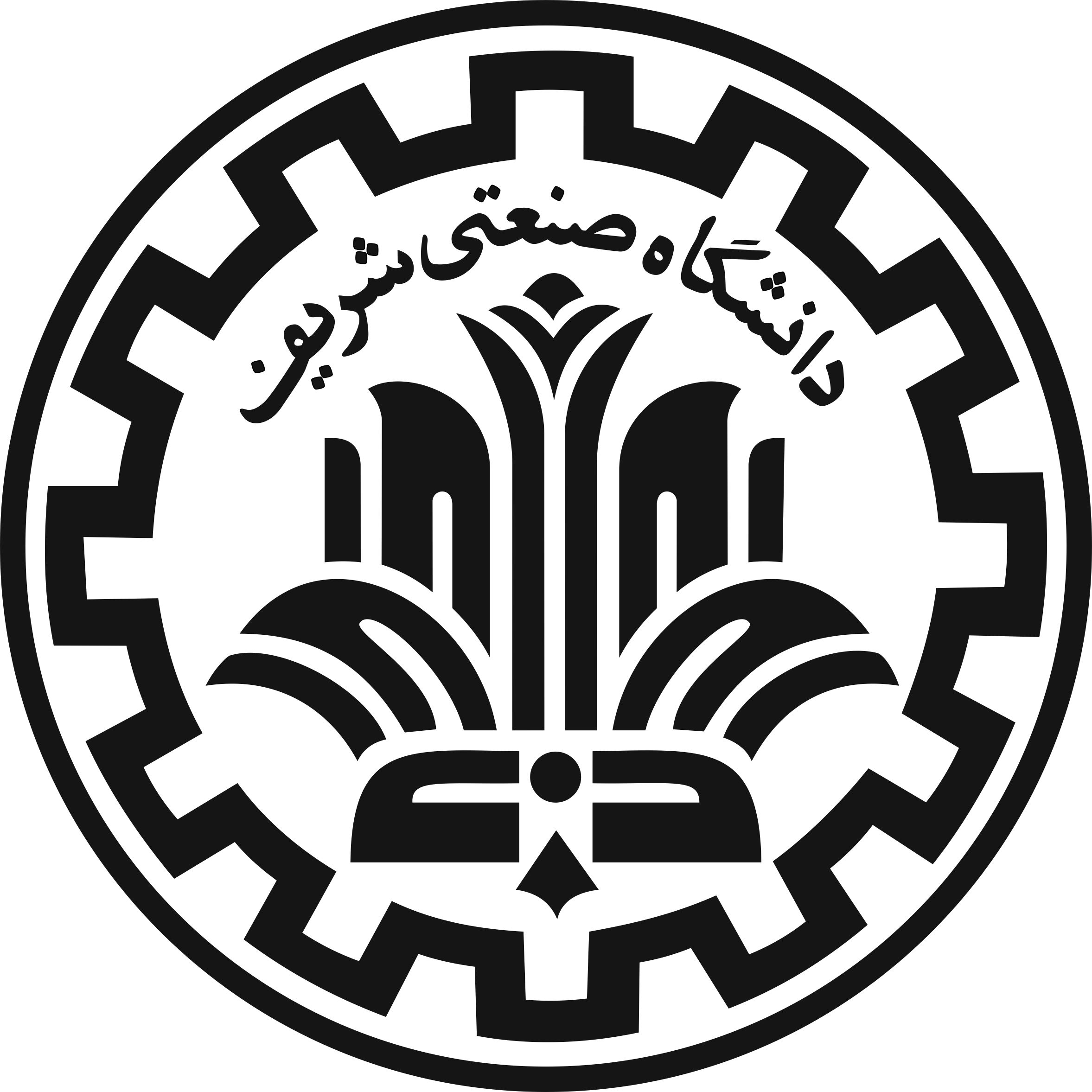}
\end{center}
\vspace*{-0.5cm}
\captionof*{figure}{\small Sharif University\\ of Technology}
\end{minipage}
\hspace*{4.7cm}
\begin{minipage}{0.4\textwidth}
\vspace*{0.2cm}
\begin{center}
\includegraphics[width=0.3\textwidth]{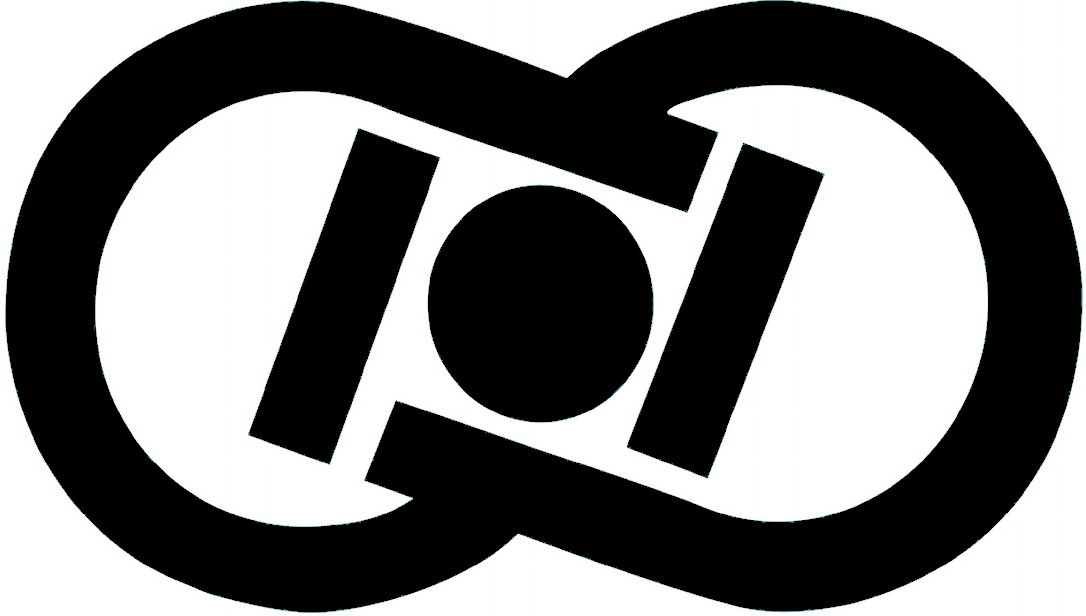}
\end{center}
\vspace*{-0.4cm}
\captionof*{figure}{\small Institute for Research in \\ Fundamental Sciences}
\end{minipage}
\\[2cm]
\textsc{\Large Doctoral Thesis}\\[0.5cm] 

\HRule \\[0.4cm] 
{\huge \bfseries \ttitle}\\[0.4cm] 
\HRule \\[1.5cm] 
 
\begin{minipage}{0.46\textwidth}
\begin{flushleft} \large
\emph{Author:}\\
\textcolor{blue}{\authornames} 
\end{flushleft}
\end{minipage}
\begin{minipage}{0.46\textwidth}
\begin{flushright} \large
\emph{Supervisor:} \\
\textcolor{blue}{\supname}\\
\emph{Advisor:}\\
\textcolor{blue}{Prof. \textsc{M. Golshani}}\\
\emph{Collaborators:}\\
\textcolor{blue}{\textsc{A. Seraj}}\\
\textcolor{blue}{\textsc{G. Comp\`ere}}
\end{flushright}
\end{minipage}\\[3cm]
 
\large \textit{A thesis submitted in fulfilment of the requirements\\ for the degree of \degreename}\\[0.3cm] 
\textit{in Physics}\\[2cm]
 
{\large July 2015}\\[0.5cm] 
{\large Tehran, IRAN} 

\vfill
\end{center}
\end{titlepage}


\abstract{\addtocontents{toc}{\vspace{1em}} 

Near Horizon Extremal Geometries (NHEG), are geometries which may appear in the near horizon region of the extremal black holes. These geometries have $SL(2,\mathbb{R})\!\times\!U(1)^n$ isometry, and constitute a family of solutions to the theory under consideration.  In the first part of this report, their thermodynamic properties are reviewed, and their three universal laws are derived. In addition, at the end of the first part, the role of these laws in black hole thermodynamics is presented. 

In the second part of this thesis, we review building their classical phase space in the Einstein-Hilbert theory. The elements in the NHEG phase space manifold are built by appropriately chosen coordinate transformations of the original metric. These coordinate transformations are generated by some vector fields, dubbed ``symplectic symmetry generators". To fully specify the phase space, we also need to identify the symplectic structure. In order to fix the symplectic structure, we use the formulation of Covariant Phase Space method. The symplectic structure has two parts, the Lee-Wald term and a boundary contribution. The latter is fixed requiring  on-shell vanishing of the symplectic current, which guarantees the conservation and integrability of the symplectic structure, and leads to the new concept of ``symplectic symmetry". Given the symplectic structure, we construct the corresponding conserved charges, the ``symplectic symmetry generators".  We also specify the explicit expression of the charges as a functional over the phase space. These symmetry generators constitute the ``NHEG algebra", which is an infinite dimensional algebra (may be viewed as a generalized Virasoro), and admits a central extension which is equal to the black hole entropy.

This report is a review of the subjects presented in \cite{HSS:2013lna,HSS:2014twa,CHSS:2015mza,CHSS:2015bca}, however some proofs and calculations are presented in new ways.
}

\clearpage 


\setstretch{1.3} 

\acknowledgements{
This report is prepared to be my Ph.D thesis, an academic outcome of eleven years of my educations at Sharif University of Technology. I really thank my teachers who provided me enough materials needed to study the subject. Specifically I am using this opportunity to express my gratitude to prof. Golshani, teacher of many of important courses in my B.Sc, M.Sc and Ph.D. I also thank him for his kind supports as supervisor/advisor of my M.Sc and Ph.D programs. Without his supports, collaboration with high energy group at the Institute for Research in Fundamental Sciences (IPM) was impossible for me, the very existence of this document.

I would also thank the high energy group at IPM for providing opportunity for me to be student researcher there. This thesis has been mainly progressed in that group. Specifically I do acknowledge prof. M.M. Sheikh-Jabbari (Shahin), the supervisor of the project. His kind scientific and financial supports, in addition to his nice lectures on related subjects, empowered me to begin, continue and finish my Ph.D program. Undoubtedly, His role as the supervisor of the project has been very significant. In addition, his hard working and courtesy would be an exemplary for me. Also, I appreciate Ali Seraj as the collaborator of this research. Results of this project are outcomes of our three years of teamwork, and any citation would refer to him and Shahin as well. I am sincerely grateful to them for sharing their truthful and illuminating views. For sure, the very existence of this thesis is due to the wonderful scientific supports from Shahin and critical contributions from Ali. What I mostly remember are Shahin's determination to the right points, and Ali's rigorous calculations augmented by innovations. I would also like to thank Geoffrey Comper\'e for his nice collaboration in the second part of the thesis.

I would like to emphasize the role of  IPM HPC cluster in my computations. Without it, some of my calculations could not be carried out. I would like to thank IPM staff, for their hospitality. Also, I would gratefully thank International Center for Theoretical Physics (ICTP) for its kind supports, providing opportunities for me to attend international conferences and interact more with physics society. Again, I would like to thank Shahin, for his prominent role in making connections between young researchers in Iran and international physics communities. I wanted to offer my grateful thanks to Allameh Tabatabaii Prize Grant of Boniad Melli Nokhbegan of Iran.  

Finally, an acknowledgement in advance! I would really appreciate a careful reader who would find mistakes in the report, and would inform me about them.
}
\clearpage 
\pagestyle{fancy} 

\tableofcontents 

\clearpage 

\setstretch{1.5} 

\listofsymbols{ll} 
{
ADM & \textbf{A}rnowitt \textbf{D}eser \textbf{M}isner\\
AdS & \textbf{A}nti \textbf{d}e \textbf{S}itter\\
BC & \textbf{B}oundary \textbf{C}ondition\\
BH & \textbf{B}lack \textbf{H}ole \\
BTZ & \textbf{B}añados \textbf{T}eitelboim \textbf{Z}anelli\\
CP & \textbf{C}arter \textbf{P}enrose\\
dim & \textbf{dim}ension\\
EH & \textbf{E}instein \textbf{H}ilbert\\
EM & \textbf{E}instein \textbf{M}axwell\\
EMD & \textbf{E}instein \textbf{M}axwell \textbf{D}ilaton\\
e.o.m & \textbf{e}quation \textbf{o}f \textbf{m}otion\\
EPL & \textbf{E}ntropy \textbf{P}erturbation \textbf{L}aw\\
EVH & \textbf{E}xtremal \textbf{V}anishing \textbf{H}orizon\\
KK & \textbf{K}aluza \textbf{K}lein\\
l.e.o.m & \textbf{l}inearized \textbf{e}quation \textbf{o}f \textbf{m}otion\\
LHS & \textbf{L}eft \textbf{H}and \textbf{S}ide\\
LW & \textbf{L}ee \textbf{W}ald\\
MP & \textbf{M}yers \textbf{P}erry\\
NHE- & \textbf{N}ear \textbf{H}orizon \textbf{E}xtremal -\\
NHEG & \textbf{N}ear \textbf{H}orizon \textbf{E}xtremal \textbf{G}eometry \\
RHS & \textbf{R}ight \textbf{H}and \textbf{S}ide\\
w.r.t & \textbf{w}ith \textbf{r}espect \textbf{t}o

}


\mainmatter 

\pagestyle{fancy} 


\setstretch{1.4}
\addtotoc{${\ast}$\,\,\,\,\,Conventions}
\chapter*{Conventions}

Some conventions which are used in the report:
\begin{itemize}
\item[--] The Greek alphabets are used to indicate spacetime indices, \eg \,$x^\nu$.
\item[--] The Einstein summation convention is used; same indices which one is up and other one is down, would be summed over. \eg there is a summation in $A_\mu A^\mu$ but there is not in $A^\mu=B^\mu$.
\item[--] The metric signature is chosen to be diag$(-,+,+,\cdots)$.
\item[--] The convention $c=\hbar=1$ is used, but the Newton constant $G$ is kept manifest for clarity. 
\item[--] The Newton constant in any dimension has been shown by $G$, without any index denoting dimensions.
\item[--] Citations which refer to a pedagogical reference (compared to original reference)  are put in brackets, \eg [[\textcolor{darkgreen}{23}]].
\item[--] Putting indices in \emph{parentheses} denotes symmetrization of those indices. For example  $A_{(\mu\nu)}=\dfrac{1}{2!}(A_{\mu\nu}+A_{\nu\mu})$. Putting them in \emph{brackets} means anti-symmetrization, \eg $A_{[\mu\nu]}=\dfrac{1}{2!}(A_{\mu\nu}-A_{\nu\mu})$.
\item[--] The standard notation for the metric $g_{\mu\nu}$, Riemann tensor $R^\alpha_{\,\,\,\beta\mu\nu}$, Ricci tensor $R_{\mu\nu}$, Ricci scalar $R$, Einstein tensor $G_{\mu\nu}$,  gauge field $A_{\mu}$ and field strength $F_{\mu\nu}$ are used. 
\item[--] The notation $\delta g^{\mu\nu}\equiv g^{\mu\alpha}g^{\nu\beta}\delta g_{\alpha\beta}$ is used. On the other hand, variations of $g^{\mu\nu}$ are denoted by $\delta (g^{\mu\nu})$.
\item[--] The sign ``$\approx$" is used to show \emph{on-shell} equations. The sign ``$\sim$" is used to identify a coordinate by a period, \eg $\,\varphi\sim\varphi+2\pi$ means that the coordinate $\varphi$ is periodic with the period $2\pi$.
\item[--] The Roman ``$\mathrm{d}$" shows exterior derivation in spacetime, while ``$\delta$" is used to show exterior derivative in phase space. The normal integration sign ``$\int$" shows integration on spacetime environment, while 

\vspace{-0.5cm}
{\small$$\uint$$}

\vspace{-0.8cm}
is used to denote integrations on the phase space. 
\end{itemize}

Some general issues about the report:
\begin{itemize}
\item[--] Naturally, some of the expressions in this report are author's personal opinions, and subjective.
\item[--] This report has been provided in a way that be most useful for undergraduate and graduate students. In order to follow the main subjects, a two-semester course in the general relativity would suffice. Therefore, from the point of view of an expert, this report might be seen as a long, boring and inaccurate paper.
\item[--] Considering intuition and examples vs. rigour and proofs, the former has been chosen. The rigorous reader is expected to refer the original papers on the subjects, if vague or inaccurate expressions are found. Nonetheless, some technical calculations and some crucial proofs are included.
\item[--] This report is an intuitive and pedagogical review of the series of works \cite{HSS:2013lna,HSS:2014twa,CHSS:2015mza,CHSS:2015bca}. More specifically, Part \ref{part I} is based on the papers \cite{HSS:2013lna,HSS:2014twa}, and Part \ref{part II} is a review of \cite{CHSS:2015mza,CHSS:2015bca}.  
\item[--] In order to follow discussions in this report, a necessary language is the language of \emph{differential form}s. We have provided a short but shallow introduction in Appendix \ref{app diff forms} for readers who are not familiar with this language. It would be ``Chapter $0$" for them.
\item[--] The Persian version of this document is available on the link below:\\
\href{https://drive.google.com/open?id=0B8AqjYFfBKw0RDk2QmwxOUdpeVU}{
https:$//$drive.google.com$/$open?id=0B8AqjYFfBKw0RDk2QmwxOUdpeVU
}
\end{itemize}
        
\addtocontents{toc}{\protect\setcounter{tocdepth}{1}}
\part{On thermodynamics of Near Horizon Extremal Geometries}\label{part I}

\addtotoc{${\ast}$\,\,\,\,\,Motivations and Outline}
\chapter*{Motivations and Outline}
Black holes (BH) are among very massive compact objects discovered to date. They first appeared theoretically, as solutions to the Einstein-Hilbert gravity, at the end of the second decade of the twentieth century. The specific feature of these solutions is the event horizon. It is a one-way surface; things can fall in to it, but can not escape to the outside. Although they are some classical exact solutions to formerly believed fundamental theories, but they also behave as thermodynamic systems. The origin of these thermodynamic properties has been a big question for black hole physicists. To be more specific, they have temperature \cite{Hawking:1976rt} and entropy \cite{Bekenstein:1973ft}, properties of thermodynamic systems. Also,  analogous to the well-known thermodynamic laws, \ie
\begin{itemize}
\item[${(0)}$] on a system in thermal equilibrium, temperature (and other chemical potentials) are constant,
\item[${(1)}$] $\delta E=T \delta S-P \delta V+\mu \delta N$,
\item[${(2)}$] in a closed thermodynamic system, entropy never decreases,
\item[${(3)}$] \begin{itemize}\item $T=0$ is not physically achievable, \item If $T\to 0$ then $S\to 0$,\end{itemize}
\end{itemize}  
they fulfil some laws. Specifically, if the BH has mass $M$, some angular momenta $J_i$ and electric charges $Q_p$ (ignoring other types of charges for simplicity in this thesis), one can associate Hawking temperature $T_{_\mrH}$ and chemical potentials $\Omega_{_\mrH}^i$ and $\Phi_{_\mrH}^p$ to it. $\Omega_{_\mrH}^i$ are the angular velocities of the BH on its event horizon, and $\Phi_{_\mrH}^p$ are its electric potentials on that horizon. Then the BH thermodynamic laws would be \cite{Bardeen:1973gd}
\begin{itemize}
\item[$\mathbf{(0)}$] $T_{_\mrH}$, $\Omega_{_\mrH}^i$ and $\Phi_{_\mrH}^p$ are constant over the horizon,
\item[$\mathbf{(1)}$] $\delta M=T_{_\mrH} \delta S+\Omega_{_\mrH}^i \delta J_i+\Phi_{_\mrH}^p \delta Q_p$ ,
\item[$\mathbf{(2)}$] For a closed thermodynamic system (including the black hole), entropy never decreases,
\item[$\mathbf{(3)}$] $T_\mrH=0$ is not physically achievable.
\end{itemize}

BHs at zero temperature are called \emph{extremal} BHs. An interesting property of the extremal BHs is that their entropy is not usually equal to zero. This property makes their thermodynamic behaviours  non-trivial. For example, keeping $T_{_\mrH}\!=\!0$, the entropy would be a non-trivial function of $J_i$ and $Q_p$. This issue makes them worth studying.  Considering BH laws of thermodynamics, two natural questions arises.
\begin{enumerate}
\item Putting $T_{_\mrH}=0$ in the first law, leads to $\delta M=\Omega_{_\mrH}^i \delta J_i+\Phi_{_\mrH}^p \delta Q_p$. Then, what universal relation does $\delta S$ satisfy? Variating the $J_i$ and $Q_p$, while keeping $T_{_\mrH}=0$, the first law is blind to the answer of this question.
\item For the daily thermodynamic systems, by the third law,  if $T\to 0$ then $S\to 0$. What is the analogue of this law for BHs? If $T_{_\mrH}\!\to\! 0$ then $S\to ?$ Is there a universal law which answers this question? 
\end{enumerate}
Answering these questions would be some motivations for studying Part \ref{part I} of the report.

Another motivation for Part \ref{part I} originates from the works of Iyer and Wald \cite{Wald:1993nt,Iyer:1994ys}. They have shown that entropy of the stationary BHs is the conserved charge associated to the Killing horizon vector, calculated on the horizon. But their analysis is heavily based on  some assumptions, which fail for the extremal BHs. Question would be whether one can carry out similar analysis for the extremal BHs. If it would be possible, then the entropy would be the conserved charge of which Killing vector? and calculated on which surface?

The approach which is used in this report, to study the thermodynamics of extremal BHs, is based on their near horizon geometries. They are geometries which are found by a limiting process towards the horizon of the extremal BHs, and are solutions to the same theory. The thermodynamic quantities (including conserved charges, entropy, temperature and chemical potentials) are encoded in the near horizon geometries. Therefore, by studying those geometries, one would be hopeful to answer the above questions. We hope that the reader would find the answers in Part \ref{part I}. In addition, Part \ref{part I} deals with some other interesting issues, like  introducing the Near Horizon Extremal Geometries \cite{Bardeen:1999ds}, their properties \cite{Kunduri:2007vf,Kunduri:2008rs,Kunduri:2013gce} and dynamical laws \cite{HSS:2013lna}. It extends the domain of thermodynamic behaviours to some solutions which are not BH. 

Outline of Part \ref{part I} is as follows. Chapter \ref{chap some grav}, will provide us the general context which we will study the physical systems.  Specifically, specifications of gravitational theories that we will study in Part \ref{part I}, are discussed. Then, two concepts are distinguished; symmetry and isometry, which are important for our discussions. Finally, some well-known BHs are provided as examples, which will be with us, till the end of Part \ref{part I}. Chapter \ref{chap Covariant Phase Space method} will enable us to calculate conserved charges, in covariant gravitational theories. It is based on a method, known as Covariant Phase Space Method. Having introduced the BH solutions, and knowing how to calculate their conserved charges, in Chapter \ref{chap BH thermo}, the thermodynamic variables of BHs are calculated. Then the laws of BH thermodynamics are reviewed. At the end of Chapter \ref{chap BH thermo}, we pose some questions to the thermodynamic laws, in the case of the extremal BHs. The questions are the ones which were quickly mentioned in Motivation. 

To answer those questions, we study the extremal BHs in their near horizon region. Chapter \ref{chap extremal near horizon} will introduce the near horizon geometry of the extremal BHs. They are some solutions, with their own isometries, without event horizon, and can be treated as a new family of solutions (in contrast with BHs). So in Chapter \ref{chap NHEG}, they are treated in this way, \ie are studied independent of the original extremal BHs. That chapter is devoted to geometrical properties of them. Finally, in Chapter \ref{chap NHEG thermo}, their conserved charges and other thermodynamic quantities are calculated. Focusing on the mentioned family of solutions, three universal (thermo) dynamic laws are then derived. At the end of Chapter \ref{chap NHEG thermo}, the role of those laws are presented in BH thermodynamics, answering the posed questions.

The last sentence, providing a big picture for the reader: Part \ref{part I} studies extremal BHs and their near horizon geometry, at the \emph{thermodynamic} level, while Part \ref{part II}, tries to go deeper, into their phase space, hopefully towards their \emph{microstates}.

\fancyhead[L]{\leftmark}
\chapter{Quick review on gravitational theories and their black hole solutions}\label{chap some grav}
The aim of this chapter is to provide the basic materials and some examples needed for later discussions. In the first section, Lagrangian, action and equation of motion are reviewed. In the second section,  symmetry and isometry are defined and distinguished. In the last section, black holes are briefly introduced and a couple of them are presented explicitly. We will follow those explicit black holes, providing examples in our analysis till the end of Part \ref{part I}.


\section{Action and equation of motion}
Our analysis in this report is based on the Lagrangian formulation of gravitational theories. The Lagrangian density $\mathcal{L}$ can depend on different dynamical fields, for example the metric $g_{\mu\nu}$, some Maxwellian gauge fields $A^{(p)}_\mu$ (distinguished by the label $p$), some scalar fields $\phi^I$ (distinguished by the label $I$) etc.
In general, the only request on the Lagrangian density $\mathcal{L}$ would be \emph{general covariance} and \emph{locality}. 
\begin{itemize}
\item \emph{General covariance:} $\mathcal{L}$ would be a scalar built covariantly from the dynamical fields; for example $\mathcal{L}=\nabla_\mu \phi \nabla^\mu\phi$ is a covariant one, but $\mathcal{L}=\partial_\mu g_{\alpha\beta}\partial^\mu g^{\alpha\beta}$ is not.
\item \emph{Locality:} 1) the fields in the Lagrangian are written in \emph{one} point of spacetime and  2) there would \emph{not} be infinite numbers of a field and its derivatives multiplied to each other. As an example $\mathcal{L}=\nabla_\mu \phi(x_1) \nabla^\mu\phi(x_2)$ is not a local Lagrangian, but $\mathcal{L}=\nabla_\mu \phi(x_1) \nabla^\mu\phi(x_1)$ is.
\end{itemize}  
Given a Lagrangian density $\mathcal{L}$ in $d$ dimensional spacetime, the \emph{action} which is denoted by $\mathcal{S}$ would be
\begin{equation}
\mathcal{S}=\int d^dx \sqrt{-g}\mathcal{L}\,.
\end{equation}
Assuming the dynamical fields to be $g_{\mu\nu}$, $A_\mu$, $\phi$ \ldots, variation of these fields are denoted as $\delta g_{\mu\nu}$, $\delta A_\mu$, $\delta \phi$ \ldots, and are referred to  \emph{dynamical field perturbations}. This nomenclature emphasises that these perturbations do not change the coordinates $x^\mu$, derivatives $\partial_\mu$ or any other non-dynamical entities. In order to find equations of motion (e.o.m), one might use the recipe of \emph{principle of least action}. Variation of the action w.r.t the dynamical fields, leads to   
\begin{align}\label{delta action}
\delta \mathcal{S}= \int \mrd^dx \Big(E_g^{\mu\nu}\delta g_{\mu\nu}+E_A^\mu\delta A_{\mu}+E_\phi\delta \phi\Big)
+\int  \mrd^dx \sqrt{-g}\nabla_\mu \Theta^\mu \,.
\end{align}
Putting $\delta \mathcal{S}=0$ and dropping the surface term, results $E_g^{\mu\nu}=0$, $E_A^\mu=0$, $E_\phi=0$, the e.o.m's. $\Theta^\mu$ denotes a surface term which appears in the variation, and is usually dropped by some boundary conditions (BCs)\footnote{It turns out that in gravitational theories with scalar Lagrangians constructed from Riemann curvature and its contractions with metric, one can not discard the surface term by a consistent boundary condition. There are some tricks on the remedy of this problem [\cite{padmanabhan2010gravitation}] and we don't discuss here. We will use the standard e.o.m and $\Theta^\mu$.}.

A more efficient way of writing the equations above is the language of forms; one can write the volume element $\mrd^dx \sqrt{-g}$ as a covariant $d$-form
\begin{equation}\label{Volume d-form}
\boldsymbol{\epsilon}\equiv \frac{\sqrt{-g}}{d!}\,\,\epsilon_{\mu_1\mu_2\cdots \mu_d}\,\,\mrd x^{\mu_1}\wedge \mrd x^{\mu_2}\wedge\cdots\wedge \mrd x^{\mu_d}\,,
\end{equation}
where $\epsilon_{\mu_1\mu_2\cdots \mu_d}$ is the Levi-Civita symbol, $\epsilon_{_{012\cdots d-1}}=+1$ and changes sign according to the permutations of indices. In this language (language of differential forms) action is written as
\begin{equation}
\mathcal{S}=\int \boldsymbol{\epsilon} \mathcal{L}=\int \mathbf{L}\,.
\end{equation}
The $d$-form $\mathbf{L}$ is Hodge dual to the $\mathcal{L}$ by the definition of Hodge duality \eqref{Hodge duality}, as
\begin{equation}\label{Lagrangian top form}
\mathbf{L}=\frac{\sqrt{-g}}{d!}\,\,\epsilon_{\mu_1\mu_2\cdots \mu_d}\,\,\mrd x^{\mu_1}\wedge \mrd x^{\mu_2}\wedge\cdots\wedge \mrd x^{\mu_d} \,\mathcal{L}\equiv \star \mathcal{L}\,.
\end{equation}
We can denote all dynamical fields (\eg $g_{\mu\nu}$, $A_{\mu}$, etc) by a single symbol $\Phi$. Then \eqref{delta action} can be written more economically as
\begin{equation}\label{delta S}
\delta \mathcal{S}=\int \delta\mathbf{L}=\int \mathbf{E}_\Phi\delta \Phi +\mrd\mathbf{\Theta}\,,
\end{equation} 
where the $(d-1)$-form $\mathbf{\Theta}$ is Hodge dual to the $1$-form $\Theta$, as
\begin{equation}\label{Theta d-1-form}
\mathbf{\Theta}=\frac{\sqrt{-g}}{(d-1)!}\,\,\epsilon_{\mu\mu_1\cdots \mu_{d-1}}\,\,\mathrm{d}x^{\mu_1}\wedge \cdots\wedge \mathrm{d}x^{\mu_{d-1}} \,\Theta^{\mu}\equiv \star \Theta\,.
\end{equation}
The \emph{on-shell} condition means $\mathbf{E}_\Phi=0$ for any one of the fields $\Phi$. Note that in order to find solutions to the e.o.m, some boundary conditions (or some isometry conditions) are needed.


\section{Symmetry vs. Isometry}\label{sec sym iso}
\emph{Symmetry} is a transformation which leaves the Lagrangian density invariant, up to a total divergence,
\begin{equation}
\mathcal{L}\to \mathcal{L}+\nabla_{\mu}\mathcal{K}^\mu\,,
\end{equation}
which using \eqref{Hodge duality} and \eqref{coderivative id} can be written as
\begin{equation}\label{symmetry condition}
\mathbf{L}\to \mathbf{L}+\mathrm{d}\boldsymbol{\mathcal{K}}\,,
\end{equation}
where $\boldsymbol{\mathcal{K}}=\star \mathcal{K}$. In other words, symmetries are transformations which leave the e.o.m intact.

\emph{Isometry} is a transformation which does not change a given field configuration, \ie\, $\delta \Phi=\Phi'(x^\mu)-\Phi(x^\mu)=0$. If that transformation is generated by some vector field $\xi$, then isometry condition is $\delta_\xi\Phi=\mathscr{L}_\xi\Phi=0$, where $\mathscr{L}_\xi$ is Lie derivative. For the specific case that $\Phi$ is  metric $g_{\mu\nu}$, the isometry is called \emph{Killing}, and $\xi$ is called \emph{Killing vector} satisfying the condition
\begin{equation}\label{Killing condition}
\mathscr{L}_\xi\, g_{\mu\nu}=2\nabla_{(\mu} \xi_{\nu)}=0\,.    
\end{equation}    
Comparing symmetry and isometry, the former is a property of a theory identified by a Lagrangian, but the latter is a property of a given field configuration irrespective to any theory.

\newpage
\vspace*{-2cm}
\dotfillb
\Example{\emph{Symmetry vs. Isometry:}\\
Assuming that planet Saturn is a spherically isometric planet, then the Newtonian theory of gravity around it would be spherically symmetric. But, having a gravitational theory with spherical symmetry, does not necessarily lead to solutions with similar isomerty. For example, in the same sense that motion of moons of Saturn are solutions to the theory, dynamics of its rings are also solutions. These solutions have not spherical isometry, but instead have cylindrical isometry. 
}
\begin{figure}[h!]
\begin{center}
\includegraphics[width=1\textwidth]{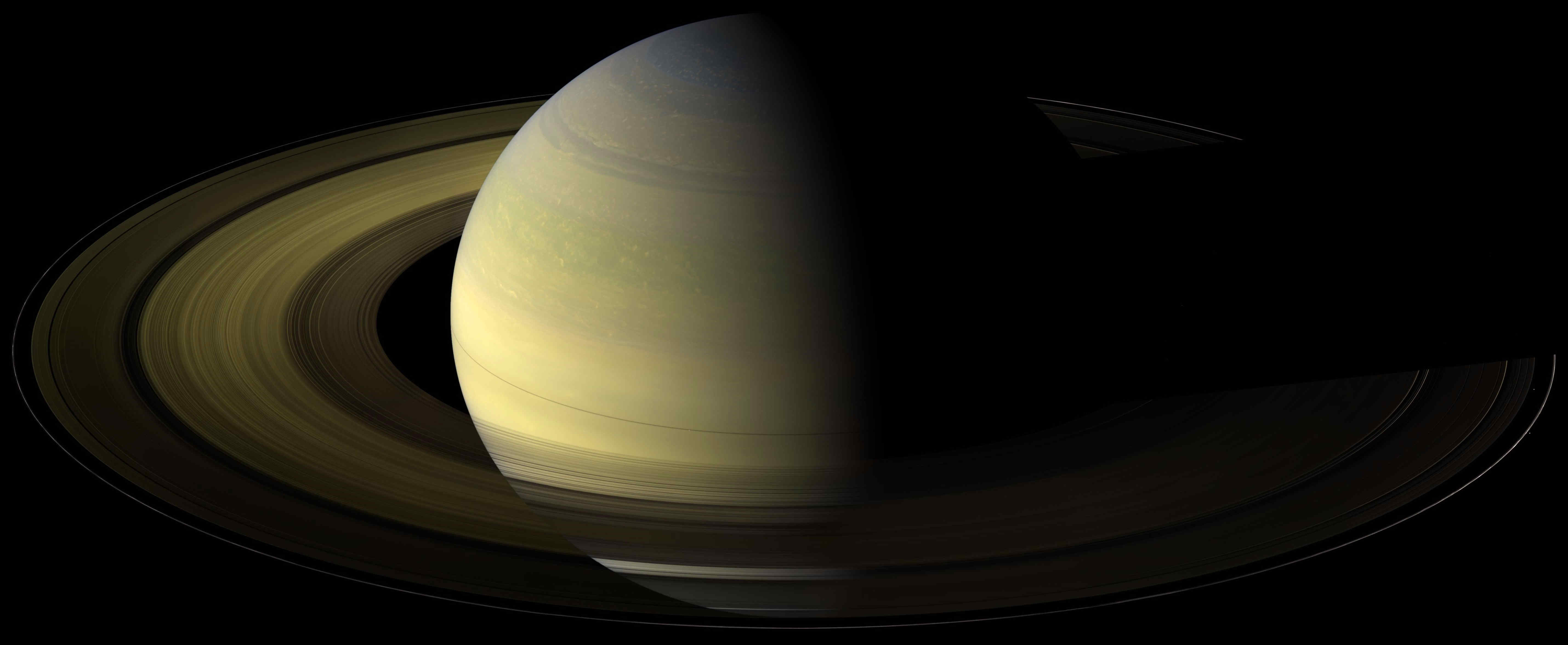}
\end{center}
\caption{Saturn is a planet with spherical isometry. The Newtonian gravity around it has spherical symmetry, while the dynamics of its rings are some solutions to that theory with cylindrical isometry. (A real image of Saturn taken by Cassini in 2009, http:$//$photojournal.jpl.nasa.gov$/$catalog$/$PIA11667)}
\end{figure}
\vspace*{-1cm}
\dotfille


\section{Some black hole solutions}\label{sec BHs}
Solutions which have topologically sphere event horizons are called \emph{black holes}. Blackness is related to the event horizon, a $d-1$ dimensional null hypersurface which act as a one-way membrane; things can go inside, but there would not be any possibility for classical return. Hole-ness is related to the sphere topology of the event horizon. There are other topologies, \eg planar/non-compact topology for black branes and $S^2\times S^1$  topology for black rings in $5$-dim, etc. An event horizon $\mrH$, is generically a \emph{Killing horizon} which is defined by the following conditions: 
\begin{enumerate}
\item $\mrH$ is a $d\!-\!1$ dimensional null hypersurface.
\item There is a Killing vector field $\zeta_{_\mrH}$ which becomes null on the $\mrH$, \ie \,$\zeta_{_\mrH}^\alpha \zeta_{_\mrH\alpha} \big|_\mrH=0$.
\end{enumerate}
Killing horizons can be \emph{bifurcating} Killing horizons. Literally it means that there is a $(d\!-\!2)$-dim surface, where the Killing horizon bifurcates on that surface (see Figure \ref{fig Kerr BH} for an example). Its concrete definition is provided in chapter \ref{chap BH thermo}.  On the bifurcation surface which is also usually denoted as $\mrH$, the $\zeta_{_\mrH}$ vanishes. We will use $\mrH$ for the horizon or its bifurcation surface interchangeably.

In the next sections, we provide a couple of examples of known BHs which would serve as our examples in later discussions. They have been chosen in a way to be examples for some different theories and dimensions. Also they are chosen to have the $\mathbb{R}\times U(1)^n$ isometry for $0<n$ as is described below.
\begin{itemize}
\item We choose the black holes to be \emph{stationary}, \ie having a timelike Killing vector field outside the horizon. Stationarity is time translation isometry in curved spacetimes, so has real numbers $\mathbb{R}$ as its parameter of transformation.
\item We choose the black holes to have at least one direction of axial isometry. \emph{Axial isometry}\footnote{We prefer using the word ``axial isometry" instead of commonly used ``axi-symmetry", emphasizing the difference between symmetry as a property of theory, and isometry as a property of a field configuration.} as the name suggests, is the isometry of rotating a cylinder around its axis. It is isomorphic to the $1\times 1$ unitary complex matrices, so is  denoted by $U(1)$. The range of its parameter of transformation is $\mathbb{R}$, but the matrix would be periodic by $2\pi$.  BHs in $d$ dimensional spacetime can have different axial isometries, denoted as $U(1)^n$ isometry. The Killing vectors are chosen to be denoted by $\mathrm{m}_i$ for $i=1,\dots,n$, for generically $n\leq d\!-\!2$. Because the $U(1)$'s commute, one can choose $n$ number of coordinates in the direction of the Killings. Those coordinates are shown by $\psi^i$ with the property $\psi^i\sim \psi^i+2\pi$, so $\mathrm{m}_i=\partial_{\psi^i}$.    
\end{itemize}
In Part \ref{part I} of this report, we confine the analysis to theories with metric $g_{\mu\nu}$ as a dynamical field. There can also be $p$ number of Abelian gauge fields $A^{(p)}_\mu$, and $I$ number of scalar fields $\phi^I$, for arbitrary number of $p$ and $I$. In this context, Our examples are selected to be solutions to the Einstein-Maxwell-Dilaton theories with a cosmological constant $\Lambda$ (EMD-$\Lambda$). These families of theories have the Einstein-Hilbert gravity (EH) as their gravity part of Lagrangian.    
 
\subsection{Kerr black hole}
\emph{Kerr} BH is a 4-dim solution to the EH gravity, with asymptotic flat BC \cite{Kerr:1963ud}. The Lagrangian density for this theory is $\mathcal{L}\!=\!\frac{1}{16\pi G}R$, where $R$ is the Ricci scalar. The only dynamical field is the metric $g_{\mu\nu}$ which should satisfy the e.o.m $G_{\mu\nu}\!\equiv\! R_{\mu\nu}\!-\!\dfrac{1}{2}R g_{\mu\nu}\!=\!0$. Coordinates $(\mt,\mr,\theta,\psi)$ can be chosen in a way that the metric would be explicitly as
\begin{align}\label{Kerr metric}
\hspace{-0.23cm}\mathrm{d}s^2= -&(1\!-\!f)\mathrm{d}\mt^2+\frac{\rho ^2}{\Delta}\mathrm{d}\mr^2+\rho ^2 \mathrm{d}\theta ^2 -2fa\sin ^2 \theta\,\mathrm{d}\mt \mathrm{d}\psi+\left( \mr^2+a^2+fa^2\sin ^2\theta \right)\sin ^2\theta\,\mathrm{d}\psi ^2\,,
\end{align}

\vspace*{-1cm}
where
\vspace*{-0.5cm}
\begin{align}
\rho^2 &\equiv \mr^2+a^2 \cos^2 \theta\,,\nonumber\\
\Delta &\equiv \mr^2-2Gm\mr+a^2\,,\nonumber\\
f&\equiv\frac{2Gm\mr}{\rho ^2}\,.
\end{align}
There are two free parameters $0\!\leq\!m$ and $a$ in it. Two horizons of this metric are at roots of $\Delta$, \ie\,$\mr_{\pm}=Gm\pm \sqrt{(Gm)^2-a^2}$. By that square root, $|a|\leq Gm$ is necessary to have horizons at real radii. Isometries of this solution are $\mathbb{R}\times U(1)$, generated by $\partial_\mt$ and $\partial_\psi$.

\subsection{Kerr-Newman black hole}
\emph{Kerr-Newman} BH is a 4-dim solution to the EM theory, with asymptotic flat BC \cite{Newman:1965tw,Newman:1965my}. The Lagrangian density for this theory is $\mathcal{L}\!=\!\frac{1}{16\pi G}(R-F^2)$, where $F$ is the electromagnetic field strength $2$-form, \ie $F\!=\!\mrd A$, and $F^2\equiv F_{\alpha\beta}F^{\alpha\beta}$. The dynamical fields are the metric $g_{\mu\nu}$ and $A_{\mu}$ which should satisfy the e.o.m's $G_{\mu\nu}\!=\!8\pi G\, T_{\mu\nu}$ and $\nabla_\alpha F^{\alpha\mu}=0$. The energy-momentum tensor $T_{\mu\nu}$ is defined as 
\begin{equation}
T_{\mu\nu}\equiv\frac{-2}{\sqrt{-g}} \frac{\delta (\sqrt{-g}\mathcal{L}_M)}{\delta{g^{\mu\nu}}}=\frac{1}{4\pi G}(F_{\mu\alpha}F^{\,\,\alpha}_{\nu}-\frac{1}{4}F^2 g_{\mu\nu})
\end{equation}
in which  $\mathcal{L}_M$ is Lagrangian density of matter, \ie whole Lagrangian except the EH gravity sector. Coordinates $(\mt,\mr,\theta,\psi)$ can be chosen in a way that the metric would be explicitly as
\begin{align}\label{Kerr-Newman metric}
\hspace{-0.23cm}\mathrm{d}s^2= -&(1\!-\!f)\mathrm{d}\mt^2+\frac{\rho ^2}{\Delta}\mathrm{d}\mr^2+\rho ^2 \mathrm{d}\theta ^2 -2fa\sin ^2 \theta\,\mathrm{d}\mt \mathrm{d}\psi+\left( \mr^2+a^2+fa^2\sin ^2\theta \right)\sin ^2\theta\,\mathrm{d}\psi ^2\,,
\end{align}

\vspace*{-1cm}
where
\vspace*{-0.5cm}
\begin{align}
\rho^2 &\equiv \mr^2+a^2 \cos^2 \theta\,,\nonumber\\
\Delta &\equiv \mr^2-2Gm\mr+a^2+q^2\,,\nonumber\\
f&\equiv\frac{2Gm\mr-q^2}{\rho ^2}\,,
\end{align}
and the gauge field $1$-form
\begin{equation}\label{Kerr-Newman gauge}
A=\frac{q\,\mr}{\rho^2}\mrd t-\frac{q\,\mr a\sin^2\theta}{\rho^2}\mrd\psi\,.
\end{equation}
There are three free parameters $0\!\leq\!m$, $a$ and $q$ in it. Two horizons of this metric are at $\mr_{\pm}=Gm\pm \sqrt{(Gm)^2-a^2-q^2}$. By that square root, $a^2+q^2\leq (Gm)^2$ is necessary  in order to have horizons at real radii. Isometries of this solution are $\mathbb{R}\times U(1)$, generated by $\partial_\mt$ and $\partial_\psi$.

\subsection{Kerr-AdS black hole}
\emph{Kerr-AdS} BH is a 4-dim solution to the EH theory with negative cosmological constant $\Lambda$, subject to the asymptotic $AdS_4$ BC \cite{Carter:1968ks}. The Lagrangian density for this theory is $\mathcal{L}\!=\!\frac{1}{16\pi G}(R-2\Lambda)$,  with the metric as its only dynamical field  $g_{\mu\nu}$, which should satisfy the e.o.m $G_{\mu\nu}\!+\Lambda g_{\mu\nu}=0$. Coordinates $(\mt,\mr,\theta,\psi)$ can be chosen in a way that the metric would be non-rotating w.r.t the infinity, and is explicitly as
\begin{align}\label{Kerr-AdS metric}
\hspace{-0.23cm}\mathrm{d}s^2= -&\Delta_\theta(\frac{1+\frac{\mr^2}{l^2}}{\Xi}\!-\!\Delta_\theta f)\mathrm{d}\mt^2+\frac{\rho ^2}{\Delta_\mr}\mathrm{d}\mr^2+\frac{\rho ^2}{\Delta_\theta} \mathrm{d}\theta ^2 -2\Delta_\theta fa\sin ^2 \theta\,\mathrm{d}\mt \mathrm{d}\psi\nonumber \\
+&\left( \frac{\mr^2+a^2}{\Xi}+fa^2\sin ^2\theta \right)\sin ^2\theta\,\mathrm{d}\psi ^2\,,
\end{align}
where
\begin{align}
\rho^2 &\equiv \mr^2+a^2 \cos^2 \theta\,,\nonumber\\
\Delta_r &\equiv (\mr^2+a^2)(1+\frac{\mr^2}{l^2})-2Gm\mr\,,\nonumber\\
\Delta_\theta&\equiv 1-\frac{a^2}{l^2}\cos ^2\theta\,,\nonumber\\
f&\equiv\frac{2Gm\mr}{\rho ^2}\,,\nonumber\\
\Xi&\equiv 1-\frac{a^2}{l^2}\,.
\end{align}

There are two free parameters $0\!\leq\!m$, $a$ in it. Radius of the $AdS_4$ has been denoted by $l$, which is related to the $\Lambda$ by $\Lambda=\dfrac{-3}{l^2}$. Horizons of this metric are at roots of $\Delta_\mr$. In order to prevent complex numbers or divergences, $|a| < l$. Isometries of this solution are $\mathbb{R}\times U(1)$, generated by $\partial_\mt$ and $\partial_\psi$.

\subsection{BTZ black hole}
\emph{BTZ} BH is a 3-dim solution to the EH gravity with cosmological constant $\Lambda\!<\!0$ and asymptotic $AdS_3$ BC \cite{Banados:1992wn}[\cite{Carlip:1995qv}]. The Lagrangian density for this theory is $\mathcal{L}\!=\!\frac{1}{16\pi G}(R-2\Lambda)$. The dynamical field is the metric $g_{\mu\nu}$ which should satisfy the e.o.m $G_{\mu\nu}+\Lambda g_{\mu\nu}=0$. Coordinates $(\mt,\mr,\psi)$ can be chosen in a way that the metric would be explicitly as
\begin{equation}\label{BTZ metric}
\mrd s^2=-\Delta\mrd{\mt}^2+\frac{\mrd \mr^2}{\Delta}+\mr^2(\mrd{\psi}-\omega \mrd{\mt})^2\,,
\end{equation}
where
\begin{align}
\Delta &\equiv -Gm+\frac{\mr^2}{l^2}+\frac{j^2}{4\mr^2}\,,\nonumber\\
\omega &=\frac{j}{2\mr^2}\,,
\end{align}
and $l=\sqrt{\frac{-1}{\Lambda}}$ is the radius of $AdS_3$.

Considering $\Lambda$ as determined by the Lagrangian, there are two free parameters $0\!\leq\!m$ and $j$ in this metric. Positive roots of $\Delta$ are the horizons of this metric. Explicitly
\begin{equation}
\Delta=\frac{(\mr^2-\mr_+^2)(\mr^2-\mr_-^2)}{l^2\mr^2}\,,
\end{equation}
so $2\mr_\pm^2\!=\!l^2(Gm\!\pm\!\sqrt{(Gm)^2-\frac{j^2}{l^2}})$, or writing in reverse,
\begin{equation}
Gm=\frac{\mr_+^2+\mr_-^2}{l^2}\,, \qquad j=\frac{2\mr_+\mr_-}{l}\,.
\end{equation}
Having real horizons necessitates $|\frac{j}{l}|\!\leq\! Gm$. Isometries of this solution are $\mathbb{R}\times U(1)$, generated by $\partial_\mt$ and $\partial_\psi$. 

\subsection{5-dim Myers-Perry black hole}
\emph{$5$-dim Myers-Perry} BH (MP for short) is a 5-dim solution to the EH gravity, with asymptotic flat BC \cite{Myers:1986un}. The Lagrangian density for this theory is $\mathcal{L}\!=\!\frac{1}{16\pi G}R$, where $R$ is the Ricci scalar. The only dynamical field is the metric $g_{\mu\nu}$ which should satisfy the e.o.m $G_{\mu\nu}\!=\!0$. Coordinates $(\mt,\mr,\theta,\psi^1,\psi^2)$ can be chosen in a way that the metric would be explicitly as 
\begin{align}\label{MP metric}
\mathrm{d}s^2= -&(\frac{-\Delta+a^2\sin^2\theta+b^2\cos^2\theta+\frac{a^2b^2}{\mr^2}}{\rho^2})\mathrm{d}\mt^2+\frac{\rho ^2}{\Delta}\mathrm{d}\mr^2+\rho^2\mathrm{d}\theta ^2 \nonumber\\
&+2\left(\Delta-(\mr^2+a^2)-\frac{b^2(\mr^2+a^2)}{\mr^2}\right)\frac{a\sin^2\theta}{\rho^2}\,\mathrm{d}\mt\, \mathrm{d}\psi^1\nonumber\\
&+2\left(\Delta-(\mr^2+b^2)-\frac{a^2(\mr^2+b^2)}{\mr^2}\right)\frac{b\cos^2\theta}{\rho^2}\,\mathrm{d}\mt\, \mathrm{d}\psi^2\nonumber\\
&+\left(-\Delta a^2\sin^2\theta+(\mr^2+a^2)^2+\frac{b^2(\mr^2+a^2)^2\sin^2\theta}{\mr^2}\right)\frac{\sin^2\theta}{\rho^2}\,\mathrm{d}\psi^1 \mathrm{d}\psi^1\nonumber\\
&+\left(-\Delta b^2\cos^2\theta+(\mr^2+b^2)^2+\frac{a^2(\mr^2+b^2)^2\cos^2\theta}{\mr^2}\right)\frac{\cos^2\theta}{\rho^2}\,\mathrm{d}\psi^2 \mathrm{d}\psi^2\nonumber\\
&+2\left(-\Delta +\frac{(\mr^2+a^2)(\mr^2+b^2)}{\mr^2}\right)\frac{ab\sin^2\theta\cos^2\theta}{\rho^2}\,\mathrm{d}\psi^1 \mathrm{d}\psi^2\,,
\end{align}
where
\begin{align}
\rho^2 &\equiv \mr^2+a^2 \cos^2 \theta+b^2\sin^2\theta\,,\nonumber\\
\Delta &\equiv \frac{(\mr^2+a^2)(\mr^2+b^2)}{\mr^2}+2Gm\,.
\end{align}
Range of $\theta$ in the chosen coordinate is $\theta\in [0,\frac{\pi}{2}]$. There are three free parameters $0\leq m$, $a$ and $b$ in this metric. The horizons of this metric are at positive roots of $\Delta$,
\begin{equation}
\mr_\pm^2=Gm-\frac{a^2+b^2}{2}\pm \sqrt{(Gm-\frac{(a-b)^2}{2})(Gm-\frac{(a+b)^2}{2})}\,.
\end{equation}
Isometries of this solution are $\mathbb{R}\times U(1)^2$ generated by $\partial_\mt$, $\partial_{\psi^1}$ and $\partial_{\psi^2}$.

\subsection{$4$-dim rotating Kaluza-Klein black hole}
\emph{Kaluza-Klein} theory of gravity (KK for short) in $4$ dimensions has the metric $g_{\mu\nu}$, a gauge field $A_\mu$, and a scalar field $\phi$ called \emph{dilaton}, as its dynamical fields. Hence the KK theory is an example of EMD theories.  Its action is as
\begin{equation}\label{KK action}
 \mathcal{S}=\frac{1}{16\pi G}\int \mrd^4 x \,(R-2\nabla^\alpha \phi \nabla_\alpha \phi -e^{-2\sqrt{3}\phi}F^2)\,.
\end{equation} 
The e.o.m derived by variation w.r.t the metric, gauge field and dilaton are 
\begin{align}
&G_{\mu\nu}=8\pi G (T_{\mu\nu}^A+T_{\mu\nu}^\phi)\,,\\
&\nabla_\mu (e^{-2\sqrt{3}\phi}F^{\mu\nu})=0\,,
\end{align}
\begin{equation}
\Box \phi=-\frac{\sqrt{3}}{2}e^{-2\sqrt{3}\phi}F^2
\end{equation}
respectively. The energy-momentum tensors are explicitly as
\begin{align}
&T_{\mu\nu}^A=\frac{1}{4\pi G} e^{-2\sqrt{3}\phi}(F_{\mu\alpha}F^{\,\,\alpha}_{\nu}-\frac{1}{4}F^2 g_{\mu\nu})\,,\\
&T_{\mu\nu}^\phi=\frac{1}{4\pi G}(\nabla_\mu\phi \nabla_\nu \phi-\frac{1}{4}R g_{\mu\nu})\,.
\end{align}
Coordinates $(\mt,\mr,\theta,\psi)$ can be chosen such that the KK BH solution to the e.o.m would be as \cite{Horne:1992zy}
\begin{align}
&\mrd s^2=-\frac{1-Z}{B}\,\mrd \mt^2+\frac{B\rho^2}{\Delta}\,\mrd \mr^2 +B\rho^2 \,\mrd \theta^2 - \frac{2Za\sin^2\theta}{B\sqrt{1-v^2}}\,\mrd \mt \,\mrd \psi \nonumber\\
&\hspace{3cm}+\left( B(\mr^2+a^2)+\frac{Z}{B}a^2\sin^2\theta\right)\sin^2\theta \,\mrd \psi^2\,,\label{KK metric}\\
&A_\mu \mrd x^\mu= \frac{Zv}{2B^2(1-v^2)}\,\mrd \mt - \frac{Zv}{2B^2\sqrt{1-v^2}}a \sin^2\theta\, \mrd \psi\,,\label{KK gauge}\\
&\phi=-\frac{\sqrt{3}}{2}\ln B\,,\label{KK scalar}
\end{align}
where
\begin{align}
&\rho^2=\mr^2+a^2\cos^2\theta\,,\nonumber\\
&\Delta=\mr^2+a^2-2Gmr\,,\nonumber\\
&Z=\frac{2Gmr}{\rho^2}\,,\nonumber\\
&B=\sqrt{1+\frac{Zv^2}{1-v^2}}\,.
\end{align}
There are three free parameters $0\leq m$, $a$ and $0\leq v<1$ in this metric. The horizons of this BH are situated at the roots of $\Delta$, \ie
\begin{equation}
\mr_\pm=Gm\pm \sqrt{G^2m^2-a^2}\,.
\end{equation}
Isometries of this solution are $\mathbb{R}\times U(1)$ generated by $\partial_\mt$ and $\partial_{\psi}$. In order to have horizons at real radius, $|a|\leq m$. In the limit $v\to 0$, one returns to the Kerr solution to the EH gravity. For generalization of KK BH to include magnetic monopole, interested reader can see \cite{Larsen:1999pp} and references in it.

\chapter{Covariant Phase Space method}\label{chap Covariant Phase Space method}
This chapter plays a very important role in this report. It deals with the concept of \emph{conserved charge}, which is the corner stone of our analysis. Introduction of this concept in our analysis is based on \emph{Covariant Phase Space method}, which is described in this chapter. In the first section, a specific family of symmetries of covariant theories is explored; the diffeomorphisms. In the second section, a short review on the generic \emph{phase spaces} is provided. In the third section, Covariant Phase Space method is introduced, which provides us symplectic structure of the field theories discussed in the previous chapter, and enables us to associate Hamiltonian generators to diffeomorphisms. In the two last sections, Noether-Wald conserved charge method is reviewed and compared with the Covariant Phase Space method.


\section{Diffeomorphism as symmetry}

Given a covariant Lagrangian density $\mathcal{L}$, then any variation ``generated" by an arbitrary vector field $\xi^\mu$ would be a symmetry, the \emph{diffeomorphism} symmetry. The ``generation" means that for infinitesimal coordinate transformation one has $x^\mu\to x^\mu-\xi^\mu$. The diffeomorphism symmetry is a local symmetry and the infinitesimal variation of dynamical fields would be the Lie derivative $\delta_\xi \Phi=\mathscr{L}_{\xi}\Phi$. Because there is not any non-dynamical field in the Lagrangian, variation of Lagrangian would be Lie derivative too,
\begin{equation}\label{Lagrangian var sym}
\delta_\xi \mathbf{L}= \mathscr{L}_{\xi}\mathbf{L}=\xi\cdot \mathrm{d} \mathbf{L}+\mathrm{d}(\xi\cdot \mathbf{L})=\mathrm{d}(\xi\cdot \mathbf{L})\,.
\end{equation}
It matches \eqref{symmetry condition} via $\boldsymbol{\mathcal{K}}=\xi\cdot \mathbf{L}$, so satisfying the symmetry condition. In \eqref{Lagrangian var sym} two identities are used. The first identity is 
\begin{equation}
\mathscr{L}_\xi\mathbf{L}=\xi\cdot \mathrm{d} \mathbf{L}+\mathrm{d}(\xi\cdot \mathbf{L})\,,
\end{equation}
which is correct for any differential form $\mathbf{L}$ and is called \emph{Cartan magic formula} \eqref{Cartan magic}. The second identity used is $\mathrm{d}\mathbf{L}\!=\!0$ correct for any $d$-form, described in \eqref{d top form zero}. 

There have been so many inconclusive efforts to associate local conserved charge to the diffeomorphism symmetry in the last century [\cite{Compere:2012hr}]. Specifically the notion of local energy and momentum is missed in general relativity or other covariant theories of gravity. Although local conserved charges seems to be ill-defined in general relativity, but for some specific solutions it is possible to associate quasi-local conserved charges. Quasi-local conserved charges are defined w.r.t a specific region of spacetime.  The Komar integral \cite{Komar:1958wp}(which associates conserved charge to some Killings by an integral formula), the ADM formalism [\cite{Gourgoulhon:2007ue}] (which associates mass and angular momentum to the asymptotic flat spacetimes), local cohomology methods \cite{Barnich:2001jy},[\cite{Barnich:2007bf,Compere:2007az}] and Covariant Phase Space method  are some successful attempts towards this goal. Our work is based on the \emph{Covariant Phase Space method} developed in \cite{Lee:1990gr,Wald:1993nt,Iyer:1994ys}, and is described in this chapter. But first we would have a review on the concepts related to a phase space.


\section{Reviewing the phase space}\label{sec phase space}
\begin{figure}[!h]
\centering
		\begin{tikzpicture}[scale=1.2]
    \begin{scope}
    \fill[even odd rule,light gray] plot [smooth] coordinates {(2,0) (4,1) (5,1.5) (9,1.5) (8,0.5) (7,0) (2.5,-0.1) (2,0)};
    \draw (5,0) node[anchor=north]{$\mathcal{M}$};
       
    \draw[thin,white,shift={(0cm,0.2cm)}] plot [smooth] coordinates {(0.5,0) (4,-0.1) (8,0)};
    \draw[thin,white,shift={(0cm,0.4cm)}] plot [smooth] coordinates {(0.5,0) (4,-0.08) (8,0)};
    \draw[thin,white,shift={(0.5cm,0.6cm)}] plot [smooth] coordinates {(0.5,0) (4,-0.03) (8,0)};
    \draw[thin,white,shift={(1.5cm,0.8cm)}] plot [smooth] coordinates {(0.5,0) (4,0.03) (8,0)};
    \draw[thin,white,shift={(2cm,1cm)}] plot [smooth] coordinates {(0.5,0) (4,0.08) (8,0)};
    \draw[thin,white,shift={(2.5cm,1.2cm)}] plot [smooth] coordinates {(0.5,0) (4,0.1) (8,0)};
    \draw[thin,white,shift={(3cm,1.4cm)}] plot [smooth] coordinates {(0.5,0) (4,0.12) (8,0)};
    
    \draw[thin,white,shift={(-0.5cm,0.0cm)}] plot [smooth] coordinates {(1.8,-0.2) (4.8,1) (5,1.5)};
    \draw[thin,white,shift={(0cm,0.0cm)}] plot [smooth] coordinates {(1.8,-0.2) (4.6,1) (5,1.5)};
    \draw[thin,white,shift={(0.5cm,0.0cm)}] plot [smooth] coordinates {(1.8,-0.2) (4.5,1) (5,1.5)};
    \draw[thin,white,shift={(1cm,0.08cm)}] plot [smooth] coordinates {(1.8,-0.2) (4.4,1) (5,1.5)};
    \draw[thin,white,shift={(1.5cm,0.08cm)}] plot [smooth] coordinates {(1.8,-0.2) (4.3,1) (5,1.5)};
    \draw[thin,white,shift={(2cm,0.08cm)}] plot [smooth] coordinates {(1.8,-0.2) (4.2,1) (5,1.5)};
    \draw[thin,white,shift={(2.5cm,0.1cm)}] plot [smooth] coordinates {(1.8,-0.2) (4.0,1) (5,1.5)};
    \draw[thin,white,shift={(3cm,0.1cm)}] plot [smooth] coordinates {(1.8,-0.2) (3.8,1) (5,1.5)};
    \draw[thin,white,shift={(3.5cm,0.1cm)}] plot [smooth] coordinates {(1.8,-0.2) (3.6,1) (5,1.5)};
    \draw[thin,white,shift={(4cm,0.1cm)}] plot [smooth] coordinates {(1.8,-0.2) (3.4,1) (5,1.5)};
    \draw[thin,white,shift={(4.5cm,0.1cm)}] plot [smooth] coordinates {(1.8,-0.2) (3.3,1) (5,1.5)};
    \draw[thin,white,shift={(5cm,0.1cm)}] plot [smooth] coordinates {(1.8,-0.2) (3.2,1) (5,1.5)};
    \draw[thin,white,shift={(5.5cm,0.1cm)}] plot [smooth] coordinates {(1.8,-0.2) (3.1,1) (5,1.5)};
    \draw[thin,white,shift={(6cm,0.1cm)}] plot [smooth] coordinates {(1.8,-0.2) (3.0,1) (5,1.5)};
    \draw[thin,white,shift={(6.5cm,0.1cm)}] plot [smooth] coordinates {(1.8,-0.2) (2.9,1) (5,1.5)};
    \draw[thin,white,shift={(7cm,0.1cm)}] plot [smooth] coordinates {(1.8,-0.2) (2.8,1) (5,1.5)};
    \draw (6,1) node[anchor=north]{$\Omega_{_{AB}}$};
    \draw[->,shift={(0.6,0.1)}](2,0)--(2.8,0);	
    \draw[->,shift={(0.6,0.1)}](2,0)--(2.5,0.25);
    \draw (2.5,0) node[anchor=south]{\footnotesize $X^{^A}$};
   	\end{scope};
	\end{tikzpicture}
	\hspace*{-2cm}
	\caption{\footnotesize Phase space is a manifold $\mathcal{M}$ equipped with a symplectic 2-form $\Omega_{_{AB}}$}	
\label{fig phase space}
\end{figure}
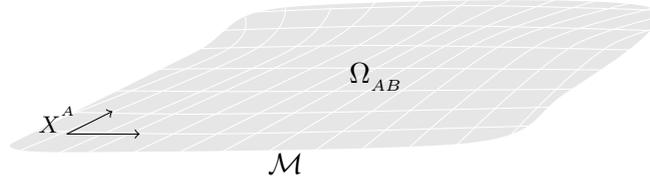

\emph{Phase space} is a manifold $\mathcal{M}$, which is equipped with a $2$-form $\Omega$, usually called \emph{symplectic} form.  If manifold is parametrized by some coordinates\footnote{$X^{^A}$ should not be confused with the spacetime coordinates $x^\mu$. We denote the exterior derivative on the phase space by $\delta$ instead of $\mathrm{d}$, to prevent confusions.} $X^{^A}$, inducing basis for the cotangent space of manifold as $\delta X^{^A}$, then $\Omega=\Omega_{_{AB}}\delta X^{^A}\delta X^{^B}$ should have the following properties.
\begin{enumerate}
\item $\Omega_{_{AB}}=-\Omega_{_{BA}}\,,$
\item $\delta \Omega=0\,,$
\item $\Omega_{_{AB}}V^{^B}=0 \,\; \Leftrightarrow \,\; V^{^B}=0\,,$
\end{enumerate}
where $V^{^B}$ is a vector in the tangent space of the manifold. $\delta$ denotes exterior derivative (see Appendix \ref{app diff forms} for definition) on the phase space. The second condition is read as ``$\Omega$ is closed", and the third one as ``$\Omega$ is non-degenerate". As a result of the latter, $\Omega$ is invertible and one can define its inverse $\Omega^{^{AB}}=(\Omega^{-1})_{_{AB}}$. The matrices $\Omega_{_{AB}}$ and $\Omega^{^{AB}}$ are responsible for lowering and raising the indices, respectively. Figure \ref{fig phase space} is a schematic illustration of a symplectic structure. 
\vspace*{1cm}

Considering a function $f=f(X^{^A})$ on the phase space, one can associate a vector field to it (usually called \emph{Hamiltonian vector field}) as
\begin{equation}
V_f\equiv \delta f=\partial_{_A}f\, \delta X^{^A}\,,
\end{equation}
where the differentiability of $f$ is assumed. In reverse, given a vector field $V^{^A}(X^{^B})$, in certain conditions, one can associate a function $H_{_V}$ (usually called \emph{Hamiltonian generator}) to it such that
\begin{equation}\label{V integrability}
\delta H_{_V}=V=\Omega_{_{AB}}\delta X^{^A} V^{^B}\,,
\end{equation}
in which $\delta H_{_V}=\partial_{_A}H_{_V}\, \delta X^{^A}$. The conditions necessary for the existence of $H_{_V}$ are called \emph{integrability}. If $V$ is not integrable, then $\deltabar$ is used instead of $\delta$, \ie $V=\deltabar H_{_V}$. It means that there is not any function $H_{_V}$ which its exterior derivative would be equal to $V$ all over the phase space.
\begin{figure}[h!]
    \begin{center}\hspace*{-1.5cm}
\subfigure[]{
     \begin{tikzpicture}[scale=1]
    \begin{scope}[shift={(-1cm,0cm)}]
    \fill[even odd rule,light gray] plot [smooth] coordinates {(2,0) (4,1) (5,1.5) (9,1.5) (8,0.5) (7,0) (2.5,-0.1) (2,0)};
    \draw (5,0) node[anchor=north]{$\mathcal{M}$};
       
    \draw[thin,white,shift={(0cm,0.2cm)}] plot [smooth] coordinates {(0.5,0) (4,-0.1) (8,0)};
    \draw[thin,white,shift={(0cm,0.4cm)}] plot [smooth] coordinates {(0.5,0) (4,-0.08) (8,0)};
    \draw[thin,white,shift={(0.5cm,0.6cm)}] plot [smooth] coordinates {(0.5,0) (4,-0.03) (8,0)};
    \draw[thin,white,shift={(1.5cm,0.8cm)}] plot [smooth] coordinates {(0.5,0) (4,0.03) (8,0)};
    \draw[thin,white,shift={(2cm,1cm)}] plot [smooth] coordinates {(0.5,0) (4,0.08) (8,0)};
    \draw[thin,white,shift={(2.5cm,1.2cm)}] plot [smooth] coordinates {(0.5,0) (4,0.1) (8,0)};
    \draw[thin,white,shift={(3cm,1.4cm)}] plot [smooth] coordinates {(0.5,0) (4,0.12) (8,0)};
    
    \draw[thin,white,shift={(-0.5cm,0.0cm)}] plot [smooth] coordinates {(1.8,-0.2) (4.8,1) (5,1.5)};
    \draw[thin,white,shift={(0cm,0.0cm)}] plot [smooth] coordinates {(1.8,-0.2) (4.6,1) (5,1.5)};
    \draw[thin,white,shift={(0.5cm,0.0cm)}] plot [smooth] coordinates {(1.8,-0.2) (4.5,1) (5,1.5)};
    \draw[thin,white,shift={(1cm,0.08cm)}] plot [smooth] coordinates {(1.8,-0.2) (4.4,1) (5,1.5)};
    \draw[thin,white,shift={(1.5cm,0.08cm)}] plot [smooth] coordinates {(1.8,-0.2) (4.3,1) (5,1.5)};
    \draw[thin,white,shift={(2cm,0.08cm)}] plot [smooth] coordinates {(1.8,-0.2) (4.2,1) (5,1.5)};
    \draw[thin,white,shift={(2.5cm,0.1cm)}] plot [smooth] coordinates {(1.8,-0.2) (4.0,1) (5,1.5)};
    \draw[thin,white,shift={(3cm,0.1cm)}] plot [smooth] coordinates {(1.8,-0.2) (3.8,1) (5,1.5)};
    \draw[thin,white,shift={(3.5cm,0.1cm)}] plot [smooth] coordinates {(1.8,-0.2) (3.6,1) (5,1.5)};
    \draw[thin,white,shift={(4cm,0.1cm)}] plot [smooth] coordinates {(1.8,-0.2) (3.4,1) (5,1.5)};
    \draw[thin,white,shift={(4.5cm,0.1cm)}] plot [smooth] coordinates {(1.8,-0.2) (3.3,1) (5,1.5)};
    \draw[thin,white,shift={(5cm,0.1cm)}] plot [smooth] coordinates {(1.8,-0.2) (3.2,1) (5,1.5)};
    \draw[thin,white,shift={(5.5cm,0.1cm)}] plot [smooth] coordinates {(1.8,-0.2) (3.1,1) (5,1.5)};
    \draw[thin,white,shift={(6cm,0.1cm)}] plot [smooth] coordinates {(1.8,-0.2) (3.0,1) (5,1.5)};
    \draw[thin,white,shift={(6.5cm,0.1cm)}] plot [smooth] coordinates {(1.8,-0.2) (2.9,1) (5,1.5)};
    \draw[thin,white,shift={(7cm,0.1cm)}] plot [smooth] coordinates {(1.8,-0.2) (2.8,1) (5,1.5)};
    \draw[thick] plot [smooth] coordinates {(4,2) (6,3) (8,2)};
    \foreach \i in {1,2,...,5}
    \foreach \j in {1,2,...,5}
    \draw[->] (0.8*\i+3.5,0.2*\j)--(0.8*\i+0.05*\j+4,0.2*\j+0.05*\i+0.2);
    \draw (6,3) node[anchor=south]{$f$};
    \draw (6,1.2) node[anchor=south]{$V_f$};
    \draw[->,shift={(0.6,0.1)}](2,0)--(2.8,0);	
    \draw[->,shift={(0.6,0.1)}](2,0)--(2.5,0.25);
    \draw (2.5,0) node[anchor=south]{\footnotesize $X^{^A}$};
    \draw[ultra thick, dgreen,->,shift={(4.6,1.7)}](1,1) to [out=210,in=120](1,0);
    \end{scope};
	\end{tikzpicture}
}\hspace*{-5cm}
\subfigure[]{
    \begin{tikzpicture}[scale=1]
    \begin{scope}[shift={(-1cm,0cm)}]
    \fill[even odd rule,light gray] plot [smooth] coordinates {(2,0) (4,1) (5,1.5) (9,1.5) (8,0.5) (7,0) (2.5,-0.1) (2,0)};
    \draw (5,0) node[anchor=north]{$\mathcal{M}$};
       
    \draw[thin,white,shift={(0cm,0.2cm)}] plot [smooth] coordinates {(0.5,0) (4,-0.1) (8,0)};
    \draw[thin,white,shift={(0cm,0.4cm)}] plot [smooth] coordinates {(0.5,0) (4,-0.08) (8,0)};
    \draw[thin,white,shift={(0.5cm,0.6cm)}] plot [smooth] coordinates {(0.5,0) (4,-0.03) (8,0)};
    \draw[thin,white,shift={(1.5cm,0.8cm)}] plot [smooth] coordinates {(0.5,0) (4,0.03) (8,0)};
    \draw[thin,white,shift={(2cm,1cm)}] plot [smooth] coordinates {(0.5,0) (4,0.08) (8,0)};
    \draw[thin,white,shift={(2.5cm,1.2cm)}] plot [smooth] coordinates {(0.5,0) (4,0.1) (8,0)};
    \draw[thin,white,shift={(3cm,1.4cm)}] plot [smooth] coordinates {(0.5,0) (4,0.12) (8,0)};
    
    \draw[thin,white,shift={(-0.5cm,0.0cm)}] plot [smooth] coordinates {(1.8,-0.2) (4.8,1) (5,1.5)};
    \draw[thin,white,shift={(0cm,0.0cm)}] plot [smooth] coordinates {(1.8,-0.2) (4.6,1) (5,1.5)};
    \draw[thin,white,shift={(0.5cm,0.0cm)}] plot [smooth] coordinates {(1.8,-0.2) (4.5,1) (5,1.5)};
    \draw[thin,white,shift={(1cm,0.08cm)}] plot [smooth] coordinates {(1.8,-0.2) (4.4,1) (5,1.5)};
    \draw[thin,white,shift={(1.5cm,0.08cm)}] plot [smooth] coordinates {(1.8,-0.2) (4.3,1) (5,1.5)};
    \draw[thin,white,shift={(2cm,0.08cm)}] plot [smooth] coordinates {(1.8,-0.2) (4.2,1) (5,1.5)};
    \draw[thin,white,shift={(2.5cm,0.1cm)}] plot [smooth] coordinates {(1.8,-0.2) (4.0,1) (5,1.5)};
    \draw[thin,white,shift={(3cm,0.1cm)}] plot [smooth] coordinates {(1.8,-0.2) (3.8,1) (5,1.5)};
    \draw[thin,white,shift={(3.5cm,0.1cm)}] plot [smooth] coordinates {(1.8,-0.2) (3.6,1) (5,1.5)};
    \draw[thin,white,shift={(4cm,0.1cm)}] plot [smooth] coordinates {(1.8,-0.2) (3.4,1) (5,1.5)};
    \draw[thin,white,shift={(4.5cm,0.1cm)}] plot [smooth] coordinates {(1.8,-0.2) (3.3,1) (5,1.5)};
    \draw[thin,white,shift={(5cm,0.1cm)}] plot [smooth] coordinates {(1.8,-0.2) (3.2,1) (5,1.5)};
    \draw[thin,white,shift={(5.5cm,0.1cm)}] plot [smooth] coordinates {(1.8,-0.2) (3.1,1) (5,1.5)};
    \draw[thin,white,shift={(6cm,0.1cm)}] plot [smooth] coordinates {(1.8,-0.2) (3.0,1) (5,1.5)};
    \draw[thin,white,shift={(6.5cm,0.1cm)}] plot [smooth] coordinates {(1.8,-0.2) (2.9,1) (5,1.5)};
    \draw[thin,white,shift={(7cm,0.1cm)}] plot [smooth] coordinates {(1.8,-0.2) (2.8,1) (5,1.5)};
    \draw[thick] plot [smooth] coordinates {(4,2) (6,3) (8,2)};
    \foreach \i in {1,2,...,5}
    \foreach \j in {1,2,...,5}
    \draw[->] (0.8*\i+3.5,0.2*\j)--(0.8*\i+0.05*\j+4,0.2*\j+0.05*\i+0.2);
    \draw (6,3) node[anchor=south]{$H_{_V}$};
    \draw (6,1.2) node[anchor=south]{$V$};
    \draw[->,shift={(0.6,0.1)}](2,0)--(2.8,0);	
    \draw[->,shift={(0.6,0.1)}](2,0)--(2.5,0.25);
    \draw (2.5,0) node[anchor=south]{\footnotesize $X^{^A}$};
    \draw[ultra thick, dgreen,<-,shift={(4.6,1.7)}](1,1) to [out=210,in=120](1,0);
    \end{scope};
	\end{tikzpicture}
}
\end{center}
\caption{\footnotesize (a) $V_f$ is the vector field associated to the function $f$ by differentiability. (b) $H_{_V}$ is the function associated to the vector filed $V$ by integrability. }
\label{fig function-vector}
\end{figure}
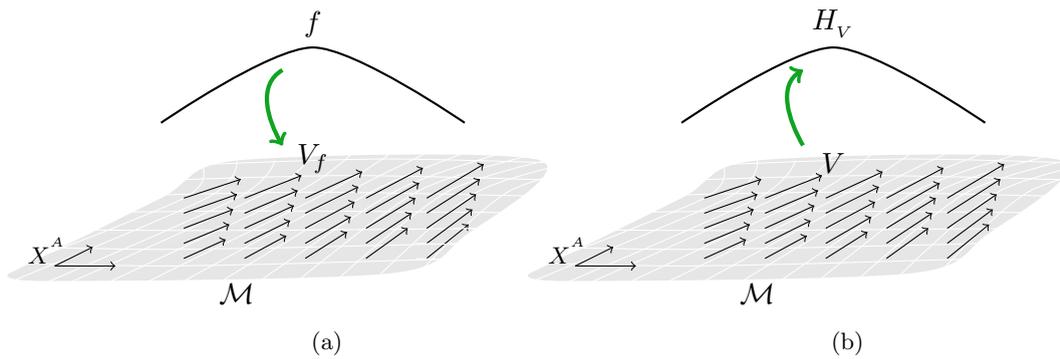

The Poisson bracket of two functions $f$ and $g$ is defined to be
\begin{align}\label{Poisson Lie}
\{f,g\}&\equiv\Omega^{^{AB}}(V_f)_{_A} (V_g)_{_B}\\
&=(V_f)^{^B}\partial_{_B}g=\mathscr{L}_{_{V_f}}g\,,
\end{align}
in which $\mathscr{L}_{_{V}}$ denotes Lie derivative in direction of $V$ on the phase space. The $\Omega$ is an anti-symmetric matrix, so \eqref{Poisson Lie} is also equal to $-\mathscr{L}_{_{V_g}}f$.

\dotfillb
\Example{\emph{1-dim motion of a particle:}\\
The simplest example, is the one-dimensional motion of a particle, with position $q$ and momentum $p$. Then by the choice of $X^1=q$ and $X^2=p$,
\begin{equation}\label{one-dim Omega}
\Omega_{_{AB}}=\begin{pmatrix}
0 & 1\\
-1 & 0
\end{pmatrix},\qquad \{q,p\}=1, \qquad\{q,q\}=\{p,p\}=0\,.
\end{equation}
The basis for the tangent and cotangent spaces are $\partial_{{X^A}}$ and $\delta X^{^A}$ respectively, which are related to each other by the matrix of the $\Omega$ in \eqref{one-dim Omega}, as $\partial_{{X^A}}=\Omega_{_{AB}}\delta X^{^B}$, \eg $\partial_q=\delta p$.\\
As an example, one can choose the function $f$ to be momentum $p$ or Hamiltonian $H$. Then
\begin{itemize}
\item $f=p \qquad\,\; \Rightarrow \qquad V_p= \delta p =\partial_q$\,.
\item $f=H \qquad \Rightarrow \qquad V_H=\delta H=\frac{\partial H}{\partial q}\delta q+\frac{\partial H}{\partial p}\delta p=\dot{p}\partial p+\dot{q}\partial_q=\partial_t$\,,
\end{itemize} 
in which the Hamiltonian equations of motion are also used. These are the manifestation of ``momentum is generator of translations in space" and ``Hamiltonian is generator of translations in time".
\begin{figure}[h!]
\vspace*{-0.5cm}
    	\centering
\begin{tikzpicture}[scale=0.8]
\fill[even odd rule,light gray]
    (0,0) to (5,1.5) to  (11,1.5) to
    (6,0) to (0,0);
    \draw[very thin,white] (1,0) to (6,1.5);
    \draw[very thin,white] (2,0) to (7,1.5);
    \draw[very thin,white] (4,0) to (9,1.5);
    \draw[very thin,white] (5,0) to (10,1.5);
    \draw[very thin,white] (0.8,0.25) to (6.8,0.25);
    \draw[very thin,white] (1.6,0.5) to (7.6,0.5);
    \draw[very thin,white] (3.3,1) to (9.3,1);
    \draw[very thin,white] (4.1,1.25) to (10.1,1.25);
    \draw[dashed,thick,->,shift={(-0.5,-0.1)}] (3,0) to (9,6*1.5/5) node[anchor=west] {$p$};
    \draw[dashed,thick,->,shift={(-0.8,0)}] (2.5,0.75) to (10.5,0.75)node[anchor=north] {$q$};
    \draw[thick] plot [smooth] coordinates {(4,2) (6,3) (8,2)};
    \draw (6,3) node[anchor=south]{$H$};
	\end{tikzpicture}
	\caption{Phase space of a particle in $1$-dim. Hamiltonian is a function on the phase space, and is the generator of translations in time.}
\label{fig 1-dim phase space}
\end{figure}
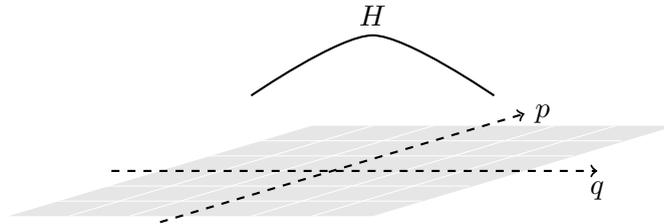
}
\vspace*{-1.5cm}
\dotfille

If $\mathcal{L}(\Phi,\partial\Phi,\dots)$ is the Lagrangian density for some classical fields $\Phi(x^\mu)$, the canonical phase space for this system is built by a specific choice of \emph{constant time} hypersurfaces. After foliating the spacetime with those hypersurfaces, then the fields $\Phi(\vec{x})$ and their conjugates $\Pi (\vec{x})\equiv \dfrac{\partial \mathcal{L}}{\partial \dot{\Phi}}$ would constitute the manifold for the phase space, where \emph{dot} is differentiation w.r.t the time. The Poisson bracket would be
\begin{equation}
\{\Phi(\vec{x}_1),\Pi(\vec{x}_2)\}=\delta(\vec{x}_1-\vec{x}_2)\,,\qquad \{\Phi(\vec{x}_1),\Phi(\vec{x}_2)\}=\{\Pi(\vec{x}_1),\Pi(\vec{x}_2)\}=0\,.
\end{equation}
Choosing a specific time foliation breaks the covariance of the theory. In order to have a covariant phase space, instead of canonical approach, the  \emph{Covariant Phase Space method} can be used. It is a covariant way of defining phase space and symplectic 2-form for (gauge)field theories, and would be the basis of our analysis in this report.


\section{Covariant Phase Space method}

In this section, we provide a brief introduction to the Covariant Phase Space method. The interested reader can refer to \cite{Lee:1990gr,Wald:1993nt,Iyer:1994ys} for original rigorous works on the subject. First, we introduce the Lee-Wald symplectic structure, which provides a covariantly built manifold $\mathcal{M}$, and equips it with a covariant symplectic $2$-form. Then using the symplectic structure, for a given vector field $\xi$, its  Hamiltonian generator variation $\delta H_\xi$ will be deduced. After that, in the next sections, the Noether-Wald conserved charge $\mathcal{Q}_\xi$ is analysed and compared to the Lee-Wald Hamiltonian generators $H_\xi$.
\subsection{Lee-Wald symplectic structure}\label{sec Lee-Wald symplectic}
 Consider a diffeomorphism covariant theory with a Lagrangian density $\mathcal{L}$ and the corresponding action in $d$-dimensional spacetime
\begin{equation}
\mathcal{S}[\Phi]=\int \mathbf{L}
\end{equation}
in which $\Phi$ denotes all of dynamical fields of the system. In the Covariant Phase Space method, the manifold $\mathcal{M}$ is constituted of the field configurations all over the spacetime, $\Phi(x^\mu)$. On the other hand, there would not be any need to conjugate fields. In order to define an appropriate symplectic $2$-form, one can use the $\mathbf{\Theta}$ in \eqref{delta S}. Explicitly, if $\delta \mathbf{L}[\Phi]=\mathbf{E}_{\Phi}\delta\Phi+\mathrm{d}\mathbf{\Theta}(\delta\Phi,\Phi)$, then the Lee-Wald symplectic $2$-form is defined as
\begin{equation}\label{Omega LW}
\Omega(\delta_1\Phi,\delta_2\Phi,\Phi)=\int_\Sigma \boldsymbol{\omega}_{_\text{LW}}(\delta_1\Phi,\delta_2\Phi,\Phi)\, 
\end{equation}
where
\begin{equation}\label{omega LW}
\boldsymbol{\omega}_{_\text{LW}}(\delta_1\Phi,\delta_2\Phi,\Phi)\equiv\delta_1\mathbf{\Theta}(\delta_2\Phi,\Phi)-\delta_2\mathbf{\Theta}(\delta_1\Phi,\Phi)\,,
\end{equation}
in which $\delta_{1}\Phi$ and $\delta_2\Phi$ are some field perturbations  as members of the tangent space of $\mathcal{M}$. It is  assumed (here and in the whole of this report) that these perturbations satisfy the linearized equations of motion (l.e.o.m, which will be described in Section \ref{chap BH thermo}). These perturbations commute with each other
\begin{equation}\label{del del Phi}
(\delta_1\delta_2-\delta_2\delta_1)\Phi=0\,.
\end{equation}
Also $\mrd$ and $\delta$ commute,
\begin{equation}\label{d delta com}
\mathrm{d}\delta\Phi=\delta \mathrm{d}\Phi\,.
\end{equation}
The $(d-1)$ surface $\Sigma$ is a Cauchy surface in the spacetime. Some issues are important to be mentioned about the proposed symplectic $2$-form.
\begin{itemize}
\item[--] \textbf{Ambiguities:} The LW $2$-form \eqref{Omega LW}, has twofold ambiguities: 
\begin{enumerate}
\item By $\mathbf{L}\to\mathbf{L}+\mathrm{d}\boldsymbol{\boldsymbol{\mu}}$ for some $(d\!-\!1)$-form $\boldsymbol{\mu}$, e.o.m does not change, but the $\mathbf{\Theta}$ changes as $\mathbf{\Theta}\to \mathbf{\Theta}+\delta \boldsymbol{\mu}$. Nonetheless, the $\boldsymbol{\omega}_{_\text{LW}}$ remains unchanged because of $(\delta_1\delta_2-\delta_2\delta_1) \boldsymbol{\mu}=0$. 
\item The $\mathbf{\Theta}$ is ambiguous by addition of an exact form, as
\begin{equation}\label{LW Y ambiguity}
\mathbf{\Theta}(\delta\Phi,\Phi)\to \mathbf{\Theta}(\delta\Phi,\Phi)+\mathrm{d}\mathbf{Y}(\delta\Phi,\Phi)\,.
\end{equation}
Hence, by adding such ambiguity,
\begin{equation}
\boldsymbol{\omega}_{_\text{LW}}\to \boldsymbol{\omega}\equiv\boldsymbol{\omega}_{_\text{LW}}+\mrd \big(\delta_1\mathbf{Y}(\delta_2\Phi,\Phi)-\delta_2\mathbf{Y}(\delta_1\Phi,\Phi)\big)\,.
\end{equation}
\end{enumerate}
\item[--] \textbf{Conservation:} The $\Omega$ should not depend on the choice of $\Sigma$, otherwise, it would break the general covariance. In order to fulfil this condition, it is enough to have two conditions:
\begin{enumerate}
\item[i)] $\mathrm{d}\boldsymbol{\omega}\approx 0$,
\item[ii)] the flow of $\boldsymbol{\omega}$ out of the boundaries of $\Sigma$ should vanish.
\end{enumerate}
In other words, $\Omega$ needs to be a \emph{conserved} entity, \ie be constant on all chosen slices $\Sigma$. The intuition for these conditions are clear; (i) means there would not be any sink or source in the bulk $\Sigma$, (ii) means there would not be any leakage from the boundaries of $\Sigma$. In Appendix \ref{app d omega 0} it is shown that the on-shell conditions leads to $\mathrm{d}\boldsymbol{\omega}\approx 0$. On the other hand, vanishing of the flow on boundary would be a criterion for appropriate fixing of ambiguities. 
\item[--]\textbf{Non-degeneracy:} The manifold $\mathcal{M}$ is supposed to be chosen such that $\Omega$ would be non-degenerate. Other two conditions, \ie anti-symmetry and closeness of $\Omega$ are guaranteed by \eqref{omega LW}.
\end{itemize} 
Assume that a vector field $\xi$ is given all over the spacetime. It induces a vector field over the phase space  by Lie derivative as $\delta_\xi \Phi=\mathscr{L}_\xi\Phi$. Motivated by the RHS of  \eqref{V integrability}, one can associate a Hamiltonian variation to $\xi$ as
\begin{align}\label{delta H xi}
\delta H_{\xi}&\equiv \Omega(\delta\Phi,\delta_\xi\Phi,\Phi)\\
&=\int_\Sigma \boldsymbol{\omega}(\delta\Phi,\delta_\xi\Phi,\Phi)\\
&=\int_\Sigma \big(\delta\mathbf{\Theta}(\mathscr{L}_\xi\Phi,\Phi)-\mathscr{L}_\xi\mathbf{\Theta}(\delta\Phi,\Phi)\big)\,. 
\end{align} 
A fundamental relation for the Covariant Phase Space method, which originates from the on-shell condition $\mathrm{d}\boldsymbol{\omega}(\Phi,\delta\Phi,\delta_\xi\Phi)\approx 0$, is that for the $\delta \Phi$ which satisfy l.e.o.m, one can find a $(d\!-\!2)$-form $\boldsymbol{k}_\xi$ such that
\begin{equation}\label{omega dk}
\boldsymbol{\omega}(\delta\Phi,\delta_\xi\Phi,\Phi)\approx\mrd\boldsymbol{k}_\xi(\delta\Phi,\Phi)\,.
\end{equation}
The general formula for $\boldsymbol{k}_\xi$ will be derived in Section \ref{sec H vs Q}, through Appendix \ref{app H v Q}. However some examples are presented earlier, in the examples \ref{example EH}-\ref{example KK}.
 
According to \eqref{omega dk} and by the Stoke's theorem \eqref{Stoke's theorem}, the definition of Hamiltonian variations for $\xi$ can be rewritten as
\begin{equation}\label{delta H k}
\delta H_{\xi}= \oint_{\mathscr{S}}\boldsymbol{k}_\xi(\delta\Phi,\Phi)\,, 
\end{equation}
where $\mathscr{S}$ is the $d\!-\!2$ dimensional boundary of the region $\Sigma$. 

It is not guaranteed that \eqref{delta H xi} be conserved, because $\delta_\xi\Phi$ might not be a member of tangent space of $\mathcal{M}$. But in the specific cases which $\Phi$ be a solution and $\xi$ be its isometry, \ie $\delta_\xi \Phi=0$, then  \eqref{delta H xi} is conserved, explained in the lemma below. 

\newpage
\vspace*{-1.5cm}
\dotfillb
\Lemma{\label{Lemma conserv H}\emph{For an isometry generator $\xi$, $\delta H_\xi$ is conserved on-shell,  independent of $\Sigma$:} \\
According to Appendix \ref{app d omega 0}, by the on-shell condition $\mrd \boldsymbol{\omega}\approx 0$. Hence the condition (i) in the conservation discussion is satisfied. On the other hand, by isometry condition $\delta_\xi\Phi=0$, which results 
\begin{equation}\label{Killing omega 0}
\boldsymbol{\omega}(\delta\Phi,\delta_\xi\Phi,\Phi)\approx 0
\end{equation}
It is true because $\boldsymbol{\omega}$ in \eqref{omega LW} is bilinear in its arguments. So the condition (ii) is satisfied for any choice of $\Sigma$, leading to conservation of $\delta H_\xi$ independent of $\Sigma$.
}
\Lemma{\label{Lemma independ H}\emph{For an isometry generator $\xi$, $\delta H_\xi$ is independent of the choice of $\mathscr{S}$ surrounding the BH:} \\
Because of \eqref{Killing omega 0} for a Killing vector $\xi$, the Hamiltonian generator $\delta H_\xi$ would be the same for any chosen $(d\!-\!2)$-dim surfaces surrounding the BH. Explicitly, using the Stoke's theorem \eqref{Stoke's theorem} for the $(d\!-\!1)$-dim region $\Sigma$ encompassed by two boundaries $\mathscr{S}_1$ and $\mathscr{S}_2$;
\begin{equation}
\int_\Sigma \boldsymbol{\omega}(\delta\Phi,\delta_\xi\Phi,\Phi)=\oint_{\mathscr{S}_2} \boldsymbol{k}_\xi(\delta\Phi,\Phi)-\oint_{\mathscr{S}_1} \boldsymbol{k}_\xi(\delta\Phi,\Phi)\,.
\end{equation}
The LHS vanishes, so by \eqref{delta H k} one has
\begin{equation}
\delta H_\xi\Big|_{\mathscr{S}_1}=\delta H_\xi\Big|_{\mathscr{S}_2}.
\end{equation}
}
\vspace{-0.5cm}
\dotfille

In brief, the property \eqref{Killing omega 0}, (which as we will see in Section \ref{sec symplectic symm}  that it is not a property of Killings exclusively) leads to conservation and $\mathscr{S}$ independence of $\delta H_\xi$. The following examples, provide explicit examples of the constructions discussed above, in addition that they would provide us necessary tools for calculations in later discussions.

\newpage
\vspace*{-2cm}
\dotfillb
\Example{\label{example EH}\emph{EH-$\Lambda$ gravity:} \\
For the EH-$\Lambda$ theory of gravity, one has
\hspace*{-1cm}
\begin{align}
&\Phi(x^\alpha)\to g_{\mu\nu}(x^\alpha),\\
&\mathbf{L}\to \frac{d^dx}{16 \pi G} \sqrt{-g}(R-2\Lambda),\label{EH Lagrangian}\\
&\mathbf{E}_{\Phi}\delta\Phi\to \frac{d^dx}{16\pi G}\sqrt{-g}\Big((G_{\mu\nu}+\Lambda g_{\mu\nu})\delta(g^{\mu\nu})\Big),\\ 
&{\Theta}^\mu(\delta\Phi,\Phi)\to\frac{1}{16\pi G}(\nabla_\alpha \delta g^{\mu\alpha}-\nabla^\mu\delta g^\alpha_{\,\,\alpha})\,, \label{EH Theta}\\
& \big(k_\xi^{\text{EH}}(\delta\Phi,\Phi)\big)^{\mu\nu}\to\dfrac{1}{16 \pi G}\Big([\xi^\nu\nabla^\mu h
-\xi^\nu\nabla_\sigma h^{\mu\sigma}+\xi_\sigma\nabla^{\nu}h^{\mu\sigma}+\frac{1}{2}h\nabla^{\nu} \xi^{\mu}-h^{\rho\nu}\nabla_\rho\xi^{\mu}]-[\mu\leftrightarrow\nu]\Big)\,\label{EH k}
\end{align}
in which $\mathbf{\Theta}=\star \Theta$, $\boldsymbol{k}_\xi^{\text{EH}}=\star k_\xi^{\text{EH}}$, $h_{\alpha\beta}\equiv \delta g_{\alpha\beta}$ and $h\equiv h^\alpha_{\,\,\alpha}$.
}
\Example{\emph{EM theory:}\\
For the EM theory, one has 
\begin{align}
&\Phi(x^\alpha)\to g_{\mu\nu}(x^\alpha) \quad \&\quad  A_{\mu}(x^\alpha)\,,\\
&\mathbf{L}\to \frac{d^dx}{16 \pi G}\sqrt{-g}(R-F^2)\,,\label{EM Lagrangian}\\
&\mathbf{E}_{\Phi}\delta\Phi\to \frac{d^dx}{16\pi G}\sqrt{-g}\Big((G_{\mu\nu}-8\pi GT_{\mu\nu})\delta( g^{\mu\nu})+4\nabla_\mu F^{\mu\nu}\delta A_\nu\Big),\\ 
&{\Theta}^\mu(\delta\Phi,\Phi)\to\frac{1}{16\pi G}(\nabla_\alpha \delta g^{\mu\alpha}-\nabla^\mu\delta g^\alpha_{\,\,\alpha}-4F^{\mu\nu}\delta A_{\nu})\,,\\
& (k_\xi^{\text{EM}})^{\mu\nu}\to(k_\xi^{\text{EH}})^{\mu\nu}+\frac{1}{8 \pi G}\Big(\big[(\frac{-1}{2}h F^{\mu\nu}+2F^{\mu\alpha}h_\alpha^{\;\;\nu}-\delta F^{\mu\nu})({\xi}^\beta A_\beta)\hspace*{3cm}\nonumber\\
 &\qquad \qquad \quad -F^{\mu\nu}\xi^\alpha \delta A_\alpha-2F^{\alpha\mu}\xi^\nu \delta A_\alpha\big]-[\mu\leftrightarrow\nu]\Big)\,.\label{EM k}
\end{align}
}

\Example{\label{example KK}\emph{KK theory:}\\
For the KK theory, one has
\begin{align}
&\Phi(x^\alpha)\to g_{\mu\nu}(x^\alpha) \quad \&\quad  A_{\mu}(x^\alpha)\quad \&\quad \phi(x^\alpha)\,,\\
&\mathbf{L}\to \frac{d^dx}{16 \pi G}\sqrt{-g}(R-2\nabla^\alpha\phi\nabla_\alpha\phi-e^{-2\sqrt{3}\phi}F^2)\,,\label{KK Lagrangian}\\
&\mathbf{E}_{\Phi}\delta\Phi\to \frac{d^dx}{16\pi G}\sqrt{-g}\Big((G_{\mu\nu}\!-\!8\pi GT_{\mu\nu})\delta (g^{\mu\nu})\!+\!4\nabla_\mu F^{\mu\nu}\delta A_\nu\!+\!4(\Box\phi\!+\!\frac{\sqrt{3}}{2}e^{-2\sqrt{3}\phi}F^2)\delta\phi\Big),\\ 
&{\Theta}^\mu(\delta\Phi,\Phi)\to\frac{1}{16\pi G}(\nabla_\alpha \delta g^{\mu\alpha}-\nabla^\mu\delta g^\alpha_{\,\,\alpha}-4e^{-2\sqrt{3}\phi}F^{\mu\nu}\delta A_{\nu}-4\nabla^\mu\phi\delta\phi)\,,\\
& (k_\xi^{\text{KK}})^{\mu\nu}\to(k_\xi^{\text{EH}})^{\mu\nu}+\frac{e^{-2\sqrt{3}\phi}}{8 \pi G}\Big(\big[(\frac{-1}{2}h F^{\mu\nu}+2F^{\mu\alpha}h_\alpha^{\;\;\nu}-\delta F^{\mu\nu}+2\sqrt{3}F^{\mu\nu}\delta\phi)({\xi}^\beta A_\beta)\hspace*{3cm}\nonumber\\
 &\qquad \qquad \quad -F^{\mu\nu}\xi^\alpha \delta A_\alpha-2F^{\alpha\mu}\xi^\nu \delta A_\alpha\big]-[\mu\leftrightarrow\nu]\Big)+\frac{1}{8\pi G}\Big(\xi^\nu\nabla^\mu\phi\,\delta\phi-[\mu\leftrightarrow\nu]\Big)\,.\label{KK k}
\end{align}
}
\vspace{-0.5cm}
\dotfille

\subsection{Integrability}
For any given vector field $\xi$, the $\delta H_\xi$ can be found by \eqref{delta H xi}, although it might not be finite. But existence of $H_\xi$ is not guaranteed. The existence of $H_\xi$ is called \emph{integrability}. The sufficient condition for a Hamiltonian variation to be integrable is \cite{Wald:1999wa}
\begin{equation}\label{integrability cond H}
(\delta_1\delta_2 -\delta_2\delta_1)H_\xi=0
\end{equation}
for any chosen members $\delta_1\Phi$ and $\delta_2\Phi$ of the tangent bundle of $\mathcal{M}$. In Appendix \ref{app Wald-Zoupas integrability} it is shown that noticing $\delta \xi=0$, \eqref{integrability cond H} is equivalent to \cite{Wald:1999wa}
\begin{equation}\label{integ cond omega}
\oint_\mathscr{S}\xi\cdot \boldsymbol{\omega}(\delta_1\Phi,\delta_2\Phi,\Phi)\approx 0\,,
\end{equation}
which is the criterion for integrability. If \eqref{integ cond omega} is satisfied for any legitimate  $\delta_1 \Phi$ and $\delta_2\Phi$, then $\delta H_\xi$, introduced in \eqref{delta H xi}, would be integrable. Hence one can find a conserved Hamiltonian generator for $\xi$ as
\begin{equation}\label{finite H xi}
H_\xi[\Phi]=\uint_{\Phi_0}^{\Phi} \delta H_\xi+H_\xi[\Phi_0]\,,
\end{equation} 
in which the integration is performed over arbitrary integral curves which connect a reference field configuration $\Phi_0$ to the $\Phi$ on the phase space. The $H_\xi[\Phi_0]$ is the reference point for the $H_\xi$ defined on the reference field configuration $\Phi_0$.

It is worth mentioning that following the steps in \ref{app Wald-Zoupas integrability}, the $\mathbf{Y}$ ambiguity has been considered implicitly in \eqref{integ cond omega}, through the $\mathbf{\Theta}$ dependency of the $\boldsymbol{\omega}$. We can keep this way of deduction in mind, by casting it in the lemma below. 

\dotfillb
\Lemma{\label{Lemma Y independence}\emph{$\mathbf{Y}$ independence of results:} \\
If any deduction has been carried out using the abstract form of $\mathbf{\Theta}$, and if the results  are explained in similar way, then the results would be correct, irrespective of any chosen $\mathbf{Y}$ ambiguity. 
}
\dotfille 

\section{Noether-Wald analysis}\label{sec Noether-Wald}
In the previous section, a (probable) conserved Hamiltonian generator $H_\xi$ was introduced for a given vector field $\xi$. In this section, the usual Noether method for definition of a conserved charge $\mathcal{Q}_\xi$ in gravitational theories is followed. We found suitable name for it to be Noether-Wald conserved charge. Then we compare it with $H_\xi$, and find that they are generally different. Eventually, the ambiguities of Noether-Wald charge $\mathcal{Q}_\xi$ is discussed.

\subsection{Noether-Wald conserved charge}
Associated to any infinitesimal diffeomorphism as a symmetry of the theory, one can find a Noether-Wald current and the corresponding Noether-Wald charge.  Following \cite{Iyer:1994ys} we take generator of  diffeomorphism symmetry to be a vector field $\xi$. Variation of Lagrangian under associated diffeomorphism is
\begin{equation}\label{lagrangian deviation}
\delta _\xi \mathbf{L}=\mathbf{E}_{\Phi} \delta _\xi\Phi +\mathrm{d}\mathbf{\Theta} (\delta_\xi\Phi,\Phi)\,,
\end{equation}
where summation on different dynamical fields should be understood in $\mathbf{E}_\Phi \delta _\xi\Phi$. The on-shell condition would be $\mathbf{E}_\Phi=0$ which is the e.o.m for $\Phi$. The $(d\!-\!1)$-form $\mathbf{\Theta}$ is the surface term generated by the variation.

According to the identity $\delta _\xi \mathbf{L}= \xi  \cdot  \mathrm{d} \mathbf{L} +\mathrm{d} (\xi  \cdot \mathbf{L})$ and noting that $\mathrm{d} \mathbf{L}=0$, we can replace the LHS of \eqref{lagrangian deviation} by $\mathrm{d} (\xi  \cdot \mathbf{L})$, so
\begin{equation}
\mathrm{d}\mathbf{\Theta}(\delta_\xi\Phi,\Phi)-\mathrm{d} (\xi  \cdot \mathbf{L}) =-\mathbf{E}_\Phi \delta _\xi\Phi \,.
\end{equation}
Now, it is possible to introduce a Noether $(d\!-\!1)$-form  current $\mathbf{J}_\xi$ as
\begin{equation}\label{Noether Wald current}
{\mathbf{J}_\xi \equiv \mathbf{\Theta}(\delta_\xi\Phi,\Phi)-\xi \! \cdot \! \mathbf{L}}\,.
\end{equation}
Therefore $\mathrm{d}\mathbf{ J}_\xi=-\mathbf{E}_\Phi \delta _\xi\Phi$,  hence $\mathrm{d} \mathbf{J}_\xi\approx 0$ whenever e.o.m is satisfied. According to the Poincar\'e's lemma, since $\mathbf{J}_\xi$ is closed, it would be exact on-shell, and can be written as
\begin{equation}\label{Q-form}
{\mathbf{J}_\xi\approx\mathrm{d} \mathrm{\mathbf{Q}}}_\xi\,,
\end{equation}
where $\mathrm{\mathbf{Q}}_\xi$ is a $(d\!-\!2)$-form, the \emph{Noether-Wald charge density}. Now one can define a conserved charge $\mathcal{Q}_\xi$ as
\begin{equation}\label{Noether Wald cons charge}
\mathcal{Q}_\xi=\int_\Sigma \mathbf{J}_\xi=\oint_{\mathscr{S}} \mathbf{Q}_\xi\,.
\end{equation}
To have $\mathcal{Q}_\xi$ as a conserved quantity, in addition to $\mrd \mathbf{J}_\xi\approx 0$, the flow of $\mathbf{J}_\xi$ out of $\mathscr{S}$ should vanish.

\dotfillb
\Lemma{\label{Lemma EH Q conserv}\emph{For $\xi$ as Killing vector of an EH gravity solution, $\mathcal{Q}_\xi$  is conserved independent of $\mathscr{S}$.} \\
Assuming that $\xi$ is Killing vector of a given solution of the EH gravity, then $\mathcal{Q}_\xi$ is conserved. It is because of (i) background has been chosen as a solution, so $\mrd \mathbf{J}_\xi\approx 0$, (ii)  $\mathbf{L}\approx 0$, and in addition, $\mathbf{\Theta}=0$ because of Killing-ness of $\xi$. So in \eqref{Noether Wald current}, $\mathbf{J}_\xi\approx 0$, preventing any flow on $\mathscr{S}$. 
}
\Lemma{\emph{For $\xi$ as Killing vector of an EH gravity solution, $\mathcal{Q}_\xi$ is independent of $\mathscr{S}$ surrounding the BH.} \\
Assuming that $\xi$ is Killing vector of a given solution of the EH gravity, then $\mathcal{Q}_\xi$ is independent of $\mathscr{S}$ surrounding the BH. It can be proved by $\mathbf{J}_\xi\approx 0$ discussed in Lemma \ref{Lemma EH Q conserv}, and by similar reasoning as in the Lemma \ref{Lemma independ H}. Hence
\begin{equation}
\mathcal{Q}_\xi\Big|_{\mathscr{S}_1}=\mathcal{Q}_\xi\Big|_{\mathscr{S}_2}\,.
\end{equation}
}
\vspace{-0.5cm}
\dotfille


\subsection{Ambiguities of Noether-Wald charge}\label{sec ambiguities}

The $(d\!-\!2)$-form $\mathbf{Q}_\xi$ in \eqref{Q-form} has threefold ambiguities: 
\begin{enumerate}
\item One ambiguity comes from freedom in the definition of Lagrangian by addition of an exact $d$-form, as
\begin{equation}
\mathbf{L} \to \mathbf{L}+\mrd \boldsymbol{\mu}\,,
\end{equation}
which according to Appendix \ref{app change of J} leads to $\mathbf{J}_\xi \to \mathbf{J}_\xi +\mrd (\xi \cdot \boldsymbol{\mu})$.
\item The other ambiguity comes from the freedom in specifying $\mathbf{J}$ itself (for a given  Lagrangian); The $\mathbf{\Theta}$ is ambiguous up to an exact $(d-1)$-form $\mrd \mathbf{Y}(\delta_\xi \Phi,\Phi)$. Therefore, the Noether-Wald current $\mathbf{J}_\xi$ is defined up to $\mathbf{J}_\xi \to \mathbf{J}_\xi+\mrd \mathbf{Y}( \delta_\xi\Phi,\Phi)$.
\item In \eqref{Q-form}, $\mathbf{Q}_\xi$ is defined up to an exact $(d-2)$-form $\mrd \mathbf{Z}$.
\end{enumerate}
So accumulating all the ambiguities, we have the freedom of choosing the Noether-Wald charge density 
\begin{equation}
\mathbf{Q}_\xi \to \mathbf{Q}_\xi +\xi \cdot \boldsymbol{\mu} +\mathbf{Y}+\mrd\mathbf{Z}\,,
\end{equation}
\dotfillb
\Example{\emph{Explicit $\mathcal{Q}_\xi$ in EH theory:} \\
Ignoring the ambiguities, in the EH theory it can be shown (Appendix \ref{app Kommar int}) that $\mathcal{Q}_\xi$ for a generic vector field $\xi$ is equal to the Komar integral,
\begin{align}\label{Komar}
\mathcal{Q}_\xi=\frac{-1}{16\pi G}\oint_\mathscr{S}\star \mrd \xi\,.
\end{align}
}
\Example{\emph{What about EM theory?} \\
For the EM theory and a Killing $\xi$, by on-shell condition $\mrd \mathbf{J}_\xi\approx0$. But one has not necessarily vanishing flows on the boundaries because of  $\mathbf{J}_\xi \not\approx 0$, which is a result of  $\mathbf{L}\not\approx 0$. So $\mathcal{Q}_\xi$ is not necessarily conserved. Therefore, in order to find a conserved charge, one needs to choose specific boundary $\mathscr{S}$, usually the $\mr\to \infty$ for asymptotic flat solutions. 
}
\dotfille

As discussed, the Noether-Wald charge density $\mathbf{Q}_\xi$ is not unique. Its most general structure is described by the following decomposition theorem\cite{Iyer:1994ys}.
\begin{theorem}
The most generic Noether-Wald conserved charge is as
\begin{equation}\label{decomposition}
\mathbf{Q}_\xi=\mathbf{W}_\mu(\Phi)\xi^\mu+\mathbf{E}^{\mu \nu}(\Phi)\nabla_{[\mu}\xi _{\nu ]}+\mathbf{Y}( \delta_\xi \Phi,\Phi)+\mrd \mathbf{Z}(\xi,\Phi),
\end{equation}
where $\mathbf{W}_\mu$ and $\mathbf{E}^{\mu \nu}$ and $\mathbf{Y}$ and $\mrd\mathbf{Z}$ are covariant $d-2$-forms which are locally constructed from fields and their derivatives, $\mathbf{Y}$ is linear in $\delta _\xi \Phi$, $\mathbf{Z}$ is linear in $\xi$ and,
\begin{equation}\label{E-four-index}
(\mathbf{E}^{\mu\nu})_{\alpha_3 \dots \alpha_d}=-\sqrt{-g}{E}^{\alpha \beta \mu \nu}\epsilon_{\alpha \beta\alpha_3 \dots \alpha_d }\,,\qquad {E}^{ \alpha \beta \mu \nu}\equiv \frac{\partial \mathcal{L}}{\partial R_{\alpha \beta \mu \nu}}\,.
\end{equation}
\end{theorem}
It is important to note that the above decomposition is not unique (except the term containing $\mathbf{E}^{\mu \nu}$). 
In order to fix/remove these ambiguities, some physical reasoning (like introducing reference point for defining the charges) are needed.

\newpage
\vspace*{-1.5cm}
\dotfillb
\Lemma{\label{Lemma ambig vanish}\emph{For a Killing vector $\xi$ which vanishes on the integration surface $\mathscr{s}$, all ambiguities vanish.} \\
Considering $\xi$ to be a Killing vector, then $\delta_\xi\Phi=0$, so the $\mathbf{Y}$ ambiguity in \eqref{decomposition} which is linear in $\delta_\xi\Phi$ vanishes. Besides, if $\xi\big|_\mathscr{S}=0$, then the terms containing $\mathbf{W}$ and $\mathbf{Z}$ would also vanish, because they are linear in $\xi$.
}
\Example{\emph{Re-deriving $\mathcal{Q}_\xi$ for EH gravity:} \\
For the EH gravity
\begin{equation}\label{EH E}
{E}^{ \alpha \beta \mu \nu}=\frac{1}{32 \pi G}(g^{\mu\alpha}g^{\nu\beta}-g^{\mu\beta}g^{\nu\alpha})\,,
\end{equation}
so by ignoring the $\mathbf{W}$, $\mathbf{Y}$ and $\mathbf{Z}$ ambiguities in \eqref{decomposition},
\begin{align}
\mathbf{Q}_\xi &=\mathbf{E}^{\mu \nu}(\Phi)\nabla_{[\mu}\xi _{\nu ]}\\
&=\Big(\frac{-1}{32 \pi G}(g^{\mu\alpha}g^{\nu\beta}-g^{\mu\beta}g^{\nu\alpha})\frac{\sqrt{-g}}{(d-2)!}\epsilon_{\alpha \beta\alpha_3 \dots \alpha_d }\,\mrd x^{\alpha_3}\wedge \cdots \wedge \mrd x^{\alpha_d}\Big) \nabla_{[\mu}\xi _{\nu ]}\\
&=\frac{-\sqrt{-g}}{16 \pi G (d-2)!}\epsilon_{\alpha \beta\alpha_3 \dots \alpha_d }\nabla^{[\alpha}\xi ^{\beta ]}\,\mrd x^{\alpha_3}\wedge \cdots \wedge \mrd x^{\alpha_d}\\
&=\frac{-1}{16 \pi G}\star \mrd\xi\,.\label{Komar 2}
\end{align}}
\dotfille

\section{Hamiltonian generators vs. Noether-Wald charges}\label{sec H vs Q}
The $\mathcal{Q}_\xi$ does not necessarily coincide with $H_\xi$ in \eqref{finite H xi}. This section is provided to make their relation clear. Ignoring the ambiguities, in Appendix \ref{app H v Q} it is shown that 
\begin{equation}\label{omega vs Q}
\boldsymbol{\omega}(\delta\Phi,\delta_\xi\Phi,\Phi)\approx \mrd \big(\delta \mathbf{Q}_\xi-\xi \cdot \mathbf{\Theta}(\delta \Phi,\Phi)\big)\,.
\end{equation}\label{k-xi general formula}
Hence, confirming the claimed \eqref{omega dk}, the $\boldsymbol{k}_\xi$ is read as
\begin{equation}
\boldsymbol{k}_\xi=\delta \mathbf{Q}_\xi-\xi \cdot \mathbf{\Theta}(\delta \Phi,\Phi)\,.
\end{equation}
As a simple example, for the EH-$\Lambda$ theory it is calculated in Appendix \ref{app k EH proof}, resulting the \eqref{EH k}.

Integrating \eqref{omega vs Q} over $\Sigma$, and using the Stoke's theorem, 
\begin{equation}\label{H v Q}
\delta H_\xi=\delta \mathcal{Q}_\xi-\oint_{\mathscr{S}}\xi\cdot \mathbf{\Theta}(\delta\Phi,\Phi)\,.
\end{equation}
So $\delta H_\xi$ and $\delta \mathcal{Q}_\xi$ coincide if the last term would vanish. The $\delta \mathcal{Q}_\xi$ by its definition is integrable; because it is variation of $\mathcal{Q}_\xi$. As a result, $\delta H_\xi$ would be integrable if and only if the last term in \eqref{H v Q} would be integrable. Besides, it is worth mentioning that according to Appendix \ref{app H v Q} and Lemma \ref{Lemma Y independence}, the \eqref{H v Q} is respected by any $\mathbf{Y}$ ambiguity.

\chapter{Laws of black holes thermodynamics}\label{chap BH thermo}

As it was briefly explained in Motivations, BHs are equipped with some laws analogous to the laws of thermodynamics. In this chapter, we explain more on this issue. In the first section, two useful concepts are introduced; \emph{linearized equation of motion} and \emph{parametric variations}. In the second section, the role of Hamiltonian generators as thermodynamic variables is reviewed. The third section, deals with temperature and other chemical potentials. In the fourth section, entropy is introduced as a Hamiltonian generator. Gathering all the materials, fifth section would present thermodynamic laws for BHs. Finally, in the sixth section, extremal BHs are introduced and are questioned about their thermodynamic laws.


\section{Linearized equation of motion and parametric variations}

In chapter \ref{chap some grav}, the e.o.m's have been denoted collectively by $\mathbf{E}_\Phi=0$. These equations are in general non-linear partial differentials of dynamical fields $\Phi$. Assuming that an exact solution to these equations is given as $\bar{\Phi}$, then it is possible to study perturbations around that solution at the linear approximations. The equations which the perturbations satisfy is called linearized equations of motion (l.e.o.m), and are found by the first order expansion of the original e.o.m's as
\begin{equation}\label{l.e.o.m}
\delta \mathbf{E}_\Phi=0\,.  
\end{equation} 

\newpage
\vspace*{-1.5cm}  
\dotfillb
\Example{\emph{l.e.o.m for EH gravity:} \\
The e.o.m for the EH gravity is $G_{\mu\nu}=0$. The dynamical field is the metric $g_{\mu\nu}$. The l.e.o.m would be $G^{(lin)}_{\mu\nu}\equiv\frac{\delta G_{\mu\nu}}{\delta g_{\alpha \beta}}|_{\bar{g}}\delta g_{\alpha\beta}=0$. The term $\frac{\delta G_{\mu\nu}}{\delta g_{\alpha \beta}}|_{\bar{g}}$ is a differential operator built of background metric $\bar{g}$ which acts linearly on $\delta g_{\alpha\beta}$. Explicitly
\begin{equation}\label{EH linearized}
G^{(lin)}_{\mu\nu}=\nabla_\alpha \nabla_\nu h^\alpha_{\; \mu}+\nabla_\alpha \nabla_\mu h^\alpha_{\; \nu}-\Box h_{\mu\nu}-\nabla_\nu \nabla_\mu h-\bar{g}_{\mu\nu}[\nabla_\alpha \nabla_\beta h^{\alpha \beta}-\Box h],
\end{equation}
in which $h_{\alpha\beta}\equiv \delta g_{\alpha\beta}$ and $h\equiv h^\alpha_{\,\,\alpha}$. Notice that $\nabla$'s are calculated using $\bar{g}$. Having non-zero cosmological constant, would add a $\Lambda h_{\mu\nu}$ to \eqref{EH linearized}.
}
\vspace*{-0.5cm}
\dotfille

There are two families of field variations which satisfy the l.e.o.m trivially:
\begin{enumerate}
\item Parametric variations,
\item Lie variations.
\end{enumerate} 
Parametric variations are the ones which we use frequently in Part \ref{part I}, and will be described here. The Lie variations are just the Lie derivatives of background fields $\mathscr{L}_\xi\bar\Phi$, and would be our main interest in Part \ref{part II}. 

Considering a solution $\bar{\Phi}$ which depends on some parameters $\bar{\Phi}(m,a,\dots)$, it can be variated w.r.t its parameters to the first order, as
\begin{equation}
\hat{\delta}\Phi\equiv \frac{\partial \bar{\Phi}}{\partial m}\delta m+ \frac{\partial \bar{\Phi}}{\partial a}\delta a+\dots\,.
\end{equation}  
We can call $\hat{\delta}\Phi$ as \emph{parametric variations} \cite{HSS:2014twa}. They satisfy l.e.o.m, because they are linearized difference of two solutions $\bar{\Phi}_1$ and $\bar{\Phi}_2$ for which $\mathbf{E}_{\bar{\Phi}_1}=0$ and $\mathbf{E}_{\bar{\Phi}_2}=0$. So to linear order,  $\hat{\delta}\mathbf{E}_\Phi=0$.

\newpage
\vspace*{-1.5cm} 
\dotfillb
\Example{\emph{Parametric variations for Kerr BH:} \\
Denoting the metric \eqref{Kerr metric} with parameters $\{m,a\}$ as $\bar{g}$, then the $\hat{\delta}g_{\mu\nu}$ would be
\begin{equation}
\hat{\delta}g_{\mu\nu}=\frac{\partial \bar{g}_{\mu\nu}}{\partial m}\delta m+ \frac{\partial \bar{g}_{\mu\nu}}{\partial a}\delta a
\end{equation}
for some constant numbers $\delta m$ and $\delta a$. For other BHs in section \ref{sec BHs} similar relation can be written depending on their parameters.
}
\dotfille


\section{Hamiltonian generators as thermodynamic variables} \label{sec BH conserved charges}
Among many possible Hamiltonian generators associated to diffeomorphisms, there are some which are used to label the BH solutions. According to the previous section, in order to guarantee the conservation of $\delta H_\xi$, $\xi$ is chosen to be Killing vector. According to the isometry $\mathbb{R}\times U(1)^n$ which we have requested for BHs, \emph{mass} and \emph{angular momenta} are defined as follows.

\subsection{Mass}
\emph{Mass} is the Hamiltonian generator $H_\xi$ associated to the stationarity, and is denoted by $M$. The coordinates are adopted to have the associated Killing vector as $\xi=\partial_\mt$. Concerning the conservation, although by Killing-ness the $H_{\partial_\mt}$ is conserved, but it is usual that $\mathscr{S}$ be chosen at infinity, in order to have conservation for $\mathcal{Q}_{\partial_\mt}$ too. The $H_{\partial_\mt}$ is different than $\mathcal{Q}_{\partial_\mt}$, because of the extra terms in \eqref{H v Q}. So if one wants to use Komar integral \eqref{Komar}, care is needed. Usually a factor of $2$ would make the Komar integral to produce the correct mass \cite{Iyer:1994ys}. In calculating the charges, normalization of Killing vectors matters. Normalization for $\xi$ is chosen in a way to coincide with the generator of time translations at asymptotics, and it is chosen to be future directed. 

\subsection{Angular momentum}
\emph{Angular momenta} are Hamiltonian generators $H_{\mathrm{m}_i}$ associated to the axial isometries, and are denoted by $J_i\equiv -H_{\mrm_i}$. The coordinates are adopted to have the associated Killing vectors as $\mathrm{m}_i=\partial_{\psi^i}$. Concerning the conservation, although by Killing-ness the $H_{\mathrm{m}_i}$ is conserved, but it is usual that $\mathscr{S}$ be chosen at infinity, in order to have conservation for $\mathcal{Q}_{\mathrm{m}_i}$ too. The $H_{\mathrm{m}_i}$ is different than $\mathcal{Q}_{\mathrm{m}_i}$, because of the extra terms in \eqref{H v Q}. But if the surface $\mathscr{S}$ is chosen to be surface of constant time and radius, then pull-back of the extra term to $\mathscr{S}$ vanishes. So $H_{\mathrm{m}_i}=\mathcal{Q}_{\mathrm{m}_i}$. Normalization for $\mathrm{m_i}$ is chosen in a way to coincide with the generator of axial isometry at asymptotics. Then $J_i$ would be defined by an extra minus sign w.r.t the $H_{\mathrm{m}_i}$.

\dotfillb
\Example{\label{example Kerr M J}\emph{Mass and angular momentum for Kerr BH:} \\
Denoting the Kerr metric \eqref{Kerr metric} with parameters $\{m,a\}$ as $\bar{g}$, then
\begin{itemize}
\item[--] One can calculate $\hat{\delta}H_{\partial_\mt}$ using $\boldsymbol{k}^{\mathrm{EH}}$ from \eqref{EH k}, and choosing $\mathscr{S}$ to be any surface of constant time and radius. Then he finds 
\begin{equation}\label{kerr delta M}
\hat{\delta}H_{\partial_\mt}=\delta m\,.
\end{equation}
According to \eqref{kerr delta M}, the $\hat{\delta}H_{\partial_\mt}$ is integrable over the parameters, so choosing the reference point of mass $H_{\partial_\mt}[\Phi_0]=0$ on the metric with $m=a=0$ (\ie the Minkowski spacetime), we would have
\begin{equation}
M\equiv H_{\partial_\mt}=m\,.
\end{equation}
\item[--] Similar calculation for the Killing $\mathrm{m}=\partial_{\psi}$, on any surface of constant time and radius surrounding the BH, leads to (remembering the additional minus sign)
\begin{equation}
-\hat{\delta}H_{\mathrm{m}}=a\delta m+m\delta a\,,
\end{equation}
which is also integrable over the parameters,
\begin{equation}\label{Kerr J}
J\equiv - H_{\mathrm{m}}=ma\,.
\end{equation}
Reference point for the angular momentum  $H_\mrm[\Phi_0]$ is chosen to vanish on the solutions with $a=0$.
\end{itemize}
 }
\Example{\emph{Mass and angular momentum for Kerr-Newman BH:} \\
Denoting the Kerr-Newman metric \eqref{Kerr-Newman metric} with parameters $\{m,a,q\}$ as $\bar{g}$, then similar analysis to the Kerr BH, but using $\boldsymbol{k}^{\mathrm{EM}}$ from \eqref{EM k} instead of \eqref{EH k}, leads to
\begin{align}
\hat{\delta}H_{\partial_\mt}=\delta m \quad &\Rightarrow\quad M=H_{\partial_\mt}=m\,,\\
-\hat{\delta}H_{\mathrm{m}}=a\delta m+m\delta a \quad &\Rightarrow \quad J= -H_{\mathrm{m}}=ma\,.
\end{align}
Reference points are chosen similarly to the analysis of the Kerr BH. Explicitly, $H_{\partial_\mt}[\Phi_0]=0$ when $m=a=q=0$. Also, the angular momentum $-H_\mrm[\Phi_0]$ is chosen to vanish on the solutions with $a=0$. 
}
\Example{\emph{Mass and angular momentum for Kerr-AdS BH:} \\
Denoting the Kerr-AdS metric \eqref{Kerr-AdS metric} with parameters $\{m,a\}$ as $\bar{g}$, then similar analysis to the Kerr BH, using $\boldsymbol{k}^{\mathrm{EH}}$ from \eqref{EH k}, leads to \cite{Gibbons:2004ai}
\begin{align}
\hat{\delta}H_{\partial_\mt}=\frac{\delta m}{\Xi^2} \quad &\Rightarrow\quad M=H_{\partial_\mt}=\frac{m}{\Xi^2}\,,\\
-\hat{\delta}H_{\mathrm{m}}=\frac{a}{\Xi^2}\delta m-\frac{m(1+3\frac{a^2}{l^2})}{\Xi^3}\delta a \quad &\Rightarrow \quad J= -H_{\mathrm{m}}=\frac{ma}{\Xi^2}\,.
\end{align}
Reference points have been chosen naturally;  For the solution with the parameters set as $m=a=0$, \ie the pure $AdS_4$ spacetime, the mass is chosen to vanish, $H_{\partial_\mt}[\Phi_0]=0$. In addition, the angular momentum $-H_\mrm[\Phi_0]$ is chosen to vanish on the solutions with $a=0$. 
}
\Example{\emph{Mass and angular momentum for BTZ BH:} \\
Denoting the BTZ metric \eqref{BTZ metric} with parameters $\{m,j\}$ as $\bar{g}$, then similar analysis to the Kerr BH, using $\boldsymbol{k}^{\mathrm{EH}}$ from \eqref{EH k} leads to \cite{Banados:1992wn}
\begin{align}
\hat{\delta}H_{\partial_\mt}=\frac{\delta m}{8} \quad &\Rightarrow\quad M=H_{\partial_\mt}=\frac{m}{8}\,,\\
-\hat{\delta}H_{\mathrm{m}}=\frac{\delta j}{8G} \quad &\Rightarrow\quad J=-H_{\mathrm{m}}=\frac{j}{8G}\,.
\end{align}
The reference point for the mass $H_{\partial_\mt}[\Phi_0]$ is chosen to vanish on the solution with $m=j=0$. For the angular momentum, $H_\mrm [\Phi_0]$ is chosen to vanish on the solutions with $j=0$.
}
\Example{\emph{Mass and angular momenta for $5$-dim MP BH:} \\
Denoting the $5$-dim MP metric \eqref{MP metric} with parameters $\{m,a,b\}$ as $\bar{g}$, then using $\boldsymbol{k}^{\mathrm{EH}}$ from \eqref{EH k} leads to 
\begin{align}
\hat{\delta}H_{\partial_\mt}=\delta m \quad &\Rightarrow\quad M=H_{\partial_\mt}=m\,,\\
-\hat{\delta}H_{\mathrm{m}_1}=\frac{\pi}{2}(a\delta m+m\delta a) \quad &\Rightarrow\quad J_1=-H_{\mathrm{m}_1}=\frac{\pi}{2} ma\,,\\
-\hat{\delta}H_{\mathrm{m}_2}=\frac{\pi}{2}(b\,\delta m+m\delta b) \quad &\Rightarrow\quad J_2=-H_{\mathrm{m}_2}=\frac{\pi}{2} mb\,.
\end{align}
The reference point $H_{\partial_\mt}[\Phi_0]$ is chosen to vanish on the solution with $m=a=b=0$, \ie the $5$-dim Minkowski spacetime. For the angular momenta, $H_{\mrm_1}$ and $H_{\mrm_2}$, are  chosen to vanish for the solutions with $a=0$ and $b=0$ respectively.
}
\Example{\label{example KK M J}\emph{Mass and angular momentum for KK BH:} \\
Denoting the KK metric \eqref{KK metric} with parameters $\{m,a,v\}$ as $\bar{g}$, then using $\boldsymbol{k}^{\mathrm{KK}}$ from \eqref{KK k} one finds \cite{Horne:1992zy}
\begin{align}
&\hat{\delta}H_{\partial_\mt}=\left(1+\frac{v^2}{2(1-v^2)}\right)\delta m + \frac{mv}{(1-v^2)^2}\delta v\,,\quad \Rightarrow\quad M=H_{\partial_\mt}=m\left(1+\frac{v^2}{2(1-v^2)}\right)\,,\\
&-\hat{\delta}H_{\mathrm{m}}=\frac{a}{\sqrt{1-v^2}}\delta m + \frac{m}{\sqrt{1-v^2}}\delta a+ \frac{mav}{(1-v^2)^{\frac{3}{2}}}\delta v\,,\quad \Rightarrow \quad J=-H_{\mathrm{m}}=\frac{ma}{\sqrt{1-v^2}}\,.
\end{align}
The $H_{\partial_\mt}[\Phi_0]$ has been chosen to vanish on the solution with $m=a=0$. For the angular momentum, $H_\mrm [\Phi_0] $ vanishes on the solutions with $a=0$.
}
\dotfille

\subsection{Electric charge}
In addition to the mass and angular momenta, for the solutions of EM theory, the electric charge is another conserved quantity used to label the solutions. It is a Noether-Wald charge, which originates from the gauge symmetry of the theories containing the fields $A^{(p)}_\mu$. The Lagrangian of gravitational theories which include these fields are chosen to be invariant under
\begin{equation}\label{gauge trans}
A^{(p)}_\mu \to A^{(p)}_\mu+\partial_\mu \Lambda^{^p}
\end{equation}
for arbitrary scalar functions $\Lambda^{^p}$. So $\mathbf{L}\to \mathbf{L}$, fulfilling the symmetry condition in Section \ref{sec sym iso}; it is called the \emph{gauge symmetry}. Performing similar analysis similar to Noether-Wald analysis in section \ref{sec Noether-Wald}, but considering $\delta \mathbf{L}=0$ results (see Appendix \ref{app electric charge}) to the on-shell conserved electric charge $Q_p$ as
\begin{equation}\label{electric charge Q}
Q_p=\oint_{\mathscr{S}}\star\, \mathrm{Q}^{(p)}\,,
\end{equation}
where
\begin{equation}\label{electric charge Q-form}
 \mathrm{Q}^{(p)\mu\nu}=\left(\frac{\partial \mathcal{L}}{\partial F^{(p)}_{\mu\nu}}\right)\,.
\end{equation}

\dotfillb
\Example{\emph{Electric charge in EM theory:} \\
By \eqref{EM Lagrangian} and \eqref{electric charge Q-form}, for the EM theory one finds $ \mathrm{Q}^{(p)}_{\mu\nu}=\dfrac{-1}{4\pi G}F^{(p)}_{\mu\nu}$. Therefore
\begin{equation}\label{EM electric charge}
Q_p=\frac{-1}{4\pi G}\oint_{\mathscr{S}}\star \mrd A^{(p)}\,.
\end{equation}
As an example in example, for the Kerr-Newman BH, putting the metric \eqref{Kerr-Newman metric} and the gauge field \eqref{Kerr-Newman gauge} into \eqref{EM electric charge}, electric charge is found to be 
\begin{equation}
Q=\frac{q}{G}\,.
\end{equation}
The surface $\mathscr{S}$ can be chosen any $(d\!-\!2)$-dim surface surrounding the BH, \eg surfaces of constant time and radius. 
}
\Example{\emph{Electric charge in KK theory:} \\
By \eqref{KK Lagrangian} and \eqref{electric charge Q-form}, for the KK theory one finds $ \mathrm{Q}^{(p)}_{\mu\nu}=\dfrac{-e^{-2\sqrt{3}\phi}}{4\pi G}F^{(p)}_{\mu\nu}$. Therefore
\begin{equation}\label{KK electric charge}
Q_p=\frac{-e^{-2\sqrt{3}\phi}}{4\pi G}\oint_{\mathscr{S}}\star \mrd A^{(p)}\,.
\end{equation}
As an example in example, for the KK BH, putting the metric \eqref{KK metric} and the gauge field  \eqref{KK gauge} into \eqref{KK electric charge}, electric charge is found to be
\begin{equation}
Q=\frac{mv}{G(1-v^2)}\,.
\end{equation}
}
\vspace*{-0.5cm}
\dotfille


\section{Temperature and chemical potentials}\label{sec BH chemicals}
The temperature which the BHs radiate particles has been derived in semi-classical approximation by Hawking, so it is called Hawking temperature $T_{_\mrH}$ \cite{Hawking:1976rt}. Although it has its origins in the quantum realm, but there is a classical entity which by calculating it, one can find the Hawking temperature. It is the \emph{surface gravity} denoted by $\kappa$, and calculated by the following relation
\begin{equation}\label{surface gravity}
\kappa^2=\frac{-1}{8}(\mrd\zeta_{_\mrH})_{\alpha\beta}(\mrd\zeta_{_\mrH})^{\alpha\beta}\Big|_\mrH\,.
\end{equation}
In the above relation, the normalization of $\zeta_{_\mrH}$ matters, which is chosen to coincide with the stationarity Killing vector $\partial_\mt$ when there is not any angular momentum present. Then the temperature $T_{_\mrH}$ can be found as
\begin{equation}\label{Hawking temp}
T_{_\mrH}=\frac{\kappa}{2\pi}\,.
\end{equation}

The chemical potentials conjugate to the angular momenta are \emph{angular velocities} of the BH,  calculated on the horizon. They are denoted as $\Omega^i_{_\mrH}$, index $i$ for $J_i$ and index $\mrH$ for the horizon. A simple, but non-covariant way of finding them out of the metric components, is solving the following system of $n$ unknowns and $n$ equations
\begin{equation}
g_{\mt i}=-g_{ij}\Omega_{_\mrH}^j\,, 
\end{equation} 
where metric components are calculated on the horizon. For sure, for this quick method to work, the metric should have been written in appropriate coordinates. Having the definition of the $\Omega^i_{_\mrH}$'s, the horizon Killing vector for different BHs would be explicitly
\begin{equation}\label{horizon Killing}
\zeta_{_\mrH}=\partial_\mt+\Omega_{_\mrH}^i\partial_{\psi^i}\,.
\end{equation}
Usually, if we know the Killing vectors for which the Hamiltonians are mass and angular momenta, the equation \eqref{horizon Killing} can be used to define $\Omega^i_{_\mrH}$ covariantly. 
 
The chemical potential conjugate to the electric charges are \emph{electric potentials} of BH on the horizon. They are denoted as $\Phi^p_{_\mrH}$ and calculated by
\begin{equation}
\Phi^p_{_\mrH}=A^{(p)}_\alpha \zeta_{_\mrH}^\alpha\Big|_\mrH\,.
\end{equation}
The sign convention is chosen such that $\Phi^p$ would be positive, if electric charge is positive.


\section{Entropy as a Hamiltonian generator}\label{sec Iyer-Wald entropy}
One of the profound progresses in thermodynamics of BHs has been introduction of their entropy as a Noether-Wald charge \cite{Wald:1993nt,Iyer:1994ys}. The Killing vector responsible for this conservation is the normalized horizon Killing vector 
\begin{equation}\label{normal Killing Horizon}
\hat{\zeta}_{_\mrH}\equiv\frac{\zeta_{_\mrH}}{\kappa}\,.
\end{equation}
The entropy is defined as
\begin{equation}
\frac{S}{2\pi}\equiv \mathcal{Q}_{\hat{\zeta}_{_\mrH}}\Big|_\mrH\,.
\end{equation}
There is not any ambiguity in this definition, because considering $\mrH$ to be the bifurcation point of the horizon, then
\begin{equation}\label{zeta vanish}
\hat{\zeta}_{_\mrH}\Big|_\mrH=0 \,,
\end{equation} 
so all kinds of ambiguities in the \eqref{decomposition} vanish, by the Lemma \ref{Lemma ambig vanish}. It results the following for the entropy
\begin{equation}\label{entropy Wald E}
\frac{S}{2\pi}=\oint_\mrH\mathbf{E}^{\mu \nu}\nabla_{[\mu}\hat{\zeta}_{_\mrH\nu ]}\,,
\end{equation}
in which $\mathbf{E}^{\mu \nu}$ is defined in \eqref{E-four-index}. It is called the Iyer-Wald entropy. By decomposition of Levi-Civita symbol to tangent and binormal components w.r.t to the horizon, it can also be written as
\begin{equation}
\frac{S}{2\pi}=-\oint_\mrH\boldsymbol{\epsilon}_{_\mrH} \epsilon_{_\perp\alpha\beta}\,{E}^{ \alpha \beta \mu \nu}\,\nabla_{[\mu}\hat{\zeta}_{_\mrH\nu ]}\,,
\end{equation}
in which $\boldsymbol{\epsilon}_{_\mrH}$ is the volume element of $(d-2)$-dim horizon
\begin{equation}
\boldsymbol{\epsilon}_{_\mrH}=\text{Vol(H)}\,{\epsilon}_{\alpha_1\dots\alpha_{d-2}}(\,\mathrm{d} x^{\alpha_1}\wedge \dots \wedge \,\mathrm{d} x^{\alpha_{d-2}})\,,
\end{equation}
and $\boldsymbol{\epsilon}_{_{\!\perp}}$ is its binormal $2$-form [\cite{Townsend:1997ew}]. Another mostly used form of the Iyer-Wald entropy, using $\nabla_{[\mu}\hat{\zeta}_{_\mrH\nu ]}\Big|_\mrH=\epsilon_{{_{\!\perp}}\mu\nu}$ \cite{Wald:1993nt,Iyer:1994ys}, can be written as 
\begin{equation}\label{entropy Wald binormal}
\frac{S}{2\pi}=-\oint_\mrH\boldsymbol{\epsilon}_{_\mrH} \epsilon_{_\perp\alpha\beta}\,\epsilon_{{_{\!\perp}}\mu\nu}\,{E}^{ \alpha \beta \mu \nu}\,.
\end{equation}

Although the entropy has been originally defined by Noether-Wald conserved charge, but it is equal to the Hamiltonian generator of $\hat{\zeta}_{_\mrH}$ calculated on the  bifurcation point $\mrH$. It is because by \eqref{zeta vanish}, the last term in \eqref{H v Q} vanishes on $\mrH$. Hence, by the Lemma \ref{Lemma independ H},  entropy is the $H_{\hat{\zeta}_{_\mrH}}$ irrespective of the choice of $\mathscr{S}$ surrounding the BH.

\dotfillb
\Example{\emph{Entropy in EH gravity is proportional to the area of the horizon.} \\
Putting ${E}^{ \alpha \beta \mu \nu}$ from \eqref{EH E} into \eqref{entropy Wald binormal}, yields
\begin{equation}
\frac{S}{2\pi}=-\frac{1}{16\pi G}\oint_\mrH\boldsymbol{\epsilon}_{_\mrH}\epsilon_{{_{\!\perp}}\alpha\beta}\,\epsilon_{_{\!\perp}}^{\alpha\beta}\,.
\end{equation}
The binormal tensor is normalized such that $\epsilon_{{_{\!\perp}}\alpha\beta}\,\epsilon_{_{\!\perp}}^{\alpha\beta}=-2$, so
\begin{equation}\label{entropy area}
\frac{S}{2\pi}=\frac{1}{8\pi G}\oint_\mrH\boldsymbol{\epsilon}_{_\mrH}\,.
\end{equation}
which is just $\frac{A_\mrH}{8\pi G}$, the area of the horizon divided by $8\pi G$, the Bekenstein-Hawking entropy \cite{Bekenstein:1973ft}. The \eqref{EH E} is true for EMD-$\Lambda$ theories, hence \eqref{entropy area} is true generically for these theories.
}
\Example{\emph{Temperature, chemical potentials and entropy for different BHs:} \\
\begin{itemize}
\item[--] \textit{Kerr BH:} 
\begin{flalign}\label{Kerr chemicals}
T_{_\mrH}&=\frac{\mr_+-\mr_-}{4\pi(\mr_+^2+a^2)}\,, \qquad \Omega_{_\mrH}=\frac{a}{\mr_+^2+a^2}\,,\qquad \frac{S}{2\pi}=\frac{\mr_+^2+a^2}{2G}\,.&
\end{flalign}
\item[--] \textit{Kerr-Newman BH:} 
\begin{flalign}
T_{_\mrH}&=\frac{\mr_+-\mr_-}{4\pi(\mr_+^2+a^2)}\,, \qquad \Omega_{_\mrH}=\frac{a}{\mr_+^2+a^2}\,, \qquad \Phi_{_\mrH}=\frac{q\mr_+}{\mr_+^2+a^2}\,,\qquad \frac{S}{2\pi}=\frac{\mr_+^2+a^2}{2G}\,.&
\end{flalign}
\item[--] \textit{Kerr-AdS BH:} 
\begin{flalign}
T_{_\mrH}&=\frac{\mr_+(1+\frac{a^2}{l^2}+3\frac{\mr_+^2}{l^2}-\frac{a^2}{\mr_+^2})}{4\pi(\mr_+^2+a^2)}\,, \qquad \Omega_{_\mrH}=\frac{a(1+\frac{\mr_+^2}{l^2})}{\mr_+^2+a^2}\,, \qquad \qquad \frac{S}{2\pi}=\frac{\mr_+^2+a^2}{2G\,\Xi}\,.&
\end{flalign}
\item[--] \textit{BTZ BH:} 
\begin{flalign}
T_{_\mrH}&=\frac{\mr_+^2-\mr_-^2}{2\pi l^2 \mr_+}\,, \qquad \Omega_{_\mrH}=\frac{\mr_-}{l\mr_+}\,, \qquad \frac{S}{2\pi}=\frac{\mr_+}{4G}\,.&
\end{flalign}
\item[--] \textit{$5$-dim MP BH:} 
\begin{flalign}
T_{_\mrH}&=\frac{\mr_+^4-a^2b^2}{2\pi \mr_+ (\mr_+^2+a^2)(\mr_+^2+b^2)}\,, \qquad \Omega_{_\mrH}^1=\frac{a}{\mr_+^2+a^2}\,,
\qquad \Omega_{_\mrH}^2=\frac{b}{\mr_+^2+b^2}\,, &\nonumber\\
 \frac{S}{2\pi}&=\frac{\pi(\mr_+^2+a^2)(\mr_+^2+b^2)}{4G \mr_+ }\,.&
\end{flalign}
\item[--] \textit{KK BH:} 
\begin{flalign}
T_{_\mrH}&=\frac{(\mr_+-\mr_-)\sqrt{1-v^2}}{4\pi(\mr_+^2+a^2)}\,, \qquad \Omega_{_\mrH}=\frac{a\sqrt{1-v^2}}{\mr_+^2+a^2}\,,\qquad \Phi_{_\mrH}=\frac{v}{2}\,,\qquad \frac{S}{2\pi}=\frac{\mr_+^2+a^2}{2G\sqrt{1-v^2}}\,.&
\end{flalign}
\end{itemize}
}
\vspace*{-0.5cm}
\dotfille


\section{Laws of black hole thermodynamics}
There are four well-known thermodynamic laws for thermodynamic systems, \eg a box of hydrogen gas. Denoting the internal energy by $E$, pressure by $P$, volume by $V$, number of particles by $N$, and its chemical potential by $\mu$, the laws are
\begin{itemize}
\item[${(0)}$] In a system in thermal equilibrium, temperature (and other chemical potentials) are constant over the system,
\item[${(1)}$] $\delta E=T \delta S-P \delta V+\mu \delta N$ ,
\item[${(2)}$] In a closed thermodynamic system, entropy never decreases,
\item[${(3)}$] \begin{itemize}\item $T=0$ is not physically achievable, \item If $T\to 0$ then $S\to 0$.\end{itemize}
\end{itemize}  
Analogous to these four laws, stationary BHs fulfil four laws, known as \emph{laws of BH thermodynamics}. These laws are as below \cite{Bardeen:1973gd}:
\begin{itemize}
\item[$\mathbf{(0)}$] $T_{_\mrH}$, $\Omega_{_\mrH}^i$ and $\Phi_{_\mrH}^p$ are constant over the horizon,
\item[$\mathbf{(1)}$] $\delta M=T_{_\mrH} \delta S+\Omega_{_\mrH}^i \delta J_i+\Phi_{_\mrH}^p \delta Q_p\,,$
\item[$\mathbf{(2)}$] For a closed thermodynamic system (including the black hole), entropy never decreases,
\item[$\mathbf{(3)}$] $T_{_\mrH}=0$ is not physically achievable.
\end{itemize}
A simple proof of the zeroth law for temperature is provided in Appendix \ref{app zeroth law} [\cite{padmanabhan2010gravitation}]. For the proof of constancy of $\Omega_{_\mrH}^i$ see [\cite{Frolov:2012}] and references therein.

It is easy to check the first law by parametric variations, explained in the example below. The Iyer-Wald proof of the law is presented after that.

\dotfillb
\Example{\emph{Checking the first law of BH thermodynamics by parametric variations:} \\
One can use the parametric variations to check the first law of BH thermodynamics. For example, for the Kerr BH, the $J$ and $S$ are known in terms of parameters $\{m,a\}$, through \eqref{Kerr J} and \eqref{Kerr chemicals}. Hence one has simply
\begin{equation}\label{Kerr delta S}
\frac{\hat{\delta}S}{2\pi}= \frac{2\mr_+^2}{\mr_+-\mr_-}\delta m-\frac{2ma}{\mr_+-\mr_-}\delta a\,,\qquad \hat{\delta}J=a\delta m+m \delta a\,.
\end{equation}
Putting \eqref{Kerr delta S} in the $T_{_\mrH} \hat{\delta} S+\Omega_{_\mrH} \hat{\delta} J$, and using \eqref{Kerr chemicals} to replace $T_{_\mrH}$ and $\Omega_{_\mrH}$, the result would confirm the first law. Similar analysis can be done for other BHs in our sample.
}
\Example{\label{example Iyer-Wald 1st law}\emph{Iyer-Wald proof of the first law (for theories without gauge fields):} \\
Assume that a stationary BH in $d$-dim with the $\mathbb{R}\times U(1)^n$ isometry is given as a solution to some gravitational theory \cite{Iyer:1994ys}. The field of that solution can be denoted by $\bar{\Phi}$. Consider the $(d\!-\!1)$-dim spacelike hypersurface $\Sigma$ connecting the bifurcation point of the horizon to the asymptotic infinity (see Figure \ref{fig Kerr non-deg}). For the Killing vector $\zeta_{_\mrH}$, and for the perturbations $\delta \Phi$ which satisfy the l.e.o.m, one can integrate the pull-back of $\boldsymbol{\omega}(\delta\Phi,\delta_{\zeta_{_\mrH}}\Phi,\bar{\Phi})$ to the $\Sigma$. By \eqref{omega vs Q} and the  Stoke's theorem,
\begin{equation}
\int_\Sigma \boldsymbol{\omega}(\delta\Phi,\delta_{\zeta_{_\mrH}}\Phi,\bar{\Phi})=\oint_{\partial\Sigma} \big(\delta \mathbf{Q}_{\zeta_{_\mrH}}-\zeta_{_\mrH} \cdot \mathbf{\Theta}(\delta \Phi,\bar{\Phi})\big)\,.
\end{equation} 
The LHS vanishes for isometry generator $\zeta_{_\mrH}$, because of $\delta_{\zeta_{_\mrH}}\Phi=0$. On the other hand, $\Sigma$ has two boundaries, the bifurcation point of the horizon and the infinity. So
\begin{equation}
0=\oint_\infty  \big(\delta \mathbf{Q}_{\zeta_{_\mrH}}-\zeta_{_\mrH} \cdot \mathbf{\Theta}(\delta \Phi,\bar{\Phi})\big)-\oint_\mrH \big(\delta \mathbf{Q}_{\zeta_{_\mrH}}-\zeta_{_\mrH} \cdot \mathbf{\Theta}(\delta \Phi,\bar{\Phi})\big)\,.
\end{equation}
According to the section \ref{sec Iyer-Wald entropy}, the last term would be $-\kappa \frac{\delta S}{2\pi}$ unambiguously (see Appendix \ref{app 1st ambig} for the reason). The first term, noticing \eqref{H v Q} and \eqref{horizon Killing} would be $\delta M-\Omega^i_{_\mrH}\delta J_i$. By the Hawking temperature \eqref{Hawking temp} then
\begin{equation}
\delta M=T_{_\mrH}\delta S+\Omega^i_{_\mrH}\delta J_i\,.
\end{equation}
}
\vspace*{-0.5cm}
\dotfille

The second law of BH thermodynamics extends the usual second low of thermodynamics to include BHs \cite{Hawking:1972vs},[\cite{Hawking:1973la,Wald:2010ge}]. It is usually called as \emph{generalized second law.}  

The third law of BH thermodynamics, which is also known as \emph{cosmic censorship conjecture} does not have a rigorous proof \cite{Bardeen:1973gd}. The conjecture claims that naked singularity, \ie a singularity which does not hide behind horizon is physically impossible [\cite{Wald:1999bl}]. The naked singularities are literally equivalent to negative temperatures $T_{_\mrH}<0$, as will be discussed in the next section. However see Figure \ref{fig Kerr naked} for their typical Carter-Penrose (CP) diagram.


\section{Extremal black holes}\label{sec Extremal BH}
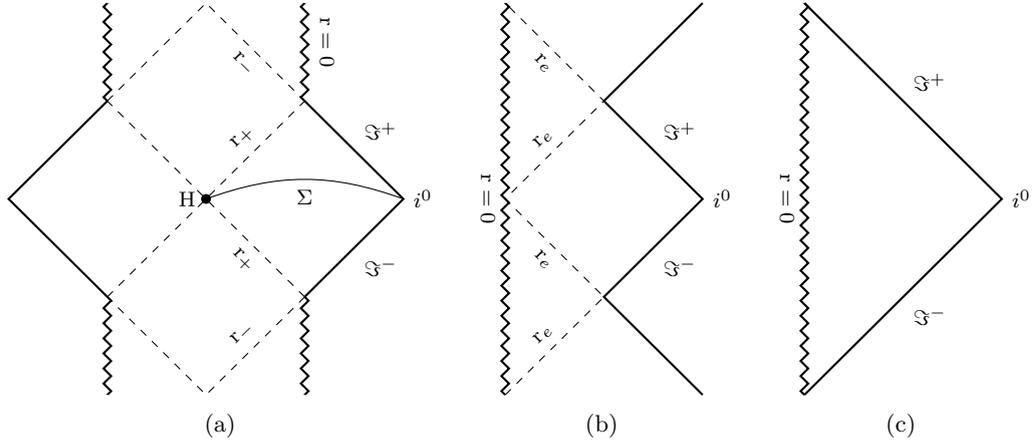
\begin{figure}[t!]
\begin{center}\hspace*{0.5cm}
\subfigure[]{\label{fig Kerr non-deg}
	\begin{tikzpicture}[scale=1.3]
    \draw[dashed] (1,1) -- (0,0) -- (-1,1); 
    \draw[thick] (-1,1) -- (-2,2) -- (-1,3);
    \draw[dashed] (-1,3) -- (0,4) -- (1,3); 
    \draw[thick] (1,3) -- (2,2) -- (1,1);
    \draw[dashed] (-1,1) -- (1,3);
    \draw[dashed] (1,1) -- (-1,3); 
    \draw[thick, decoration = {zigzag,segment length = 2mm, amplitude = 0.5mm},decorate] (1,1)--(1,0);
    \draw[thick, decoration = {zigzag,segment length = 2mm, amplitude = 0.5mm},decorate] (-1,1)--(-1,0);
    \draw[thick, decoration = {zigzag,segment length = 2mm, amplitude = 0.5mm},decorate] (1,3)--(1,4);  
    \draw[thick, decoration = {zigzag,segment length = 2mm, amplitude = 0.5mm},decorate] (-1,3)--(-1,4); 
    \draw (1.5,1.5) node[anchor=north west]{\scriptsize $\Im^-$};
    \draw (1.5,2.5) node[anchor=south west]{\scriptsize $\Im^+$};
    \draw (2,2) node[anchor=west]{\scriptsize $i^0$};
    \fill (0,2) circle(0.05cm) node[anchor=east]{\scriptsize $\mathrm{H}$};
    \draw (0.5,0.5) node[rotate=45,anchor=south]{\scriptsize $\mr_-$};
    \draw (0.5,1.5) node[rotate=-45,anchor=north]{\scriptsize $\mr_+$};
    \draw (0.5,2.5) node[rotate=45,anchor=south]{\scriptsize $\mr_+$};
    \draw (0.5,3.5) node[rotate=-45,anchor=north]{\scriptsize $\mr_-$};
    \draw (1.4,3.6) node[rotate=-90,anchor=north]{\scriptsize $\mr=0$};
    \draw[thin] (0,2) to [out=20,in=160] node[midway,below]{\scriptsize $\Sigma$}(2,2);
	\end{tikzpicture}
}
\subfigure[]{\label{fig Kerr deg}
	\begin{tikzpicture}[scale=1.3]
	\draw[thick, decoration = {zigzag,segment length = 2mm, amplitude = 0.5mm},decorate] (0,0)--(0,4);
	\draw[thick] (2,4) -- (1,3) -- (2,2) -- (1,1) -- (2,0);
	\draw[dashed] (0,0) -- (1,1) -- (0,2) -- (1,3)  -- (0,4);
	\draw (0.5,0.5) node[rotate=45,anchor=south]{\scriptsize $\mr_e$};
	\draw (0.5,1.5) node[rotate=-45,anchor=north]{\scriptsize $\mr_e$};
	\draw (0.5,2.5) node[rotate=45,anchor=south]{\scriptsize $\mr_e$};
	\draw (0.5,3.5) node[rotate=-45,anchor=north]{\scriptsize $\mr_e$};
	\draw (1.5,1.5) node[anchor=north west]{\scriptsize $\Im^-$};
	\draw (1.5,2.5) node[anchor=south west]{\scriptsize $\Im^+$};
    \draw (2,2) node[anchor=west]{\scriptsize $i^0$};
    \draw (0,2) node[rotate=-90,anchor=north]{\scriptsize $\mr=0$};
	\end{tikzpicture}
}
\subfigure[]{\label{fig Kerr naked}
	\begin{tikzpicture}[scale=1.3]
	\draw[thick, decoration = {zigzag,segment length = 2mm, amplitude = 0.5mm},decorate] (0,0)--(0,4);
	\draw[thick] (0,4)  -- (2,2) --  (0,0);
	\draw (1,1) node[anchor=north west]{\scriptsize $\Im^-$};
	\draw (1,3) node[anchor=south west]{\scriptsize $\Im^+$};
    \draw (2,2) node[anchor=west]{\scriptsize $i^0$};
    \draw (0,2) node[rotate=-90,anchor=north]{\scriptsize $\mr=0$};
	\end{tikzpicture}
}
\end{center}
\caption{\footnotesize (a) CP diagram for the Kerr BH at $T_{_\mathrm{H}}\!>\!0$. Horizons are depicted by dashed lines.  The point $\mathrm{H}$ is the bifurcation point of the horizon. The spacelike $d\!-\!1$-dim surface $\Sigma$ connects the $\mathrm{H}$ to the spatial infinity $i^0$. (b) CP diagram for Kerr BH at $T_{_\mathrm{H}}\!=\!0$, the extremal Kerr. The horizon is at $\mr_e$, and is a degenerate horizon, without bifurcation point. (c) CP diagram for the Kerr BH at $T_{_\mathrm{H}}\!<\!0$, the naked singularity. There is not any horizon.}
\label{fig Kerr BH}
\end{figure}

\emph{Extremal} BHs, are BHs with zero temperature, $T_{_\mrH}\!=\!0$, which according to \eqref{Hawking temp} is equivalent to $\kappa=0$. The Killing horizons with this property are called \emph{degenerate} horizons. The nomenclature originates from the fact that extremal BHs have their two outermost horizons at the same radius $\mr_+=\mr_-\equiv \mr_e$. It is in contrast to the Killing horizons with $\kappa>0$, which are called \emph{bifurcating} Killing horizons. Extremality constraints the parameters of the BH. Usually mass is expressed in terms of other parameters. Therefore the extremal BHs have one parameter less than the similar BHs at non-zero temperature. Emphasizing again, in contrast to the BHs at non-zero temperature which have bifurcation horizon, extremal BHs have not bifurcation horizons. Illustrating by an example, the CP diagrams for the Kerr and extremal Kerr are depicted in Figures \ref{fig Kerr non-deg} and \ref{fig Kerr deg}. 
An interesting property of the extremal BHs is that they have usually non-zero entropy. This is in contradiction to the third law of thermodynamics which states that if $T\to 0$, then $S\to 0$. A natural question would be generalization of this version of the third law to the BHs thermodynamics. In other words, the question is whether there is a law which determines that if $T_{_\mrH}\to 0$ then $S\to \,?$. In the next chapters of Part \ref{part I}, the answer will be presented.

\newpage
\vspace*{-1.5cm} 
\dotfillb
\Example{\label{example extremality}\emph{Entropy of the extremal BHs is not generally zero.} \\
The  extremality condition constraints one of the parameters (usually chosen to be $m$) to the other parameters of the BH. It can be found simply by the constraint $\mr_+=\mr_-$. Calling this degenerate radius as $\mr_e$, then the entropy would be read simply by inserting $\mr_e$ to \eqref{Kerr chemicals} etc. The extremality condition, $\mr_e$ and the entropy for the BHs in our examples, as a function of BHs parameters are as follows.
\begin{itemize}
\item[--] Extremal Kerr BH:
\begin{flalign}
Gm&=a\,, \qquad \mr_e=a\,, \qquad \frac{S}{2\pi}=\frac{a^2}{G}\,. &
\end{flalign}
\item[--] Extremal Kerr-Newman BH:
\begin{flalign}
Gm&=\sqrt{a^2+q^2}\,, \qquad \mr_e=\sqrt{a^2+q^2}\,, \qquad \frac{S}{2\pi}=\frac{2a^2+q^2}{2G}\,. &
\end{flalign}
\item[--] Extremal Kerr-AdS BH:
\begin{flalign}
Gm&=\frac{\mr_e(1+\frac{\mr_e^2}{l^2})^2}{1-\frac{\mr_e^2}{l^2}}\,, \qquad \mr_e^2=\frac{\sqrt{l^4+14l^2a^2+a^4}-a^2-l^2}{6}\,, \qquad \frac{S}{2\pi}=\frac{\mr_e^2+a^2}{2G\,\Xi}\,. &
\end{flalign}
\item[--] Extremal BTZ BH:
\begin{flalign}
Gm&=\frac{j}{l}\,, \qquad \mr_e=\sqrt{\frac{jl}{2}}\,, \qquad \frac{S}{2\pi}=\frac{1}{4G}\sqrt{\frac{jl}{2}}\,. &
\end{flalign}
\item[--] Extremal $5$-dim MP BH:
\begin{flalign}\label{EVH example}
Gm&=\frac{(a+b)^2}{2}\,, \qquad \mr_e=ab\,, \qquad \frac{S}{2\pi}=\frac{\pi\sqrt{ab}\, (a+b)^2}{4G}\,. &
\end{flalign}
\item[--] Extremal KK BH:
\begin{flalign}
Gm&=a\,, \qquad \mr_e=a\,, \qquad \frac{S}{2\pi}=\frac{a^2}{G\sqrt{1-v^2}}\,. &
\end{flalign}
\end{itemize}
}
\Example{\emph{Extremal Vanishing Horizons (EVH):} \\
Although Example \ref{example extremality} shows that the entropy is not usually zero at $T_{_\mrH}\!=\!0$, but there are some specific BHs which do have zero entropy, at zero temperature. They are called \emph{Extremal Vanishing Horizons} (EVHs for short) \cite{Sheikh-Jabbari:2011hj}. In a sense, they behave as the usual thermodynamic systems. One example which can be found by the above equations, is the extremal $5$-dim MP BH, when only one of the parameters $a$ or $b$ is zero.  Following the \eqref{EVH example}, the entropy vanishes if $a=0$ \emph{or} $b=0$. It is important that in this situation, the mass would be finite \emph{and} non-zero.
}
\dotfille

Another interesting thermodynamic property which extremal BHs posses, is that they can have non-trivial dynamics at zero temperature. The entropy at extremality is a non-trivial function of other thermodynamic variables. So although the first law of BH thermodynamics at $T_{_\mrH}=0$ would become $\delta M=\Omega^i_{\text{ext}}\delta J_i+\Phi^p_{\text{ext}}\delta Q_p$, but it is blind to the variations of the entropy $\delta S$. In other words, there is another natural question that if $T_{_\mrH}$ is kept to be zero, then $\delta S=?$. This question will also be addressed in the next chapters. 

A careful reader might have noticed that at exactly zero temperature, \ie $\kappa=0$, the Iyer-Wald analysis for the entropy breaks down. It is because $\hat{\zeta}_{_\mrH}$ in \eqref{normal Killing Horizon} is ill-defined. In other word, having a bifurcation horizon is a key condition for the Iyer-Wald analysis \cite{Iyer:1994ys}. Therefore, a question arises is whether at zero temperature, entropy is Hamiltonian generator for a Killing vector. 

The key step towards studying the extremal BHs and answering the questions above, is to study them at their near horizon region, which will be described in the next chapter.

\chapter{Near horizon geometry of the extremal black holes}\label{chap extremal near horizon}
In this chapter we will study the extremal BHs in the regions near their horizons. There are several reasons for conducting this study; BHs are compact objects localized in some portion of the space. In addition, thermodynamic variables of the BHs are encoded in their near horizon region; Temperature and chemical potentials are calculated on the horizon, while entropy and other Hamiltonian generators can be calculated on the horizon. Another reason comes from the zeroth law of BH thermodynamics, which requests the temperature and chemical potentials to be constant \emph{over the horizon}. Hence, it is natural to study BH thermodynamic behaviours in the regions around them. As it was described in the end of the last chapter, we will focus on the extremal BHs. Interestingly, studying their near horizon region also makes the study simpler and more fruitful.  There are some specific features for the extremal ones, making them more manageable in the ``near horizon" approach of studying; one is the ``isometry enhancement" described in this chapter.  The other one is ``uniqueness of their dynamical perturbations", described in Part \ref{part II}.  

In the first section, the closer look at the horizon of the extremal BHs is put into mathematical language. In the second section, isometries of the resulted geometries are studied, and ``isometry enhancement" is introduced.  


\section{Near horizon limit}
Approaching to the near horizon region of BHs is a well-known procedure in the literature of BHs. The physical meaning of this procedure is taking a closer look at the horizon. More precisely, it is studying the field configurations around $\mr \to \mr_+$. In order to perform this procedure, it is customary to introduce a dimensionless radius $r$ by the coordinate transformation 
\begin{equation}\label{Sch'd near horizon coord}
\mr=\mr_+(1+\lambda r)\,.
\end{equation}
Then, assuming the constant $\lambda$ to be infinitesimal, the approximation can be performed to the desired order.

\dotfillb
\Example{\emph{Near horizon approximation of Schwarzschild BH:} \\
\emph{Schwarzschild} BH is the Kerr BH without rotation, \ie with $a=0$. The metric is explicitly as
\begin{equation}
\mrd s^2= -f \mrd \mt^2+\frac{\mrd \mr^2}{f}+\mr^2\mrd \theta^2+\mr^2\sin^2 \theta\mrd \psi^2\,, \qquad f=1-\frac{2Gm}{\mr}\,.
\end{equation}
The horizon of this BH is at $\mr_+=2Gm$. The near horizon approximation of this metric can be found by the transformation $\mr=\mr_+(1+\lambda r)$ and keeping relevant terms up to the order $\mathcal{O}(\lambda)$. The result would be
\begin{equation}\label{Sch'd NHG}
\mrd s^2= -\lambda r \mrd \mt^2+\frac{4\lambda G^2m^2}{r}\mrd r^2 +\mr_+^2\mrd \theta^2+\mr_+^2\sin^2 \theta\mrd \psi^2\,.
\end{equation}
The $(\mt,r)$ sector of this near horizon metric is a Rindler space with acceleration equal to $\frac{1}{4Gm}$. It can be made explicit by the change of coordinate $\rho^2 \equiv \lambda (4Gm)^2 r$, which makes the metric of that sector to be $-\frac{\rho^2}{(4Gm)^2}\mrd \mt^2+\mrd \rho^2$. The $(\theta,\psi)$ sector is just a $2$-sphere with radius $\mr_+$.
}
\vspace*{-0.5cm}
\dotfille

The geometry which is derived by the mentioned approximation is called \emph{near horizon geometry}. The near horizon geometries are not necessarily a solution to the theory. For example, although Schwarzschild BH is a solution to the EH gravity, but its near horizon geometry, the \eqref{Sch'd NHG}, is not a solution. It is because the procedure of performing the near horizon approximation is composed of two steps; 1) a coordinate transformation \eqref{Sch'd near horizon coord} 2) an approximation procedure. The former can not ruin the metric (and other probable dynamical fields) to be a solution, but the latter can.

Studying the near horizon geometry of the extremal BHs is conceptually similar, but with an important mathematical difference, and some subtleties. The mathematical difference is that in order to derive the near horizon geometry of the extremal BHs, a \emph{limiting} process can be used instead of approximation.

The first subtle issue is that in addition to the transformation \eqref{Sch'd near horizon coord}, one needs to choose coordinates in which the time is in the direction of the Killing horizon $\zeta_{_\mrH}$. It is done by the transformation $\psi^i\to \psi^i-\Omega^i_{_\mrH}\mt$. This transformation is necessary because the horizon is behind the ergosphere, so the usual time generator $\partial_\mt$ would not be timelike. On the other hand, $\zeta_{_\mrH}$ defines a natural time outside but close to the horizon. The second subtle issue is that taking the near horizon limit of the extremal BHs leads to divergences in time components of the metric. Preventing it, one needs to rescale the time by a factor $\lambda$. Taking the mentioned two issues into account, the coordinate transformations from original extremal BH coordinates $(\mt,\mr,\theta^\alpha,\psi^i)$ to the near horizon geometry coordinates $(t,r,\theta^\alpha,\varphi^i)$ would be \cite{Bardeen:1999ds}[\cite{Compere:2012hr}] 
\begin{equation}\label{NH coordinate}
\mt=\frac{c_0 \mr_e t}{\lambda}\,, \qquad \mr= \mr_e(1+\lambda r)\,,\qquad  \psi^i=\varphi^i+\Omega^i_{_\mrH}\frac{c_0 \mr_e t}{\lambda}\,.
\end{equation}
in which the constant $c_0$ is an unimportant factor\footnote{however see section 5 of \cite{HSS:2013lna} to find out how to determine it appropriately.}. The last subtlety is that in the presence of some gauge fields $A^{(p)}$, in addition to the \eqref{NH coordinate}, a gauge transformation would also be necessary to prevent divergences in their components as
\begin{equation}\label{NH gauge trans}
A^{(p)}\to A^{(p)}+\mrd \Lambda^p \,, \qquad \Lambda^p=-\Phi^p_{_\mrH}\frac{c_0 \mr_e t}{\lambda}\,.
\end{equation}
After the above transformations, taking the limit $\lambda\to 0$ would result the near horizon geometry of the extremal BHs. There are some interesting remarks on this limit which is described below.
\begin{itemize}
\item The transformations are chosen in a way that all of the coordinates $(t,r,\theta^\alpha,\varphi^i)$ are dimensionless.
\item The process of taking near horizon limit for the extremal BHs, is composed of two steps; 1) the  transformations \eqref{NH coordinate}  2) the \emph{exact} limit $\lambda\to 0$. Both of these steps keep a solution to remain a solution. So the near horizon geometry would be a new solution to the theory itself.
\item It is proved \cite{Kunduri:2007vf,Kunduri:2008rs},[\cite{Kunduri:2013gce}] that for a generic family of theories which we will focus on implicitly in Part \ref{part I}, the fields of the resulted near horizon geometries have the following generic shapes.
\begin{align}\label{NHEG-metric}
	{\mrd s}^2&=\Gamma\big[-r^2\mrd t^2+\frac{\mrd r^2}{r^2}+\tilde{\gamma}_{\alpha\beta} \mrd\theta^\alpha\mrd \theta^\beta+\gamma_{ij}(\mrd\varphi^i+k^ir\,\mrd t)(\mrd\varphi^j+k^jr\,\mrd t)\big]\,,\\
\label{NHEG-gauge}
A^{(p)}&=\sum_{i=1}^n f^{(p)}_i(\mrd\varphi^i+k^ir\,\mrd t)-e^pr\,\mrd t\,,\\
\phi^I&=\phi^I(\theta^\alpha)\,.\label{NHEG-Scalar}
\end{align}
where summation of $i,j$ is from $1$ to $n$, and  $\alpha,\beta$ go from $1$ to $d\!-\!2\!-\!n$. The $\Gamma, \tilde{\gamma}_{\alpha\beta}, \gamma_{ij}, f^{(p)}_i$ are some functions of the coordinate $\theta^\alpha$ whose explicit functionality are fixed by the limiting process. The $k^i$ and $e^p$ are some constants determined similarly.   
\end{itemize}

\dotfillb
\Example{\label{Kerr NHEG}\emph{Near horizon geometry of the extremal Kerr BH:} \\
Beginning from \eqref{Kerr metric} in its extremal form, and by the coordinate transformations \eqref{NH coordinate} choosing $c_0=2$, and taking the limit $\lambda\to 0$ one finds
\begin{align}
\Gamma=a^2(1+\cos^2\theta)\,,\qquad \tilde{\gamma}=1\,,\qquad \gamma=\left(\frac{2\sin\theta}{1+\cos^2\theta}\right)^2\,, \qquad k=1\,.
\end{align}
}
\Example{\label{KerrNewman NHEG}\emph{Near horizon geometry of the extremal Kerr-Newman BH:} \\
Beginning from \eqref{Kerr-Newman metric} in its extremal form, and by the transformations \eqref{NH coordinate} and \eqref{NH gauge trans} choosing $c_0=\frac{q^2+2a^2}{q^2+a^2}$, and taking the limit $\lambda\to 0$, one finds
\begin{align}
\Gamma&=q^2+a^2(1+\cos^2\theta)\,, \qquad \tilde{\gamma}=1\,, \qquad \gamma=\left(\frac{q^2+2a^2\sin\theta}{q^2+a^2(1+\cos^2\theta)}\right)^2\,,\qquad k=\frac{2a\sqrt{q^2+a^2}}{q^2+2a^2}\,,\nonumber\\
f&=\frac{-\sqrt{q^2+a^2}\,qa\sin^2\theta}{q^2+a^2(1+\cos^2\theta)}\,,\qquad e=\frac{q^3}{q^2+2a^2}\,.
\end{align}
}
\Example{\label{KerrAdS NHEG}\emph{Near horizon geometry of the extremal Kerr-AdS BH:} \\
Beginning from \eqref{Kerr-AdS metric} in its extremal form, and by the transformations \eqref{NH coordinate}, by defining the $\Delta_0\equiv 1+\frac{a^2}{l^2}+\frac{6\mr_e^2}{l^2}$ and choosing $c_0=\frac{\mr_e^2+a^2}{\mr_e^2 \Delta_0}$, and taking the limit $\lambda\to 0$, one finds
\begin{align}
\Gamma&=\frac{\rho_e^2}{\Delta_0}\,, \qquad \tilde{\gamma}=\frac{\Delta_0}{\Delta_{\theta}}\,, \qquad \gamma=\frac{\Delta_0}{\Delta_\theta}\,\frac{(\mr_e^2+a^2)^2\sin^2\theta}{(\rho_e^2)^2 \,\Xi^2}\,,\qquad k=\frac{2a\mr_e\Xi}{\Delta_0 (\mr_e^2+a^2)}\,,
\end{align}
in which
\begin{align}
\rho_e^2 &\equiv \mr_e^2+a^2 \cos^2 \theta\,,\nonumber\\
\Delta_\theta&\equiv 1-\frac{a^2}{l^2}\cos ^2\theta\,,\nonumber\\
\Xi&\equiv 1-\frac{a^2}{l^2}\,.
\end{align}
}
\Example{\emph{Near horizon geometry of the extremal BTZ BH:} \\
The coordinates of the BTZ BH would be $(t,r,\varphi)$, \ie there is not any $\theta$. So $\Gamma$ and $\gamma$ are expected to be some constants. By the coordinate transformations \eqref{NH coordinate} choosing $c_0=\frac{l}{2j}$, and taking the limit $\lambda\to 0$, one finds
\begin{align}
\Gamma&=\frac{l^2}{4}\,,\qquad \gamma=\frac{2j}{l}\,,\qquad k=\sqrt{\frac{l}{2j}}\,.
\end{align}
}
\Example{\emph{Near horizon geometry of the extremal $5$-dim MP BH:} \\
By the coordinate transformations \eqref{NH coordinate} choosing $c_0=\frac{(a+b)^2}{4ab}$, and taking the limit $\lambda\to 0$, one finds
\begin{equation}
\begin{split}
\Gamma &= \frac{1}{4} (a+b) (a\cos^2\theta+b\sin^2\theta),\qquad \tilde{\gamma}=4\,,\qquad k^1=\frac{1}{2}\sqrt{\frac{b}{a}},\qquad k^2=\frac{1}{2}\sqrt{\frac{a}{b}},\\\\
\gamma_{ij}&=\dfrac{4}{(a\cos^2\theta+b\sin^2\theta)^2}
\begin{pmatrix}
 a (a+b\sin^2\theta) \sin ^2\theta &  a b \cos ^2\theta \sin ^2\theta \\ \ \ & \ \ \\
 a b \cos ^2\theta \sin ^2\theta &  b (b+a\cos^2\theta)\cos ^2 \theta\\
\end{pmatrix}\,.
\end{split}
\end{equation}
}
\Example{\label{KK NHEG}\emph{Near horizon geometry of the extremal KK BH:} \\
By the coordinate transformations \eqref{NH coordinate} and \eqref{NH gauge trans} choosing $c_0=\dfrac{2}{\sqrt{1-v^2}}$ and taking the limit $\lambda\to 0$, one finds
\begin{equation}
\begin{split}
\Gamma &= \sqrt{\frac{{(1+\cos^2\theta)(1+\cos^2\theta+v^2\sin^2\theta)}}{1-v^2}}\,a^2,\qquad \tilde{\gamma}=1\,,\qquad k=1,\\
\gamma&=\frac{4\sin^2\theta}{(1+\cos^2\theta)(1+\cos^2\theta+v^2\sin^2\theta)}\,,\qquad f=\frac{-av\sqrt{1-v^2}\sin^2\theta}{1+\cos^2\theta+v^2\sin^2\theta}\,,\qquad e=0\,,\\\\
\phi&=-\frac{\sqrt{3}}{4}\ln \left(\frac{1+\cos^2\theta+v^2\sin^2\theta}{(1-v^2)(1+\cos^2\theta)}\right)\,.
\end{split}
\end{equation}
}
\vspace*{-0.5cm}
\dotfille

One can explicitly check that the geometries derived in the examples above, are themselves solutions to the theory which the extremal BH has been. An important property of the near horizon geometries of the extremal BHs is the isometry enhancement described in the next section.


\section{Isometry enhancement}
In the process of near horizon limit for the extremal BHs, something unexpected but important and helpful happens; the \emph{isometry enhancement}. Note that this isometry enhancement is crucial in giving us a good handle and control over the geometry (compared to usual BHs). In fact, one of the major obstacles in understanding the BH physics is the strong (nonlinear) gravitational effects. Having more isometries will ameliorate the problem. Before describing ``isometry enhancement", we emphasize that approaching to the surface of an object, can change isometries of the region under study. It is done by a simple but intuitive example. After that, the ``isometry enhancement" is described.

\dotfillb
\Example{\emph{Changing of the isometries in the limit to the surface of an object:} \\
\begin{figure}[H]
\input{fig-isometry-earth}
\label{fig-isometry-earth}
\end{figure}
Consider the planet earth, as a spherical lonely object in the $3$-dim space. It has spherical isometry. The group which is responsible for this isometry is known as $SO(3)$. Its algebra is a three dimensional non-commutative algebra. Now, assume that one approaches to the surface of the earth. The surface of the earth would be seen as a flat plane. Its isometries would be translations tangent to the plane, and  rotation around an axis perpendicular to the plane. The group would be $\mathbb{R}^2\!\times\!U(1)$, which has also a $3$-dimensional algebra. But the translations commute. In other words, approaching to the surface of the earth, changes the isometries of the region under consideration from $SO(3)$ to $\mathbb{R}^2\!\times\!U(1)$.
}
\vspace*{-0.2cm}
\dotfille

Following the example above, assume a more realistic planet earth, which is flattened towards its equator. It would have the $U(1)$ isometry instead of  $SO(3)$. Again, approaching its surface would change the isometry to the $\mathbb{R}^2\!\times\!U(1)$. In this case an \emph{isometry enhancement} has happened. The isometry algebra gets larger.

In the case of near horizon limit of the extremal BHs, a similar phenomenon happens. The isomety changes. It gets larger; the isometry enhancement. The $\mathbb{R}\times U(1)^n$ isometry of the extremal BH enhances to $SL(2,\mathbb{R})\!\times\!U(1)^n$ isometry, described in the remaining of the section.
\begin{figure}[h!]
\input{fig-isometry-BH}
\label{fig-isometry-BH}
\end{figure}

The extremal BHs under consideration have $\mathbb{R}\times U(1)^n$ isometry as described in chapter \ref{chap some grav}. But their near horizon geometries (\ref{NHEG-metric}-\ref{NHEG-Scalar}) have $SL(2,\mathbb{R})\!\times\!U(1)^n$. The  $SL(2,\mathbb{R})$ isometries are generated by Killing vectors $\xi_a$ with $ a\in \{-,0,+\}$ and 
\begin{equation}\label{xi1-xi2}
\begin{split}
\xi_- &=\partial_t\,,\qquad \xi_0=t\partial_t-r\partial_r,\qquad	\xi_+ =\dfrac{1}{2}(t^2+\frac{1}{r^2})\partial_t-tr\partial_r-\frac{1}{r}{k}^i{\partial}_{\varphi^i}\,.
\end{split}
\end{equation}
The  $U(1)^{n}\;$ isometries are generated by Killing vectors $\mathrm{m}_i$ with $i\in \{1,\dots, n \}$ and 
\begin{align}\label{U(1)-generators}
\mathrm{m}_i=\partial_{\varphi^i}.
\end{align}
The isometry algebra is then
\begin{align}\label{commutation relation}
[\xi_0,\xi_-]=-\xi_-,\qquad [\xi_0,\xi_+]=\xi_+, \qquad [\xi_-,\xi_+]=\xi_0\,,\qquad [\xi_a,\mathrm{m}_i]=0.
\end{align}
Notice that the above vector fields are Killings, irrespective to the specific functionality of undetermined functions in (\ref{NHEG-metric}-\ref{NHEG-Scalar}). To be more precise, the form of NHEG ``ansatz" given in (\ref{NHEG-metric}-\ref{NHEG-Scalar}) is the most general field configuration with these isometries. One subtle issue worth mentioning is that $\xi_+$ is not an isometry generator of the gauge fields, but it leaves them invariant up to a gauge transformation, 
\begin{equation}\label{xi_+ A}
\mathscr{L}_{\xi_+}A^{(p)}=\mrd (\frac{-e^p}{r})\equiv \mrd \Lambda^p \,.
\end{equation}
As it is expected, the $\mathrm{m}_i$ vectors in \eqref{U(1)-generators} are the original $\partial_{\psi^i}$ in the near horizon limit. Also the original stationarity Killing $\partial_\mt$ goes to a linear combination of \eqref{U(1)-generators}. On the other hand, it can be checked that $\zeta_{_\mrH}$ would vanish in the near horizon limit as $\lambda \partial_t$.

Although the geometries  which were described in this chapter are derived by a limiting process from the extremal BHs, but they are some solutions with their own isometries, uniqueness theorems, Killing horizon structures etc,  which some of them will be discussed in the next chapter. It makes studying their properties, an interesting subject on its own. Specifically we will study thermodynamic behaviours of these geometries in chapter \ref{chap NHEG thermo}. In addition to deriving those properties,  it would be expected finding the answers to the questions posed in Section \ref{sec Extremal BH} on the thermodynamics of the extremal BHs.

\chapter{Near Horizon Extremal Geometries}\label{chap NHEG}

In the previous chapter, by taking the near horizon limit of extremal BHs, we found new geometries as solutions to the considered theory. Isometries of the new geometries are enhanced to $SL(2,\mathbb{R})\!\times\!U(1)^n$. The goal of this chapter is delving into the interesting mathematical properties of these geometries, which are necessary tools for later chapters. We will deal the geometries, found at near horizons of extremal BHs, as geometries having independent personality, and call them \emph{Near Horizon Extremal Geometries}, NHEGs for short. 

As mentioned, taste of this chapter is more mathematical rather than physical.  In the first section, NHEGs are defined. In the second section, the $AdS_2$ sector of NHEGs are studied. Third section deals with causal structure of NHEGs, and identifies infinite numbers of Killing horizons in it. Fourth section studies discrete isometries of NHEGs. Finally, in the fifth section, modular covariance of these geometries are introduced.  


\section{Introducing NHEGs} \label{sec NHEG geometry} 
The NHEGs can be viewed as an independent family of solutions by the definition below.
\begin{Definition}
NHEGs are the most general solutions, with local $SL(2,\mathbb{R})\!\times\!U(1)^n$ isometry group and $U(1)^p$ gauge group\footnote{In this thesis, we ignore the topological issues for simplicity. An interested reader can refer to the original papers, for more discussions on NHEG topological issues.}.
\end{Definition}
The coordinates can be chosen to make the isometry manifest. In those coordinates, the fields would have the generic shape of [\cite{Kunduri:2013gce}]
\begin{align}\label{NHEG metric}
	{\mrd s}^2&=\Gamma\big[-r^2\mrd t^2+\frac{\mrd r^2}{r^2}+\tilde{\gamma}_{\alpha\beta} \mrd\theta^\alpha\mrd \theta^\beta+\gamma_{ij}(\mrd\varphi^i+k^ir\,\mrd t)(\mrd\varphi^j+k^jr\,\mrd t)\big]\,,\\
\label{NHEG gauge}
A^{(p)}&= f^{(p)}_i(\mrd\varphi^i+k^ir\,\mrd t)-e^pr\,\mrd t\,,\\
\phi^I&=\phi^I(\theta^\alpha)\,.\label{NHEG Scalar}
\end{align}
where summation of $i,j$ is from $1$ to $n$, and  $\alpha,\beta$ go from $1$ to $d-2-n$. The $\Gamma,  \tilde{\gamma}_{\alpha\beta}, \gamma_{ij}, f^{(p)}_i$ are some functions of the coordinate $\theta^\alpha$ whose explicit functionality are fixed by the e.o.m. The $k^i$ and $e^p$ are some constants determined similarly. The isometry group is generated by Killing vectors
\begin{equation}\label{NHEG isometry}
\begin{split}
\xi_- &\!=\!\partial_t\,,\qquad \, \xi_0\!=\!t\partial_t\!-\!r\partial_r,\qquad\,	\xi_+\! =\!\dfrac{1}{2}(t^2\!+\!\frac{1}{r^2})\partial_t\!-\!tr\partial_r-\frac{{k}^i}{r}{\partial}_{\varphi^i}\,, \qquad\, \mathrm{m}_i\!=\!\partial_{\varphi^i},
\end{split}
\end{equation}
whose algebra is
\begin{align}\label{NH commutation relation}
[\xi_0,\xi_-]=-\xi_-,\qquad [\xi_0,\xi_+]=\xi_+, \qquad [\xi_-,\xi_+]=\xi_0\,,\qquad [\xi_a,\mathrm{m}_i]=0.
\end{align}
in which $ a\in \{-,0,+\}$ and $i\in\{1,\dots,n\}$.

\dotfillb
\Example{\emph{Near horizon extremal Kerr (NHE-Kerr):} \\
Choosing the theory to be EH gravity in $d=4$, and by the request of $SL(2,\mathbb{R})\!\times\!U(1)$ one finds an NHEG solution, as
\begin{align}
\Gamma=a^2(1+\cos^2\theta)\,,\qquad \tilde{\gamma}=1\,,\qquad \gamma=\left(\frac{2\sin\theta}{1+\cos^2\theta}\right)^2\,, \qquad k=1\,,
\end{align}
known as ``near horizon extremal Kerr". It is a solution with one parameter $a$. Other examples of NHEGs can be examples \ref{KerrNewman NHEG}-\ref{KK NHEG} which we do not repeat.
}
\dotfille

Although the NHEGs are not asymptotic flat solutions, but they can be labelled by their angular momenta $J_i$ and electric charges $Q_p$. Instead of asymptotics, one can choose any $(d\!-\!2)$-dim surface of constant $(t,r)$ for charge integrations. The next two sections will provide us mathematical tools, necessary for explaining these issues in the next chapter.


\section{$AdS_2$ sector and its $SL(2,\mathbb{R})$ isometry}\label{sec AdS2 sector}
Keeping $\theta^\alpha,\varphi^i$ constant, there is an $AdS_2$ sector in the metric \eqref{NHEG metric}, which is proportional to
\begin{equation}\label{AdS_2 metric}
\mrd s^2=-r^2 \mrd t^2 +\frac{\mrd r^2}{r^2} \,.
 \end{equation} 
$SL(2,\mathbb{R})$ isometry of NHEGs is basically related to this sector. Forgetting about the $g_{t\varphi^i}$ terms in \eqref{NHEG metric}, which means dropping the $\dfrac{k^i}{r}\partial_{\varphi^i}$ in $\xi_+$, the $\xi_a$'s are isometries of the $AdS_2$ sector. Having a review on the $SL(2,\mathbb{R})$ group and $AdS_2$ geometry is useful for later discussion.

$SL(2,\mathbb{R})$ is the group composed of $2\!\times\!2$ real valued matrices with unit determinant. It is a Lie group, and has a Lie algebra denoted by $sl(2,\mathbb{R})$. The algebra is a $3$-dim algebra generated by $\xi_a$ basis which satisfy the commutation relations \eqref{NH commutation relation}. The commutation relations can be written efficiently using the \emph{structural coefficients}, $f_{ab}^{\;\;\ c}$ as
\begin{align}\label{structure coeff}
[\xi_a,\xi_b]=f_{ab}^{\,\,\,\;c}\xi_c\,.
\end{align}
Killing form of the $sl(2,\mathbb{R})$ algebra can be found from structural coefficients as
\begin{equation}\label{Killing form def}
K_{ab}=\frac{-1}{2}f_{ac}^{\;\;\ d}f_{bd}^{\;\;\ c}\,,
\end{equation}
which in the basis chosen for the algebra, it is
\begin{equation}\label{Killing form}
K_{ab}=\begin{pmatrix}
0&0&1\\
0&-1&0\\
1&0&0
\end{pmatrix}\,.
\end{equation}
See Appendix \ref{app NHEG global} for the reason of conventional factor $\dfrac{-1}{2}$. Its inverse $K^{ab}=(K^{-1})_{ab}$ has the same components as itself (in the chosen basis). Using the Killing form one can raise or lower the indices, \eg $f_{abc}=K_{cd}f_{ab}^{\;\;\;d}$. An algebra can have different representations, so $sl(2,\mathbb{R})$ also can. The set of vectors $\xi_a$ in \eqref{NHEG isometry} is one of the representations of it. Another one is given in the example below.

\newpage
\vspace*{-1.5cm} 
\dotfillb
\Example{\emph{Adjoint representation of $sl(2,\mathbb{R})$:} \\
$sl(2,\mathbb{R})$ is a $3$-dim algebra. Every algebra is a linear vector space. So consider $\mathbb{R}^3$ with the standard basis $\{\hat{i},\hat{j},\hat{k}\}$ to be used for the representation of $sl(2,\mathbb{R})$. We can simply choose the basis as $|\xi_-\rangle=\hat{i}$, $|\xi_0\rangle=\hat{j}$ and $|\xi_+\rangle=\hat{k}$. The \emph{adjoint} representation of $sl(2,\mathbb{R})$ would be members of the endomorphisms of $\mathbb{R}^3$, \ie the $3\!\times\!3$  matrices,  with the property $\xi_a|\xi_b\rangle=|[\xi_a,\xi_b]\rangle$. They are explicitly
\begin{equation}
\xi_-=
\begin{pmatrix}
\,0 \, & \, 1\, & \, 0 \, \\ 0&0&1\\ 0&0&0
\end{pmatrix}\,\qquad
\xi_0=
\begin{pmatrix}
-1&\, 0\, &\, 0 \, \\ 0&0&0\\ 0&0&1
\end{pmatrix}\,\qquad
\xi_+=
\begin{pmatrix}
0 & 0 &\, 0 \, \\ -1&0&0\\ 0&-1&0
\end{pmatrix}\,.
\end{equation}
The Killing form can be read by $K_{ab}=\dfrac{-1}{2}\text{Tr}(\xi_a\,\xi_b)$.
}
\vspace*{-0.5cm}
\dotfille

$SL(2,\mathbb{R})$ is the isometry group of the $AdS_2$ geometry. $AdS_2$ is a $2$-dim manifold with constant negative Ricci scalar. The $AdS_2$ can be embedded in $\mathbb{R}^{2,1}$, the $3$-dim spacetime with signature $(-,-,+)$ and coordinates $(X^1,X^2,X^3)$, as
\begin{equation}
-(X^1)^2-(X^2)^2+(X^3)^2=-l^2.
\end{equation}
$l$ is called $AdS$ radius, and we will choose it to be unit radius. The coordinate system which covers the manifold of $AdS_2$ is called \emph{global} coordinate.  The reader can have a glimpse on it in Appendix \ref{app NHEG global}. There are another coordinates used for parametrizing the geometry, \eg \emph{Gaussian null} coordinate, or \emph{Kruskal type} coordinate described in \cite{CHSS:2015bca}. Coordinates which are used widely in this report, are $(t,r,\theta^\alpha,\varphi^i)$ such that $AdS_2$ metric would be \eqref{AdS_2 metric}. It is called the Poincar\'e \emph{patch}, because it does not cover all the manifold. In the CP diagram for $AdS_2$, which is depicted in Figure \ref{fig AdS2 CP 1}, the patches are illustrated. The range of $t$ is $(-\infty,\infty)$, and the range of $r$  can be $(0,\infty)$ or $(-\infty,0)$ depending on which patches are expected to be covered. The $r=0$ is the Poincar\'e horizon in $AdS_2$ manifold, which is a degenerate Killing horizon, corresponding to the Killing vector $\xi_-$. The $r=-\infty$ and $r=\infty$ are conformal boundaries of the manifold. The $AdS_2$ metric, in these coordinates, would be \eqref{AdS_2 metric}.

Embedding $AdS_2$ in $\mathbb{R}^{2,1}$, the vector which connects origin to the point of $(t,r)$ is denoted as $\mrn^a$ where \cite{HSS:2013lna}
\begin{equation}\label{n_a up}
\mrn^-=-\frac{t^2r^2-1}{2r}\,, \qquad \mrn^0=tr\,,\qquad \mrn^+=-r\,.
\end{equation}
Appendix \ref{app NHEG global} is added to provide a rigorous introduction for this vector. However an insight is given in Figure \ref{fig AdS2 embed}. Indices can be lowered by the Killing form \eqref{Killing form} as
\begin{equation}\label{n_a down index}
\mrn_-=-r\,, \qquad \mrn_0=-tr\,,\qquad \mrn_+=-\frac{t^2r^2-1}{2r}\,.
\end{equation}
The $\mrn^a$ has interesting properties as
\begin{align}
& \mrn_a \mrn^a=-1\,,\label{n_a norm}\\
&\mathcal{L}_{\xi_a}\mrn_b=f_{ab}^{\,\,\,c}\,\mrn_c\,, \label{del_xi n_a}\\
&\mrn^a\xi_a=k^i \mrm_i\,.\label{n xi vs k m}
\end{align}
Paying attention to the \eqref{del_xi n_a}, $\mrn_a$ is a vector representation of the $sl(2,\mathbb{R})$.


\section{Causal structure}\label{sec causal structure}
In the previous section we had a look at $(t,r)$ sector of the NHEG metric \eqref{NHEG metric}. Now we would have a look at the other sector. At each point of the $AdS_2$ sector identified by some constant $(t,r)$, \eg $(t_{_\mathcal{H}},r_{_\mathcal{H}})$, there is a $(d\!-\!2)$-dimensional closed smooth surface $\mathcal{H}$. It is parametrized by  the coordinates $(\theta^\alpha, \varphi^i)$. The induced metric on $\mathcal{H}$ is
\begin{equation}\label{H-metric}
{\mrd s}^2_{_\mathcal{H}}=\Gamma\left(\tilde{\gamma}_{\alpha\beta} \mrd\theta^\alpha\mrd \theta^\beta+\gamma_{ij}\mrd\varphi^i\mrd\varphi^j\right)\,,
\end{equation}
which induces the volume $(d\!-\!2)$-form 
\begin{equation}\label{H volume form}
\boldsymbol{\epsilon}_{_\mathcal{H}}=\frac{\sqrt{-g}}{\Gamma}\mrd \theta^1\wedge \dots\wedge\mrd \theta^{d-n-2}\wedge \mrd \varphi^1\wedge\dots\wedge\mrd\varphi^n\,.	
\end{equation}	
Then the binormal $2$-form of the $\mathcal{H}$ would be
\begin{equation}\label{binormal form}
\boldsymbol{\epsilon_{_\perp}}=\Gamma \mrd t\wedge \mrd r\,.
\end{equation}
Volume $d$-form of the NHEG geometry is related to the forms above as
\begin{align}\label{NHEG volume d-form decomp}
\boldsymbol{\epsilon}&=\boldsymbol{\epsilon_{_\perp}}\wedge \boldsymbol{\epsilon}_{_\mathcal{H}}\\
&=\frac{\sqrt{-g}}{d\,!}{\epsilon}_{\mu_1\mu_2\cdots\mu_d}\mrd x^{\mu_1}\wedge \mrd x^{\mu_2}\wedge\cdots \wedge \mrd x^{\mu_d}.\label{NHEG volume d-form}
\end{align}
In addition that each point $(t_{_\mathcal{H}},r_{_\mathcal{H}})$ on $AdS_2$ identifies a surface $\mathcal{H}$, it also identifies associated vector $\mrn_{_\mathcal{H}}^a$ as
 \begin{equation}\label{n-a vector up}
\mrn^-_{_\mathcal{H}}=-\frac{t_{_\mathcal{H}}^2r_{_\mathcal{H}}^2-1}{2r_{_\mathcal{H}}}\,, \qquad \mrn^0_{_\mathcal{H}}=t_{_\mathcal{H}}r_{_\mathcal{H}}\,,\qquad \mrn^+_{_\mathcal{H}}=-r_{_\mathcal{H}}\,.
\end{equation} 
Figure \ref{fig AdS2 embed} summarizes the  geometrical objects described in this chapter, intuitively.
\begin{figure}[h!]
\input{fig-AdS2}
\label{fig AdS2 embed}
\end{figure}

\begin{figure}[h!]
\begin{center}
\subfigure[]{\label{fig AdS2 CP 1}
		\begin{tikzpicture}[scale=0.6]
		\draw (-2,8) --node[midway, below, sloped] {\footnotesize $r=-\infty$} (-2,0);
		\draw (2,8) -- (2, 4);
		\draw [ultra thick, draw=black, fill=gray!20] (-2,0) -- node[midway, above, sloped] {\footnotesize $r=0$} (2,-4) -- node[midway, below, sloped] {\footnotesize $r=\infty$} (2,4)  -- node[midway, below, sloped] {\footnotesize $r=0$}  (-2,0) -- cycle;
		\draw (2,4) -- node[midway, above, sloped] {\footnotesize $r=0$} (-2,8);
		\draw (-2,0) -- (-2,-4);
		\node at (.5,0) {${\mathbf{I}}$};
		\node at (-0.5,4) {${\mathbf{II}}$};
		\draw[>=stealth,->] (-2,5.1)--(-2,5);
		\draw[>=stealth,->] (-2,1.1)--(-2,1);
		\draw[>=stealth,->] (-2,-2.9)--(-2,-3);
		\draw[>=stealth,->] (2,4.9)--(2,5);
		\draw[>=stealth,->,ultra thick] (2,0.9)--(2,1);
		\draw[>=stealth,->, ultra thick] (2,-3.1)--(2,-3);
		\end{tikzpicture}
}
\hspace*{1cm}
\subfigure[]{\label{fig AdS2 CP 2}
		\begin{tikzpicture}[scale=0.6]
		\draw (-2,8) -- (-2,0);
		\draw (2,8) -- (2,4);
		\draw [ultra thick, draw=black, fill=gray!20] (-2,0) --  (2,-4) --  (2,4)  --   (-2,0) -- cycle;
		\draw (2,4) -- (-2,8);
		\draw (-2,0) -- (-2,-4);
		\draw[>=stealth,->] (-2,5.1)--(-2,5);
		\draw[>=stealth,->] (-2,1.1)--(-2,1);
		\draw[>=stealth,->] (-2,-2.9)--(-2,-3);
		\draw[>=stealth,->] (2,4.9)--(2,5);
		\draw[>=stealth,->,ultra thick] (2,0.9)--(2,1);
		\draw[>=stealth,->, ultra thick] (2,-3.1)--(2,-3);
		\begin{scope}[shift={(0cm,0.3cm)}]	
		\draw[dashed,shift={(0cm,1cm)}] (-2,-2) -- (2,2);
		\draw[dashed,shift={(0cm,-1cm)}] (-2,-2) -- (2,2);
		\draw[dashed,shift={(0cm,-2cm)}] (-2,-2) -- (2,2);
		\draw[dashed] (-2,-2) -- (2,2);	
		\draw[dashdotted] (-2,2) -- (2,-2);
		\draw[dashdotted,shift={(0cm,1cm)}] (-2,2) -- (2,-2);
		\draw[dashdotted,shift={(0cm,2cm)}] (-2,2) -- (2,-2);
		\draw[dashdotted,shift={(0cm,3cm)}] (-2,2) -- (2,-2);
		\node at (-0.5,-3.3cm) {\footnotesize $\mathcal{N}_+$};
		\node at (-0.5,4.3cm) {\footnotesize $\mathcal{N}_-$};
		\end{scope}
		\end{tikzpicture}
}
\hspace*{1.3cm}
\subfigure[]{\label{fig AdS2 CP 3}
		\begin{tikzpicture}[scale=0.6]
		\draw (-2,8) --(-2,0);
		\draw (2,8) -- (2,4);
		\draw [ultra thick, draw=black, fill=gray!20] (-2,0) --  (2,-4) --  (2,4)  --   (-2,0) -- cycle;
		\draw (2,4) -- (-2,8);
		\draw (-2,0) -- (-2,-4);
		\draw[>=stealth,->] (-2,5.1)--(-2,5);
		\draw[>=stealth,->] (-2,1.1)--(-2,1);
		\draw[>=stealth,->] (-2,-2.9)--(-2,-3);
		\draw[>=stealth,->] (2,4.9)--(2,5);
		\draw[>=stealth,->,ultra thick] (2,0.9)--(2,1);
		\draw[>=stealth,->, ultra thick] (2,-3.1)--(2,-3);
		\begin{scope}[shift={(0cm,0.3cm)}]	
		\draw[dashed] (-2,-2) -- (2,2);	
		\draw[dashdotted] (-2,2) -- (2,-2);
		\fill (0,0) circle (0.1cm) node[anchor=south]{\footnotesize $\mathcal{H}$};
		\node at (1,-1.7cm) {\footnotesize $\mathcal{N}_{_\mathcal{H}}$};
		\end{scope}
		\end{tikzpicture}
}
\end{center}
\caption{\footnotesize (a) CP diagram for the $AdS_2$ sector of an NHEG. In the Poincar\'e coordinates, the region (I) is parametrized by $0<r$ and the region (II) is parametrized by $r<0$. The $r\to \pm\infty$ are conformal boundaries. Arrows show the positive direction of coordinate $t$, ranging in the interval $(-\infty,\infty)$, in both I and II. (b) A couples of null surfaces $\mathcal{N}_+$ are sketched by dashed lines, and $\mathcal{N}_-$ are sketched by dashdotted lines. (c) NHEG has infinite numbers of Killing horizons $\mathcal{N}_{_\mathcal{H}}$, for infinite numbers of choices for bifurcation point $\mathcal{H}$. One of them is depicted.}
\label{fig-AdS2-CP}
\end{figure}
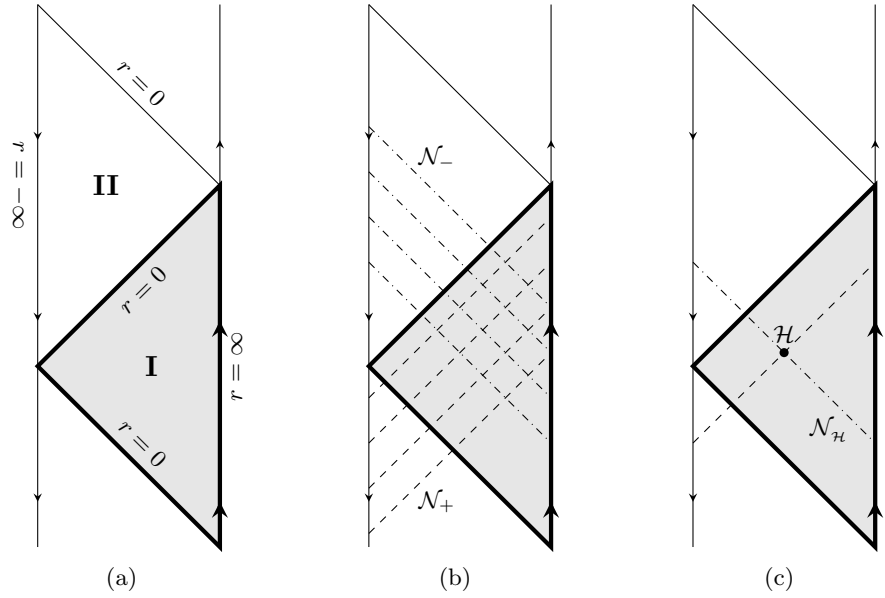

Returning back to the full NHEG geometry, there are two null $1$-form fields all over the spacetime,
\begin{equation}
\ell_+=\mrd t-\frac{\mrd r}{r^2}\,,\qquad \ell_-=\mrd t +\frac{\mrd r}{r^2}\,,
\end{equation}
which are normal to the $(d-1)$-dim hypersurfaces
\begin{equation}\label{N+ N-}
\mathcal{N}_+:\,\,t+\frac{1}{r}=\text{const}\,,\qquad \mathcal{N}_-:\,\,t-\frac{1}{r}=\text{const}\,.
\end{equation}
respectively. One can check that 
\begin{equation}\label{ell enn null}
\ell_+\cdot\ell_+=\ell_-\!\cdot\ell_-=0 \,, \qquad \ell_+\cdot \ell_-=\frac{-2}{\Gamma r^2}\,,
\end{equation}
hence $\mathcal{N}_{\pm}$ are null hyper surfaces. A couple of them are depicted in Figure \ref{fig AdS2 CP 2}.  For any $(t_{_\mathcal{H}},r_{_\mathcal{H}})$, \eqref{N+ N-}  defines associated $\mathcal{N}_{_{\mathcal{H}\pm}}$ as
\begin{equation}
\mathcal{N}_{_{\mathcal{H}+}}:\,\,t+\frac{1}{r}=t_{_\mathcal{H}}\!+\!\frac{1}{r_{_\mathcal{H}}}\,,\qquad \mathcal{N}_{_{\mathcal{H}-}}:\,\,t-\frac{1}{r}=t_{_\mathcal{H}}\!-\!\frac{1}{r_{_\mathcal{H}}}\,.
\end{equation}
The $\mathcal{H}$ is at the bifurcation point of $\mathcal{N}_{_{\mathcal{H}+}}$ and $\mathcal{N}_{_{\mathcal{H}-}}$ as is depicted in Figure \ref{fig AdS2 CP 3}. The binormal tensor of $\mathcal{H}$, \ie \eqref{binormal form}, can be written in terms of $\ell_\pm$ as 
\begin{equation}\label{epsilon ell}
\boldsymbol{\epsilon}_{_\perp}=\frac{\Gamma r^2}{2} \ell_+\wedge\ell_-\,,
\end{equation}
where by inserting \eqref{ell enn null} in it,  one can find $\epsilon_{_{\perp}\alpha\beta}\,\epsilon_{_\perp}^{\alpha\beta}=-2$.

Summarizing in a single sentence, $\mathcal{H}$ is bifurcation point of null hypersurface 
\begin{equation}\label{NHEG Killing horizon}
\mathcal{N}_{_\mathcal{H}}\equiv \mathcal{N}_{_{\mathcal{H}+}}\!\cup \mathcal{N}_{_{\mathcal{H}-}}\,.
\end{equation}
It would be nice that for any chosen $\mathcal{H}$, there would be a Killing vector $\zeta_{_\mathcal{H}}$, which would become null on $\mathcal{N}_{_\mathcal{H}}$. Then, according to the definition of Killing horizons in Section \ref{sec BHs}, $\mathcal{N}_{_\mathcal{H}}$ would be a Killing horizon. Fortunately such a Killing vector exists and is \cite{HSS:2013lna}
\begin{equation}\label{zeta H}
\zeta_{_\mathcal{H}}=\mrn^a_{_\mathcal{H}}\xi_a-k^i\mrm_i\,. 
\end{equation} 
with the expected properties
\begin{equation}\label{zeta properties}
\zeta_{_\mathcal{H}\alpha}\zeta_{_\mathcal{H}}^\alpha\Big|_{\mathcal{N}_{_\mathcal{H}}}=0\,, \qquad \zeta_{_\mathcal{H}}\Big|_\mathcal{H}=0\,.
\end{equation}
Notice that $\mrn^a_{_\mathcal{H}}$ are some constants, \ie they are \eqref{n-a vector up}. Hence  $\zeta_{_\mathcal{H}}$ is a linear combination of $SL(2,\mathbb{R})\!\times\!U(1)^n$ Killings vectors, so it is a Killing vector. 

Concluding in brief, $\zeta_{_\mathcal{H}}$ is a Killing vector which makes $\mathcal{N}_{_\mathcal{H}}$ a Killing horizon with bifurcation point $\mathcal{H}$. There are infinite choices for $(t_{_\mathcal{H}},r_{_\mathcal{H}})$, so there are infinite numbers of such Killing horizons. One is depicted in Figure \ref{fig AdS2 CP 3}.


\section{$Z_2$ isometries}\label{sec Z2 isometry}
In addition to the $SL(2,\mathbb{R})\!\times\!U(1)^n$, NHEGs have some discrete isometries which will be explained in this section. 
\paragraph{Inversion of $(t,\varphi^i)$:} By the $t\to -t$ and $\varphi^i\to -\varphi^i$, the NHEG  (\ref{NHEG metric}-\ref{NHEG Scalar}) does not change. This isometry is reminiscent of the same isometry in the extremal BHs which NHEGs are their near horizon region \cite{Schiffrin:2015yua}. By this isometry,
\begin{equation}
\xi_0\to \xi_0\,, \qquad \xi_{\pm}\to -\xi_\pm \,, \qquad \mrm_i\to -\mrm_i\,, \qquad \ell_\pm\to -\ell_\mp\,, \qquad \zeta_{_\mathcal{H}}\to -\zeta_{_\mathcal{H}}\,.
\end{equation} 
\paragraph{Inversion of $(r,\varphi^i)$:} By the $r\to -r$ and $\varphi^i\to -\varphi^i$ the NHEG (\ref{NHEG metric}-\ref{NHEG Scalar}) does not change. This isometry exchanges the regions I and II in Figure \ref{fig AdS2 CP 1}. Hence, it exchanges the conformal boundaries. By this isometry,
\begin{equation}
\xi_a\to \xi_a\,, \qquad \mrm_i\to -\mrm_i\,, \qquad \ell_\pm\to \ell_\pm\,,\qquad \zeta_{_\mathcal{H}}\to -\zeta_{_\mathcal{H}}\,.
\end{equation}


\section{$SL(n,\mathbb{Z})$ covariance}
NHEGs have continuous $SL(2,\mathbb{R})\!\times\!U(1)^n$ and some discrete isometries, described above. There are another transformations which, although are not isometry, but keep the generic shape of NHEG dynamical fields (\ref{NHEG metric}-\ref{NHEG Scalar}) intact. In other words, the NHEG generic shape changes covariantly by those transformations. This section introduces these transformations; the $SL(n,\mathbb{Z})$ transformations of coordinates $\varphi^i$. 
 
The group $SL(n,\mathbb{Z})$ consists of $n\times n$ matrices of unit determinant with integer-valued elements. Its members map a vector on a $n$-dimensional square lattice, to another vector on the same lattice. Explicitly, if a vector on the $n$-dimensional square lattice is specified by $\vec{m}$ such that $m^i\in \mathbb{Z}$, then
\begin{equation}
m'^i=P^i_{\,\,j}m^j\,,
\end{equation}
where $P^i_{\,\,j}$ is a member of the $SL(n,\mathbb{Z})$ group. The inverse of these transformations is simply found by inverting the matrix $P$. It can be denoted simply as
\begin{equation}\label{SLnZ 1}
P^{\,\,i}_{j}=(P^{-1})^i_{\,\,j}\,, \qquad \qquad P^i_{\,\,j}P^{\,\,j}_{k}=\delta^i_{\,\,k}\,.
\end{equation}

Returning back the the NHEG metric, the coordinates $\varphi^i$ have periodicity $\varphi^i\sim \varphi^i +2\pi$.  We can transform these coordinates by members $P$ of $SL(n,\mathbb{Z})$ as 
\begin{equation}\label{SLnZ 3}
\varphi'^i=P^i_{\,\,j}\varphi^j\,, \qquad \qquad P^i_{\,\,j}\in SL(n,\mathbb{Z})\,,
\end{equation}
and the inverse transformation
\begin{equation}\label{SLnZ 6}
\varphi^i=P^{\,\,i}_{j}\varphi'^j\,.
\end{equation}
The specific feature of these transformations is that they keep the periodicity of $\varphi'^i$ to be $2\pi$, \textit{i.e.} $\varphi'\sim \varphi'+2\pi$. A proof for the simple case $n=2$ is provided in Appendix \ref{Sl2Z period}. Also from \eqref{SLnZ 3} we immediately have
\begin{equation}
\mrd\varphi'^i=P^i_{\,\,j}\mrd\varphi^j\,.
\end{equation}
These transformation can be viewed as diffeomorphisms on the NHEG, so for the metric we would have
\begin{equation}\label{SLnZ 2}
g'_{\mu\nu}=\frac{\partial x^\alpha}{\partial x'^\mu}\frac{\partial x^\beta}{\partial x'^\nu}g_{\alpha\beta}\,.
\end{equation}
Using the transformations \eqref{SLnZ 6} in \eqref{SLnZ 2}, then $g'_{\mu\nu}$ turns out to be of the same generic shape of the original NHEG in \eqref{NHEG metric}, in which the following changes are realized:
\begin{align}
\gamma'_{ij}&=P^{\,\,k}_{i}P^{\,\,l}_{j}\gamma_{kl}\,,\label{SLnZ 4}\\
k'^i&=P_{\,\,j}^{i}k^j\,.\label{SLnZ 5}
\end{align}
The transformations \eqref{SLnZ 4} do not ruin the smoothness of the surfaces $\mathcal{H}$, because smoothness is eventually related to the eigenvalues of the $\gamma_{ij}$, which are not affected by \eqref{SLnZ 4}. Finally, for the transformations of $\mathrm{m}_i=\partial_{\varphi^i}$, according to $\mathrm{m}_i=\Gamma \gamma_{ij}(\mrd\varphi^j+k^jr\,\mrd t)$, and the transformations \eqref{SLnZ 4} and \eqref{SLnZ 5} we find
\begin{equation}\label{SLnZ axi gen tran}
\mathrm{m}'_i=P^{\,\,j}_{i}\mathrm{m}_j\,.
\end{equation}
Some care is needed to distinguish the lattice vectors $m^i$,  from the Killing vectors $\mrm_i=\partial_{\varphi^i}$, specifically in Part \ref{part II} of the report, where $m_i$ also would appear.

Motivated by the $SL(n,\mathbb{Z})$ covariance, we can adopt the notation of $n$-dim vectors as 
\begin{align}
&\vec{\varphi}\equiv (\varphi^1, \dots ,\varphi^n)\,, \qquad \vec{\mrm}=\vec{\partial}_{\varphi}\equiv (\partial_{\varphi^1},\dots,\partial_{\varphi^n})\,, \qquad \vec{k}\equiv (k^1,\dots,k^n)\,,\nonumber\\
& \vec{J}\equiv (J_1,\dots,J_n)\,,\qquad \vec{m}\equiv (m_1,\dots,m_n)\,.
\end{align}
A dot product notation can also be adopted as contraction of indices of two $n$-vectors, one with up index and the other with down index, \eg 
\begin{equation}
 \vec{k}\cdot\vec{J}\equiv\sum_{i=1}^{n}k^i J_i\,.
\end{equation} 

\dotfillb
\Example{\emph{$\zeta_{_\mathcal{H}}$ is invariant under $SL(n,\mathbb{Z})$.} \\
Using the new notation, we can rewrite $\zeta_{_\mathcal{H}}$ in \eqref{zeta H} as
\begin{equation}\label{zeta invariant}
\zeta_{_\mathcal{H}}=\mrn^a_{_\mathcal{H}} \xi_a-\vec{k}\cdot\vec{\mrm}\,.
\end{equation}
By our new conventions, it would be easy to check that this vector is invariant under  $SL(n,\mathbb{Z})$ transformations; $\mrn^a\xi_a$ does not have any $\vec{\varphi}$ dependency, except through $\xi_+$. According to the explicit components of this vector in \eqref{NHEG isometry}, that dependency would be a term proportional to $\vec{k}\cdot \vec{\mathrm{m}}$, which is invariant, because of \eqref{SLnZ 5} and \eqref{SLnZ axi gen tran}. Similarly the explicit $\vec{k}\cdot\vec{\mrm}$  in \eqref{zeta invariant} is invariant. 
}
\vspace*{-0.5cm}
\dotfille

\chapter{Thermodynamic laws of NHEGs}\label{chap NHEG thermo}

In recent decades, there have been motivations to search for thermodynamic behaviours in geometries not necessarily  having event horizon. As a prototype of these attempts, one can mention the generalization to the Rindler spacetime \cite{Rindler:1966zz}, leading to the Unruh effect \cite{Unruh:1976db}. Hence, studying thermodynamics of the NHEGs can be an independent and interesting line of research, although it has been inspired by the near horizon limit of extremal BHs.  In this chapter, we elaborate the generalization of thermodynamic behaviours to the NHEGs. Specifically for these geometries which have not event horizon, we would introduce entropy, and three universal laws (\ie laws which are independent of theory and dimensions). Nonetheless, at the end of the chapter, we will discuss their relation to the BH thermodynamics. 

This chapter has the same logical steps as chapter \ref{chap BH thermo}; In the first and second sections, the thermodynamic variables of NHEG are described. In the third section, NHEG entropy is defined as a Hamiltonian generator. In the forth section, which is the heart of Part \ref{part I}, the NHEG dynamical laws are introduced and proved. Finally, in the last section, their role in the BH thermodynamics is investigated.

\section{NHEG Hamiltonian generators}\label{sec NHEG conserved charges}
This section is analogous to Section \ref{sec BH conserved charges}, but this time for NHEGs. First the surface of integration are studied, and then the angular momenta and electric charges are calculated by the Covariant Phase Space method introduced in chapter \ref{chap Covariant Phase Space method}.
 
\subsection{Similarity of surfaces $\mathcal{H}$}\label{sec similar H}
Similarly to the BHs, one can use Covariant Phase Space method to find $\delta H_\xi$ for a given vector field $\xi$. In order to find it, in addition to $\boldsymbol{k}_\xi(\delta\Phi,\Phi)$ introduced in \eqref{delta H k}, one needs to introduce appropriate $(d\!-\!2)$-dim surface $\mathscr{S}$ as the surface of integration. Notice that NHEGs have neither event horizon, nor flat asymptotics. Therefore in definition of charges, one needs to introduce appropriate surfaces. We have denoted surfaces of constant time and radius $(t=t_{_\mathcal{H}},r=r_{_\mathcal{H}})$ by $\mathcal{H}$. These surfaces are $(d\!-\!2)$-dim bifurcation point of the Killing horizon  $\mathcal{N}_{_\mathcal{H}}$ in \eqref{NHEG Killing horizon}. An important property of the surfaces $\mathcal{H}$ is that one can map them to each other using the group action associated to the subalgebra $\{\xi_-,\xi_0\}$. Appendix \ref{app H to H by xi} provides a proof for this proposition. Noticing that $\{\xi_-,\xi_0\}$ are isometry generators, it literally means that different surfaces $\mathcal{H}$ are physically similar surfaces. 

An implication of this similarity, is that although $\boldsymbol{\epsilon}_{_\mathcal{H}}$ and $\boldsymbol{\epsilon_{_\perp}}$ are defined for an $\mathcal{H}$, \ie for the constant time and radius $(t_{_\mathcal{H}},r_{_\mathcal{H}})$, but they are independent of that choice. Because $(t_{_\mathcal{H}},r_{_\mathcal{H}})$ does not appear explicitly and implicitly in \eqref{H volume form} and \eqref{binormal form}.  Another implication of the similarity of these surfaces, is that asymptotics would not play an important role, concerning thermodynamic behaviours of NHEGs. Hence the surfaces $\mathcal{H}$, which are bifurcation point of Killing horizons of NHEGs, are the best candidates to be the surfaces of integration $\mathscr{S}$.

Two useful lemmas are mentioned below, which are useful for checking $\mathcal{H}$ independence of conserved charges of NHEGs \cite{HSS:2014twa}. 

\vspace*{-0.5cm}
\dotfillb
\Lemma{\label{Lemma Hodge xi inv}\emph{If a $p$-form is $\{\xi_-,\xi_0\}$ isometric, then its Hodge dual is also $\{\xi_-,\xi_0\}$ isometric.} \\
For the proof, it is enough to note that in the definition of Hodge duality \eqref{Hodge duality}, in addition to the $p$-form under considerations, the volume element \eqref{NHEG volume d-form} and the metric \eqref{NHEG metric} appear which both of them are also  $\{\xi_-,\xi_0\}$ isometric.
}
\Lemma{\label{Lemma NHEG H independence}\emph{If $\mathbf{Q}$ is a $(d\!-\!2)$-form which is independent of constants $(t_{_\mathcal{H}},r_{_\mathcal{H}})$, and with $\{\xi_-,\xi_0\}$ isometry, then $\oint_{\mathcal{H}}\mathbf{Q}$ is independent of $\mathcal{H}$.} \\
It would suffice to show that the pull-back of $\mathbf{Q}$ to the $\mathcal{H}$ is $(t,r)$ independent; Remembering that $\mathcal{H}$ is parametrized by coordinates $(\theta^\alpha,\varphi^i)$, then the $\mathrm{Q}^{tr}$ component of $\mathbf{Q}=\star \mathrm{Q}$ should be $(t,r)$ independent. By the Lemma \ref{Lemma Hodge xi inv}, $\mathrm{Q}$ is $\{\xi_-,\xi_0\}$ isometric. In Appendix \ref{app t r dependences}, $(t,r)$ dependency of tensors of rank $2$, is determined by their $\{\xi_-,\xi_0\}$ isometry. It proves that $\mathrm{Q}^{tr}$ is $(t,r)$ independent. Hence the lemma is proved.
}
\dotfille

\subsection{Angular momenta and electric charges}
By the Covariant Phase Space method, one can associate Hamiltonian generators $\delta J_i\equiv -\delta H_{\mrm_i}$ to the Killing vectors of the $U(1)^n$ isometry, \ie to the  $\mrm_i=\partial_{\varphi^i}$. These Hamiltonian generators are angular momenta. For a given theory and its NHEG solution, 
\begin{equation}\label{NHEG angular momentum}
\delta J_i=-\delta H_{\mrm_i}=-\oint_\mathcal{H}\boldsymbol{k}_{\mrm_i}(\delta\Phi,\Phi)
\end{equation}
provides the variation of the angular momenta for a given $\delta \Phi$ in the tangent space of the phase space of the NHEG (which will be discussed in Part \ref{part II}). Using similar analysis as in  the examples \ref{example Kerr M J}-\ref{example KK M J}, one can use parametric variations to find $\hat{\delta} J_i$, and then integrate them over the parameters of the solutions, to find $J_i$.   

If there are gauge fields $A^{(p)}$ in the theory, electric charges are simply found by \eqref{electric charge Q} and \eqref{electric charge Q-form}, in which integration is done on the $\mathcal{H}$. 

For the specific NHEG examples in \ref{Kerr NHEG}-\ref{KK NHEG} , the angular momenta and electric charges will be mentioned in Example \ref{example NHEG thermo entities}. By the Lemma \ref{Lemma NHEG H independence},  $J_i$ and $Q_p$ are independent of the surfaces $\mathcal{H}$. It is basically because of $[\xi_{a},\mrm_i]=0$ and $\mathscr{L}_{\xi_{a}}A^{(p)}=0$ for $a\in\{-,0\}$. For the independence of angular momenta, of the surfaces $\mathcal{H}$, a similar analysis as in Lemma \ref{Lemma independ H} could also be used.

\subsection{$SL(2,\mathbb{R})$ conserved charges}
Although NHEGs are stationary solutions, but they lack a Killing vector which would be timelike globally. To be more specific, $\xi_-\cdot\xi_-$ (or norm of any other combination of NHEG Killings, as candidate for generator of stationarity) change sign when one sweeps the $\theta^\alpha$ in the allowed ranges. So introducing mass for these geometries seems to be ill-defined. It is not surprising, because remembering their relation to the extremal BHs, extremality removes the mass as an independent conserved charge. But on the other hand, the isometry enhancement has endowed the NHEGs extra isometries. Hamiltonian generators associated to the $SL(2,\mathbb{R})$ isometry, denoted by $H_{\xi_a}$, and Noether-Wald charges denoted by $\mathcal{Q}_{\xi_a}$, would play an important role in the analysis of NHEG thermodynamics. 

One can use the standard Covariant Phase Space method for studying $H_{\xi_a}$. But having bookkeeping in mind,  we will deal the $SL(2,\mathbb{R})$ charges in such a way to be useful for the proofs of NHEG thermodynamic laws; We will not study each one of the three $H_{\xi_a}$ individually, but we will consider a specific linear combination of them, $\mrn^a_{_\mathcal{H}}H_{\xi_a}$, in which $\mrn_{_\mathcal{H}}^a$ is \eqref{n-a vector up}. There are two main reason for this specific choice; the first reason is that in studying NHEG thermodynamic laws, the $H_{\xi_a}$ appear only in this combination. The second reason is that by \eqref{n xi vs k m},
\begin{equation}
\mrn^a_{_\mathcal{H}}\xi_a\Big|_{\mathcal{H}}=\vec{k}\cdot\vec{\mrm}\,,
\end{equation}
hence in the calculation of $\mrn^a_{_\mathcal{H}}H_{\xi_a}$, the last term of \eqref{H v Q} drops simply because of vanishing of the pull-back of $(\vec{k}\cdot\vec{\mrm})\cdot \mathbf{\Theta}$ to the $\mathcal{H}$. Therefore 
\begin{equation}\label{n-a  H vs n-a Q}
\mrn^a_{_\mathcal{H}}H_{\xi_a}=\mrn^a_{_\mathcal{H}}\mathcal{Q}_{\xi_a}\,.
\end{equation}
Interestingly the RHS can be calculated explicitly \emph{independent} of the details of theory. In the Appendix \ref{app n-h Q-a proof} it is shown that the result is
\begin{equation}\label{n-a Qxi-a}
\mrn^a_{_\mathcal{H}}\mathcal{Q}_{\xi_a}=e^pQ_p-\oint_\mathcal{H} \sqrt{-g} \,\mathcal{L} 
\end{equation}
in which $\mathcal{L}$ is the $0$-form Lagrangian density of the theory. According to Lemma \ref{Lemma NHEG H independence} and following Appendix \ref{app n-h Q-a proof}, \eqref{n-a  H vs n-a Q} is an $\mathcal{H}$ independent combination of conserved charges. 

\section{NHEG chemical potentials}

According to Section \ref{sec causal structure}, although NHEGs have not event horizons, but they have infinite number of Killing horizons, with bifurcation points $\mathcal{H}$. In the same way as the chemical potentials of BHs  are defined in Section \ref{sec BH chemicals}, we can define three types of chemical potentials for NHEGs, as conjugates to the entropy $S$, angular momenta $J_i$ and electric charges $Q_p$. 

Chemical potential conjugate to the entropy can be found by the surface gravity $\kappa$ on the surface $\mathcal{H}$ as
\begin{equation}\label{NHEG kappa}
\kappa^2=\frac{-1}{8}(\mrd\zeta_{_\mathcal{H}})_{\alpha\beta}(\mrd\zeta_{_\mathcal{H}})^{\alpha\beta}\Big|_\mathcal{H}\,.
\end{equation}
Analogous to BHs, the positive root of the above is considered. It turns out to be $\kappa=1$ for all NHEGs, \ie by just using the generic shape \eqref{NHEG metric}. Motivated by \eqref{Hawking temp}, the conjugate to the entropy $S$ would be
\begin{equation}
\frac{\kappa}{2\pi}=\frac{1}{2\pi}\,,
\end{equation}
\ie conjugate to the $\frac{S}{2\pi}$ is always equal to unity.

Conjugate to the angular momenta $J_i$ and electric charges $Q_p$ can not be found by tricks in the section \ref{sec BH chemicals}, because those methods are heavily based on some appropriate coordinates and gauges. But analyzing the NHEG dynamical laws reveals that the thermodynamic conjugates of $J_i$ and $Q_p$ are $k^i$ and $e^p$ respectively. So we will postpone their discussion to the next sections.

\section{NHEG entropy as a Hamiltonian generator}\label{sec NHEG entropy}
Similarly to the BHs, the NHEG entropy can be defined as a Noether-Wald charge \cite{HSS:2013lna}. The Killing vector responsible for this conservation is the horizon Killing vector $\zeta_{_\mathcal{H}}$. Noting the $\kappa=1$, its normalization would be trivial. The entropy can be defined as
\begin{equation}
\frac{S}{2\pi}\equiv \mathcal{Q}_{\zeta_{_\mathcal{H}}}\Big|_\mathcal{H}\,.
\end{equation}
There is not any ambiguity in this definition, because by \eqref{zeta properties} and Lemma \ref{Lemma ambig vanish},  all kinds of ambiguities in the \eqref{decomposition} vanish. It results the following for the entropy
\begin{equation}\label{NHEG entropy E}
\frac{S}{2\pi}=\oint_\mathcal{H}\mathbf{E}^{\mu \nu}\nabla_{[\mu}{\zeta}_{_\mathcal{H}\nu ]}\,,
\end{equation}
in which $\mathbf{E}^{\mu \nu}$ is defined in \eqref{E-four-index}. By decomposition  \eqref{NHEG volume d-form decomp}, it can also be written as
\begin{equation}
\frac{S}{2\pi}=-\oint_\mathcal{H}\boldsymbol{\epsilon}_{_\mathcal{H}} \epsilon_{_{\!\perp\alpha\beta}}\,\epsilon_{_{\!\perp\mu\nu}}\,{E}^{ \alpha \beta \mu \nu}\,.
\end{equation}
All of the terms appearing in the RHS are $\{\xi_-,\xi_0\}$ isometric. So by the Lemma \ref{Lemma NHEG H independence}, NHEG entropy is independent of the choice of $\mathcal{H}$. 

The entropy defined above is equal to the Hamiltonian generator of ${\zeta}_{_\mathcal{H}}$. It is because by \eqref{zeta properties}, the last term in \eqref{H v Q} vanishes on $\mathcal{H}$.

\dotfillb
\Example{\label{example NHEG thermo entities}\emph{Thermodynamic entities for different NHEGs:} 
\begin{itemize}
\item[--] Near horizon extremal Kerr (NHE-Kerr):
\begin{flalign}\label{NHE-Kerr entities}
&J=\frac{a^2}{G}\,, \qquad Q=0\,, \qquad k=1\,, \qquad e=0\,,\qquad \frac{S}{2\pi}=\frac{a^2}{G}\,, \qquad \oint_\mathcal{H}\sqrt{-g}\mathcal{L}=0\,.&
\end{flalign}
\item[--] NHE-Kerr-Newman:
\begin{flalign}
&J=\frac{\sqrt{a^2+q^2}}{G}a\,, \qquad Q=\frac{q}{G}\,,\qquad k=\frac{2a\sqrt{a^2+q^2}}{q^2+2a^2}\,, \qquad e=\frac{q^3}{q^2+2a^2}\,,&\nonumber\\
&\frac{S}{2\pi}=\frac{2a^2+q^2}{2G}\,, \qquad \oint_\mathcal{H}\sqrt{-g}\mathcal{L}=\frac{q^4}{2G(q^2+2a^2)}\,. &
\end{flalign}
\item[--] NHE-Kerr-AdS:
\begin{flalign}
&J=\frac{\mr_e (1+\frac{\mr_e^2}{l^2})^2}{G\,\Xi^2 (1-\frac{\mr_e^2}{l^2})}a\,, \qquad Q=0\,, \qquad k=\frac{2a\mr_e\Xi}{\Delta_0 (\mr_e^2+a^2)}\,, \qquad e=0\,,&\nonumber\\
&\frac{S}{2\pi}=\frac{\mr_e^2+a^2}{2G\,\Xi}\,, \qquad \oint_\mathcal{H}\sqrt{-g}\mathcal{L}=\frac{2a^2\mr_e^2(1+\frac{\mr_e^2}{l^2})^2}{G\,\Delta_0 \Xi(\mr_e^2+a^2)(1-\frac{\mr_e^2}{l^2})}-\frac{\mr_e^2+a^2}{2G\,\Xi}\,. &
\end{flalign}
\item[--] NHE-BTZ:
\begin{flalign}
&J=\frac{j}{\mathrm{8G}}\,, \qquad Q=0\,, \qquad k=\sqrt{\frac{l}{2j}}\,, \qquad e=0\,, \qquad\frac{S}{2\pi}=\frac{1}{4G}\sqrt{\frac{jl}{2}}\,,&\nonumber\\
& \oint_\mathcal{H}\sqrt{-g}\mathcal{L}=\frac{-1}{8G}\sqrt{\frac{jl}{2}}\,. &
\end{flalign}
\item[--] NHE-MP in $5$-dim:
\begin{flalign}
&J_1=\frac{\pi(a+b)^2}{4G}a\,, \qquad J_2=\frac{\pi(a+b)^2}{4G}b\,, \qquad Q=0\,, \qquad  k^1=\frac{1}{2}\sqrt{\frac{b}{a}},\qquad k^2=\frac{1}{2}\sqrt{\frac{a}{b}}\,,&\nonumber\\
&e=0\,,\qquad \frac{S}{2\pi}=\frac{\pi \sqrt{ab}\,(a+b)^2}{4G}\,, \qquad \oint_\mathcal{H}\sqrt{-g}\mathcal{L}=0\,. &
\end{flalign}
\item[--] NHE-KK in $4$-dim:
\begin{flalign}
&J=\frac{a^2}{G\sqrt{1-v^2}}\,, \qquad Q=\frac{av}{\sqrt{1-v^2}}\,, \qquad k=1\,, \qquad e=0\,,\qquad \frac{S}{2\pi}=\frac{a^2}{G\sqrt{1-v^2}}\,,&\nonumber\\
&\oint_\mathcal{H}\sqrt{-g}\mathcal{L}=0\,.&
\end{flalign}
\end{itemize}
}
\vspace*{-0.5cm}
\dotfille

\section{Laws of NHEG thermodynamics}
In this section, the dynamical laws for NHEGs are described. These laws are in parallel to BH thermodynamic laws, as will be described in Section \ref{sec Revisiting laws}.

\subsection{NHEG Zeroth law}
NHEG zeroth law can be explained as
\begin{center}
\emph{NHEG chemical potentials, $\kappa$, $k^i$ and $e^p$ are constant over the whole geometry.}
\end{center}
\dotfillb
\vspace*{-0.5cm}
\begin{proof} \small{Proof for the constancy of $\kappa$ has two steps; the first step is showing that $\kappa$ is constant over $\mathcal{N}_{_\mathcal{H}}$. This step can be elaborated similar to the one for BHs, \ie by replacing $\zeta_{_\mrH}\to \zeta_{_\mathcal{H}}$ in Appendix \ref{app zeroth law}. The second step is to show that $\kappa$ is constant over the whole geometry. According to \eqref{NHEG kappa},  $\kappa^2$ is a  scalar, which inherits $\{\xi_-,\xi_0,\mrm_i\}$ isometry from the $\mrd \zeta_{_\mathcal{H}}\Big|_{\mathcal{H}}=2\boldsymbol{\epsilon}_{_\perp}$. According to Appendix \ref{app t r dependences}, any scalar having these isometries, can be at most a function of $\theta^\alpha$.  But according to the first step, $\kappa^2$ is  constant over $\mathcal{H}$, so can not be a function of $\theta^\alpha$. So it would be constant over the whole geometry.

For the constancy of $k^i$ and $e^p$, as it was explained in the beginning of Section \ref{sec NHEG geometry}, the $SL(2,\mathbb{R})\!\times\!U(1)^n$ isometry of NHEG leads to the constancy of $k^i$ and $e^p$. For the precise discussion, the reader can refer to \cite{Kunduri:2013gce}. Nonetheless, in Appendix \ref{app constant k} a simple proof for the case of $d\!=\!4$ is provided. 
}
\end{proof}
\normalsize
\dotfille

\subsection{NHEG Entropy law}
\emph{NHEG Entropy law} is a universal law in parallel to the third law of BH thermodynamics, relating the entropy of NHEGs to their other thermodynamic entities as
\begin{equation}\label{entropy law}
\frac{S}{2\pi}=\vec{k}\cdot\vec{J}+e^pQ_p-\oint_\mathcal{H}\sqrt{-g}\,\mathcal{L}\,. 
\end{equation} 

\dotfillb
\vspace*{-0.5cm}
\begin{proof} \small{By taking covariant derivative of both sides of \eqref{zeta H}, 
\begin{equation}
\nabla_\mu\zeta_{_\mathcal{H}\nu}=\mrn_{_\mathcal{H}}^a \nabla_\mu \xi_{a\nu}-k^i \nabla_\mu \mrm_{i\nu}\,.
\end{equation} 
Multiplying both sides by \eqref{E-four-index} and integrating over $\mathcal{H}$ yields
\begin{equation}\label{entropy law proof 1}
\oint_\mathcal{H}\mathbf{E}^{\mu \nu}\nabla_{[\mu}{\zeta}_{_\mathcal{H}\nu ]}=\mrn_{_\mathcal{H}}^a\oint_\mathcal{H}\mathbf{E}^{\mu \nu}\nabla_{[\mu}\xi_{a\nu ]}-k^i\oint_\mathcal{H}\mathbf{E}^{\mu \nu}\nabla_{[\mu}\mrm_{i\nu ]}\,.
\end{equation}
According to \eqref{NHEG entropy E}, the LHS is unambiguously the $\dfrac{S}{2\pi}$. The first and second terms of the RHS, using the \eqref{decomposition}, are $\mrn_{_\mathcal{H}}^a\mathcal{Q}_{\xi_a}$ and $-k^i \mathcal{Q}_{\mrm_i}$ respectively, upto their individual ambiguities. But their ambiguities are not independent;  summation of each type of  their ambiguities vanishes (via Lemma \ref{Lemma ambig vanish} and thanks to the vanishing of $\mrn_{_\mathcal{H}}^a \xi_a-k^i\mrm_i$ on the $\mathcal{H}$ and its Killingness). As a result fixing the ambiguities of one of them suffices. By vanishing of the last term in \eqref{H v Q}, $\mathcal{Q}_{\mrm_i}=H_{\mrm_i}$, so we can fix the ambiguities in RHS of \eqref{entropy law proof 1} by demanding $\mathcal{Q}_{\mrm_i}=-J_i$. It turns out that by this fixation of ambiguities, the first term in \eqref{entropy law proof 1} would have its ambiguities fixed such that it is equal to \eqref{n-a Qxi-a}. Therefore $\dfrac{S}{2\pi}=e^pQ_p-\oint_\mathcal{H}\sqrt{-g}\,\mathcal{L}+\vec{k}\cdot\vec{J}$, 
the entropy law. 
}
\end{proof}
\normalsize
\dotfille

Entropy law provides a universal relation between entropy of NHEGs and its other thermodynamic variables. This relation has been firstly observed in \cite{Astefanesei:2006dd,Sen:2008jk}. Its realization as a universal thermodynamic law of NHEGs (\textit{c.f.} Smarr formulas [\cite{padmanabhan2010gravitation}]), plus its proof based on the definition of entropy as a conserved charge and isometry relation \eqref{zeta H}, has been carried out in \cite{HSS:2013lna}. Realization of this law as a thermodynamic law which parallels the third law of BHs will be discussed in section \ref{sec Revisiting laws}.  

\dotfillb
\Example{\emph{Checking the entropy law:} \\
For the NHE-Kerr:
\begin{flalign}
&J=\frac{a^2}{G}\,, \qquad Q=0\,, \qquad k=1\,, \qquad e=0\,,\qquad \frac{S}{2\pi}=\frac{a^2}{G}\,, \qquad \oint_\mathcal{H}\sqrt{-g}\mathcal{L}=0\,.\nonumber&
\end{flalign}
which simply confirms the entropy law. Checking the entropy law for other NHEGs in Example \ref{example NHEG thermo entities} is also very simple task to do. 
}
\dotfille

\subsection{NHEG Entropy perturbation law}\label{sec EPL}
Assuming the NHEG dynamical fields as background fields $\bar{\Phi}$, they can be perturbed by $\delta\Phi$ which satisfy l.e.o.m. Satisfying l.e.o.m is necessary if $\delta \Phi$ is supposed to be an acceptable member of the linear space, tangent to the presumed NHEG phase space. As a result of l.e.o.m, $\delta\Phi$ would respect symplectic structure properties discussed in Chapter \ref{chap Covariant Phase Space method}. 

The perturbations $\delta \Phi$ would vary the NHEG Hamiltonian generators by $\delta J_i$, $\delta Q_p$, $\delta H_{\xi_a}$ and $\delta S$.  Although by Lemma \ref{Lemma independ H}, $\delta J_i$ and $\delta H_{\xi_a}$ are independent of the surface of integration $\mathcal{H}$,  but this issue is not guaranteed for $\delta S$ and $\delta Q_p$. Specifically, although $\zeta_{_\mathcal{H}}$ is a Killing, but  because of its explicit $(t_{_\mathcal{H}},r_{_\mathcal{H}})$ dependency, $\delta S$ can depend on the choice of $\mathcal{H}$. In Appendix \ref{app NHEG entropy H indep}, it is shown that requesting $\{\xi_-,\xi_0\}$  isometry of $\delta\Phi$ would necessarily lead to  $\mathcal{H}$ independence of $\delta S$ and $\delta Q_p$. We will postpone providing more reasons for justifying the request of $\{\xi_-,\xi_0\}$  isometry of $\delta\Phi$ to Part \ref{part II}.

Assuming that perturbations are $\{\xi_-,\xi_0\}$ isometric, \emph{entropy perturbation law} (EPL) is a universal law in parallel to the first law of BH thermodynamics, relating the perturbations of the entropy of NHEGs to perturbations of other conserved charges as
\begin{equation}\label{EPL}
\frac{\delta S}{2\pi}=\vec{k}\cdot\delta\vec{J}+e^p \delta Q_p\,.
\end{equation}

\dotfillb
\begin{proof} \small{The idea of the proof is similar to the Iyer-Wald proof for the first law of BHs, explained in Example \ref{example Iyer-Wald 1st law}. Beginning from \eqref{H v Q proof 1} for the Killing vector $\zeta_{_\mathcal{H}}$, 
\begin{equation}
\boldsymbol{\omega}(\delta \Phi,\delta_{\zeta_{_\mathcal{H}}}\Phi,\bar{\Phi})  =\mathrm{d}\Big(\delta  \mathbf{Q}_{\zeta_{_\mathcal{H}}}- \zeta_{_\mathcal{H}} \! \cdot \! \mathbf{\Theta} (\delta \Phi,\bar{\Phi})\Big)\,,
\end{equation}
by integration over $(d\!-\!1)$-dim surface $\Sigma$ with boundaries at $\mathcal{H}$ and $r\to \infty$ we have 
\begin{align}\label{EPL proof 1}
\Omega(\delta \Phi,\delta_{\zeta_{_\mathcal{H}}}\Phi,\bar{\Phi})=\oint_\infty \Big(\delta  \mathbf{Q}_{\zeta_{_\mathcal{H}}}- \zeta_{_\mathcal{H}} \! \cdot \! \mathbf{\Theta} (\delta \Phi,\bar{\Phi})\Big) -\oint_\mathcal{H} \Big(\delta  \mathbf{Q}_{\zeta_{_\mathcal{H}}}- \zeta_{_\mathcal{H}} \! \cdot \! \mathbf{\Theta} (\delta \Phi,\bar{\Phi})\Big)\,.
\end{align} 
where $\Omega(\delta \Phi,\delta_{\zeta_{_\mathcal{H}}}\Phi,\bar{\Phi}) \equiv \int_\Sigma\boldsymbol{\omega}(\delta \Phi,\delta_{\zeta_{_\mathcal{H}}}\Phi,\bar{\Phi})$. The $\Omega$ in the LHS is \emph{not} identically zero, because $\mathscr{L}_{\zeta_{_\mathcal{H}}}A_\mu^{(p)}\!\neq \!0$. In Appendix \ref{app Omega xi3 calc} the LHS is calculated to be $-e^p\delta Q_p$. The last integration in RHS is unambiguously equal to $\dfrac{-\delta S}{2\pi}$. For the reason, in Appendix \ref{app 1st ambig} make the replacement $\zeta_{_\mrH}\to \zeta_{_\mathcal{H}}$. Noticing Appendix \ref{app NHEG entropy H indep}, by $\{\xi_-,\xi_0\}$ isometry of perturbations, $\delta Q_p$ and $\delta S$ are independent of the chosen $\mathcal{H}$. Gathering the results so far, \eqref{EPL proof 1} can be rewritten as
\begin{equation}\label{EPL proof 2}
\frac{\delta S}{2\pi}=e^p\delta Q_p -k^i\oint_\infty \Big(\delta  \mathbf{Q}_{\mrm_i}- \mrm_i \! \cdot \! \mathbf{\Theta} (\delta \Phi,\bar{\Phi})\Big) +\mrn^a_{_\mathcal{H}}\oint_\infty \Big(\delta  \mathbf{Q}_{\xi_a}- \xi_a \! \cdot \! \mathbf{\Theta} (\delta \Phi,\bar{\Phi})\Big)\,.  
\end{equation}
According to \eqref{NHEG angular momentum} and \eqref{k-xi general formula}, the first integral is the definition of $\delta H_{\mrm_i}\equiv -\delta J_i$. Notice that it is explicitly independent of $\mathcal{H}$, because the integration is taken at $r=\infty$. So
\begin{equation}\label{EPL proof 3}
\frac{\delta S}{2\pi}=e^p\delta Q_p +k^i\delta J_i +\mrn^a_{_\mathcal{H}}\delta H_{\xi_a}\,,  
\end{equation}
where the notation $\delta H_{\xi_a}$ is adapted for individual integrals in the last term of \eqref{EPL proof 2}, just for bookkeeping.

As the final step, we will show that the last term in \eqref{EPL proof 3} vanishes. The key point is that all terms in \eqref{EPL proof 3}, except that last term, are shown to be $\mathcal{H}$-independent. So by \eqref{EPL proof 3}, the last term has to be $\mathcal{H}$-independent. Paying attention that the integrals in $\delta H_{\xi_a}$ are taken at $\infty$, the $\delta H_{\xi_a}$ have not any $r_{_\mathcal{H}}$-dependency. So the $r_{_\mathcal{H}}$ dependency only comes from the factors $\mrn_{_\mathcal{H}}^a$. Using \eqref{n-a vector up}, requesting that the coefficients of different powers of $r_{_\mathcal{H}}$ to vanish, leads to
\begin{equation}\label{EPL proof 4}
\delta H_{\xi_-}=0\,, \qquad t_{_\mathcal{H}}\delta H_{\xi_0}-\delta H_{\xi_+}=0. 
\end{equation}
Putting \eqref{EPL proof 4} back into the \eqref{EPL proof 3}, makes the last term of \eqref{EPL proof 3} to vanish, hence the proof would finish. 
}
\end{proof}
\normalsize
\dotfille

Initial works on the EPL and its proof based on the near horizon limit of extremal BHs have been carried out in \cite{Hartman:2009ge,Compere:2012hr,HSS:2013lna} and \cite{Johnstone:2013er}, while the proof above which is based on $\{\xi_-,\xi_0\}$ isometry of perturbations has been introduced and elaborated in \cite{HSS:2014twa}. 

\dotfillb
\Example{\emph{Checking the EPL by parametric variations:} \\
EPL can be checked using the parametric variations introduced in Chapter \ref{chap BH thermo}, but with caring a subtle issue. The Killing vectors $\xi_+$ and $\zeta_{_\mathcal{H}}$ are parameter dependant. But in the proof of EPL it is assumed that the perturbations do not change any non-dynamical vector field, including $\xi_+$ and $\zeta_{_\mathcal{H}}$. So it is not guaranteed that EPL would be respected by the parametric variations. In \cite{HSS:2014twa} it is proved that this subtle issue does not ruin the EPL for parametric variations. Hence one can use parametric variations to check EPL. The NHE-Kerr example is provided here. For the NHE-Kerr using \eqref{NHE-Kerr entities}, one finds
\begin{equation}
\hat{\delta}J=\frac{2a}{G}\delta a\,,\qquad \frac{\hat{\delta}S}{2\pi}=\frac{2a}{G}\delta a\,,
\end{equation}
which by considering the $k=1$, simply satisfies the EPL 
\begin{equation}
\frac{\hat{\delta}S}{2\pi}= k \hat{\delta} J\,.
\end{equation}
Other examples gathered in Example \ref{example NHEG thermo entities} are easy to be worked out similarly.
}
\dotfille


\section{Revisiting laws of black hole thermodynamics}\label{sec Revisiting laws}

In the end of Chapter \ref{chap BH thermo} we posed three main questions concerning the thermodynamics of the extremal BHs, summarized as: 
\begin{enumerate}
\item How can one extend Iyer-Wald definition of the entropy, \ie \, as a Hamiltonian generator, to the extremal BHs?
\item What is the universal law of BH thermodynamics analogous to the ``as $T\to 0$ then $S\to 0$" version of the third law of thermodynamics?
\item Keeping the $T_{_\mrH}=0$, \ie for the extremal family of BH solutions, what is the universal law which determines $\delta S$? The first law of BH thermodynamics is blind to the answer\footnote{For additional information which is needed to reproduce the $\delta S$ by taking the limit $T_{_\mrH}\to 0$ of the first law, see \cite{Johnstone:2013er}.}.
\end{enumerate}
 The answer to the first question was provided in Section \ref{sec NHEG entropy}. Answers to the second and third question would be the \emph{NHEG entropy law} and EPL respectively, if the relation of NHEG chemical potentials $k^i$ and $e^p$  with the original extremal BHs would be determined. After a series of works \cite{Guica:2009cd,Hartman:2009ge,Azeyanagi:2009sd,Johnstone:2013er}, this relation is analytically proved to be as \cite{HSS:2013lna}
\begin{equation}\label{k e BH relation}
k^i=-\frac{1}{2\pi}\frac{\partial \Omega^i}{\partial T_{_\mrH}}\Big|_{T_{_\mrH}=0}\,, \qquad e^p =-\frac{1}{2\pi}\frac{\partial \Phi^p}{\partial T_{_\mrH}}\Big|_{T_{_\mrH}=0}\,.
\end{equation}
As a result, the NHEG dynamical laws are suggestive for enhancing the BHs thermodynamic laws as follows.
\begin{itemize}
\item[$\mathbf{(0)}$] $T_{_\mrH}$, $\Omega_{_\mrH}^i$ and $\Phi_{_\mrH}^p$ are constant over the horizon.\\
 If $T_{_\mrH}=0$, then  $k^i$ and $e^p$ in \eqref{k e BH relation} would also be constant over the horizon. 
\item[$\mathbf{(1)}$] $\delta M=T_{_\mrH} \delta S+\Omega_{_\mrH}^i \delta J_i+\Phi_{_\mrH}^p \delta Q_p$.\\
 If $T_{_\mrH}=0$, then $\delta M=\Omega_{_\mrH}^i \delta J_i+\Phi_{_\mrH}^p \delta Q_p$, and $\frac{\delta S}{2\pi}=k^i\delta J_i+e^p\delta Q_p$.
\item[$\mathbf{(2)}$] For a closed thermodynamic system (including the black hole), entropy never decreases.
\item[$\mathbf{(3)}$] \begin{itemize}\item $T_{_\mrH}=0$ is not physically achievable, \item If $T_{_\mrH}\to 0$ then $
\frac{S}{2\pi}\to \vec{k}\cdot\vec{J}+e^pQ_p-\oint_\mrH\sqrt{-g}\,\mathcal{L}$.
\end{itemize}
\end{itemize}

 \addtocontents{toc}{\vspace{1em}}
\part{On phase space of Near Horizon Extremal Geometries}\label{part II}

\addtotoc{${\ast}$\,\,\,\,\,Motivations and Outline}
\chapter*{Motivations and Outline}
In Part \ref{part I}, based on the Covariant Phase Space method, thermodynamics of BHs were reviewed and enhanced. Thermodynamics of BHs at non-zero temperature could be directly analyzed, while for the analysis of the extremal ones, NHEGs were employed. A natural question arises is about the phase space on which the Hamiltonian generators were introduced. In other words, it was implicitly assumed that there is a well-defined phase space on which those Hamiltonian generators are defined. In Part \ref{part II}, we will try to identify that phase space explicitly. So,  Part \ref{part II} is a natural continuation of Part \ref{part I}. Nonetheless, passing from Part \ref{part I} to Part \ref{part II} is in some sense similar to passing from thermodynamics to statistical mechanics. 

Since the recognition of thermodynamic behaviours  of the BHs, underlying statistical mechanics of those behaviours has been a big question for BH physicists. Following this question, seminal explanations have been proposed for this question, as far as some specific supersymmetric BHs are concerned \cite{Strominger:1996sh,Dabholkar:2014ema}. Unfortunately those explanations are not generalizable to include non-supersymmetric BHs, \eg the Kerr BH. Providing satisfactory explanations for the microstate counting of the Kerr-type black holes is a goal of BH physics. Interestingly, there are some evidences that microstates of BHs, are some degrees of freedom around their horizons. For example, the temperature $T_{_\mathrm{H}}$, chemical potentials $\Omega_{_\mrH}^i$ and $\Phi_{_\mrH}^p$ are determined by the field configurations around the horizon. Entropy and other conserved charges as Hamiltonian generators can be calculated similarly. Hence, although the phase space which will be discussed in this part,  would be NHEG phase space, but it might open new gates for microstate counting of the corresponding extremal BHs. In the same way that studying classical phase space of ideal gases sheds light on their statistical mechanics, one is motivated to study NHEG phase space, in order to search for the underlying microscopic system responsible for thermodynamic behaviours of the extremal BHs. 

Another motivation for studying NHEG phase space, is to provide a setup for understanding proposed holographic dualities between gravity theories, and some other (usually conformal symmetric) theories on the boundaries of spacetime (\eg \cite{Guica:2009cd,Hartman:2009ge,Loran:2009cr}). There has been a hope that by these dualities, the entropy of BHs could be reproduced using the field theories on the boundary. Investigating the NHEG phase space, can enable us to have new insights on that hope.  

The outline of Part \ref{part II} is as follows. Chapter \ref{chap Prerequisites} at first, reviews the role of phase space in thermodynamics. Specifically it emphasizes that information about the entropy of a thermodynamic system is usually encoded in the phase space of that system. Then it provides some generic properties which NHEG phase space is expected to own. Those properties confine the manifold of the NHEG phase space greatly. Chapter \ref{chap NHEG phase space} will add some more specific constraints, to fix the manifold completely. Hence, the NHEG phase space manifold is completely identified. In order to determine the symplectic $2$-form, one needs to fix the $\mathbf{Y}$ ambiguity in the Lee-Wald symplectic structure. Remaining sections of Chapter \ref{chap NHEG phase space} are provided to do this job. As a result, at the end of that chapter, the NHEG phase space is completely built. Chapter \ref{chap NHEG algebra} deals with symmetries of the built phase space, in order to find a way to extract information about the entropy of NHEG as a statistical system. In addition, algebra of the symplectic structure is studied, opening a gate towards a probable dual field theory description of the system. Finally, Chapter \ref{chap outlook} will summarize the discussions in both Part \ref{part I} and \ref{part II}. In addition, some interesting lines of research are gathered at the end.

\chapter{Prerequisites for NHEG phase space}\label{chap Prerequisites}

In Part \ref{part II}, we focus on the EH gravity, in $4$ and higher dimensions, with the action
\begin{equation}
\mathcal{S}=\int \mrd^d x \sqrt{-g} R\,.
\end{equation}
 In addition, the isometry of NHEGs would be chosen to be $SL(2,\mathbb{R})\!\times\!U(1)^{n}$ for $n\!=\!d\!-\!3$. Hence there would be just one coordinate $\theta$, and the only dynamical field would be the metric
\begin{align}\label{NHEG metric d-3}
	{\mrd s}^2&=\Gamma\big[-r^2\mrd t^2+\frac{\mrd r^2}{r^2}+ \mrd\theta^2+\gamma_{ij}(\mrd\varphi^i+k^ir\,\mrd t)(\mrd\varphi^j+k^jr\,\mrd t)\big]\,.
\end{align} 
All of the discussions in Part \ref{part II} are classical, unless any quantization is mentioned explicitly. This part is mainly based on the works \cite{CHSS:2015mza,CHSS:2015bca}.

The first section of the present chapter provides a quick review on the role of phase space in thermodynamics. The sections after that, would emphasize basic and generic properties which are expected for NHEG phase space. They provide us an overview on the analysis which will be worked out in later chapters.


\section{Revisiting the phase space}\label{sec revisit phase space}
As described in Chapter \ref{chap Covariant Phase Space method}, phase space is a manifold $\mathcal{M}$, equipped to a symplectic $2$-form $\Omega$. Studying phase space of thermodynamic systems, literally means studying those systems at the level of statistical mechanics. Hence one expects to find the entropy, using some kind of microstate counting through the phase space. Below, two famous  examples are reminded.

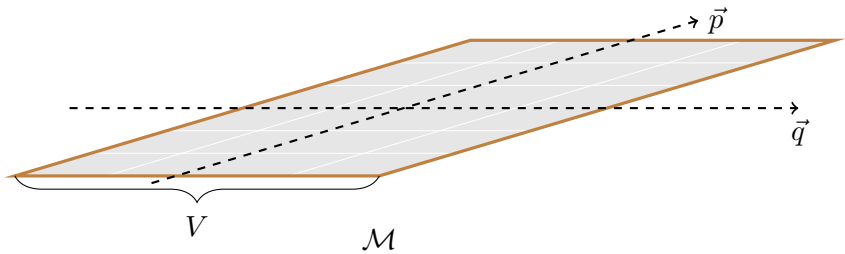
\begin{figure}[H]
	\centering
\begin{tikzpicture}[scale=1.2]
\fill[even odd rule,light gray]
    (1,0) to (6,1.5) to  (10,1.5) to
    (5,0) to (1,0);
    \draw[very thin,white] (1,0) to (6,1.5);
    \draw[very thin,white] (2,0) to (7,1.5);
    \draw[very thin,white] (4,0) to (9,1.5);
    \draw[very thin,white] (5,0) to (10,1.5);
    \draw[very thin,white] (0.8,0.25) to (6.8,0.25);
    \draw[very thin,white] (1.6,0.5) to (7.6,0.5);
    \draw[very thin,white] (3.3,1) to (9.3,1);
    \draw[very thin,white] (4.1,1.25) to (10.1,1.25);
    
    \draw[very thick,brown] (1,0) to  (5,0) to  (10,1.5) to (6,1.5) -- cycle;
    
    \draw[dashed,thick,->,shift={(-0.5,-0.08)}] (3,0) to (9,6*1.5/5) node[anchor=west] {$\vec{p}$};
    \draw[dashed,thick,->,shift={(-0.9,0)}] (2.5,0.75) to (10.5,0.75)node[anchor=north] {$\vec{q}$};
    \draw (5,-0.7) node {$\mathcal{M}$};
    \draw [decorate,decoration={brace,amplitude=10pt}]
    (5,0) -- (1,0) node [black,midway,below,shift={(0cm,-0.4cm)}] {$V$};
    \end{tikzpicture}
    \caption{A schematic of the manifold $\mathcal{M}$ of the phase space describing one free particle in a micro-canonical ensemble. The position is confined in the volume $V$. By the micro-canonical definition, the momentum is also confined in some region. The manifold of the phase space describing an ideal gas consisted of $N$ particles would be the Cartesian product of $N$ number of $\mathcal{M}$s, denoted as $\mathcal{M}^N$.}
    \label{fig ideal gas micro}
\end{figure}

\dotfillb
\Example{\label{example ideal gas micro}\emph{Calculating entropy of an ideal gas using its phase space; micro-canonical ensemble:} \\
Considering a box of ideal gas, consisted of $N$ number of non-interacting particles enclosed in $3$-dim volume $V$.  The manifold of the phase space for this system would be the Cartesian product of the phase spaces of each individual particle. The phase space of each particle, denoted by $\mathcal{M}$, is a $6$-dim  manifold of position $\vec{q}$ and momentum $\vec{p}$. Hence the phase space of the whole gas would be $\mathcal{M}^N$. The position $\vec{q}$ is confined to be inside the volume $V$. Also, by micro-canonical definition, the $\vec{p}$ is also confined to be in a region which would provide the gas an energy in the range $[E,E+\delta E]$. A schematic of the phase space of one particle is provided in Figure \ref{fig ideal gas micro}. Having the fundamental principle of statistical mechanics in mind, any point of the phase space is as probable as other points. Denoting the $6N$-dim volume of the phase space by $\text{Vol}(\mathcal{M}^N)$, then the entropy of the system would be
\begin{equation}
S=-\mathrm{k}_\mathrm{B}\ln \Big(\text{Vol}(\mathcal{M}^N)\Big)
\end{equation}
where $\mathrm{k}_\mathrm{B}$ is Boltzmann constant. 
}

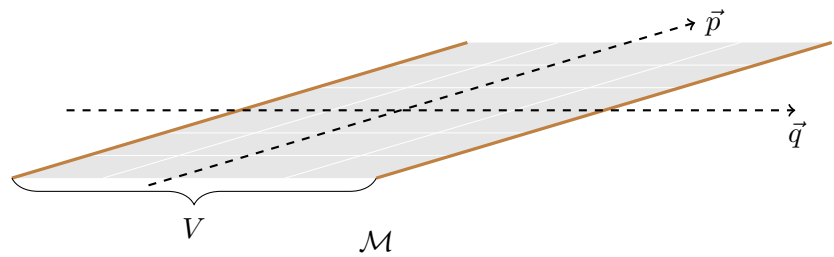
\begin{figure}[H]
	\centering
\begin{tikzpicture}[scale=1.2]
\fill[even odd rule,light gray]
    (1,0) to (6,1.5) to  (10,1.5) to
    (5,0) to (1,0);
    \draw[very thin,white] (1,0) to (6,1.5);
    \draw[very thin,white] (2,0) to (7,1.5);
    \draw[very thin,white] (4,0) to (9,1.5);
    \draw[very thin,white] (5,0) to (10,1.5);
    \draw[very thin,white] (0.8,0.25) to (6.8,0.25);
    \draw[very thin,white] (1.6,0.5) to (7.6,0.5);
    \draw[very thin,white] (3.3,1) to (9.3,1);
    \draw[very thin,white] (4.1,1.25) to (10.1,1.25);
    
    \draw[very thick,brown] (5,0) to  (10,1.5);
    \draw[very thick,brown] (1,0) to  (6,1.5);
    
    \draw[dashed,thick,->,shift={(-0.5,-0.08)}] (3,0) to (9,6*1.5/5) node[anchor=west] {$\vec{p}$};
    \draw[dashed,thick,->,shift={(-0.9,0)}] (2.5,0.75) to (10.5,0.75)node[anchor=north] {$\vec{q}$};
    \draw (5,-0.7) node {$\mathcal{M}$};
    \draw [decorate,decoration={brace,amplitude=10pt}]
    (5,0) -- (1,0) node [black,midway,below,shift={(0cm,-0.4cm)}] {$V$};
    \end{tikzpicture}
    \caption{A schematic of the manifold $\mathcal{M}$ of the phase space describing one free particle in a canonical ensemble. The position is confined in the volume $V$. The manifold of the phase space describing an ideal gas consisted of $N$ particles would be $\mathcal{M}^N$.}
\label{fig ideal gas canonic}
\end{figure}

\Example{\label{example ideal gas canonic}\emph{Calculating entropy of an ideal gas using its phase space; canonical ensemble:} \\
Considering a similar system of ideal gas as Example \ref{example ideal gas micro}, but as a canonical ensemble. Therefore, there would be $N$ particles in volume $V$, this time in a temperature $T$. The phase space is similar to the one in Example \ref{example ideal gas micro}, but there would not be any constraints on the momentums $\vec{p}$. A schematic of the phase space of one particle is provided in Figure \ref{fig ideal gas canonic}. Different points of the phase space can describe the state of the system with the  probability proportional to $e^{-\beta E}$ where $\beta=\dfrac{1}{\mathrm{k}_\mathrm{B}T}$. One can find the entropy of the system through the partition function $Z$ as
\begin{align}
Z=\int_{\mathcal{M}_N}e^{-\beta E}\,, \qquad S=\mathrm{k}_\mathrm{B}(1-\beta \frac{\partial}{\partial \beta})\ln Z\,.
\end{align}
}
\vspace*{-0.5cm}
\dotfille

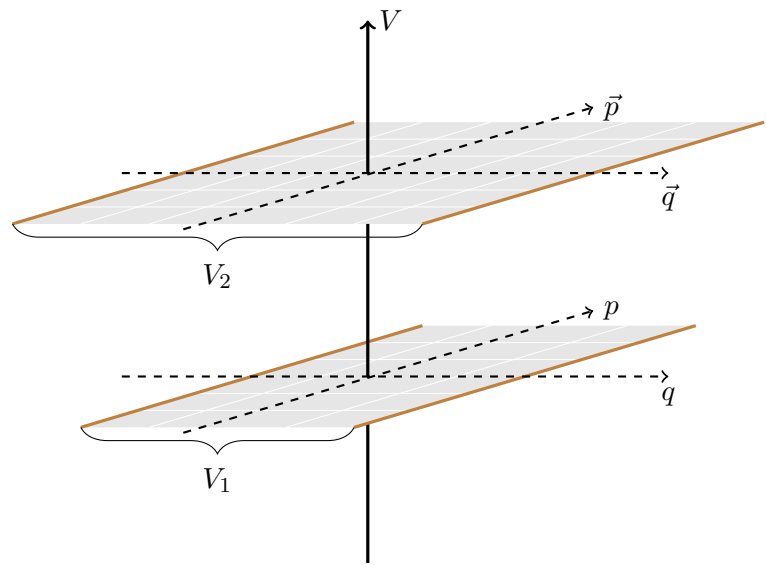
\begin{figure}[!h]
	\centering
\begin{tikzpicture}[scale=0.9,decoration={%
   markings,%
   mark=at position 0.25 with {\arrow[black]{stealth};},%
   mark=at position 0.5 with {\arrow[black]{stealth};},%
   mark=at position 0.75 with {\arrow[black]{stealth};}}]
    \draw[very thick] (5.2,-2) to (5.2,0.5);
    \fill[even odd rule,light gray]
    (1,0) to (6,1.5) to  (10,1.5) to
    (5,0) to (1,0);
    \draw[very thin,white] (1,0) to (6,1.5);
    \draw[very thin,white] (2,0) to (7,1.5);
    \draw[very thin,white] (4,0) to (9,1.5);
    \draw[very thin,white] (5,0) to (10,1.5);
    \draw[very thin,white] (0.8,0.25) to (6.8,0.25);
    \draw[very thin,white] (1.6,0.5) to (7.6,0.5);
    \draw[very thin,white] (3.3,1) to (9.3,1);
    \draw[very thin,white] (4.1,1.25) to (10.1,1.25);
    
    \draw[very thick,brown] (1,0) to (6,1.5);
    \draw[very thick,brown] (5,0) to (10,1.5);
    
    \draw[dashed,thick,->,shift={(-0.5,-0.08)}] (3,0) to (9,6*1.5/5) node[anchor=west] {$p$};
    \draw[dashed,thick,->,shift={(-0.9,0)}] (2.5,0.75) to (10.5,0.75)node[anchor=north] {$q$};
    \draw [decorate,decoration={brace,amplitude=10pt}]
    (5,0) -- (1,0) node [black,midway,below,shift={(0cm,-0.4cm)}] {$V_1$}; 
    
    \draw[very thick] (5.2,0.75) to (5.2,4);
    
    \begin{scope}[shift={(0,3)}]
    \fill[even odd rule,light gray]
    (0,0) to (5,1.5) to  (11,1.5) to
    (6,0) to (0,0);
    \draw[very thin,white] (1,0) to (6,1.5);
    \draw[very thin,white] (2,0) to (7,1.5);
    \draw[very thin,white] (4,0) to (9,1.5);
    \draw[very thin,white] (5,0) to (10,1.5);
    \draw[very thin,white] (0.8,0.25) to (6.8,0.25);
    \draw[very thin,white] (1.6,0.5) to (7.6,0.5);
    \draw[very thin,white] (3.3,1) to (9.3,1);
    \draw[very thin,white] (4.1,1.25) to (10.1,1.25);
    
    \draw[very thick,brown] (0,0) to (5,1.5);
    \draw[very thick,brown] (6,0) to (11,1.5);
    
    \draw[dashed,thick,->,shift={(-0.5,-0.08)}] (3,0) to (9,6*1.5/5) node[anchor=west] {$\vec{p}$};
    \draw[dashed,thick,->,shift={(-0.9,0)}] (2.5,0.75) to (10.5,0.75)node[anchor=north] {$\vec{q}$};
    \draw [decorate,decoration={brace,amplitude=10pt}]
    (6,0) -- (0,0) node [black,midway,below,shift={(0cm,-0.4cm)}] {$V_2$};    
    
    \end{scope}
    \draw[very thick,->] (5.2,3.75) to (5.2,6) node[anchor=west]{$V$};
    \end{tikzpicture}
    \caption{Depending on the volume chosen for the box of gas in Example \ref{example ideal gas canonic}, manifold of the phase space would be different.}
\label{fig different volumes}
\end{figure}

\begin{figure}[H]
	\centering
\begin{tikzpicture}[scale=1]

    \draw[very thick] (5.5,1) to (5.5,3.3);
    \begin{scope}[shift={(0,3)}]
        \fill[even odd rule,light gray]
    (0,0) to (5,1.5) to  (11,1.5) to
    (6,0) to (0,0);
    \draw[very thin,white] (1,0) to (6,1.5);
    \draw[very thin,white] (2,0) to (7,1.5);
    \draw[very thin,white] (3,0) to (8,1.5);
    \draw[very thin,white] (4,0) to (9,1.5);
    \draw[very thin,white] (5,0) to (10,1.5);
    \draw[very thin,white] (0.7,0.25) to (6.9,0.25);
    \draw[very thin,white] (1.5,0.5) to (7.7,0.5);
    \draw[very thin,white] (2.2,0.75) to (8.6,0.75);
    \draw[very thin,white] (3.3,1) to (9.3,1);
    \draw[very thin,white] (4.0,1.25) to (10.2,1.25);
    \end{scope}
    \draw[very thick,->] (5.5,3.75) to (5.5,6) node [anchor=west,align=center]{\footnotesize Thermodynamic \\   \footnotesize variables};
    \draw (4.5,3.5) node[anchor=south]{$\Omega$};
    \draw (4,2.3) node[anchor=south]{$\mathcal{M}$};
    \end{tikzpicture}
    \caption{A helpful figure which illustrates phase space schematically, in terms of thermodynamic variables.}
\label{fig phase space vs thermo}
\end{figure}
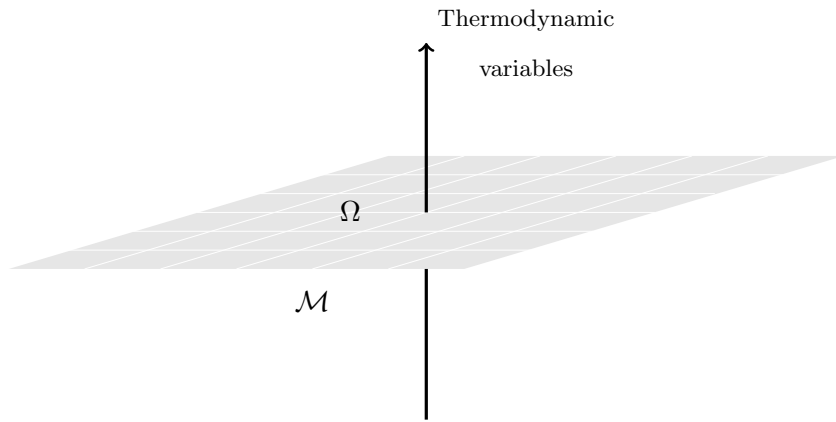

Given a thermodynamic system in different thermodynamic variables, its phase space would be different. As an example, in Example \ref{example ideal gas canonic} considering the volume as one of the thermodynamic variables, then the manifold of the phase space would depend on the $V$. Figure \ref{fig different volumes} depicts it schematically. 
So it is helpful to introduce figures which illustrate the phase space, for determined thermodynamic variables. See Figure \ref{fig phase space vs thermo}. Notice that all points of the phase space which are depicted, correspond to the same thermodynamic variables.


\section{Uniqueness of NHEGs}
\emph{Uniqueness theorems} are propositions which for a given theory, under some specific conditions and isometries, determine solutions uniquely. As the prototypes of uniqueness theorems, two examples are reminded below.

\dotfillb
\Example{\emph{Uniqueness of Schwarzschild solution (Birkhoff's theorem):} \\
Considering EH gravity in $d\!=\!4$, spherical isometric solutions are uniquely Schwarzschild solutions.
}

\Example{\emph{Uniqueness of Kerr solution:} \\
Considering EH gravity in $d\!=\!4$, stationary solutions with flat asymptotics are uniquely Kerr solutions.
}
\vspace*{-0.5cm}
\dotfille

Notice that although it is not usually mentioned explicitly, uniqueness theorems determine solutions uniquely \emph{upto coordinate transformations}. 

NHEGs also enjoy some uniqueness theorems proved for different theories, and  different conditions and topologies. An interested reader can refer to \cite{Kunduri:2007vf,Kunduri:2008rs,Hollands:2009ng},[\cite{Kunduri:2013gce}] for more details. Here we mention the NHEG uniqueness theorem which is relevant to our later discussions \cite{Kunduri:2007vf}. 
\begin{theorem}\label{theorem NHEG uniqueness}
Considering EH-$\Lambda$ theories in $d$ dimension, the $SL(2,\mathbb{R})\!\times\!U(1)^{d-3}$ isometric solutions  are uniquely NHEGs labelled by $d\!-\!3$ number of angular momenta $J_i$.
\end{theorem}

\dotfillb
\Example{\emph{Uniqueness of NHE-Kerr solution:} \\
Considering EH gravity in $d\!=\!4$, $SL(2,\mathbb{R})\!\times\!U(1)$ isometric solutions is NHE-Kerr solution. It is labelled by its angular momentum $J$.
}

\Example{\emph{Uniqueness of NHE-MP solution in $d\!=\!5$:} \\
Considering EH gravity in $d\!=\!5$, $SL(2,\mathbb{R})\!\times\!U(1)^2$ isometric solutions is NHE-MP solution. It is labelled by its angular momenta $J_1$ and $J_2$.
}
\vspace*{-0.5cm}
\dotfille

\begin{figure}[H]
	\centering
\begin{tikzpicture}[scale=1]

    \draw[very thick] (5.5,1) to (5.5,3.3);
    \begin{scope}[shift={(0,3)}]
        \fill[even odd rule,light gray]
    (0,0) to (5,1.5) to  (11,1.5) to
    (6,0) to (0,0);
    \draw[very thin,white] (1,0) to (6,1.5);
    \draw[very thin,white] (2,0) to (7,1.5);
    \draw[very thin,white] (3,0) to (8,1.5);
    \draw[very thin,white] (4,0) to (9,1.5);
    \draw[very thin,white] (5,0) to (10,1.5);
    \draw[very thin,white] (0.7,0.25) to (6.9,0.25);
    \draw[very thin,white] (1.5,0.5) to (7.7,0.5);
    \draw[very thin,white] (2.2,0.75) to (8.6,0.75);
    \draw[very thin,white] (3.3,1) to (9.3,1);
    \draw[very thin,white] (4.0,1.25) to (10.2,1.25);
    \end{scope}
    \draw[very thick,->] (5.5,3.75) to (5.5,6) node [anchor=west]{$J_i$};
    \draw (4.5,3.5) node[anchor=south]{$\Omega$};
    \draw (7,3.5) node[anchor=south]{\footnotesize $g_{\mu\nu}$};
    \draw (5.5,3.75) node[anchor=north]{\footnotesize $\bar{g}_{\mu\nu}$};
    \fill (5.5,3.75) circle (0.08cm);
    \draw (4,2.3) node[anchor=south]{$\mathcal{M}$};
    \end{tikzpicture}
    \caption{A schematic of NHEG phase space, in terms of angular momenta $\vec{J}$. The manifold $\mathcal{M}$ is comprised of some metric configurations. $\bar{g}_{\mu\nu}$ is the known NHEG solution, \ie \eqref{NHEG metric}. Symplectic $2$-form $\Omega$, is the LW form \eqref{Omega LW}, upto $\mathbf{Y}$ ambiguities.}
\label{fig NHEG phase space pre}
\end{figure}
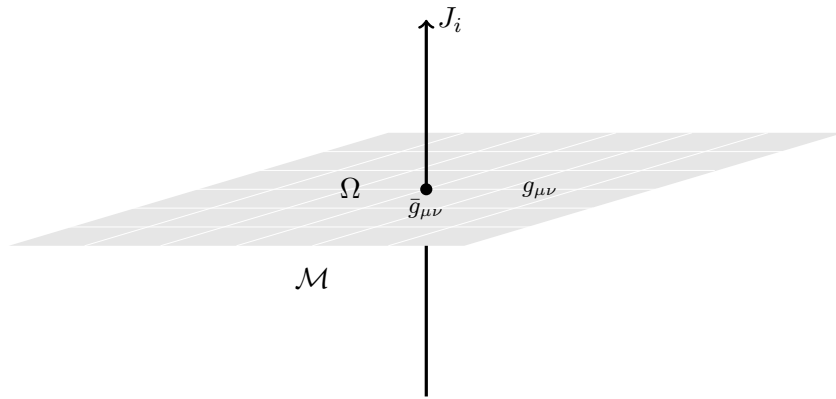

According to the Theorem \ref{theorem NHEG uniqueness}, $J_i$ would be thermodynamic variables which identify the NHEG as a thermodynamic system. So a schematic of the NHEG phase space would be as Figure \ref{fig NHEG phase space pre}. The aim of Part \ref{part II} is to determine $\mathcal{M}$ and $\Omega$ in that figure.

Fortunately Covariant Phase Space method provides us some of the steps, hence we would not reinvent the wheel. The phase space would be constructed of dynamical field configurations, all over the spacetime. Therefore, for the EH theory it would be $g_{\mu\nu}(x^\alpha)$. We will denote the NHEG solution in our hand, as the reference point of the phase space, and  denote it by $\bar{g}_{\mu\nu}$. We adopt this notation for any other entities related to the $\bar{g}_{\mu\nu}$, \eg $\bar{\xi}_a$, $\bar{\mrm}_i$ as its Killing vectors etc. 
In addition, the symplectic structure $\Omega$ would be the LW $2$-form \eqref{Omega LW}, upto some $\mathbf{Y}$ ambiguities discussed in Section \ref{sec Lee-Wald symplectic} (see \eqref{LW Y ambiguity}). 

Summarizing, what which remains to be found, is the identification of $\mathcal{M}$ and fixation of $\mathbf{Y}$. Although those are just two tasks, but the main body of Part \ref{part II} would be devoted to them. In the remaining of this chapter, we continue discussing general expectations from the NHEG phase space.


\section{Isometries of NHEG phase space}\label{sec Isometries of NHEG phase space}
According to the Lemma \ref{Lemma conserv H}, if a vector field $\xi$ is Killing, then conservation of its Hamiltonian generator $\delta H_\xi$ is guaranteed on-shell. The isometries of NHEG solution are $SL(2,\mathbb{R})\!\times U(1)^{d-3}$, generated by the Killings $\bar{\xi}_a$ and $\bar{\mrm}_i$. Their Hamiltonian generators are conserved on the $\bar{g}_{\mu\nu}$. But their Hamiltonian generators might not be conserved on arbitrary point of phase space. In order to guarantee their conservation all over the phase space, it is sufficient to request that all points of the phase space would be $SL(2,\mathbb{R})\!\times U(1)^{d-3}$ isometric. We can denote the Killing vectors generating isometries of an arbitrary point of the phase space $g_{\mu\nu}$, by $\xi_a$ and $\mrm_i$.


\section{Constancy of NHEG thermodynamic variables over phase space}\label{sec constancy of delJ}
Considering that NHEG phase space has $SL(2,\mathbb{R})\!\times U(1)^{d-3}$ isometry generated by $\xi_a$ and $\mrm_i$, as mentioned in the previous section, $\delta H_{\xi_a}$ and $\delta H_{\mrm_i}$ are conserved. They need to be finite and integrable, in order to associate $H_{\xi_a}$ and  $H_{\mrm_i}$ to any point of the phase space. Remembering Section \ref{sec revisit phase space} and Figure \ref{fig NHEG phase space pre}, the thermodynamic variables should be constant over the phase space focused on. The consistent choice of the reference points $H_{\mrm_i}[\bar{g}]$ is naturally the $H_{\mrm_i}[\bar{g}]=-J_i$. Hence the constancy of it over the phase space means 
\begin{equation}
H_{\mrm_i}[{g}]=-J_i
\end{equation}
for all points of the phase space. On the other hand, one can calculate parametric variations of the $SL(2,\mathbb{R})$ Hamiltonians using \eqref{EH k}, and find that $\hat{\delta}H_{\xi_a}=0$, so the natural choice for the $H_{\xi_a}[\bar{g}]$ would be zero. Therefore all over the phase space one expects to have
\begin{equation}
H_{\xi_a}[{g}]=0\,.
\end{equation}


\section{Uniqueness of NHEG perturbations}

A guideline towards determination of the manifold $\mathcal{M}$ is to study its tangent space. Specifically it is usual to study field perturbations $\delta \Phi$ around $\bar{g}_{\mu\nu}$, which satisfy l.e.o.m. As described in Chapter \ref{chap Covariant Phase Space method}, on-shell-ness of dynamical fields $\Phi$, and satisfying l.e.o.m by $\delta \Phi$, lead to have a closed  symplectic structure. In order to define appropriate field perturbations, in addition to the l.e.o.m, some other conditions, which are usually some BCs, are needed. Assuming some conditions including fall-off conditions asymptotics to the $\bar{g}_{\mu\nu}$, it has been shown \cite{Amsel:2009bs,Dias:2009ev} that there is not possibility for any dynamical perturbations $\delta \Phi$ around NHEGs, for specific theories and dimensions. 

One important hint towards understanding better the dynamical perturbations, is that any chosen conditions in addition to the l.e.o.m, should not prohibit the parametric variations. However, the fall-off condition asymptotic to $\bar{g}_{\mu\nu}$ discards the parametric variations in NHEGs. Motivated by this issue, and having general covariance for any proposed BC in mind, the following theorem can be proved \cite{HSS:2014twa}.
\begin{theorem}\label{theorem NHEG pert uniqueness}
Considering EH-$\Lambda$ theories in $d$ dimension, and a NHEG with $SL(2,\mathbb{R})\!\times\!U(1)^{d-3}$ isometry  as a solution, then dynamical perturbations $\delta \Phi$ which
\vspace*{-0.3cm}
\begin{enumerate}
\item satisfy l.e.o.m,
\item are $\{\xi_-,\xi_0\}$ isometric,
\item have the full $SL(2,\mathbb{R})\!\times\!U(1)^{d-3}$ isometry asymptotically, 
\end{enumerate}
\vspace*{-0.3cm}
are uniquely determined to be the parametric variations $\hat{\delta}\Phi$. 
\end{theorem}
We refer to this theorem as \emph{uniqueness of NHEG perturbations}. There are some remarks considering this theorem:
\begin{itemize}
\item[--] Similar to the NHEG itself which is uniquely determined by $J_i$, its perturbations are also uniquely determined by $\delta J_i$.
\item[--] Although not explained explicitly, infinitesimal coordinate transformations are excluded from the above theorem.
\item[--] About the essence of $\{\xi_-,\xi_0\}$ isometry, in addition that it is sufficient condition for the proof of EPL, there are other reasons described in the next chapter. 
\item[--] Notice that according to the Section \ref{sec Isometries of NHEG phase space}, perturbations in the tangent space of $\mathcal{M}$ are supposed to have $SL(2,\mathbb{R})\!\times\!U(1)^{d-3}$ not only asymptotically, but over the whole spacetime.
\end{itemize}
Returning back to the determination of NHEG phase space manifold, Theorem \ref{theorem NHEG pert uniqueness} provides an important hint about the tangent space of $\mathcal{M}$ at $\bar{g}_{\mu\nu}$; that is if one wants to find perturbations $\delta g_{\mu\nu}$ around $\bar{g}_{\mu\nu}$ satisfying the conditions in the theorem, \emph{and} with $\delta J_i=0$, there is not possibility for any candidates, except the infinitesimal coordinate transformations. This is the basis on which the next chapter is relied on. Thanks to the covariance of Killing condition \eqref{Killing condition}, the $g_{\mu\nu}$ constructed by coordinate transformations from $\bar{g}_{\mu\nu}$ would be automatically $SL(2,\mathbb{R})\!\times \!U(1)^{d-3}$ isometric. Hence they would be in agreement with the discussion in Section \ref{sec Isometries of NHEG phase space}.

\chapter{The NHEG phase space}\label{chap NHEG phase space}

Chapter \ref{chap Prerequisites} provided us prerequisite conditions of the NHEG phase space. Summarizing it, NHEG phase space is comprised of metrics $g_{\mu\nu}$ which are derived from $\bar{g}_{\mu\nu}$ by some coordinate transformations. Those metrics are expected to have $SL(2,\mathbb{R})\!\times \!U(1)^{d-3}$ isometry, with the same angular momenta as $\bar{g}_{\mu\nu}$, and vanishing $H_{\xi_a}$. 

This chapter imposes more constraints on the presumed coordinate transformations, such that the manifold $\mathcal{M}$ would be built. Then it will be equipped to a symplectic structure by fixing the $\mathbf{Y}$ ambiguity. 

\section{Phase space manifold}

\subsection{Identifying ``symplectic symmetry generators"}\label{sec identify chi}

Each point of the NHEG phase space is a metric $g_{\mu\nu}$ constructed by some coordinate transformations from $\bar{g}_{\mu\nu}$ in \eqref{NHEG metric d-3}. Assuming that these transformations are generated by some family of vector fields $\chi$, \ie \, having the infinitesimal transformations $x^\mu\to x^\mu - \bar{\chi}^\mu$, then the field perturbations $\delta_{\bar{\chi}}\Phi=\mathscr{L}_{\bar{\chi}} \Phi$ would be members of the tangent space of $\mathcal{M}$ at the point $\bar{g}_{\mu\nu}$.  Then, by exponentiation of $\chi$, finite transformations are found. 

A very simple example of building a manifold by exponentiation of infinitesimal transformations, is building $\mathbb{R}^2$. Choosing one arbitrary point of $\mathbb{R}^2$, then  by exponentiation of translations, the whole of $\mathbb{R}^2$ can be built. What follows is another example, providing more intuitions on the process of exponentiation. 

\newpage
\vspace*{-1.5cm}
\dotfillb
\Example{\emph{Building manifold $S^2$ by exponentiation of the $so(3)$ algebra:} \\
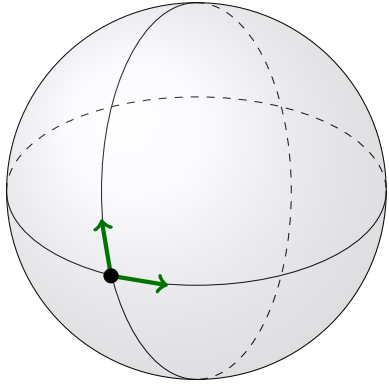
\begin{figure}[H]
	\centering
\begin{tikzpicture}[scale=2.5]
    \draw (-1,0) arc (180:360:1cm and 0.5cm);
    \draw[dashed] (-1,0) arc (180:0:1cm and 0.5cm);
    \draw (0,1) arc (90:270:0.5cm and 1cm);
    \draw[dashed] (0,1) arc (90:-90:0.5cm and 1cm);
    \draw (0,0) circle (1cm);
    \shade[ball color=blue!10!white,opacity=0.20] (0,0) circle (1cm);
    \draw[ultra thick,color=darkgreen, ->, shift={(-0.45,-0.45)}] (0,0) to (0.3,-0.05);
    \draw[ultra thick,color=darkgreen, ->, shift={(-0.45,-0.45)}] (0,0) to (-0.05,0.3);
    \fill[black,shift={(-0.45,-0.45)}] (0,0) circle (0.04cm);
\end{tikzpicture}
\vspace*{0.5cm}
    \caption{In $\mathbb{R}^3$, choosing a point at unit distance from the origin, one can exponentiate the generators of $SO(3)$ to build the manifold $S^2$.}
\label{fig so3}
\end{figure}
Consider the linear space $\mathbb{R}^3$, and a point in it at a unit distance from the origin. One can build the manifold $S^2$ by exponentiation of the $so(3)$ algebra as described below. By the choice of standard basis $\{\hat{i},\hat{j},\hat{k}\}$, a representation for the basis of $so(3)$ algebra would be as
\begin{equation}\label{so3 matrices}
\mathrm{l}_1=\begin{pmatrix}
0&1&0\\ -1&0&0\\0&0&0
\end{pmatrix}\,,\qquad 
\mathrm{l}_2=\begin{pmatrix}
0&0&-1\\ 0&0&0\\1&0&0
\end{pmatrix}\,,\qquad
\mathrm{l}_3=\begin{pmatrix}
0&0&0\\ 0&0&1\\0&-1&0
\end{pmatrix}\,.
\end{equation}
An arbitrary member of the algebra would be a linear combination of the matrices above. \\
Choosing a point in $\mathbb{R}^3$ at unit distance of the origin, marked by a vector $|v\rangle$, one can study the tangent space of $S^2$ at that point, by studying
\begin{equation}
|\delta v\rangle=(\lambda_1\mathrm{l}_1+\lambda_2\mathrm{l}_2+\lambda_3\mathrm{l}_3)|v\rangle
\end{equation}
for small constants $\lambda$s. Notice that although $|\delta v\rangle$ seems to be identified by three constants $\lambda$s, but the tangent space would be $2$-dim plane. It is because there is a $1$-dim subalgebra of $so(3)$ for which $|\delta v\rangle=0$. \\
By exponentiation the matrices \eqref{so3 matrices}, matrices of finite transformations, (\ie induced representation of $SO(3)$ group), would be found as
\begin{equation}
\mathrm{L}_1=\begin{pmatrix}
\cos \Lambda_1 &\sin \Lambda_1 &0\\ -\sin \Lambda_1&\cos \Lambda_1&0\\0&0&1
\end{pmatrix}\,,\quad 
\mathrm{L}_2=\begin{pmatrix}
\cos \Lambda_2&0&-\sin \Lambda_2\\ 0&1&0\\\sin \Lambda_2&0&\cos \Lambda_2
\end{pmatrix}\,,\quad
\mathrm{L}_3=\begin{pmatrix}
1&0&0\\ 0&\cos \Lambda_3&\sin \Lambda_3\\0&-\sin \Lambda_3&\cos \Lambda_3
\end{pmatrix}\,.
\end{equation}
$\Lambda$'s are some constants in $\mathbb{R}$. Then, acting on $|v\rangle$ by different powers of $\mathrm{L}$'s with arbitrary chosen constants $\Lambda$'s, would create (multiple covers of) the $S^2$ manifold.  Summarizing, we could build the manifold $S^2$, by picking a point $|v\rangle$, and exponentiating an algebra, the $so(3)$, illustrated in Figure \ref{fig so3}. 
}
\vspace*{-0.5cm}
\dotfille

For building the NHEG phase space, the initial point in the phase space would be $\bar{g}_{\mu\nu}$. The algebra, which is exponentiated, would be the algebra of generators $\bar{\chi}$, determined by some steps below \cite{CHSS:2015mza,CHSS:2015bca}. We can coin the name \emph{symplectic symmetry generators} for them, posteriorly motivated by Section \ref{sec symplectic symm}.  The coordinates in which $\bar{g}_{\mu\nu}$ and $g_{\mu\nu}$ are represented are denoted as $\bar{x}^\mu$ and $x^\mu$ respectively. In order to determine NHEG symplectic symmetry generators, We start with the most general vector fields
\begin{equation}
\bar{\chi}=\bar{\chi}^\mu (\bar{x})\bar{\partial}_{\mu}
\end{equation}
and constraint them through  the six conditions listed below.

\paragraph{ (1) $[\bar \chi,\bar\xi_{-}]=[\bar \chi,\bar\xi_{0}]=0$.} These conditions are supported as follows.
\begin{description}
\item[1.1] \emph{Similarity of the surfaces $\mathcal{H}$:} In Section \ref{sec similar H} it was discussed that by the $\{\bar \xi_-,\bar \xi_0\}$  isometry of $\bar {g}_{\mu\nu}$, the surfaces of constant $(\bar{t},\bar{r})$, denoted by $\mathcal{H}$, would be similar to each other. We request that the same would be true for surfaces of constant $(t,r)$, which for bookkeeping are also denoted by $\mathcal{H}$.  Then, given $g_{\mu\nu}$, any conserved charge can be defined through integrating over these $d\!-\!2$ dimensional surfaces.  However, there are infinitely many of such surfaces at any given $t_{_\mathcal H},r_{_\mathcal H}$.  We require that all such conserved charges be equal. According to a similar reasoning as in Appendix \ref{app H to H by xi}, this goal is achieved if $\xi_-=\partial_t$ and $\xi_0=t\partial_t-r\partial_r$ would be Killings of $g_{\mu\nu}$. The (1) guarantees it.
\item[1.2] \emph{Perturbations $\delta_{\chi}\Phi$ be $\{\xi_-,\xi_0\}$ isometric all over the phase space, specifically around $\bar{g}_{\mu\nu}$.} There are two main reasons for this request; 1) By this request, one would have $H_{\xi_a}[g]\!=\!0$  over the phase space (will be discussed in Section \ref{sec constant J}). 2) It can be shown \cite{HSS:2014twa} that beginning from finite perturbations in extremal BHs, and demanding them to remain finite after the near horizon limit, it automatically put the derived perturbations of $\bar{g}_{\mu\nu}$ in NHEG,  to be $\{\bar \xi_-,\bar \xi_0\}$  isometric.\\
In Appendix \ref{app isometries pert to commut} it is shown that requesting $\{\bar{\xi}_-,\bar{\xi}_0\}$ isometry for $\delta_\chi\Phi$ yields $[\bar \chi,\bar\xi_{-}]=[\bar \chi,\bar\xi_{0}]=0$, after discarding vectors $\bar\chi$ which are linear combinations of $\bar{\xi}_a$ Killings.
\item[1.3] \emph{To guarantee $\delta S=0$:} In Section \ref{sec EPL} it was shown that the necessary and sufficient condition for the EPL \eqref{EPL}, is $\bar{\xi}_-,\bar{\xi}_0$ invariance of the l.e.o.m-satisfying perturbations around $\bar{\Phi}$. On the other hand, for the perturbations in the tangent space of $\mathcal{M}$ at $\bar{g}_{\mu\nu}$,  according to Section \ref{sec constancy of delJ} it is expected to have $\delta J_i=0$. As a result, requesting \textbf{(1)} through the EPL yields  $\delta S=0$. 
\end{description}
The condition (1) fixes the $\bar t$ and $\bar r$ dependence of all components  of ${\bar \chi}$ (see Appendix \ref{app t r dependences})
\begin{equation}\label{chi-xi1,2}
{\bar \chi}=\frac{1}{\bar r}\epsilon^{\bar t}\partial_{\bar t}+\bar{r}\epsilon^{\bar r}\partial_{\bar r}+\epsilon^{\bar{\theta}}\partial_{\bar \theta}+\vec{\epsilon}\cdot\vec{\partial}_{\bar \varphi}\,.
\end{equation}
\paragraph{(2) $\bar\nabla_\mu {\bar \chi}^\mu=0$.} We require the volume element $\boldsymbol{\epsilon}$ \eqref{NHEG volume d-form}, to be the same for all elements in the phase space; i.e.  $\delta_{{\bar \chi}}\boldsymbol{\epsilon}=0$. Since $\boldsymbol{\epsilon}$ is covariant, $\delta_{{\bar \chi}}\boldsymbol{\epsilon}=\mathscr{L}_{\bar \chi} \boldsymbol{\epsilon}$. On the other hand,
\begin{equation}
\mathscr{L}_{\bar \chi} \boldsymbol{\epsilon}={\bar \chi}\cdot \mrd\boldsymbol{\epsilon}+\mrd({\bar \chi}\cdot \boldsymbol{\epsilon})=\mrd({\bar \chi}\cdot \boldsymbol{\epsilon})=\star(\bar{\nabla}_\mu {\bar \chi}^\mu)\,.
\end{equation}
Therefore, $\mathscr{L}_{\bar \chi} \boldsymbol{\epsilon}=0$ is equivalent to $\bar{\nabla}_\mu {\bar \chi}^\mu=0$.
\paragraph{(3)}\emph{$\delta_{\bar \chi}\mathbf{L}=0$}, where $\mathbf{L}=\frac{1}{16\pi G} R\boldsymbol{\epsilon}$ is the Einstein-Hilbert  Lagrangian $d$-form evaluated on the NHEG background \eqref{NHEG metric d-3} before imposing the equations of motion. The functional form of $\Gamma(\bar{\theta})$ and $\gamma_{ij}(\bar{\theta})$ is therefore arbitrary except for the regularity conditions. According to $\mathbf{L}=\star \mathcal{L}$ and the request (2), $\mathscr{L}_{\bar\chi} \mathcal{L}=0$. But $\mathcal{L}$ is a scalar density built from the metric, and it is invariant under the background $SL(2,\mathbb{R})\!\times\! U(1)^{d-3}$ isometries. So by Appendix \ref{app t r dependences} it only admits $\bar \theta$ dependence, yielding $\mathscr{L}_{\bar\chi} \mathcal{L}=\bar\chi^{\bar\theta}\partial _\theta \mathcal{L}=0$. Therefore we find
\begin{equation}
\epsilon^{\bar\theta}=0,
\end{equation}
and then, by (2) 
\begin{equation}
\epsilon^{\bar r}=-\vec{\partial}_{\bar \varphi}\cdot \vec{\epsilon}\,.
\end{equation}
\paragraph{(4)} $ \epsilon^{\bar t} = - b\, \vec{\partial}_{\bar \varphi}\cdot \vec{\epsilon}$, for $b=\pm 1$.
This condition can be motivated from two different perspectives: 
\begin{description}

\item[4.1] \emph{Invariance of one of the $\mathcal{N}_\pm$:} As discussed in Section \ref{sec causal structure}, the NHEG has two null geodesic congruences generated by $\bar{\ell}_+$ and $\bar{\ell}_-$ which are respectively normal to constant $\bar{v} = \bar{t}+\frac{1}{\bar{r}}$ and $\bar{u}=\bar{t}-\frac{1}{\bar{r}}$ surfaces, the $\bar{\mathcal{N}}_\pm$  \cite{Durkee:2010ea}. We request that either $\mathscr{L}_{\bar \chi} \bar{v}=0$ or $\mathscr{L}_{\bar \chi} \bar u=0$. It implies that each element in the phase space will have one of surfaces of constant $v=t+\frac{1}{r}$ and $u=t-\frac{1}{r}$, denoted by $\mathcal{N}_+$ and $\mathcal{N}_-$, as one branch of its bifurcate Killing horizons. This request yields the condition (4).

\item[4.2] \emph{Regularity of $\mathcal H$ surfaces:} As we will discuss in section \ref{sec finite trans}, the condition (4) ensures that constant $t,r$ surfaces $\mathcal H$ are regular without singularities at poles on each element of the phase space. 

\end{description}

\paragraph{(5)} \emph{$\vec{\epsilon}$ are $\bar\theta$-independent and periodic functions of $\bar \varphi^i$.} We impose these conditions as they guarantee (i) smoothness of the $(t,r)$ constant surfaces ${\mathcal H}$ of each element of the phase space, as we will show below in section \ref{sec finite trans}, and (ii) constancy of the area of ${\mathcal H}$  and the angular momenta $\vec{J}$ over the phase space, as we will also show in the sections \ref{sec finite trans} and \ref{sec constant J}.

\paragraph{(6)}  \emph{Finiteness, conservation and regularity of the symplectic structure.} 
These final conditions crucially depend on the definition of the symplectic structure which will be presented in Section \ref{sec symplectic structure}. Our analysis reveals that additional conditions are required in order to obtain a well-defined  symplectic structure. After fixing $\mathbf{Y}$ ambiguities in the symplectic structure, we found the generators $\vec{\epsilon} = \vec{k} \epsilon(\bar \varphi^1,\dots \bar \varphi^{d-3})$, where $\epsilon$ is arbitrary function periodic in all its $d\!-\!3$ variables. The $SL(n,\mathbb{Z})$ covariance is also used to discard other possibilities (\eg see Appendix C in \cite{CHSS:2015bca}).

As a result, we end up with the following NHEG symplectic symmetry generator 
\begin{equation}\label{ASK-1}
\bar\chi[{\epsilon}(\vec{\bar \varphi})]=-\vec{k}\cdot\vec{\partial}_{\bar\varphi}{\epsilon}\;(\dfrac{b}{\bar r}\partial_{\bar t}+\bar r\partial_{\bar r})+{\epsilon}\vec{k}\cdot\vec{\partial}_{\bar\varphi}
\end{equation}
with $b= \pm 1$ which generates the infinitesimal perturbations tangent to the phase space around the background, $\delta\Phi[{\epsilon}(\vec{\bar\varphi})]=\mathscr{L}_{\bar\chi} \bar{\Phi}$.

Each one of the two families of the generators labelled by $b=\pm 1$, constitute a closed algebra as
\begin{equation}
[\bar \chi [\epsilon_1], \bar\chi [\epsilon_2]]=\bar \chi [\epsilon_3] \qquad \epsilon_3 =\epsilon_1 \vec{k}\cdot \vec{\partial}_{\bar\varphi} \epsilon_2-\epsilon_2 \vec{k}\cdot \vec{\partial}_{\bar\varphi} \epsilon_1\,.
\end{equation}
These two family of generators $b = \pm 1$, are related to each other by $Z_2$ isometries of the background. Hence the two phase spaces built with either of these choices are mapped to each other by those $Z_2$ isometries, proved in Appendix \ref{app Z2}. Without loss of generality we choose $b=+1$.


\subsection{Exponentiation of the symplectic symmetry generators}\label{sec finite trans}

Having the point $\bar{g}_{\mu\nu}$ in the phase space, and equipped with the generators \eqref{ASK-1}, the exponentiation can be performed to build the $\mathcal{M}$ \cite{CHSS:2015bca}. Technical calculations are provided in Appendix \ref{app finite trans}. Here the results are summarized.
\begin{itemize}
\item Beginning from coordinates $\bar{x}^\mu$, the finite coordinate transformations relating it to $x^\mu$ would be as
\begin{align}\label{finite transformations}
\bar{t} =t-\frac{1}{r}(e^{\Psi}-1),  \quad\qquad \bar{r} =re^{-{\Psi}},\quad\qquad \bar{\theta}=\theta\,,\quad\qquad \bar{\varphi}^i=\varphi^i + k^i F\,,
\end{align}
with \emph{wiggle function} $F(\vec{\varphi})$ periodic in all of its arguments, and
\begin{align}\label{Psi-def}
e^{\Psi}=1+\vec{k}\cdot\vec{\partial}_\varphi F.
\end{align}
\item $g_{\mu\nu}$ as different points of the $\mathcal{M}$ are identified with the wiggle function $F(\vec{\varphi})$ as
\begin{equation}\label{finalphasespace}
\mrd s^2=\Gamma(\theta)\Big[-\left( \boldsymbol\sigma -  \mrd \Psi \right)^2+\Big(\dfrac{\mrd r}{r}-\mrd{\Psi}\Big)^2+\mrd\theta^2+\gamma_{ij}(\mrd\tilde{\varphi}^i+{k^i}\boldsymbol{\sigma})(\mrd\tilde{\varphi}^j+{k^j}\boldsymbol{\sigma})\Big], 
\end{equation}
in which
\begin{align}
\boldsymbol{\sigma}=e^{-{\Psi}}r\,\mrd(t+\frac{1}{r})+\dfrac{\mrd r}{r},\qquad \tilde{\varphi}^i=\varphi^i+k^i (F-{\Psi})\,.
\end{align}
\item The metric \eqref{finalphasespace} has $SL(2,\mathbb{R})\!\times\!U(1)^{d-3}$ isometry generated by
\begin{flalign}\label{phase space isometry}
&\xi_- =\partial_t\,,\qquad \xi_0=t\partial_t-r\partial_r, &\nonumber \\
&\xi_+ =\dfrac{1}{2}(t^2+\frac{1}{r^2})\partial_t-tr\partial_r-\frac{1}{r}{k}^i{\partial}_{\varphi^i}+\dfrac{1}{r}\vec{k}\cdot\vec{\partial}_\varphi (F-{\Psi}){\eta_+},&\nonumber\\
&\mathrm{m}_i=\left(\delta^j_i-e^{-\Psi}k^j\partial_i F\right)\partial_{\varphi^j} +(\partial_i {\Psi}-e^{-\Psi}\vec{k}\cdot\vec{\partial}_\varphi  {\Psi}\partial_i F) {\eta_+},&
\end{flalign}
where $\eta_+$ is defined as (read more about it in Appendix \ref{app eta})
\begin{equation}
\eta_+ = \frac{1}{r}\partial_t+r\partial_r\,.
\end{equation}
\item Following Appendix \ref{app finite trans}, the $\chi$ responsible for generation of $\mathcal{M}$ around any point $g_{\mu\nu}$, have the same form as \eqref{ASK-1}. More explicitly, on any given point of the $\mathcal{M}$, generically we have the family of symplectic symmetry generators as
 \begin{equation}\label{ASK}
\chi[{\epsilon}(\vec{ \varphi})]=-\vec{k}\cdot\vec{\partial}_{\varphi}{\epsilon}\;(\dfrac{b}{r}\partial_{t}+ r\partial_{r})+{\epsilon}\vec{k}\cdot\vec{\partial}_{\varphi}
\end{equation}
for $b=1$. It is the push-forward of the vector \eqref{ASK-1}, in which 
\begin{equation}\label{chi push forward}
\bar\epsilon(\vec{\bar\varphi})=(1+\vec{k}\cdot \vec{\partial}_\varphi F)\ {\epsilon}(\vec{\varphi})
\end{equation}
where $\epsilon$ in \eqref{ASK-1} is denotes by $\bar\epsilon$. It emphasizes that although we have used $g_{\mu\nu}[F\!=\!0]$ as a seed for the building the phase space, but equivalently any other points $g_{\mu\nu}[F]$ and the generators \eqref{ASK} could be used.
\end{itemize}
According to \eqref{finalphasespace}, the metric induced on the  $\mathcal{H}$s, \ie the surfaces of constant $(t,r)$, is
\begin{equation}
\mrd s^2_{_\mathcal{H}}=\Gamma(\mrd\theta^2+\gamma_{ij}\mrd\tilde{\varphi}^i \mrd\tilde{\varphi}^j)\,. 
\end{equation}
 Comparing it with the metric of the same surfaces in $g_{\mu\nu}[F\!=\!0]$, it is manifestly smooth. Following Appendix \ref{app finite trans}, this smoothness is crucially a result of $b=\pm 1$, and $\theta$-independence of $\epsilon$. In addition, it shows that the area of the surfaces $\mathcal{H}$ are invariant over the phase space, in agreement with constancy of the entropy.
 
Note that by $F\!=\!0$, $\bar{x}^\mu=x^\mu$, so we can drop the \emph{bar} from $\bar{g}_{\mu\nu}$ and $\bar{x}^\mu$ and any other things related to them. It would be simply $g_{\mu\nu}[F\!=\!0]$, with coordinates $x^\mu$. To the end, we will adopt this more convenient notation instead of \emph{bar} notation.

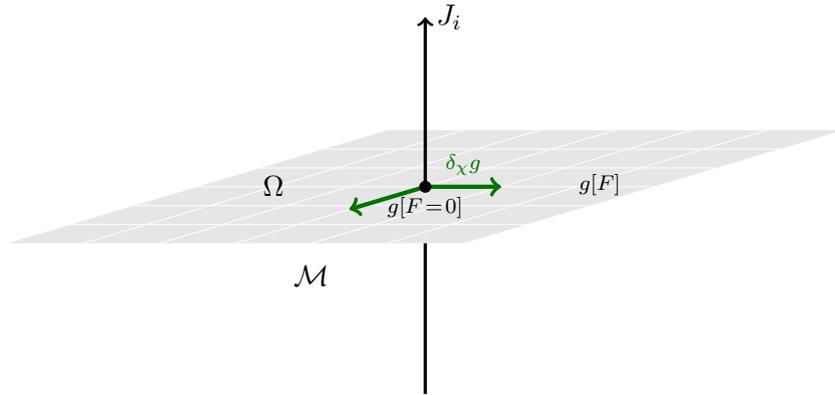
\begin{figure}[H]
	\centering
\begin{tikzpicture}[scale=1]

    \draw[very thick] (5.5,1) to (5.5,3.3);
    \begin{scope}[shift={(0,3)}]
        \fill[even odd rule,light gray]
    (0,0) to (5,1.5) to  (11,1.5) to
    (6,0) to (0,0);
    \draw[very thin,white] (1,0) to (6,1.5);
    \draw[very thin,white] (2,0) to (7,1.5);
    \draw[very thin,white] (3,0) to (8,1.5);
    \draw[very thin,white] (4,0) to (9,1.5);
    \draw[very thin,white] (5,0) to (10,1.5);
    \draw[very thin,white] (0.7,0.25) to (6.9,0.25);
    \draw[very thin,white] (1.5,0.5) to (7.7,0.5);
    \draw[very thin,white] (2.2,0.75) to (8.6,0.75);
    \draw[very thin,white] (3.3,1) to (9.3,1);
    \draw[very thin,white] (4.0,1.25) to (10.2,1.25);
    \end{scope}
    \draw[very thick,->] (5.5,3.75) to (5.5,6) node [anchor=west]{$J_i$};
    \draw (3.5,3.5) node[anchor=south]{$\Omega$};
    \draw (7.8,3.5) node[anchor=south]{\scriptsize $g[F]$};
    \draw (5.5,3.75) node[anchor=north]{\scriptsize ${g}[F\!=\!0]$};
    \draw[ultra thick, color=darkgreen, ->] (5.5,3.75) --node[midway,above]{\scriptsize $\delta_{\chi}g$} (6.5,3.75);
    \draw[ultra thick, color=darkgreen, ->] (5.5,3.75) --(4.5,3.75-1.5/5);
    \draw (4,2.3) node[anchor=south]{$\mathcal{M}$};
    \fill (5.5,3.75) circle (0.08cm);
    \end{tikzpicture}
    \caption{NHEG phase space manifold $\mathcal{M}$ is built by the exponentiation of the NHEG symplectic symmetry generators $\chi$.}
\label{fig exponentiation}    
\end{figure}

Figure \ref{fig exponentiation} summarizes the process of exponentiation of NHEG symplectic symmetry generators schematically.


\section{Symplectic structure}\label{sec symplectic structure}
In the previous section, the NHEG phase space manifold $\mathcal{M}$ was built. The aim of this section is to equip that manifold to an appropriate  symplectic $2$-form $\Omega$. Pragmatically, it would be done by fixing the $\mathbf{Y}$ ambiguity in the LW symplectic $2$-form \eqref{Omega LW}.  An important concept needed, is the concept of ``symplectic symmetry" introduced in the following. 

\subsection{Symplectic symmetry}\label{sec symplectic symm}
The manifold $\mathcal{M}$ is built by some coordinate transformations generated by $\chi$ \eqref{ASK}. Different points of the manifold are metrics which are distinguished by the wiggle function $F$. An important question would be ``how can one consider different metrics $g_{\mu\nu}[F]$, related to each other by some coordinate transformations, as \emph{physically} distinct points of the phase space?". The answer to this question is based on the Hamiltonian generator associated to the $\chi$. In order to have physically distinct points, $\delta H_\chi$ needs to be well-defined, conserved, and integrable. In addition, $H_\chi$ needs to be finite  and non-trivial (\ie not to be constant) function over the phase space. 

In general, the vector fields $\xi$ for which $H_\xi$ is constant over the phase space are called \emph{pure gauges}, while the ones for which $H_\xi$ is not finite, are not allowed. 

Generically, $\delta H_\chi$ is not guaranteed to be conserved. In Lemma \ref{Lemma conserv H} it was shown that for the exceptional case of isometry generators of any point of $\mathcal{M}$, the conservation is guaranteed on all points of the $\mathcal{M}$. The origin of this conservation is  $\boldsymbol{\omega}\!\approx\! 0$ for those generators. But for sure, $\chi$ can not be an isometry generator, otherwise, $\delta_\chi \Phi=0$, prohibiting the exponentiation process. Here, we enlarge that exceptional case, to include the vectors $\chi$.
\begin{Definition}\label{definition symplectic sym}
Considering any arbitrary $\Phi$ and $\delta \Phi$ in $\mathcal{M}$ and its tangent space, then the vector fields $\chi$ for which
\begin{enumerate}
\item $\boldsymbol{\omega}(\delta\Phi,\delta_\chi\Phi,\Phi)\approx 0$,
\item $\delta H_\chi$ be integrable,
\item $H_\chi$ be finite over the $\mathcal{M}$,
\end{enumerate}
are called \emph{symplectic symmetry generators} \cite{CHSS:2015mza}. Then, \emph{symplectic symmetry} is defined as mapping $\mathcal{M}$  to itself by the action of $H_\chi$.  
\end{Definition}
Notice that in the general covariant theories, diffeomorphism is a symmetry. Hence, symplectic symmetry is a subset of diffeomorphism symmetry. If the manifold $\mathcal{M}$ is generated by the family of $\chi$'s, then $\delta_\chi \Phi$ spans the tangent space to the manifold. So, by the same reasoning as in Lemma \ref{Lemma conserv H}, one can show that $\delta H_\chi$ is conserved on-shell, independent of the choice of $\mathscr{S}$. Besides, if $\chi$ would be a field independent vector field, then by \eqref{integ cond omega}, $\delta H_\chi$ would be automatically integrable. Let's emphasize this issue in the following lemma, for later usage.

\dotfillb
\Lemma{\label{Lemma symplectic symm integ}\emph{If $\mathcal{M}$ is built by exponentiation of some field independent symplectic symmetry generators $\chi$, then  $\delta H_\chi$ is integrable.} \\
By the integrability condition \eqref{integ cond omega}, we deduce that $\delta H_\chi$ is integrable if $\boldsymbol{\omega}(\delta_1 \Phi,\delta_2\Phi,\Phi)\approx 0$. But this condition is gauranteed by the first condition in Definition \ref{definition symplectic sym}. The reason is that the tangent bundle of $\mathcal{M}$ is spanned by the generators $\chi$. Therefore $\delta_1 \Phi$ and $\delta_2 \Phi$ have to be of type $\delta _\chi\Phi$. Hence the lemma is proved. Notice that by the ``field independent", the independency of $\chi$ from the points in $\mathcal{M}$ is understood. It should be considered, because in the derivation of \eqref{integ cond omega}, $\delta \chi=0$ has been assumed.
}
\dotfille

In order to understand the field independency of $\chi$'s in our analysis, a regression is needed; the NHEG phase space manifold $\mathcal{M}$ is composed of the metrics $g_{\mu\nu}$ in \eqref{finalphasespace}, represented in the coordinates $x^\mu$ and identified by the wiggle function $F(\vec{\varphi})$. The NHEG symplectic generators are the vector fields $\chi$ in \eqref{ASK}, which also depend on $F(\vec{\varphi})$ through \eqref{chi push forward}. At any point of the $\mathcal{M}$, the $\delta _{\chi}g_{\mu\nu}$ span the tangent space of the manifold in that point. In spite of the  $F(\vec{\varphi})$ dependence of the push-forwarded $\chi$, one can absorb the $F(\vec{\varphi})$ dependence to the $\epsilon(\vec{\varphi})$. As a result, focusing on the \emph{set} of $\chi$, there is not any privileged point on $\mathcal{M}$.  Therefore, without lose of generality, we can study the reference point $g_{\mu\nu}[F\!=\!0]$ and the $\chi$'s constituting the tangent space at that point  
 \begin{equation}
\chi[{\epsilon}(\vec{ \varphi})]=-\vec{k}\cdot\vec{\partial}_{\varphi}{\epsilon}\;(\dfrac{b}{r}\partial_{t}+ r\partial_{r})+{\epsilon}\vec{k}\cdot\vec{\partial}_{\varphi}\,.
\end{equation}
Now, this vectors can be considered as $F$ independent vectors over the whole phase space. They induce vectors $\delta_\chi \Phi$ on the tangent bundle of $\mathcal{M}$. Similar to any other field independent (here $F$ independent) vector field, one can study their Hamiltonian generators $\delta H_\chi$. Thanks to Lemma \ref{Lemma symplectic symm integ}, $\delta H_\chi$ are integrable. So we can use notation $H_\chi$ for their integrated Hamiltonians.

\subsection{Fixing ambiguities}
Now we are ready to fix the ambiguities in the LW symplectic $2$-form \eqref{Omega LW}. Notice that the tangent space to the NHEG phase space manifold $\mathcal{M}$ is spanned by the vectors $\delta_\chi\Phi$. Hence we already know the perturbations well. Reminding that the only ambiguity in the LW symplectic $2$-form is of the $\mathbf{Y}$ type, then the NHEG symplectic structure is
\begin{equation}\label{Omega NHEG}
\Omega(\delta_{\chi_1}\Phi,\delta_{\chi_2}\Phi,\Phi)=\int_\Sigma \boldsymbol{\omega}(\delta_{\chi_1}\Phi,\delta_{\chi_2}\Phi,\Phi)\, 
\end{equation}
where
\begin{equation}\label{omega NHEG}
\boldsymbol{\omega}=\delta_{\chi_1}\mathbf{\Theta}(\delta_{\chi_2}\Phi,\Phi)-\mathscr{L}_{\chi_2}\mathbf{\Theta}(\delta_{\chi_1}\Phi,\Phi)+\mrd \big(\delta_{\chi_1}\mathbf{Y}(\delta_{\chi_2}\Phi,\Phi)-\mathscr{L}_{\chi_2}\mathbf{Y}(\delta_{\chi_1}\Phi,\Phi)\big)\,,
\end{equation}
for $\mathbf{\Theta}$ to be \eqref{EH Theta}, and some $(d\!-\!2)$-form $\mathbf{Y}$ linear in $\delta_\chi\Phi$. Hence the Poisson bracket can be written in terms of $\boldsymbol{k}_\chi$ as
\begin{equation}\label{Poisson NHEG}
\{H_{\chi_2},H_{\chi_1}\}\equiv \delta _{\chi_1}H_{\chi_2}=\oint_\mathcal{H}  \boldsymbol{k}_{\chi_2}(\delta_{\chi_1}\Phi,\Phi)=\oint_\mathcal{H}\Big(\boldsymbol{k}^{\text{EH}}_{\chi_2}(\delta_{\chi_1}\Phi,\Phi)+\boldsymbol{k}^{\mathbf{Y}}_{\chi_2}(\delta_{\chi_1}\Phi,\Phi)\Big)\,,
\end{equation} 
in which $\boldsymbol{k}^{\text{EH}}$ is \eqref{EH k} and $\boldsymbol{k}^{\mathbf{Y}}$ waits to be determined.
It can be fixed by the following considerations:
\begin{enumerate}
\item According to the Section \ref{sec symplectic symm}, we request the $\chi$ to be symplectic symmetry generator, \ie make the \eqref{omega NHEG} vanish on-shell. 
\item We request that $\delta_{\chi_1} H_{\chi_2}$ be independent of the $b$ in $\chi$. 
\end{enumerate}
The first condition makes $\delta H_{\chi}$ to be conserved, in addition to regularizing the \eqref{omega NHEG} (see Appendix \ref{app regularizing} to see why LW $2$-form is singular). 

In Appendix \ref{app Y fix} it is shown that the first request fixes the $\mathbf{Y}$ upto a complementary $\mathbf{Y}_{\text{c}}$ for which $\mrd \big(\delta_{\chi_1}\mathbf{Y}_{\text{c}}(\delta_{\chi_2}\Phi,\Phi)-\mathscr{L}_{\chi_2}\mathbf{Y}_{\text{c}}(\delta_{\chi_1}\Phi,\Phi)\big)\approx 0$. The result is 
\begin{equation}\label{NHEG Y theta}
\mathbf{Y}(\delta\Phi,\Phi)=-\eta_b\cdot\mathbf{\Theta}(\delta_\chi\Phi,\Phi)+\mathbf{Y}_{\text{c}}(\delta\Phi,\Phi)\,,
\end{equation}
where $\eta_b$ is the invariant vector 
\begin{equation}
\eta_b=\dfrac{b}{r}\partial_t+r\partial_r\,.
\end{equation}
The second condition then fixes \cite{CHSS:2015bca}  $\mathbf{Y}_{\text{c}}$ to be
\begin{equation}
\mathbf{Y}_{\text{c}}=\left( \frac{1}{\Gamma} \delta g_{\alpha\beta}\;\eta_b^{\;\alpha} \eta_2^{\;\beta }\right)\;\boldsymbol{\epsilon}_{_\perp}\,,
\end{equation}
in which $\eta_2$ and $\boldsymbol{\epsilon}_{_\perp}$ are push-forwarded version of $\bar{\eta}_2\equiv\frac{1}{\bar r}\partial_{\bar t}$ and \eqref{binormal form} respectively. Hence 
\begin{equation}\label{NHEG Y}
\mathbf{Y}=-\eta_b\cdot\mathbf{\Theta}(\delta_\chi\Phi,\Phi)+\left( \frac{1}{\Gamma} \delta g_{\alpha\beta}\;\eta_b^{\;\alpha} \eta_2^{\;\beta }\right)\;\boldsymbol{\epsilon}_{_\perp}\,.
\end{equation}
Summarizing, the NHEG symplectic structure is fixed to be
\begin{equation}
\delta _{\chi_1}H_{\chi_2}=\oint_\mathcal{H}\boldsymbol{k}^{\text{EH}}_{\chi_2}(\delta_{\chi_1}\Phi,\Phi)+\oint_{\mathcal{H}}\Big(\delta_{\chi_1}\mathbf{Y}(\delta_{\chi_2}\Phi,\Phi)-\mathscr{L}_{\chi_2}\mathbf{Y}(\delta_{\chi_1}\Phi,\Phi)\Big)
\end{equation}
in which $\mathbf{Y}$ is the \eqref{NHEG Y}. Thanks to the $\{\xi_-,\xi_0\}$ isometry of $g_{\mu\nu}[F]$, $\chi$, $\eta$ and $\boldsymbol{\epsilon}_{_\perp}$, by the Lemma \ref{Lemma NHEG H independence}, $\delta_{\chi_1} H_{\chi_2}$ is an $\mathcal{H}$-independent Hamiltonian. 

\subsection{Checking constancy of NHEG charges over the phase space}\label{sec constant J}
Being equipped to a symplectic structure with symplectic symmetry, the constancy of $H_{\mrm_i}$ and $H_{\xi_a}$ can be assessed. Notice that $\mrm_i$ and $\xi_a$ are the vectors in \eqref{phase space isometry}, which are Killing vectors of all points in the $\mathcal{M}$. 

The vectors $\{\xi_{-},\xi_0\}$ are isometry of all points of $\mathcal{M}$, so
\begin{equation}\label{J constancy proof 1}
\delta_{\xi_{a}} H_\chi=\oint_\mathcal{H}\boldsymbol{k}_\chi (\delta_{\xi_a}\Phi,\Phi)=0 \,,\qquad \forall \Phi\in \mathcal{M},\quad a\in\{-,0\}\,.
\end{equation}
The last equation, is a result of the Killing condition and the linearity of $\boldsymbol{k}_\chi$ in $\delta_{\xi_a}\Phi$. On the other hand, by the Lemma \ref{Lemma symplectic symm integ}, $\delta_\chi H_{\xi_a}$ is integrable. So 
\begin{equation}
\{H_\chi,H_{\xi_a}\}=-\delta _\chi H_{\xi_a}=\delta_{\xi_{a}} H_\chi=0\,,\qquad \forall \Phi\in \mathcal{M},\quad a\in\{-,0\}\,.
\end{equation}
It vanishes because of the \eqref{J constancy proof 1}. Hence, $H_{\xi_a}$ for $a\!=\!-,0$ are constant over the whole manifold. The reference point can be chosen to put them equal to zero. So, they are zero all over the phase space.

The $\xi_+$ and $\mrm_i$ in \eqref{phase space isometry}, are field dependent, \ie depend on the wiggle function. So the integrability condition \ref{integ cond omega}, and therefore Lemma \ref{Lemma symplectic symm integ} are not necessarily true for them. But if their Hamiltonian would be integrable, similar argument as above could prove their constancy. In Appendix \ref{app constancy of J} we prove their constancy using another method. 

By the arguments above, $J_i$ and $H_{\xi_a}$ which are conserved Hamiltonian generators associated to $\mrm_i$ and $\xi_a$ on any point of $\mathcal{M}$, are constants over the phase space, \ie
\begin{equation}
\{H_\chi,J_i\}=\{H_\chi,H_{\xi_a}\}=0\,,
\end{equation}
in agreement with Section \ref{sec constancy of delJ}.

\chapter{The NHEG algebra}\label{chap NHEG algebra}

In the first section of this chapter, the Hamiltonian generators for the symplectic symmetry generators $\chi$ are calculated, and the algebra of them is introduced, which is dubbed \emph{NHEG algebra}. Then, in the second section, some discussions on the quantization and probable dual field theory description are provided.


\section{NHEG symplectic symmetry algebra}\label{sec NHEG symplectic symmetry}

In order to delve into the algebraic properties of the introduced $\chi$'s, one can expand the periodic function $\epsilon(\vec{\varphi})$ in its Fourier modes, as
\begin{equation}
\epsilon(\vec{\varphi})=- \sum _{\vec{n}} c_{\vec{n}}\,e^{- i(\vec{n}\cdot \vec{\varphi})}\,
\end{equation}
for some constants $c_{\vec{n}}$ and $\vec{n}\equiv (n_1,n_2,\dots,n_n)$, $n_i\in \mathbb{Z}$.
 Therefore the generator $\chi$ decomposes as
\begin{equation}\label{Ln expansion}
\chi=\sum _{\vec{n}} c_{\vec{n}}\chi_{_{\vec{n}}}\,,
\end{equation}
where
\begin{equation}\label{Final Ln}
\chi_{_{\vec{n}}}=-e^{-i(\vec{n}\cdot\vec{\varphi})}\bigg(i(\vec{n}\cdot\vec{k})(\frac{1}{r}\partial_t+ r\partial_r)+\vec{k}\cdot\vec{\partial_{\varphi}} \bigg)\,.
\end{equation}
The Lie bracket between two such Fourier modes is given by
\begin{equation}\label{chi-algebra}
\left[\chi_{_{\vec{m}}},\chi_{_{\vec{n}}}\right]_{L.B.}= -i\vec{k}\cdot(\vec{m}-\vec{n})\chi_{_{\vec{m}+\vec{n}}}\,.
\end{equation}
The algebra of the associated Hamiltonians would be similar to \eqref{chi-algebra}, except for a possible central extension
\begin{align}\label{central extension intro}
&\{H_{\vec{m}}, H_{\vec{n}}\} = -i\vec{k} \cdot (\vec{m}- \vec{n}) H_{\vec{m}+\vec{n}} + C_{\vec{m},\vec{n}}\,\\
&\{H_{\vec{p}}, C_{\vec{m},\vec{n}} \} =0, \qquad \forall\, \vec{p},\vec{n},\vec{m}.\label{central extension commut}
\end{align}
The aim of the remaining of this section would be calculating $H_{\vec{n}}$ and $C_{\vec{m},\vec{n}}$. 

At first, the central extension can be found. According to \eqref{central extension commut}, $\delta_\chi C_{\vec{m},\vec{n}}=0$ all over the phase space. So $C_{\vec{m},\vec{n}}$ is a constant function over the phase space. Therefore, it can be calculated on the reference point $g_{\mu\nu}[F\!=\!0]$. Details of the calculation are sent to Appendix \ref{app central extension}. The final result is that by the choice of references as
\begin{align}\label{reference of Hn}
&H_{\chi_{\vec{n}}}[F\!=\!0]=0\,, \qquad\quad \,\, \vec{n}\neq \vec{0},\cr
&H_{\chi_{\vec{0}}}[F\!=\!0]=\vec{k}\cdot\vec{J}\,, 
\end{align}
the central extension is
\begin{align}\label{Central extension}
 C_{\vec{m},\vec{n}}&= -i (\vec{k}\cdot \vec{m})^3 \frac{S}{2\pi}\,\delta_{\vec{m}+\vec{n},0},
\end{align}
in which $S$ is the entropy of the NHEG. Notice that different choices of references in \eqref{reference of Hn}, only changes the central extension by a term proportional to $(\vec{k}\cdot \vec{m})$.

For the calculation of $H_{\vec{n}}$ over the phase space, by \eqref{central extension intro} one has the followings for $\vec{n}\neq \vec{0}$
\begin{align}\label{H_n}
H_{\vec{n}}&=\frac{i}{\vec{k}\cdot\vec{n}}\{H_{\vec{n}}, H_{\vec{0}}\}\,, \\
H_{\vec{0}}&=\dfrac{i}{2\vec{k} \cdot \vec{n}}\left(\{H_{\vec{n}}, H_{-\vec{n}}\}+i(\vec{k}\cdot \vec{n})^3\frac{S}{2 \pi} \right).
\end{align}
The RHS of the equations above can be calculated on an arbitrary point of $\mathcal{M}$. The details are in Appendix \ref{app finite charges}, resulting to
\begin{align}\label{charge}
H_{\vec{n}}&= \oint_\mathcal{H} \boldsymbol{\epsilon}_\mathcal{H}\  T[\Psi] e^{-i \vec{n}\cdot \vec{\varphi} },
\end{align}
where $\boldsymbol{\epsilon}_\mathcal{H}$ is the volume form on $\mathcal H$ and
\begin{align}\label{resd4}
T[\Psi]&= \frac{ 1}{16 \pi G}  \Big( (\Psi' )^2 -2 \Psi''+ 2 e^{2 \Psi  } \Big)
\end{align}
where primes are directional derivatives along the vector $\vec{k}$, \ie $\Psi' = \vec{k}\cdot \vec{\partial}_\varphi \Psi$. 
The charges $H_{\vec{n}}$ are therefore the Fourier modes of $T[\Psi]$.


\section{More about the NHEG algebra}

\subsection{Probable Liouville field description}\label{sec Liouville field}
In order to understand the result in \eqref{charge} more, it is interesting to first note how the wiggle function $F$ transforms under a symplectic symmetry transformation generated by $\chi[\epsilon]$. To this end, we recall that by construction 
\begin{equation}
\mathscr{L}_{\chi[\epsilon]}(g_{\mu\nu}[F]) = g_{\mu\nu}[F+\delta_\epsilon F]-g_{\mu\nu}[F].
\end{equation}
We find 
\begin{align}
\delta_{\epsilon} F=  (1 +\vec{k}\cdot \vec{\partial}_\varphi F )\epsilon=e^{{\Psi}}\epsilon.
\end{align}
The field $\Psi$ then transforms as
\begin{align}\label{Psi-transform}
\delta_\epsilon {\Psi} =\epsilon\,  {\Psi}'+   \epsilon'.
\end{align}
where prime denotes again the directional derivative $\vec{k}\cdot \vec{\partial}_\varphi $. Therefore, $\Psi$ transforms like a Liouville field. In particular, note that $\delta_\epsilon e^{{\Psi}}=(e^{{\Psi}}\epsilon)'$ and hence $e^{{\Psi}}$ resembles a ``weight one operator'' in the terminology of conformal field theory. It is then natural to define the Liouville stress-tensor
\begin{align}\label{T-tensor}
T[\Psi]&= \frac{ 1}{16 \pi G}  \Big( (\Psi' )^2 -2 \Psi''+ \Lambda e^{2 \Psi  } \Big)
\end{align}
with ``cosmological constant'' $\Lambda$, which according to \eqref{resd4} is found to be $\Lambda = 2$. In addition, $T[\Psi]$ transforms as 
\begin{align}\label{T-transform}
\delta_\epsilon T = \epsilon  T' + 2  \epsilon' T - \frac{1}{8\pi G}  \epsilon'''.
\end{align}
Expanding in Fourier modes as in \eqref{charge}, it is straightforward to check from the transformations \eqref{T-transform} that the algebra \eqref{central extension intro} is recovered. 

The above resembles the transformation of the energy momentum tensor, a ``quasi-primary operator of weight two''. However, we would like to note that ${\Psi}$ and hence $T[\Psi]$ are not function of time but are functions of all coordinates $\varphi^i$, in contrast with the standard Liouville theory.

Given \eqref{charge} and \eqref{resd4}, one can immediately make the following interesting observation; the charge associated with the zero mode $\vec{n}=0$, $H_{\vec{0}}$, is positive definite over the whole phase space. This is due to the fact that the $\partial^2 \Psi$ term does not contribute to $H_{\vec{0}}$ and the other two terms in \eqref{T-tensor} give positive contributions.

\subsection{Quantization of the NHEG algebra}
One can use the Dirac quantization rules
\begin{equation}\label{quantization}
\{\quad\}\to \frac{1}{i}[\quad]\,,\qquad\mathrm{and}\qquad  H_{\vec{n}}\to  \,L_{\vec{n}},
\end{equation} 
to  promote the symmetry algebra to an operator algebra, the \emph{NHEG algebra} $\widehat{\mathcal{V}_{\vec{k},S}}$
\begin{align}\label{NHEG-algebra quantized}
[L_{\vec{m}}, L_{\vec{n}}] = \vec{k} \cdot (\vec{m}- \vec{n}) L_{\vec{m}+\vec{n}} +  \frac{S}{2\pi}(\vec{k}\cdot \vec{m})^3 \delta_{\vec{m}+\vec{n},0}\,.
\end{align}
According to Section \ref{sec constant J}, the $J_i$ and $H_{\xi_a}$ commute with $L_{\vec{n}}$, and are therefore central elements of the NHEG algebra  $\widehat{\mathcal{V}_{\vec{k},S}}$. Also by Definition \ref{definition symplectic sym}, they are symplectic symmetry generators. Hence, the \emph{full symplectic symmetry of the phase space} is
\begin{equation}\label{full-algebra}
\mathrm{NHEG \ Symplectic \ Symmetry\ Algebra}=\widehat{\mathcal{V}_{\vec{k},S}} \oplus \;\;\mathrm{sl}(2,\mathbb R) \underbrace{\oplus  \;\;\mathrm{u}(1)}_{(d-3 \;\text{times})} .
\end{equation}
We reiterate that all geometries in the phase space have vanishing $SL(2,\mathbb{R})$ charges, and $U(1)$ charges equal to $J_i$.

\subsection{Virasoro subalgebras}
The NHEG algebra has Virasoro algebras as its subalgebras, described below.
\paragraph{The case $d=4$.}
For the four dimensional {Kerr case},  $k=1$ and one obtains the familiar Virasoro algebra
\begin{align}\label{Virasoro}
[L_m, L_n  ]= (m-n) L_{m+n} + \frac{c}{12}m^3\delta_{m+n,0}
\end{align}
with central charge $c= 12 \frac{S}{2 \pi}=12J$, as in  \cite{Guica:2009cd}.
\paragraph{The cases $d>4$.} In higher dimensions, the NHEG algebra  $\widehat{\mathcal{V}_{\vec{k},S}}$ \eqref{NHEG-algebra quantized} is a more general infinite-dimensional algebra in which the entropy appears as the central extension.
For $d>4$ the NHEG algebra contains infinitely many Virasoro subalgebras. To see the latter, first we note that vectors $\vec{n}$ construct a $d-3$ dimensional lattice. $\vec{k}$ may or may not be on the lattice.  Let $\vec{e}$ be any given vector on this lattice  such that $\vec{e}\cdot \vec{k}\neq 0$. Consider the set of generators $L_{\vec{n}}$ such that $\vec{n}=n \vec{e}$ (see Figure \ref{fig NHEG algebra}). Then one may readily observe that these generators form a Virasoro algebra of the form \eqref{Virasoro}. If we define
\begin{equation}\label{Virasoro-subalgebra-generators}
\ell_n\equiv \frac{1}{\vec{k}\cdot \vec{e}} L_{\vec{n}}\,,
\end{equation}
then
\begin{align}
[\ell_m,\ell_n]&=[\frac{L_{\vec{m}}}{\vec{k}\cdot \vec{e}} ,\frac{L_{\vec{n}}}{\vec{k}\cdot \vec{e}} ]
=\frac{\vec{k}\cdot(\vec{m}-\vec{n})}{\vec{k}\cdot \vec{e}}\frac{L_{\vec{m}+\vec{n}}}{\vec{k}\cdot \vec{e}}+\frac{(\vec{k}\cdot \vec{m})^3}{(\vec{k}\cdot \vec{e})^2} \frac{S}{2\pi} \,\delta_{\vec{m}+\vec{n},0}\cr
&=(m-n)\ell_{m+n}+\frac{c_{\vec{e}}}{12}m^3 \,\delta_{m+n,0}\,.
\end{align}
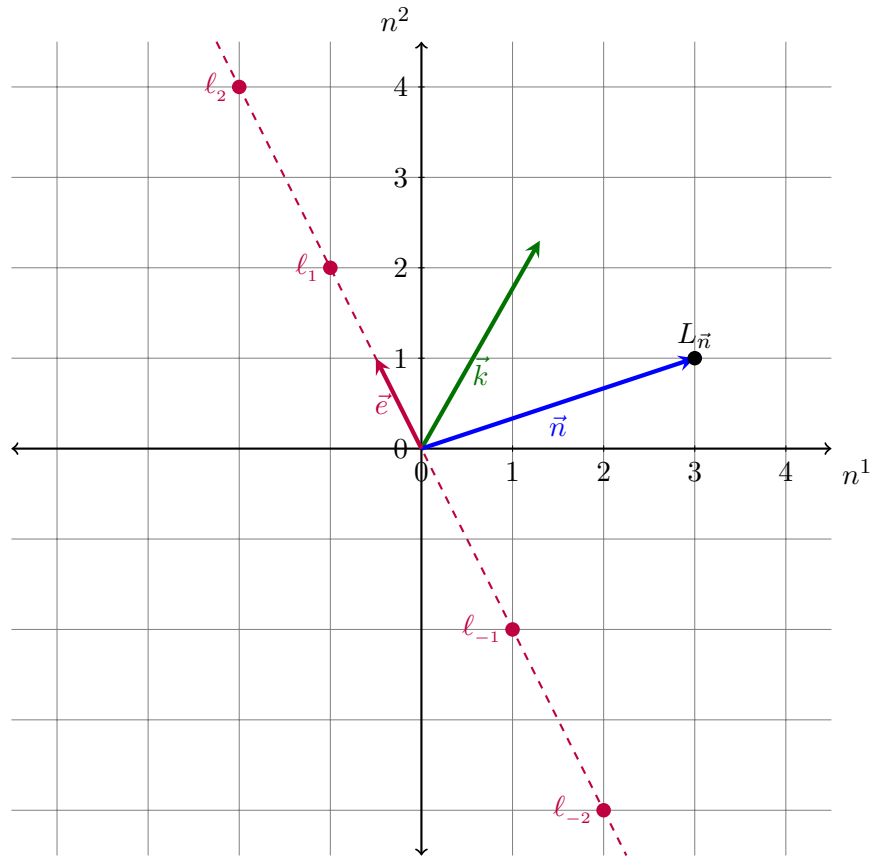
\begin{figure}[t!]
\centering
	\begin{tikzpicture}[scale=1.2]
	\begin{scope}
	\draw[step=1cm,gray,very thin] (-4.5,-4.5) grid (4.5,4.5);
	\draw[thick,<->] (-4.5,0) -- (4.5,0) node[anchor=north west] { $n^1$};
\draw[thick,<->] (0,-4.5) -- (0,4.5) node[anchor=south east] { $n^2$};
\foreach \x in {0,1,2,3,4}
    \draw (\x cm,1pt) -- (\x cm,-1pt) node[anchor=north] { $\x$};
\foreach \y in {0,1,2,3,4}
    \draw (1pt,\y cm) -- (-1pt,\y cm) node[anchor=east] { $\y$};
    \draw[ultra thick,->,>=stealth,darkgreen](0,0)--(1.3,2.3)node[midway, below] { $\vec{k}$};
    \draw[ultra thick,->,>=stealth,blue](0,0)--(3,1)node[midway, below] { $\vec{n}$};
    \fill[](3,1)circle (0.08cm)node[anchor=south] { $L_{\vec{n}}$};
    \draw[thick,purple,dashed](-2.25,4.5)--(2.25,-4.5);
    \fill[purple](-1,2)circle (0.08cm) node[anchor=east] { $\ell_{_{1}}$};
    \fill[purple](-2,4)circle (0.08cm) node[anchor=east] { $\ell_{_{2}}$};
    \fill[purple](1,-2)circle (0.08cm) node[anchor=east] { $\ell_{_{-1}}$};
    \fill[purple](2,-4)circle (0.08cm) node[anchor=east] { $\ell_{_{-2}}$};
    \draw[ultra thick,->,>=stealth,purple](0,0) to (-0.5,1);
    \draw[purple](-0.25,0.5) node[anchor=east]{ $\vec{e}$};
    \end{scope}
	\end{tikzpicture}
	\caption{In the case of $U(1)^2$ isometry, the lattice which $\vec{n}$ relies on is a two dimensional lattice. The vector $\vec{k}$ is not necessarily on the lattice. For any node of the lattice, there is an associated symplectic symmetry generator $L_{\vec{n}}$. A direction which falls on the lattice, and is not perpendicular to  $\vec{k}$  is depicted by the dashed line. This line is identified by a vector $\vec{e}$. The nodes on such lines, \ie $\ell_n$, comprise a Virasoro subalgebra. There are infinite numbers of such subalgebras.}	
\label{fig NHEG algebra}
\end{figure}
As a result, the central charge for the selected subalgebra would be
\begin{equation}\label{virasoro-subalgenra-central-charge}
c_{\vec{e}}=12 (\vec{k}\cdot \vec{e}) \frac{S}{2\pi}\,.
\end{equation}
The entropy might then be written in the suggestive form $S= \frac{\pi^2}{3} c_{\vec{e}} \,T_{F.T.}$ where
\begin{equation}
T_{F.T.}=\frac{1}{2\pi(\vec{k}\cdot \vec{e})}
\end{equation}
is the { extremal Frolov-Thorne chemical potential associated with $\vec{e}$}, as reviewed in [\cite{Compere:2012hr}].

\chapter{Summary and Outlook}\label{chap outlook}

\section{Summary}
In Part \ref{part I}, the thermodynamic properties of NHEGs were studied. The entropy for these  family of solutions was introduced as the Hamiltonian generator associated to the infinite numbers of Killing vectors:
\begin{equation}
\zeta_{_\mathcal{H}}=\mrn^a_{_\mathcal{H}}\xi_a-k^i\mrm_i\,. 
\end{equation} 
Three universal laws, describing dynamics of these solutions has been introduced as:
\begin{itemize}
\item \emph{NHEG Zeroth law:} NHEG chemical potentials, $\kappa$, $k^i$ and $e^p$ are constant over the whole geometry.
\item \emph{NHEG Entropy law:} The entropy of NHEGs is related to their other thermodynamic entities as
\begin{equation}
\frac{S}{2\pi}=\vec{k}\cdot\vec{J}+e^pQ_p-\oint_\mathcal{H}\sqrt{-g}\,\mathcal{L}\,. 
\end{equation}
\item \emph{NHEG entropy perturbation law:} Assuming that perturbations are $\{\xi_-,\xi_0\}$ isometric, the entropy of NHEGs is related to perturbations of other conserved charges as
\begin{equation}
\frac{\delta S}{2\pi}=\vec{k}\cdot\delta\vec{J}+e^p \delta Q_p\,.
\end{equation}
\end{itemize}

After the NHEG zeroth law, entropy law and EPL were elaborated, their contributions to the BH thermodynamics at zero temperature were discussed. They enhanced the BH thermodynamic laws at zero temperature, by the following laws:
\begin{itemize}
\renewcommand{\labelitemi}{$\circ$} 
\item If $T_{_\mrH}=0$, then  $k^i$ and $e^p$ in \eqref{k e BH relation} would also be constant over the horizon. 
\item If $T_{_\mrH}=0$, then $\frac{\delta S}{2\pi}=k^i\delta J_i+e^p\delta Q_p$.
\item If $T_{_\mrH}\to 0$ then $
\frac{S}{2\pi}\to \vec{k}\cdot\vec{J}+e^pQ_p-\oint_\mrH\sqrt{-g}\,\mathcal{L}$.
\end{itemize}

Part \ref{part II} has been an attempt towards building the phase space of NHEGs, hopefully towards  understanding their microstates. For EH theory of gravity, and for the NHEG solutions in $d$ dimensional spacetime, with $SL(2,\mathbb{R})\times U(1)^{d-3}$ isometry, the NHEG phase space manifold is composed of the metrics $g[F]$, with arbitrary periodic function $F$ of $\vec{\varphi}$; the wiggle function. All of the metrics $g[F]$ are diffeomorphic to the NHEG solution $g[F=0]$ itself. The coordinate transformation relating different points of the phase space, are explicitly as
\begin{align}
  \bar{t} =t-\frac{1}{r}(e^{\Psi}-1),\qquad \bar{r} =re^{-{\Psi}},\qquad \bar\theta=\theta\,, \qquad \bar{\varphi}^i=\varphi^i + k^i F,
\end{align}
with \emph{wiggle function} $F(\vec{\varphi})$ periodic in all of its arguments, and
\begin{align}
e^{\Psi}=1+\vec{k}\cdot\vec{\partial}_\varphi F.
\end{align}
The infinitesimal vector fields which generate these coordinate transformations, are 
\begin{equation}
\chi[{\epsilon}(\vec{ \varphi})]=-\vec{k}\cdot\vec{\partial}_{\varphi}{\epsilon}\;(\dfrac{b}{r}\partial_{t}+ r\partial_{r})+{\epsilon}\vec{k}\cdot\vec{\partial}_{\varphi}\,.
\end{equation}
in which $\epsilon$ is infinitesimal version of the wiggle function. Requesting these vector fields to be generators of symplectic symmetry, mainly requesting $\boldsymbol{\omega}(\delta\Phi,\delta_\chi\Phi,\Phi)\approx 0$, fixes the $\mathbf{Y}$ ambiguity of the Lee-Wald symplectic structure. Then, the Hamiltonian generators of the Fourier modes of the symplectic symmetry generators would be calculated as
\begin{align}
H_{\vec{n}}&= \oint_\mathcal{H} \boldsymbol{\epsilon}_\mathcal{H}\  T[\Psi] e^{-i \vec{n}\cdot \vec{\varphi} },
\end{align}
where 
\begin{align}
T[\Psi]&= \frac{ 1}{16 \pi G}  \Big( (\Psi' )^2 -2 \Psi''+ 2 e^{2 \Psi  } \Big)
\end{align}
and primes are directional derivatives along the vector $\vec{k}$. The Poisson bracket of the $H_{\vec{n}}\,$'s, have the same commutation relations as the Lie bracket of $\chi$'s themselves, but with a central extension. The central extension turns out to be the entropy of the NHEG, \ie
\begin{align}
&\{H_{\vec{m}}, H_{\vec{n}}\} = -i\vec{k} \cdot (\vec{m}- \vec{n}) H_{\vec{m}+\vec{n}} -i (\vec{k}\cdot \vec{m})^3 \frac{S}{2\pi}\,\delta_{\vec{m}+\vec{n},0},
\end{align} 
All of the points of the NHEG phase space have the same isometry, \ie the $SL(2,\mathbb{R})\times U(1)^{d-3}$. Their conserved charges $J_i$, $H_{\xi_a}$ are constant over the phase space. In addition, the entropy $S$ is also a constant over the phase space, in agreement with constancy of a central element. According to Definition \ref{definition symplectic sym}, the symplectic symmetry of the NHEG phase space is as
\begin{equation}
\widehat{\mathcal{V}_{\vec{k},S}} \oplus \;\;\mathrm{sl}(2,\mathbb R) \underbrace{\oplus  \;\;\mathrm{u}(1)}_{(d-3 \;\text{times})} .
\end{equation}

\section{Outlook}
There are some interesting lines of research, considering Part \ref{part I}. 
\begin{itemize}
\item[--] It can be possible to find similar laws, in the thermodynamics of materials. There have been  progresses in studying materials with non-zero entropy at zero temperature, \eg the \emph{strange metals} \cite{Sachdev:1992ab}. They are fermions with long range interactions. Their thermodynamics are studied, and there have been observed some (may be universal) relations between the entropy of these metals in terms of other thermodynamic variables (see \cite{Sachdev:2015efa} and references therein). Interestingly, in that context, one can find some roles played by $U(1)$ and $SL(2,\mathbb{R})$ isometries.  It can be an interesting topic for research, if one would be able to augment the usual thermodynamic laws at zero temperature, by some laws analogous to the NHEG laws.  
\item[--] The NHEGs are of few geometries which one can define the entropy as a conserved charge. Also there are not known geometries except the BHs, which admit a complete analogy with thermodynamic systems. Inspiring by NHEG, there might be found another geometries with such behaviours.  
\end{itemize}
In Part \ref{part II}, symplectic structure of the NHEGs has been studied. There are many interesting possible lines of research, waiting for an interested reader to think more about them.  
\begin{itemize}
\item[--] The analysis of Part \ref{part II} is limited to the EH theory. It is tempting to generalize the results to other theories, \eg the EMD-$\Lambda$ or higher derivative gravity theories.
\item[--]The NHEG phase space might be derivable by near horizon limit of the extremal BHs. So there might be possibility to find extensions of the metrics $g_{\mu\nu}[F]$ in the NHEG phase space to the extremal BH geometry.
\item[--] The NHEGs are near horizon regions of the extremal BHs. It might be possible to apply similar analysis as in Part \ref{part II} for the near horizons of \emph{near} extremal BHs. These BHs are not at exact zero temperature, but very close to it. This line of research is also motivated by the microstate counting of BHs at non-zero temperature.
\item[--] According to Section \ref{sec Liouville field}, there might be possibility for a dual description of NHEG dynamics. That description seems to be a Liouville field theory. Studying the field $\Psi$ more, might shed light on this issue.
\item[--] So far, the entropy of NHEG has appeared as a central extension of the NHEG algebra. In the same way that NHEG algebra is different than Virasoro algebra, one might be tempted to change the Cardy's analysis of $2$-dim conformal field theories \cite{Blote:1986ds,Cardy:1986ie}, in order to make a kind of microstate counting associated to the NHEG algebra.
\item[--] Purely by mathematical motivations, the NHEG algebra is a new algebra waiting to be studies mathematically. In addition to exploring its unknown geometrical origin, studying its field representations might provide a new insight over the subject. 
\item[--] Finally, the classical NHEG phase space is waiting to be quantized. It might open a new line of research, towards quantization of the gravity.     
\end{itemize}


\addtocontents{toc}{\vspace{2em}} 

\appendix

\addtocontents{toc}{\protect\setcounter{tocdepth}{0}}

\chapter{A brief introduction to differential form language} \label{app diff forms}
This appendix is written to make a reader who is not familiar with the language of \emph{forms} capable to follow discussions of this thesis. It is neither complete nor rigorous. For a short but more rigour lecture note,  [\cite{Avramidi:2003sd}] is suggested. For a complete study of the subject [\cite{Flanders:2012}] can be a good choice.

\paragraph{Differential forms:} Consider a $d$ dimensional manifold parametrized by some coordinates $x^\mu$. A $p$-form $\boldsymbol{a}$ is roughly a tensor with $p$ number of anti-symmetric indices, all down. $p$ is an integer number such that $0\!\leq p\! \leq d$. There are four usual ways to represent a $p$-form:
\begin{enumerate}
\item[--] by its components explicitly as $a_{\mu_1\mu_2\dots\mu_p}$,
\item[--] covariant representation, but contracted with ordered basis,
\begin{equation}
a_{\mu_1\mu_2\dots\mu_p}\,\mrd x^{\mu_1}\wedge \mrd x^{\mu_2}\wedge\dots\wedge \mrd x^{\mu_p}\,,\qquad \mu_1<\mu_2<\dots<\mu_p\,,
\end{equation}
\item[--] covariant representation, but contracted with unordered basis,
\begin{equation}
\frac{1}{p!}a_{\mu_1\mu_2\dots\mu_p}\,,\mrd x^{\mu_1}\wedge \mrd x^{\mu_2}\wedge\dots\wedge \mrd x^{\mu_p}\,,
\end{equation}
\item[--] abstract representation as $\boldsymbol{a}$.
\end{enumerate}
 
\newpage
\vspace*{-1.5cm}
\dotfillb
\Example{\emph{Some differential forms in different representations, in $4$-dim spacetime:} \\
Considering the 4-dim Minkowski spacetime as the manifold parametrized by the coordinates $(t,r,\theta,\varphi)$. The metric would be $\eta_{\mu\nu}=\text{diag}(-1,1,r^2,r^2\sin^2\theta)$. Assuming that a pointlike electric charge $Q$ is situated at the origin, the gauge field $A_\mu=(\dfrac{Q}{r},0,0,0)$ is a $1$-form which has other representations as
\begin{equation}
\mathbf{A}=\frac{Q}{r} \mrd t\,. 
\end{equation}
The LHS is the abstract presentation, although we have not use the bold font. Also the surface element $2$-form of the sphere $S^2$, denoted abstractly as $\mathbf{\Sigma}$ has the following different representations
\begin{align}
&\Sigma_{\mu\nu}=\begin{pmatrix}
0&0&0&0\\
0&0&0&0\\
0&0&0&\sin\theta\\
0&0&-\sin\theta&0
\end{pmatrix}\,,\qquad \mathbf{\Sigma}=\sin\theta\, \mrd \theta\wedge\mrd\varphi = \frac{1}{2!}(\sin\theta\,\mrd \theta\wedge\mrd\varphi-\sin\theta\,\mrd\varphi\wedge\mrd\theta)\,.
\end{align}  
}
\vspace*{-0.5cm}
\dotfille

There are a couple of operations on differential forms which are described below.
\paragraph{Exterior derivative:} Exterior derivative of a $p$-form which is denoted as $\mrd \boldsymbol{a}$, is a $(p\!+\!1)$-form defined as below.
\begin{align}
\left(\mrd \boldsymbol{a}\right)_{\mu_1\dots\mu_{p+1}}&=(p+1)\partial_{[\mu_1}a_{\mu_2\dots\mu_{p+1}]}\\
&=\sum_{q=1}^{p+1}(-1)^{q-1}\,\partial_{\mu_q}a_{\mu_1\dots\mu_{q-1}\mu_{q+1}\dots\mu_{p+1}}\,.
\end{align}
It has the following properties:
\begin{align}
&\mrd \boldsymbol{a}=0\,, \hspace{2cm} \text{if}\,\,p=d\,,\label{d top form zero}\\
&\mrd^2 \boldsymbol{a}\equiv \mrd (\mrd \boldsymbol{a})=0\,, \qquad \forall p\label{d2=0}\,.
\end{align}
\paragraph{Exterior product:} Exterior product (or wedge  product) of a $p$-form $\boldsymbol{a}$ by a $q$-form $\boldsymbol{b}$, which is denoted as $\boldsymbol{a}\wedge\boldsymbol{b}$, is a $(p\!+\!q)$-form defined as
\begin{align}
\left(\boldsymbol{a}\wedge\boldsymbol{b}\right)_{\mu_1\dots\mu_{p+q}}&=\frac{(p+q)!}{p!\,q!}a_{[\mu_1\dots \mu_p}\,b_{\mu_{p+1} \dots \mu_{p+q}]}\,.
\end{align}
It has the following properties:
\begin{align}
&\boldsymbol{a}\wedge\boldsymbol{b}=(-1)^{pq}\,\, \boldsymbol{b}\wedge\boldsymbol{a} \qquad (\text{anticommutativity}),\\
&(\boldsymbol{a}\wedge\boldsymbol{b})\wedge \boldsymbol{c}=\boldsymbol{a}\wedge(\boldsymbol{b}\wedge \boldsymbol{c}) \qquad (\text{associativity}),\\
&(\boldsymbol{a}+\boldsymbol{b})\wedge \boldsymbol{c}=\boldsymbol{a}\wedge\boldsymbol{c} +\boldsymbol{b}\wedge\boldsymbol{c} \qquad (\text{distributivity})\,,\\
&\boldsymbol{a}\wedge\boldsymbol{a}=0\,\, \hspace{3cm} \text{if}\,\, p=\text{odd}\,,\label{a wedge a}\\
&\boldsymbol{a}\wedge\boldsymbol{b}=0\,\, \hspace{3cm} \text{if}\,\, d<p+q\,.
\end{align}
Specifically for the basis $1$-forms $\mrd x^\mu$ we have
\begin{align}\label{dx wedge dx}
\mrd x^\mu \wedge \mrd x^\nu=-\mrd x^\nu \wedge \mrd x^\mu \quad\Rightarrow\quad \mrd x^\mu \wedge \mrd x^\mu=0\,.
\end{align}
\paragraph{Interior product:} Interior product of a vector $v$ by a $p$-form $\boldsymbol{a}$, which is denoted as $v\cdot\boldsymbol{a}$ or $i_v \boldsymbol{a}$, is a $(p\!-\!1)$-form defined as
\begin{align}
\left(v\cdot\boldsymbol{a}\right)_{\mu_1\dots\mu_{p-1}}&=\frac{1}{(p-1)!}v^\mu a_{\mu\mu_1\dots \mu_{p-1}}\,.
\end{align}
For a $p$-form $\boldsymbol{a}$ and a $q$-form $\boldsymbol{b}$ it satisfies
\begin{equation}
v\cdot(\boldsymbol{a}\wedge\boldsymbol{b})=(v\cdot\boldsymbol{a})\wedge\boldsymbol{b}+(-1)^p \boldsymbol{a}\wedge (v\cdot\boldsymbol{b})\,.
\end{equation} 
\paragraph{Hodge duality:} Hodge duality (or star operator)  of a $p$-form $\boldsymbol{a}$, which is denoted as $\star\boldsymbol{a}$, is a $(d\!-\!p)$-form defined as 
\begin{align}\label{Hodge duality}
\left(\star\boldsymbol{a}\right)_{\mu_{p+1}\dots\mu_{d}}=\frac{1}{p!}\epsilon_{\mu_1\dots\mu_p\mu_{p+1}\dots\mu_d}\sqrt{-g}g^{\mu_1\nu_1}\dots g^{\mu_p\nu_p}\,a_{\nu_1\dots\nu_p}\,.
\end{align}
$\epsilon_{\mu_1\mu_2\dots\mu_d}$ is the Levi-Civita symbol in $d$ dimensions, such that $\epsilon_{12\dots d}=+1$. $g_{\mu\nu}$ is the metric defined on the manifold. Hodge duality has the property
\begin{equation}
\star^2\boldsymbol{a}\equiv \star(\star \boldsymbol{a})=(-1)^{p(d-p)}\boldsymbol{a}\,.
\end{equation} 
Notice that if $d$ is odd, then $\star^2\boldsymbol{a}=\boldsymbol{a}$ for any $p$.

\newpage
\vspace*{-1.5cm}
\dotfillb
\Example{\emph{Operations on some differential forms in $4$-dim spacetime:} \\
Field strength $\mathbf{F}$ is a $2$-form which is defined by the exterior derivative of the gauge field. For the pointlike charge it is
\begin{equation}
\mathbf{F}=\mrd \mathbf{A}=\frac{Q}{r^2}\mrd t\wedge \mrd r\,.
\end{equation}
According to \eqref{d2=0}
\begin{equation}
\mrd \mathbf{F}=\mrd^2\mathbf{A}=0\,,
\end{equation}
which is called Bianchi identity. \\
As an example of exterior product, one can find
\begin{equation}
\mathbf{A}\wedge \mathbf{\Sigma}=\frac{Q\sin\theta}{r} \mrd t\wedge \mrd \theta\wedge\mrd \varphi\,.
\end{equation}
Also by \eqref{a wedge a},
\begin{equation}
\mathbf{A}\wedge\mathbf{A}=0\,.
\end{equation}
As an example for the Hodge duality, one can find $\star \mathbf{F}$ as
\begin{align}
\star \mathbf{F}&=\frac{\sqrt{-g}}{2!\,2!}\epsilon_{\alpha\beta\mu\nu}g^{\alpha\sigma}g^{\beta\rho}F_{\sigma\rho}\,\mrd  x^\mu\wedge\mrd x^\nu\\
&=-Q\sin\theta\, \mrd \theta\wedge\mrd \varphi\,.
\end{align}
In the absence of any electric current, the e.o.m is $\mrd(\star \mathbf{F})=0$. As a check, we have
\begin{equation}
\mrd(\star \mathbf{F})=\mrd(-Q\sin\theta\, \mrd \theta\wedge\mrd \varphi)=-\cos\theta \, \mrd \theta\wedge\mrd \theta\wedge\mrd \varphi\,,
\end{equation}
which vanishes by \eqref{dx wedge dx}. Notice that in the body of the thesis, we have denoted $\mathbf{A}$ and $\mathbf{F}$ by $A$ and $F$, respecting the standard notations. Also, if both of a form and its Hodge dual appear, we have denoted one by bold, but the other not in bold, \eg in \eqref{Lagrangian top form} and \eqref{Theta d-1-form}. 
}
\dotfille

There are a couple of concepts, identities and theorems, useful for the reader of this thesis. Here,  they are introduced  in brief.
\paragraph{Exact and closed forms:} If a $p$-form $\boldsymbol{a}$ can be written in terms of a $(p\!-\!1)$-form $\boldsymbol{b}$ as $\boldsymbol{a}=\mrd \boldsymbol{b}$, then $\boldsymbol{a}$ is called an \emph{exact} form.\\
If a $p$-form $\boldsymbol{a}$ has the property $\mrd\boldsymbol{a}=0$, then $\boldsymbol{a}$ is called a \emph{closed} form. By \eqref{d2=0}, any exact form is closed. But the reveres is not always true.

\paragraph{Pull-back of a $\boldsymbol{p}$-form to a surface:} Assume that in our $d$-dim manifold, there is a $p$-dim surface $\Sigma$, parametrized by the coordinates $(x^{\sigma_1},\dots,x^{\sigma_p})$. Then the \emph{pull-back} of a $p$-form $\boldsymbol{a}$ to that surface would be simply $a_{\sigma_1\dots\sigma_p}$ calculated on $\Sigma$, \ie
\begin{equation}
a_{\sigma_1\dots\sigma_p}\Big|_{\Sigma}.
\end{equation}

\paragraph{Cartan's magic formula:} Given a vector $\xi$ and a $p$-form $\boldsymbol{a}$, the Cartan's magic formula is the identity
\begin{equation}\label{Cartan magic}
\mathscr{L}_\xi \boldsymbol{a}=\xi\cdot \mrd \boldsymbol{a}+\mrd (\xi\cdot \boldsymbol{a})\,. 
\end{equation} 
\paragraph{Coderivative identity:} Assume that a $p$-form $\boldsymbol{a}$ and a $p\!-\!1$-form $\boldsymbol{b}$ satisfy $\boldsymbol{a}=\mrd \boldsymbol{b}$. In addition, it is always possible to consider $\boldsymbol{A}$ and $\boldsymbol{B}$  such that $\boldsymbol{a}=\star \boldsymbol{A}$ and $\boldsymbol{b}=\star \boldsymbol{B}$. Then  
\begin{equation}\label{coderivative id}
A_{\mu_1\dots\mu_{d-p}}=-\nabla_{\alpha}\left(g^{\alpha\beta}B_{\beta\mu_1\dots\mu_{d-p}}\right)\,,
\end{equation}
See [\cite{Avramidi:2003sd}] for the reason of the name and the original shape of the identity.
\paragraph{Stoke's theorem:} Consider a $p$-dim surface $\Sigma$ with $(p\!-\!1)$-dim boundary $\partial\Sigma$. Also consider an exact $p$-form $\boldsymbol{a}$, so $\boldsymbol{a}=\mrd \boldsymbol{b}$ for some $\boldsymbol{b}$. Then
\begin{equation}\label{Stoke's theorem}
\int_\Sigma \boldsymbol{a}=\int_{\partial\Sigma}\boldsymbol{b}\,. 
\end{equation} 
$\int_\Sigma \boldsymbol{a}$ means pull-backing $\boldsymbol{a}$ to the surface $\Sigma$, and integrating over it multiplied by the infinitesimal surface element $\mrd x^{\sigma_1}\dots \mrd x^{\sigma_p}$. A Similar meaning for the  $\int_{\partial\Sigma}\boldsymbol{b}$, but $p\to p-1$.


\chapter{NHEG in global coordinates}\label{app NHEG global}

By the coordinate transformations[\cite{Compere:2012hr}]
\begin{equation}\label{Poincare to global}
r=\sqrt{1\!+\!\rho^2}\cos\tau\!+\!\rho\,, \quad tr=\sqrt{1\!+\!\rho^2}\sin\tau\,,\quad \varphi^i=\phi^i\!+\!k^i\ln \bigg|\frac{\cos\tau\!+\!\rho\sin\tau}{1\!+\!\sqrt{1\!+\!\rho^2}\sin\tau}\bigg|\,,
\end{equation}
in which $\tau,\rho\in(-\infty,\infty)$ one can find NHEG in the \emph{global} coordinate for its $AdS_2$ sector, as
\begin{align}\label{NHEG global metric}
\mrd s^2&=\Gamma\left[-(1+\rho^2)\mrd\tau^2+\dfrac{\mrd \rho^2}{1+\rho^2}+ \tilde{\gamma}_{\alpha \beta}\mrd\theta^\alpha \mrd\theta^\beta\!+\!{\gamma}_{ij}(\mrd\phi^i+k^i\rho\,\mrd \tau)(\mrd\phi^i+k^j\rho\,\mrd\tau)\right]\,,\nonumber\\
A^{(p)}&= f^{(p)}_i(\mrd\phi^i+k^i\rho\mrd \tau)+e^p\rho\mrd \tau\,,\nonumber\\
\phi^I&=\phi^I(\theta^\alpha)\,.
\end{align}
where $\Gamma,\tilde{\gamma}_{\alpha \beta},\gamma_{ij},f^{(p)}_i$ are some functions of $\theta^\alpha$, specified by the equations of motion, and not necessarily similar functions of the Poincar\'e patch \eqref{NHEG metric}. Associated with this coordinate system, the $sl(2,\mathbb{R})$ Killing vector fields $\xi_\sigma$, $\sigma\in\{1,2,3\}$ are given as
\begin{align}\label{global Killings}
\xi_1&=-\cos \tau\frac{\rho}{\sqrt{1+\rho^2}}\partial_\tau- \sin \tau{\sqrt{1+\rho^2}}\partial_\rho-\cos \tau \frac{k^i}{\sqrt{1+\rho^2}}\partial_{\phi^i}\,,\nonumber\\
\xi_2&=\sin \tau\frac{\rho}{\sqrt{1+\rho^2}}\partial_\tau- \cos \tau{\sqrt{1+\rho^2}}\partial_\rho+\sin \tau \frac{k^i}{\sqrt{1+\rho^2}}\partial_{\phi^i}\,, \nonumber\\
\xi_3&=\partial_\tau\,.
\end{align}
In this basis the $sl(2,\mathbb{R})$ commutation relations and Killing form are
\begin{align}\label{global commutations}
\left[\xi_1,\xi_2\right]=-\xi_3\,,\qquad \left[\xi_1,\xi_3\right]=-\xi_2\,,\qquad\left[\xi_2,\xi_3\right]=\xi_1\,,
\end{align}
\begin{equation}\label{global Killing form}
K_{ab}=K^{ab}=\begin{pmatrix}
-1&0&\,0\,\\
0&-1&0\\
0&0&1
\end{pmatrix}\,.
\end{equation}
The $AdS_2$ sector can be immersed in $\mathbb{R}^{2,1}$, the $3$-dim spacetime with signature $(-,-,+)$ and coordinates $X^\sigma$, as
\begin{flalign}
&-(X^1)^2-(X^2)^2+(X^3)^2=-1\,,\label{AdS2 embeding}\\
&X^1=-\sqrt{1+\rho^2}\cos \tau\,,\qquad X^2=\sqrt{1+\rho^2}\sin \tau\,,\qquad X^3=-\rho\,.
\end{flalign}
One can then define the vector $\mrn^\sigma$ in the $\mathbb{R}^{2,1}$ which connects the origin to the point parametrized by $(\tau,\rho)$ as
\begin{equation}
\mrn^1=-\sqrt{1+\rho^2}\cos \tau\,,\qquad \mrn^2=\sqrt{1+\rho^2}\sin \tau\,,\qquad \mrn^3=-\rho\,.
\end{equation}
Using the Killing form \eqref{global Killing form}, or the metric of $\mathbb{R}^{2,1}$, the indices can be lowered as
\begin{equation}\label{global n_a}
\mrn_1=\sqrt{1+\rho^2}\cos \tau\,,\qquad \mrn_2=-\sqrt{1+\rho^2}\sin \tau\,,\qquad \mrn_3=-\rho\,.
\end{equation}
The conventional factor $\dfrac{-1}{2}$ in the definition of the Killing form \eqref{Killing form def} has been chosen such that makes the Killing form \eqref{global Killing form} coincide with the metric of $\mathbb{R}^{2,1}$. Accordingly we have  $\mrn_\sigma\mrn^\sigma=-1$ which is nothing but \eqref{AdS2 embeding}. 
Also one can check that
\begin{equation}
\mathscr{L}_{\xi_\rho}\mrn_\sigma=f_{\rho\sigma}^{\,\,\,\,\tau}\mrn_\tau\,,
\end{equation}
\ie $\mrn_\sigma$ is a vector representation of the $sl(2,\mathbb{R})$. Considering the $AdS_2$ in Poincar\'e patch, the Killing vectors in global coordinates \eqref{global Killings} are related to ones in Poincar\'e patch \eqref{NHEG isometry} as
\begin{equation}
\xi_-=-\xi_1+\xi_3\,,\qquad \xi_0=\xi_2\,,\qquad \xi_+=\frac{\xi_1+\xi_3}{2}\,.
\end{equation}
Accordingly we have
\begin{align}
\mrn_-&=-\mrn_1+\mrn_3=-r\\
\mrn_0&=\mrn_2=-tr\\
\mrn_+&=\frac{\mrn_1+\mrn_3}{2}=-\frac{t^2r^2-1}{2r}\,,
\end{align}
where \eqref{global n_a} and transformations \eqref{Poincare to global} are used. These are nothing but the \eqref{n_a down index} in Section \ref{sec AdS2 sector}.

\chapter{Technical proofs and calculations of part I}


\section{}\label{app d omega 0}
If the dynamical fields $\Phi$ satisfy e.o.m, and $\delta\Phi$ satisfy l.e.o.m then the Lee-Wald symplectic two form \eqref{omega LW} is closed \cite{Lee:1990gr}. Here is the proof. Using \eqref{del del Phi}, $(\delta_1\delta_2-\delta_1\delta_2)\mathbf{L}=0$. Replacing $\delta \mathbf{L}$ from \eqref{delta S} then
\begin{align}
(\delta_1\delta_2-\delta_1\delta_2)\mathbf{L}=&\delta_1\big(\mathbf{E}_\Phi\delta_2\Phi+\mrd \mathbf{\Theta}(\delta_2\Phi,\Phi)\big)-\delta_2\big(\mathbf{E}_\Phi\delta_1\Phi+\mrd \mathbf{\Theta}(\delta_1\Phi,\Phi)\big)\\
= &(\delta_1\mathbf{E}_\Phi)\delta_2\Phi+\mrd \delta_1\mathbf{\Theta}(\delta_2\Phi,\Phi)-(\delta_2\mathbf{E}_\Phi)\delta_1\Phi-\mrd \delta_2 \mathbf{\Theta}(\delta_1\Phi,\Phi)
\end{align}
where \eqref{del del Phi} and \eqref{d delta com} are used. Now if $\Phi$ satisfies the e.o.m and $\delta \Phi$ satisfy l.e.o.m, then $\delta \mathbf{E}_\Phi=0$ (see \eqref{l.e.o.m}), so
\begin{align}
(\delta_1\delta_2-\delta_1\delta_2)\mathbf{L}\approx& \mrd \delta_1\mathbf{\Theta}(\delta_2\Phi,\Phi)-\mrd \delta_2 \mathbf{\Theta}(\delta_1\Phi,\Phi)\\
=&\mrd \boldsymbol{\omega}(\delta_1\Phi,\delta_2\Phi,\Phi)\,.
\end{align}
As a result,
\begin{equation}
\mrd \boldsymbol{\omega}(\delta_1\Phi,\delta_2\Phi,\Phi)\approx 0\,.
\end{equation}

\section{}\label{app H v Q}
The Hamiltonian variation $\delta H_\xi$ introduced in \eqref{delta H xi} is related to the Noether-Wald charge variation $\delta \mathcal{Q}_\xi$ in \eqref{Noether Wald cons charge} by the relation \eqref{H v Q}. Here the steps to prove this proposition are provided \cite{Iyer:1994ys}. By variation of \eqref{Noether Wald current}, we have
\begin{equation}
\delta{\mathbf{J}_\xi = \delta\mathbf{\Theta}( \delta_\xi\Phi,\Phi)-\xi \! \cdot \! \delta\mathbf{L}}\,.
\end{equation}
Note that $\xi$ is not dynamical, so $\delta$ passes from it. Substituting in the last term from \eqref{delta S}  and by on-shell condition,
\begin{align}
\delta\mathbf{J}_\xi &\approx \delta\mathbf{\Theta}( \delta_\xi\Phi,\Phi)-\xi \! \cdot \! \mrd \mathbf{\Theta}(\delta \Phi,\Phi)\\
&=\delta\mathbf{\Theta}( \delta_\xi\Phi,\Phi)-\mathscr{L}_\xi \mathbf{\Theta}(\delta \Phi,\Phi)+\mrd \big(\xi \cdot \mathbf{\Theta}(\delta \Phi,\Phi)\big)\,.
\end{align}
By rearrangement, and by $\delta\mathbf{J}_\xi=\mrd \delta \mathbf{Q}_\xi$ which is a result of l.e.o.m,
\begin{equation}\label{H v Q proof 1}
\delta\mathbf{\Theta}( \delta_\xi\Phi,\Phi)-\mathscr{L}_\xi \mathbf{\Theta}(\delta \Phi,\Phi)\approx \mrd \big(\delta \mathbf{Q}_\xi-\xi \cdot \mathbf{\Theta}(\delta \Phi,\Phi)\big)\,.
\end{equation}
Notice that this result is correct, irrespective of any chosen $\mathbf{Y}$ ambiguity. It is because in \eqref{Noether Wald current} any chosen $\mathbf{Y}$ ambiguity for the $\mathbf{\Theta}$ would also be the same ambiguity in $\mathbf{J}_\xi$. Comparing \eqref{H v Q proof 1} with \eqref{delta H xi} and \eqref{omega dk}, the explicit general formula for $\boldsymbol{k}_\xi(\delta\Phi,\Phi)$ can be read as
\begin{equation}
\boldsymbol{k}_\xi(\delta\Phi,\Phi)=\delta \mathbf{Q}_\xi-\xi \cdot \mathbf{\Theta}(\delta \Phi,\Phi)\,.
\end{equation}
Finally, integrating \eqref{H v Q proof 1} over $\Sigma$ and using the Stoke's theorem leads to \eqref{H v Q}.

\section{}\label{app Wald-Zoupas integrability}
The integrability condition is explicitly \eqref{integrability cond H}. We can use \eqref{H v Q} to replace $\delta H_\xi$ by
\begin{equation}
\delta H_\xi=\int \boldsymbol{\omega}(\delta\Phi,\mathscr{L}_\xi\Phi,\Phi)=\oint \delta \mathbf{Q}_\xi-\xi\cdot\mathbf{\Theta}(\delta\Phi,\Phi)\,.
\end{equation}
Then noticing that $\delta \mathbf{Q}_\xi$ is by definition integrable, $(\delta_1\delta_2 -\delta_2\delta_1)\mathbf{Q}_\xi=0$, so
\begin{align}
0=(\delta_1\delta_2 -\delta_2\delta_1)H_\xi &\approx -\oint \delta_1(\xi\cdot\mathbf{\Theta}(\delta_2\Phi,\Phi))-\delta_2(\xi\cdot\mathbf{\Theta}(\delta_1\Phi,\Phi))\label{integ cond proof 1}\\
&=-\oint \xi\cdot\delta_1\mathbf{\Theta}(\delta_2\Phi,\Phi)-\xi\cdot\delta_2\mathbf{\Theta}(\delta_1\Phi,\Phi)\label{integ cond proof 2}\\
&=-\oint \xi\cdot\big(\delta_1\mathbf{\Theta}(\delta_2\Phi,\Phi)-\delta_2\mathbf{\Theta}(\delta_1\Phi,\Phi)\big)\\
&=-\oint \xi\cdot \boldsymbol{\omega}(\delta _1\Phi,\delta_2\Phi,\Phi)\,.\label{integ cond proof}
\end{align}
Vanishing of \eqref{integ cond proof} is the claimed integrability condition \eqref{integ cond omega}. An important thing to note is that in derivation of \eqref{H v Q}  which is used in this proof, and also in deriving \eqref{integ cond proof 2} from \eqref{integ cond proof 1}, we have had assumed $\delta \xi=0$. So in order to have the \eqref{integ cond omega} as a valid integrability condition, $\xi$ should not depend on the points of the manifold of the phase space.

\section{}\label{app change of J}
Here we find the effect of addition a total derivative to the Lagrangian, on the Noether-Wald current $\mathbf{J}_\xi$. By $\mathbf{L}\to \mathbf{L}+\mrd\boldsymbol{\mu}$, then $\delta_\xi \mathbf{L}\to \delta_\xi\mathbf{L}+\mrd \delta_\xi\boldsymbol{\mu}$. According to \eqref{lagrangian deviation} then $\mathbf{\Theta}\to \mathbf{\Theta}+\delta_\xi\boldsymbol{\mu}$. On the other hand, $\xi\cdot\mathbf{L}\to \xi\cdot\mathbf{L} +\xi\cdot \mrd \boldsymbol{\mu}$. Using the identity $\delta_\xi \boldsymbol{\mu}=\xi \cdot \mrd \boldsymbol{\mu}+\mrd (\xi \cdot \boldsymbol{\mu})$, then we would have
\begin{equation}
{\mathbf{J}_\xi= \mathbf{\Theta}( \delta_\xi\Phi,\Phi)-\xi \! \cdot \! \mathbf{L}} \to \mathbf{J}_\xi+\mrd (\xi \cdot \boldsymbol{\mu})\,.
\end{equation}

\section{}\label{app Kommar int}
The derivation of explicit form of Noether-Wald charge $\mathcal{Q}_\xi$ in the EH gravity is presented. Beginning from \eqref{Noether Wald current}, we can substitute the $\mathbf{L}$ and $\mathbf{\Theta}$ of the EH gravity from \eqref{EH Lagrangian} and \eqref{EH Theta} respectively. As a result, for the dual of $\mathbf{J}_\xi$ we have
\begin{align}
J^\mu_\xi &=\Theta^\mu(g,\delta_\xi g)-\mathcal{L}\xi^\mu\\
&=\frac{1}{16\pi G}(\nabla_\alpha \delta_\xi g^{\mu\alpha}-\nabla^\mu\delta_\xi g^\alpha_{\,\,\alpha})-\frac{1}{16 \pi G} R\,\xi^\mu\,.
\end{align}
Substituting $\delta_\xi g^{\mu\alpha}=\nabla^\mu \xi^\alpha+\nabla^\alpha\xi^\mu$, then
\begin{align}
J^\mu_\xi &=\frac{1}{16\pi G}\Big(\nabla_\alpha (\nabla^\mu \xi^\alpha+\nabla^\alpha\xi^\mu)-2\nabla^\mu \nabla_\alpha \xi^\alpha - R\,\xi^\mu\Big)\\
&=\frac{1}{16\pi G}\Big(\nabla_\alpha (\nabla^\alpha\xi^\mu-\nabla^\mu \xi^\alpha)+2\nabla_\alpha\nabla^\mu \xi^\alpha-2\nabla^\mu \nabla_\alpha \xi^\alpha - R\,\xi^\mu\Big)\,.
\end{align}
Now one can use the identity $(\nabla_\mu\nabla_\nu-\nabla_\nu\nabla_\mu)\xi^\alpha=R^\alpha_{\:\beta \mu\nu}\xi^\beta$ to rewrite it as
\begin{align}
J^\mu_\xi &=\frac{1}{16\pi G}\Big(\nabla_\alpha (\nabla^\alpha\xi^\mu-\nabla^\mu \xi^\alpha)+ 2R^\mu_{\,\,\alpha}\xi^\alpha- R\,\xi^\mu\Big)\\
&\approx \frac{1}{16\pi G}\Big(\nabla_\alpha (\nabla^\alpha\xi^\mu-\nabla^\mu \xi^\alpha)\Big)\,.
\end{align}
Noticing the \eqref{coderivative id} and $\mathbf{J}_\xi=\mrd\mathbf{Q}_\xi$, we have $\mathbf{Q}_\xi=\star \mathrm{Q}_\xi$ in which
\begin{equation}
\big(\mathrm{Q}_\xi)_{\mu\nu}=\frac{-1}{16\pi G}(\nabla_\mu\xi_\nu-\nabla_\nu \xi_\mu)=\frac{-1}{16\pi G}(\mrd \xi)_{\mu\nu}\,,
\end{equation}
It can be easily checked  that adding a cosmological constant $\Lambda$ to the EH theory, does not change the final result.

\section{}\label{app k EH proof}
In this section, we will derive the explicit form of the $\boldsymbol{k}_\xi^{\text{EH}}$, \ie the equation \eqref{EH k}. We can begin from the general formula of the $\boldsymbol{k}_\xi$, derived in Appendix \eqref{app Kommar int}, which is
\begin{equation}\label{k EH proof 1}
\boldsymbol{k}_\xi(\delta\Phi,\Phi)=\delta \mathbf{Q}_\xi-\xi \cdot \mathbf{\Theta}(\delta \Phi,\Phi)\,.
\end{equation}
For the EH theory, by the \eqref{Komar 2} and \eqref{EH Theta} we have 
\begin{align}
\mathbf{Q}_\xi&=\star \Big( \frac{-1}{16\pi G }\frac{1}{2!} (\nabla_\mu\xi_\nu-\nabla_\nu\xi_\mu)\,\mrd x^\mu\wedge\mrd x^\nu\Big)\\
&=\frac{-1}{16\pi G}\frac{\sqrt{-g}}{(2!(d-2)!)}\epsilon_{\mu\nu\alpha_1\dots\alpha_{d-2}}(\nabla^\mu\xi^\nu-\nabla^\nu\xi^\mu)\,\mrd x^{\alpha_1}\wedge\dots\wedge\mrd x^{\alpha_{d-2}}\label{k EH proof 2}
\end{align}
and
\begin{align}\label{k EH proof 4}
\mathbf{\Theta}(\delta \Phi,\Phi)=\star\Big(\frac{1}{16\pi G}(\nabla_\alpha \delta g_{\,\,\mu}^{\alpha}-\nabla_\mu\delta g^\alpha_{\,\,\alpha})\,\mrd x^\mu\Big)
\end{align}
respectively. So the terms in \eqref{k EH proof 1} can be calculated using them. Putting \eqref{k EH proof 2} into the first term of \eqref{k EH proof 1}

\vspace*{-0.8cm}
\small{
\begin{align}
\delta \mathbf{Q}_\xi&=\delta \Big( \frac{-1}{16\pi G}\frac{\sqrt{-g}}{(2!(d-2)!)}\epsilon_{\mu\nu\alpha_1\dots\alpha_{d-2}}g^{\mu\beta}(\nabla_\beta\xi^\nu)\,\mrd x^{\alpha_1}\wedge\dots\wedge\mrd x^{\alpha_{d-2}}\Big)-[\mu\leftrightarrow\nu]\\
&=\frac{-1}{16\pi G}\frac{1}{(2!(d-2)!)}\epsilon_{\mu\nu\alpha_1\dots\alpha_{d-2}}\Big((\delta \sqrt{-g})\,g^{\mu\beta}(\nabla_\beta\xi^\nu)-\sqrt{-g}\,\delta g^{\mu\beta}(\nabla_\beta\xi^\nu)\nonumber\\
&\hspace*{4cm}+\sqrt{-g}g^{\mu\beta}\delta (\nabla_\beta\xi^\nu)\Big)\,\mrd x^{\alpha_1}\wedge\dots\wedge\mrd x^{\alpha_{d-2}}-[\mu\leftrightarrow\nu]\,,\label{k EH proof 3}
\end{align}}\normalsize

\vspace*{-0.5cm}
in which the notation $\delta g^{\mu\nu}\equiv g^{\mu\alpha}g^{\nu\beta}\delta g_{\alpha\beta}=-\delta (g^{\mu\nu})$ has been used. Now by the relations 
\begin{align}
\delta \sqrt{-g}=\frac{1}{2\sqrt{-g}}\delta g^\alpha_{\,\,\alpha}\,, \qquad \delta \Gamma ^\lambda_{\mu\nu}&= \frac{1}{2}[g^{\lambda \sigma}\big(\nabla _\mu\delta  g_{\sigma \nu}+\nabla_\nu\delta g_{\sigma \mu}-\nabla _\sigma \delta g_{\mu\nu}]
\end{align}
the \eqref{k EH proof 3} is calculated to be
\begin{align}
\delta \mathbf{Q}_\xi&=\frac{-1}{16\pi G}\frac{\sqrt{-g}}{(2!(d-2)!)}\epsilon_{\mu\nu\alpha_1\dots\alpha_{d-2}}\Big(\frac{1}{2}\delta g^\alpha_{\,\,\alpha}(\nabla^\mu\xi^\nu)-\delta g^{\mu\beta}(\nabla_\beta\xi^\nu)\nonumber\\
&\hspace*{4cm}+\xi^\alpha\nabla^\mu \delta g^\nu_{\,\,\alpha}\Big)\,\mrd x^{\alpha_1}\wedge\dots\wedge\mrd x^{\alpha_{d-2}}-[\mu\leftrightarrow\nu]\,.\label{k EH proof 5}
\end{align}
Now, let's calculate the second term of \eqref{k EH proof 1}. Inserting \eqref{k EH proof 4} into the second term in \eqref{k EH proof 1}, then

\vspace*{-0.8cm}
\small{
\begin{align}
-\xi \cdot \mathbf{\Theta}(\delta \Phi,\Phi)&=-\xi \cdot\Big(\frac{1}{16\pi G}\frac{\sqrt{-g}}{(d-1)!}\epsilon_{\mu\alpha_{1}\dots\alpha_{d-1}}(\nabla_\alpha \delta g^{\alpha\mu}-\nabla^\mu\delta g^\alpha_{\,\,\alpha})\mrd x^{\alpha_1}\wedge\dots \wedge \mrd x^{\alpha_{d-1}}\Big)\nonumber\\
&=\frac{-1}{16\pi G}\frac{\sqrt{-g}}{(d-2)!}\epsilon_{\mu\nu\alpha_{1}\dots\alpha_{d-2}}(\nabla_\alpha \delta g^{\alpha\mu}-\nabla^\mu\delta g^\alpha_{\,\,\alpha})\xi^\nu \,\mrd x^{\alpha_1}\wedge\dots \wedge \mrd x^{\alpha_{d-2}}\\
&=\frac{-1}{16\pi G}\frac{\sqrt{-g}}{2(d-2)!}\epsilon_{\mu\nu\alpha_{1}\dots\alpha_{d-2}}(\nabla_\alpha \delta g^{\alpha\mu}-\nabla^\mu\delta g^\alpha_{\,\,\alpha})\xi^\nu \,\mrd x^{\alpha_1}\wedge\dots \wedge \mrd x^{\alpha_{d-2}}-[\mu\leftrightarrow\nu]\label{k EH proof 6}
\end{align}}\normalsize

\vspace*{-0.5cm}
Hence, having found the \eqref{k EH proof 5} and \eqref{k EH proof 6}, the $\boldsymbol{k}_\xi^{\text{EH}}$ can be read as
\begin{align}
\boldsymbol{k}_\xi^{\text{EH}}&=\frac{-1}{16\pi G}\frac{\sqrt{-g}}{(2!(d-2)!)}\epsilon_{\mu\nu\alpha_1\dots\alpha_{d-2}}\Big(\frac{1}{2}\delta g^\alpha_{\,\,\alpha}(\nabla^\mu\xi^\nu)-\delta g^{\mu\beta}(\nabla_\beta\xi^\nu)+\xi^\alpha\nabla^\mu \delta g^\nu_{\,\,\alpha}\nonumber\\
&+(\nabla_\alpha \delta g^{\alpha\mu}-\nabla^\mu\delta g^\alpha_{\,\,\alpha})\xi^\nu\Big)\,\mrd x^{\alpha_1}\wedge\dots\wedge\mrd x^{\alpha_{d-2}}-[\mu\leftrightarrow\nu]
\end{align}
By the Hodge duality, we would have $\boldsymbol{k}_\xi^{\text{EH}}=\star k_\xi^{\text{EH}}$, where
\begin{equation}
k_\xi^{\text{EH}\mu\nu}=\frac{-1}{16\pi G}\Big(\frac{1}{2}\delta g^\alpha_{\,\,\alpha}(\nabla^\mu\xi^\nu)-\delta g^{\mu\beta}(\nabla_\beta\xi^\nu)+\xi^\alpha\nabla^\mu \delta g^\nu_{\,\,\alpha}+(\nabla_\alpha \delta g^{\alpha\mu}-\nabla^\mu\delta g^\alpha_{\,\,\alpha})\xi^\nu\Big)-[\mu\leftrightarrow\nu]\,,
\end{equation}
which is the equation \eqref{EH k}.

If one adds the cosmological constant $\Lambda$ to the EH theory, the equations \eqref{k EH proof 2} and \eqref{k EH proof 4} would remain intact, resulting to the same $\boldsymbol{k}_\xi$.

\section{}\label{app electric charge}
This section is provided to introduce the electric charge as a Noether-Wald charge. By the gauge transformation \eqref{gauge trans}, the Lagrangian is invariant $\delta \mathbf{L}=0$. So noticing \eqref{delta S}, $\mrd \mathbf{\Theta}\approx 0$ in which $\mathbf{\Theta}=\star \Theta$ and
\begin{equation}\label{gauge Theta pre}
\Theta^{(p)\mu}=\frac{\partial \mathcal{L}}{\partial F^{(p)}_{\mu\nu}}\delta A^{(p)}_{\nu}\,.
\end{equation}
By the $\delta A^{(p)}_\mu=\partial_\mu \Lambda^p=\nabla_\mu \Lambda^p$, and by the e.o.m $\nabla_\alpha \frac{\partial \mathcal{L}}{\partial F^{(p)}_{\mu\alpha}}\approx 0$ we have
\begin{equation}\label{gauge Theta}
\Theta^{(p)\mu}=\nabla_\alpha(\frac{\partial \mathcal{L}}{\partial F^{(p)}_{\mu\alpha}}\Lambda^p)\,.
\end{equation}
Considering the Noether-Wald current $(d-1)$-form as $\mathbf{J}\!=\!\mathbf{\Theta}$ and $\mathbf{J}\!=\!\mrd \mathbf{Q}$, then from the \eqref{gauge Theta} one can read
\begin{equation}
\mathrm{Q}^{(p)\mu\nu}=\frac{\partial \mathcal{L}}{\partial F^{(p)}_{\mu\nu}}\Lambda^p\,,
\end{equation} 
where the notation $\mathbf{Q}^{(p)}=\star \mathrm{Q}^{(p)}$ is used. The global part of the gauge transformations is responsible for electric charge conservation, so taking the $\Lambda^p=\text{const.}=1$, would result the $(d-2)$-form electric charge density as Hodge dual to
\begin{equation}\label{electric charge 2-form}
\mathrm{Q}^{(p)\mu\nu}=\left(\frac{\partial \mathcal{L}}{\partial F^{(p)}_{\mu\nu}}\right)\,.
\end{equation}

\section{}\label{app zeroth law}
In order to show the constancy of Hawking temperature on the horizon, it is enough to show the constancy of $\kappa^2$ in \eqref{surface gravity}, \ie 
\begin{equation}\label{constant kappa}
t ^\mu \partial _\mu \kappa ^2=0
\end{equation}
for any arbitrary vector field $t^\mu$ tangent to the horizon. The proof has two steps. 

The first step is showing that it is correct for $t^\mu=\zeta_{_\mrH}^\mu$. So $\kappa^2$ is constant on the orbits of $\zeta_{_\mrH}$. To show it,
\begin{align}
\zeta_{_\mrH} ^\mu \partial _\mu \kappa ^2=\zeta_{_\mrH} ^\mu \nabla _\mu \kappa ^2 =-(\nabla ^\alpha \zeta_{_\mrH}^\beta)(\zeta_{_\mrH} ^\mu \nabla _ \mu\nabla_ \alpha \zeta_{_\mrH\beta})=-(\nabla ^\alpha \zeta_{_\mrH} ^ \beta) \zeta_{_\mrH} ^ \mu R _{\,\,\mu \alpha \beta}^ {\nu}\zeta_{_\mrH\nu}
\end{align}
where in the last equation, the identity $\nabla _ \mu\nabla_ \alpha \xi_\beta=R^\nu_{\,\,\mu\alpha\beta}\xi_\nu$ true for any Killing is used. But the last term vanished by the anti-symmetric properties of indices $\{\mu,\nu\}$ in the Riemann tensor.

The second step is to show \eqref{constant kappa} on the bifurcation point of the horizon for any chosen tangent $t^\mu$. Then, because the bifurcation point is the origin of orbits of $\zeta_{_\mrH}$, and the horizon is spanned by those orbits, $\kappa^2$ would be constant all over the horizon. So in order to show the claim,
\begin{align}
t ^\mu \partial _\mu \kappa ^2&=t ^\mu \nabla _\mu \kappa ^2 =-(\nabla ^\alpha \zeta_{_\mrH}^\beta)(t ^\mu \nabla _ \mu\nabla_ \alpha \zeta_{_\mrH\beta})=-(\nabla ^\alpha \zeta_{_\mrH} ^ \beta) t^ \mu R _{\,\,\mu \alpha \beta}^ {\nu}\zeta_{_\mrH\nu}
\end{align}
The last term vanishes because $\zeta_{_\mrH}$ vanishes on the bifurcation point of the horizon, proving the claim.

\section{}\label{app 1st ambig}
Here we show that $\oint_\mrH \delta \mathbf{Q}_{\zeta_{_\mrH}}$ is unambiguously equal to $\frac{\kappa\delta S}{2\pi}$ \cite{Iyer:1994ys}. Using the \eqref{decomposition} in Section \ref{sec Iyer-Wald entropy} it was shown that $\oint_\mrH \mathbf{Q}_{\zeta_{_\mrH}}$ is unambiguously equal to $\frac{\kappa S}{2\pi}$ because none of the $\mathbf{W}$, $\mathbf{Y}$ and $\mathbf{Z}$ ambiguities could contribute. But taking variations, this issue might not remain valid. We show that it is not the case. The variations of $\mathbf{W}$ and $\mathbf{Z}$ ambiguities would remain linear in $\zeta_{_\mrH}$, because $\delta$ does not act on $\zeta_{_\mrH}$. So according to \eqref{zeta vanish} they would vanish. For the $\mathbf{Y}$ term we have
\begin{align}\label{1st law deltaY}
\delta \mathbf{Y}(\delta_{\zeta_{_\mrH}}\Phi,\bar{\Phi})&=\mathbf{Y}(\delta\delta_{\zeta_{_\mrH}} \Phi,\bar{\Phi})\nonumber\\
&=\mathbf{Y}(\delta_{\zeta_{_\mrH}}\delta \Phi,\bar{\Phi})\nonumber\\
&=\delta_{\zeta_{_\mrH}}\mathbf{Y}(\delta \Phi,\bar{\Phi})\nonumber\\
&=\zeta_{_\mrH} \cdot \mrd\mathbf{Y}+\mrd(\mathbf{Y}\cdot \zeta_{_\mrH})\,.
\end{align}
In the above we have used the fact that since $\delta\zeta_{_\mrH}=0$, we can  interchange $\delta_{\zeta_{_\mrH}}$ and $\delta$. The last term is linear in $\zeta_{_\mrH}$, so it would not have any contribution too.

\section{}\label{Sl2Z period}
This section is provided to show that in the case of $U(1)^2$ isometry, the action of $SL(2,\mathbb{Z})$ does not change the periodicity of the coordinates $\varphi^1,\varphi^2$. Considering the following $3$ points on the plane $\mathbb{R}^2$
\begin{align}
A=(\varphi^1,\varphi^2)\,,\qquad B=(\varphi^1,\varphi^2)+(2\pi,0)\,,\qquad C=(\varphi^1,\varphi^2)+(0,2\pi)\,,
\end{align}
we identify these three points to obtain a \textit{torus}
\begin{align}\label{SLnZ A-B-C}
A\sim B\sim C\,, \qquad \forall\,(\varphi^1,\varphi^2)\,.
\end{align}
Therefore the periodicity of $\varphi^1,\varphi^2$ are equal to $2\pi$. Now suppose that we make a $SL(2,\mathbb{Z})$ coordinate transformation on the torus by
\begin{align}
\begin{pmatrix}
\varphi'^1\\
\varphi'^2
\end{pmatrix}&=
\begin{pmatrix}
a&b\\
c&d
\end{pmatrix} \begin{pmatrix}
\varphi^1\\
\varphi^2
\end{pmatrix}
\end{align}
where $a,\cdots, d\in \mathbb{Z}$ and $ad-bc=1$. In the new coordinates, the \eqref{SLnZ A-B-C} would be as
\begin{align}\label{SlnZ A'-B'-C'}
(\varphi'^1,\varphi'^2)\sim (\varphi'^1,\varphi'^2)+2\pi (a,c)\sim (\varphi'^1,\varphi'^2)+2\pi (b,d)\,,\qquad \forall\,(\varphi'^1,\varphi'^2)\,. 
\end{align}
From the above relation, it is not clear what is the periodicity of the new coordinates $\varphi'^1,\varphi'^2$. But Suppose that we move $n$ periods along $\varphi^1$, and $m$ periods along $\varphi^2$. Since $n,m$ are integers, then the endpoint is identified with the starting point, as
\begin{align}
(\varphi'^1,\varphi'^2)\sim (\varphi'^1,\varphi'^2)+2\pi n (a,c)+2\pi m (b,d)\,.
\end{align}
Now we want to find the period of $\varphi'^1$. In other words, the question is ``\textit{find the smallest values for $n,m$ such that the resulting movement is only along $\varphi'^1$}?"\\
In order that the movement be along $\varphi'^1$ only, we should have
\begin{align}
nc+md=0 \implies m=-\dfrac{c}{d}n\,.
\end{align}
Since the determinant of the transformation is 1, and the entries of $SL(n,\mathbb{Z})$ are integeres, one can show that the greatest common factor of $c,d$ is 1 (otherwise the common factor could be factor out and the determinant would be proportional to the square of that integer factor). Therefore the smallest (positive) solution to the above equation is $n=d$, and accordingly $m=-c$. Going back to \eqref{SlnZ A'-B'-C'}, it turns out that
\begin{align}
(\varphi'^1,\varphi'^2)\sim (\varphi'^1,\varphi'^2)+2\pi n(a-\dfrac{c}{d}b, 0)\,.
\end{align}
Using the fact that $ad-bc=1$, we find that 
\begin{align}
(\varphi'^1,\varphi'^2)\sim (\varphi'^1,\varphi'^2)+2\pi(1, 0)\,.
\end{align}
This means that the period of the new coordinate $\varphi'^1$ is again $2\pi$. By the same way, we can show that the period of $\varphi'^2$ is also $2\pi$. Therefore the periodicity is not changed by $SL(2,\mathbb{Z})$ transformations.

\section{}\label{app H to H by xi}
Surfaces $\mathcal{H}$ are $(d\!-\!2)$-dim surfaces which are identified by some constant time and radius $(t_{_\mathcal{H}},r_{_\mathcal{H}})$. They are mapped to each other under the action of group generated by the subalgebra $\{\xi_-,\xi_0\}$. To prove this proposition, we notice that the action of the generators $\xi_-=\partial_t$ and $\xi_0=t\partial_t-r\partial_r$ are as 
\begin{equation}
t\to t+a \,, \qquad \begin{cases}t \to bt\\ r\to \frac{r}{b}\,,\end{cases}
\end{equation}
respectively, for some constants $a$ and $b$. Now assume surfaces $\mathcal{H}_1$ and $\mathcal{H}_2$ are identified by $(t_1,r_1)$ and $(t_2,r_2)$ respectively. Defining $b\equiv \dfrac{r_1}{r_2}$, then by the action of group generated by $\xi_0$ one can map $t_1 \to bt_1$ and $r_1\to r_2$. Then, by definition of $a\equiv t_2-bt_1$, by the action of group generated by $\xi_-$ one can map $bt_1\to t_2$. Accumulating the above actions would lead to $t_1\to t_2$ and $r_1\to r_2$. Other coordinates $(\theta^\alpha,\varphi^i)$ are left unchanged, so $\mathcal{H}_1$ is mapped to $\mathcal{H}_2$ by isometry actions.

\section{}\label{app t r dependences}
Assuming that a tensor has $\{\xi_-,\xi_0\}$ isometry, its $(t,r)$ dependence is determined completely, component by component. Here we will mention it for tensors of rank $(0,1,2)$, but it can be generalized to any rank. 

For a tensor of rank $0$, denoted by $T$, $\mathscr{L}_{\xi_-}T=\partial_t T=0$, so $T$ has not any $t$ dependence. Then by $\mathscr{L}_{\xi_0}T=-r\partial_r T=0$, it would not have any $r$ dependence. Hence
\begin{equation}
T=T(\theta^\alpha,\varphi^i) \,.
\end{equation}

For a tensor of rank $1$ with down index, denoted by $T_\mu$, $\mathscr{L}_{\xi_-}T_\mu=\partial_t T_\mu=0$ so $T_\mu$ have not any $t$ dependence. Then by $\mathscr{L}_{\xi_0}T=0$ a system of differential equations follows
\begin{equation}
\begin{cases}
-r\partial_r T_t+T_t =0\\
-r\partial_r T_r-T_r=0\\
-r\partial_r T_{\theta^\alpha}=0\\
-r\partial_r T_{\varphi^i}=0
\end{cases}
\end{equation}
It is a homogeneous system of differential equations, with the answer 
\begin{equation}\label{T1 tr dependence}
T_\mu=(r\tilde{T}_t,\frac{\tilde{T}_r}{r},\tilde{T}_{\theta^\alpha},\tilde{T}_{\varphi^i})\,,  
\end{equation}  
for some functions $\tilde{T}_\mu=\tilde{T}_\mu(\theta^\alpha,\varphi^i)$. 

For a second rank tensor $T_{\mu\nu}$, by a similar analysis, one finds that by $\mathscr{L}_{\xi_-}T_{\mu\nu}=\partial_t T_{\mu\nu}=0$, components are $t$ independent, and by $\mathscr{L}_{\xi_0}T_{\mu\nu}=0$, the $r$ dependences are fixed as
\begin{align}\label{T2 tr dependence}
T_{\mu\nu}&=\begin{pmatrix}r^2\tilde{T}_{tt} &  \tilde{T}_{tr} & r\tilde{T}_{t\theta^\alpha}&r\tilde{T}_{t \varphi^i}\\
& \frac{\tilde{T}_{rr}}{r^2}&\frac{\tilde{T}_{r\theta^\alpha}}{r}& \frac{\tilde{T}_{r \varphi^i}}{r}\\
 & & \tilde{T}_{\theta^\alpha\theta^\beta}&  \tilde{T}_{\theta^\alpha\varphi^i}\\
  & & &\tilde{T}_{\varphi^i\varphi^j}
\end{pmatrix}
\end{align} 
for some functions $\tilde{T}_{\mu\nu}=\tilde{T}_{\mu\nu}(\theta^\alpha,\varphi^i)$.

For the tensors $T^\mu$ and $T^{\mu\nu}$, by similar analysis (or by raising indices using $g^{\mu\nu}$) one finds that components are $t$ independent. The $r$ dependence would be found to be $r\to \dfrac{1}{r}$ in \eqref{T1 tr dependence} and \eqref{T2 tr dependence}.

\section{}\label{app n-h Q-a proof}
In this appendix, we will show that $\mrn^a_{_\mathcal{H}}\mathcal{Q}_{\xi_a}$ is equal to $e^pQ_p-\oint_\mathcal{H}\sqrt{-g}\, \mathcal{L}$, independent of the theory under considerations. Beginning from \eqref{structure coeff}, rewritten as
\begin{equation}
[\xi_b,\xi_c]=f_{bc}^{\,\,\,\;d}\xi_d\,,
\end{equation}
and by multiplication of both sides by $f_a^{\,\,bc}$, we have
\begin{equation}
\xi_a=\frac{1}{2}f_a^{\,\,bc}[\xi_b,\xi_c]\,,
\end{equation}
where \eqref{Killing form def} is used. On the other hand by the definition of Lie bracket,
\begin{align}
[\xi_b,\xi_c]^\mu &=\xi_b^\nu\nabla_\nu\xi_c^\mu -\xi_c^\nu\nabla_\nu\xi_b^\mu\cr
&=2\nabla_\nu\left(\xi_b^{[\nu} \xi_c^{\mu]}\right)\,.
\end{align}
Therefore we find
\begin{equation}\label{xi in term brackets}
\xi_a^\mu=f_a^{\,\,bc}\nabla_\nu\left(\xi_b^{[\nu} \xi_c^{\mu]}\right)=f_a^{\,\,bc}\nabla_\nu\left(\xi_b^{\mu} \xi_c^{\nu}\right)\,, 
\end{equation}
which will be used in a moment. Returning back to the calculations of Noether-Wald charges by \eqref{Noether Wald current}, in the Hodge dual language,
\begin{equation}
{J}_{\xi_a}^\mu ={\Theta}^\mu(\delta_{\xi_a}\Phi,\Phi)-\mathcal{L}\xi_a^\mu \,.
\end{equation}
Calculating the first term in RHS: For the $\{\xi_-,\xi_0\}$, the ${\Theta}^\mu(\delta_{\xi_a}\Phi,\Phi)$ vanishes identically by the isometry condition $\delta_{\xi_a}\Phi=0$. For the $\xi_+$ and in the presence of gauge fields, according to \eqref{xi_+ A} and \eqref{gauge Theta}, we have
\begin{equation}
\Theta^{\mu}(\delta_{\xi_+}A^{(p)},A^{(p)})=\nabla_\nu\left(\frac{-e^p}{r}\frac{\partial \mathcal{L}}{\partial F^{(p)}_{\mu\nu}}\right)\,.
\end{equation}
Calculating the second term in RHS: Replacing $\xi_a^\mu$ from \eqref{xi in term brackets},
\begin{align}
-\mathcal{L}\,\xi_a^\mu &=\mathcal{L}\,f_a^{\,\,bc}\nabla_\nu\left(\xi_b^{\mu} \xi_c^{\nu}\right)\\
\approx& f_a^{\,\,bc}\nabla_\nu\left(\mathcal{L}\,\xi_b^{\mu} \xi_c^{\nu}\right)\,,
\end{align}
where in the last equation, the on-shell isometry condition $\xi_a^\nu\nabla_\nu \mathcal{L}\approx 0$ is used.\\ Accumulating the results above, for the Hodge dual of the $\mathbf{Q}_{\xi_a}$ we read
\begin{equation}\label{na Q xi proof 1}
\mathrm{Q_{\xi_a}^{\mu\nu}}=\frac{-e^p}{r} \frac{\partial \mathcal{L}}{\partial F^{(p)}_{\mu\nu}}\delta_{a+} +f_a^{\,\,bc}\mathcal{L}\xi_b^{\mu} \xi_c^{\nu}\,.
\end{equation}
Noticing that integration over $\mathcal{H}$ does not include integration over coordinate $r$, then by \eqref{electric charge Q-form}, integration of the first term in the RHS of \eqref{na Q xi proof 1} simply leads to  $\dfrac{-e^p}{r_{_\mathcal{H}}} Q_p\delta_{a+}$. The second term in the RHS of \eqref{na Q xi proof 1} is a $2$-form with its indices are raised. One needs to use \eqref{Hodge duality} to find its Hodge dual, and then pull-back it to $\mathcal{H}$ and integrate it. So we are dealing with
\begin{align}
\oint_\mathcal{H}\frac{1}{2!}\epsilon_{\mu\nu\alpha_1\dots\alpha_{d-2}} \sqrt{-g} \left(f_a^{\,\,bc}\mathcal{L}\xi_b^{\mu} \xi_c^{\nu}\right)\mrd x^{\alpha_1}\wedge \dots\wedge \mrd x^{\alpha_{d-2}}
=\oint_\mathcal{H}\,\boldsymbol{\epsilon}_\mathcal{H}\epsilon_{_\perp\mu\nu}\left(\frac{1}{2}f_a^{\,\,bc}\mathcal{L}\xi_b^{\mu} \xi_c^{\nu}\right)\label{na Q xi proof 2}
\end{align}
where \eqref{NHEG volume d-form} is used. It can be made simpler by the identity 
\begin{equation}
\epsilon_{_\perp\mu\nu}(\xi_b^{\mu} \xi_c^{\nu})=\Gamma \delta_{\xi_b}\mrn_c 
\end{equation}
contracted by $f_a^{\,\,bc}$, which is
\begin{equation}
f_a^{\,\,bc}\epsilon_{_\perp\mu\nu}(\xi_b^{\mu} \xi_c^{\nu})=f_a^{\,\,bc}\Gamma \delta_{\xi_b}\mrn_c=\Gamma f_a^{\,\,bc} f_{bc}^{\,\,\,d}\,\mrn_d=2\Gamma\mrn_a\,,
\end{equation}
in which \eqref{del_xi n_a} and \eqref{Killing form def} are also used. Putting it into \eqref{na Q xi proof 2}
and noticing that $\mrn_a$ are just functions of $(t,r)$ which are not integrated over, we end with
\begin{equation}
\mathcal{Q}_{\xi_a}= \dfrac{-e^p}{r_{_\mathcal{H}}} Q_p\delta_{a+} + \mrn_{_\mathcal{H}a}\oint_\mathcal{H}\,\boldsymbol{\epsilon}_\mathcal{H} \Gamma \mathcal{L}\,.
\end{equation} 
As a result, by  \eqref{n_a norm} and \eqref{n-a vector up}, 
\begin{equation}
\mrn_{_\mathcal{H}}^a\mathcal{Q}_{\xi_a}=e^p Q_p-\oint_\mathcal{H}\,\boldsymbol{\epsilon}_\mathcal{H} \Gamma \mathcal{L}=e^p Q_p-\oint_\mathcal{H}\sqrt{-g} \mathcal{L}\,.
\end{equation}

\section{}\label{app constant k}
It is intended to show that constancy of $k=\frac{g_{t\varphi}}{rg_{\varphi\varphi}}$ in the $4$-dim spacetime is a result of $SL(2,\mathbb{R})\!\times\!U(1)$ isometry. To begin, we note that by the $U(1)$ isometry, the coordinates $\varphi$ can be chosen in a way that metric components would be independent of this coordinate. In this coordinate, the $U(1)$ Killing vector would be $\mrm=\partial_\varphi$.   Similarly, the coordinate $t$ can be chosen in a way that $\xi_-$ would be $\partial_t$, hence the metric components would also be independent of $t$. The $r$ coordinate can be chosen in a way that makes $g_{r\varphi}=0$. Now, the $\xi_0$ would be chosen to be $t\partial_t-r\partial_r$, fulfilling the $[\xi_0,\xi_-]=-\xi_-$. Requesting $\xi_0$ to be Killing vector, fixes the metric non-zero components to have $r$ dependencies as \eqref{T2 tr dependence}. 

By solving some simple differential equations, the most generic vector $\xi_+$ with the commutation relations in \eqref{commutation relation} turns out to be
\begin{equation}
\xi_+=(\frac{t^2}{2}+\frac{a}{r^2})\partial_t+(b-tr)\partial_r+\frac{c}{r} \partial_{\theta}+\frac{d}{r}\partial_\varphi\,,
\end{equation}
where $\{a,b,c,d\}$ are some functions of $\theta$. We could define $r$ from the beginning such that make $a=\text{const.}=\dfrac{1}{2}$. Requesting the $\xi_+$ to be Killing vector of the metric, one finds that $\mathscr{L}_{\xi_+}g_{\varphi\varphi}=0$ and $\mathscr{L}_{\xi_+}g_{tt}=0$ lead to $c=0$ and $b=0$ respectively. Then $\mathscr{L}_{\xi_+}g_{\theta\varphi}=0$ results $\partial_\theta d=0$, \ie  $\,d=\text{const.}\equiv k$. Finally, $\mathscr{L}_{\xi_+}g_{t\varphi}=0$ leads to $\frac{g_{t\varphi}}{rg_{\varphi\varphi}}=k$.

\section{}\label{app NHEG entropy H indep}
For the $\{\xi_-,\xi_0\}$ isometric perturbations, $\delta S$ is $\mathcal{H}$-independent. To show it, consider the entropy perturbation associated with field perturbations $\delta\Phi$ around the NHEG background $\bar{\Phi}$, as
 \begin{align}\label{S H indep proof 1}
\frac{\delta S}{2 \pi}\bigg|_\mathcal{H} = - \oint_\mathcal{H} \ \frac{\delta(\boldsymbol{\epsilon}_{_\mathcal{H}} \epsilon_{_{\!\perp\alpha\beta}}\,\epsilon_{_{\!\perp\mu\nu}}\,{E}^{ \alpha \beta \mu \nu})}{\delta\Phi}\bigg|_{\bar{\Phi}}\ \delta\Phi\,.
\end{align}
According to Appendix \ref{app H to H by xi}, any two arbitrary $\mathcal{H}$ surfaces (defined at different values of $t_{_\mathcal{H}}, r_{_\mathcal{H}}$) are related by a diffeomorphism generated by $\{\xi_-,\xi_0\}$.  $\delta{S}$ would be $\mathcal{H}$-independent if the integrand in \eqref{S H indep proof 1} would be invariant under such diffeomorphisms. That is, if
\begin{equation}
\mathscr{L}_{\xi_a}\left(\frac{\delta(\boldsymbol{\epsilon}_{_\mathcal{H}} \epsilon_{_{\!\perp\alpha\beta}}\,\epsilon_{_{\!\perp\mu\nu}}\,{E}^{ \alpha \beta \mu \nu})}{\delta\Phi}\bigg|_{\bar{\Phi}}\ \delta\Phi\right)=\frac{\delta(\boldsymbol{\epsilon}_{_\mathcal{H}} \epsilon_{_{\!\perp\alpha\beta}}\,\epsilon_{_{\!\perp\mu\nu}}\,{E}^{ \alpha \beta \mu \nu})}{\delta\Phi}\bigg|_{\bar{\Phi}}\ \mathscr{L}_{\xi_a}(\delta\Phi)=0\,,\quad a=-,0\,,
\end{equation}
where in the second equality we used the fact that background fields $\bar{\Phi}$ and $\boldsymbol{\epsilon}$'s are $\{\xi_-,\xi_0\}$ invariant. Hence $\delta{S}$ is $\mathcal{H}$-independent if  $\mathcal{L}_{\xi_-}(\delta\Phi)=\mathcal{L}_{\xi_0}(\delta\Phi)=0$.

A similar reasoning leads to the same proposition for the electric charges, \ie for the $\{\xi_-,\xi_0\}$ isometric perturbations, $\delta Q_p$ is $\mathcal{H}$-independent. For the proof, it would suffice to replace  $\,-\epsilon_{_{\!\perp\mu\nu}}\,{E}^{ \alpha \beta \mu \nu}$ in \eqref{S H indep proof 1} by $\mathrm{Q}^{(p)\alpha\beta}$, introduced in \eqref{electric charge Q-form}, which is also $\{\xi_-,\xi_0\}$ invariant.

\section{}\label{app Omega xi3 calc}
{Here we present details of computation of the symplectic form appearing in the LHS of \eqref{EPL proof 1}. As described in Lemma \ref{Lemma independ H}, for an isometry generator $\xi$, $\boldsymbol{\omega}=0$. All of the Killing vectors of NHEG in \eqref{NHEG isometry} are isometry generators, except for $\xi_+$. Because as it acts on gauge fields $A^{(p)}$, there is a residual gauge transformation, described in \ie \eqref{xi_+ A}. For simplicity we drop all indices $(p)$, without losing generalization. To compute the effects of this residual gauge transformation, we start with the definition of $\boldsymbol{\omega}$  
\begin{align}
\boldsymbol{\omega}(\delta\Phi,\delta_{\xi_+}\Phi,\bar{\Phi})= \delta \mathbf{\Theta}(\delta_{\xi_+}\Phi,\bar{\Phi})-\mathscr{L}_{\xi_+} \mathbf{\Theta}(\delta\Phi,\bar{\Phi})\,.
\end{align}
According to \eqref{gauge Theta pre}
\begin{align*}
\Theta^\mu(\delta A)=\dfrac{\partial \mathcal{L}}{\partial F_{\mu\nu}}\delta A_\nu,
\end{align*}
so
\begin{align}
\delta_2\Theta^\mu(\delta_1 A)&=\delta_2(\dfrac{\partial \mathcal{L}}{\partial F_{\mu\nu}}\delta_1 A_\nu)\\
&=\delta_2(\dfrac{\partial \mathcal{L}}{\partial F_{\mu\nu}})\;\delta_1 A_\nu+\dfrac{\partial \mathcal{L}}{\partial F_{\mu\nu}}\;\delta_2\delta_1 A_\nu\,.
\end{align}
By \eqref{del del Phi}, \ie $\delta_1\delta_2=\delta_2\delta_1$
\begin{align}
\omega^\mu(\Phi,\delta_1\Phi,\delta_2\Phi)=\delta_2(\dfrac{\partial \mathcal{L}}{\partial F_{\mu\nu}})\delta_1 A_\nu-\delta_1(\dfrac{\partial \mathcal{L}}{\partial F_{\mu\nu}})\delta_2 A_\nu\,,
\end{align}
where $\omega^\mu$ is the vector Hodge dual to the $(d\!-\!1)$-form symplectic current $\boldsymbol{\omega}$.
The nonvanishing part of $\omega$ is hence 
\begin{align}
\omega^\mu(\Phi,\delta\Phi,\delta_{\xi_+}\Phi) &=\delta(\dfrac{\partial \mathcal{L}}{\partial F_{\mu\nu}})\delta_{\xi_+} A_\nu-\delta_{\xi_+}(\dfrac{\partial \mathcal{L}}{\partial F_{\mu\nu}})\delta A_\nu\\
&=\delta(\dfrac{\partial \mathcal{L}}{\partial F_{\mu\nu}})\delta_{\xi_+} A_\nu\,.
\end{align}
The second term on the right hand side is zero since $\xi_+$ is an isometry of Lagrangian and $F_{\mu\nu}$.
Next, recall from \eqref{xi_+ A} that
\begin{align*}
\delta_{\xi_+} A_\nu=\nabla_\nu \Lambda ,\hspace*{1cm}\Lambda=\dfrac{-e}{r}\,,
\end{align*}
therefore
\begin{align}
\nonumber\omega^\mu(\Phi,\delta\Phi,\delta_{\xi_+}\Phi) &=\delta(\dfrac{\partial \mathcal{L}}{\partial F_{\mu\nu}})\nabla_\nu \Lambda\\
\nonumber &=\nabla_\nu\Big(\Lambda\delta(\dfrac{\partial \mathcal{L}}{\partial F_{\mu\nu}})\Big)-\Lambda\nabla_\nu\delta(\dfrac{\partial \mathcal{L}}{\partial F_{\mu\nu}})\\
&=\nabla_\nu\Big(\Lambda\delta(\dfrac{\partial \mathcal{L}}{\partial F_{\mu\nu}})\Big)
\end{align}
where we have used the linearized equation of motion for the gauge field perturbations $\delta A_\mu$. Summarizing the calculations above, $\boldsymbol{\omega}(\delta\Phi,\delta_{\xi_+}\Phi,\Phi)=\mrd \mathbf{Q}(\delta\Phi,\delta_{\xi_+}\Phi,\Phi)$ in which $\mathbf{Q}=\star \mathrm{Q}$ and
\begin{equation}
\mathrm{Q}^{\mu\nu}=\Lambda\delta(\frac{\partial \mathcal{L}}{\partial F_{\mu\nu}})\,.
\end{equation}
 Therefore, we obtain
\begin{align}
\Omega(\Phi,\delta\Phi,\delta_{\xi_+}\Phi) &=\int_\Sigma \boldsymbol{\omega}(\delta\Phi,\delta_{\xi_+}\Phi,\Phi)\\ 
&=\oint_{\partial \Sigma}\mathbf{Q}=\oint_{\infty}\mathbf{Q}-\oint_\mathcal{H} \mathbf{Q}\,.
\end{align}
$\Omega$ will hence have a term at infinity and a term on $\mathcal{H}$. The term at infinity does not contribute, since by $\{\xi_-,\xi_0\}$ isometry of $\bar{\Phi}$ and $\delta \Phi$, and noticing Appendix \ref{app t r dependences}, pull-back of $\mathbf{Q}$ to $\mathcal{H}$ has the $r$-dependency similar to $\Lambda=\dfrac{-e}{r}$. Hence it vanishes at infinity. So, the only contribution is
 \begin{align}\label{Lambda-infinity}
\Omega (\delta\Phi,\delta_{\xi_+}\Phi,\bar{\Phi})&=-\oint_\mathcal{H} \mathbf{Q} \\
&=\frac{e}{r_{_\mathcal{H}}}\delta Q
\end{align}
where \eqref{electric charge Q} and \eqref{electric charge Q-form} are used. Noting that $\zeta_{_\mathcal{H}}=n_{_\mathcal{H}}^a\xi_a-k^i\mrm_i$, and that $\Omega$ is linear in $\delta_{\zeta_{_\mathcal{H}}}A=n_{_\mathcal{H}}^a\delta_{\xi_a}A-k^i\delta_{\mrm_i}A$, we obtain (returning back the index $(p)$)
\begin{align}
\Omega(\delta\Phi,\delta_{\zeta_\mrH}\Phi,\bar{\Phi})=n_{_\mathcal{H}}^+\,\Omega(\delta\Phi,\delta_{\xi_+}\Phi)=n_{_\mathcal{H}}^+\,\dfrac{e^p}{r_{_\mathcal{H}}}\delta Q_p=-e^p\delta Q_p\,.
\end{align}}
\newpage


\chapter{Technical proofs and calculations of part II}

\section{}\label{app isometries pert to commut}
We will show that requesting $\{\bar{\xi}_-,\bar{\xi}_0\}$ isometry for $\delta_\chi\Phi$ yields $[\bar \chi,\bar\xi_{-}]=[\bar \chi,\bar\xi_{0}]=0$.  Let  $\bar\Phi$ denote the NHEG background \eqref{NHEG metric} and $\mathcal{A}$ the algebra of background isometries  $sl(2,\mathbb{R})\times u(1)^{d-3}$. 
For notational convenience, we will drop all bars on vector fields in this appendix but it is understood that we are considering generators of diffeomorphisms around the background. 
First, we note 
\begin{equation}
\mathscr{L}_{\xi_{-,0}}\delta_{\chi} \bar{\Phi}=\mathscr{L}_{\xi_{-,0}}{\mathscr{L}}_{\chi} \bar{\Phi} = \mathscr{L}_{[\xi_{-,0},\chi]}\bar{\Phi},
\end{equation}
since $\xi_{-,0}$ are Killing vectors of the background. Requiring $\mathscr{L}_{\xi_{-,0}}\delta_{\chi} \bar{\Phi} = 0$ is therefore equivalent to requiring that $[\chi,\xi_{-,0}] \in \mathcal A$. 

\dotfillb
\Lemma{\label{Lemma pert isometry commut} \emph{The only vectors $\chi$ for which $[\chi,\xi_{-,0}]\in \mathcal{A}$ are linear combination of members of the $sl(2,\mathbb{R})$ algebra and the ones for which $[\chi,\xi_{-}]=0, [\chi,\xi_{0}]=\beta^i {{\mathrm m}_i}$ with $\beta^i$ fixed constants. }
\begin{proof}
$[\chi,\xi_{-,0}]\in \mathcal{A}$ means that
\begin{equation}\label{Sym of pert 2}
\begin{split}
[\chi,\xi_{-}]&=\alpha^1\xi_-+\alpha^2\xi_0+\alpha^3\xi_++\alpha^i \mathrm{m}_i,\\
[\chi,\xi_{0}]&=\beta^1\xi_-+\beta^2\xi_0+\beta^3\xi_++\beta^i \mathrm{m}_i,
\end{split}\end{equation}
for some constants $\alpha$ and $\beta$'s. By the Jacobi identity we have
\begin{equation}
[[\chi,\xi_-],\xi_0]+[[\xi_0,\chi],\xi_-]+[[\xi_-,\xi_0],\chi]=0.
\end{equation}
Inserting \eqref{Sym of pert 2}  in the above equation, and using the algebra of Killings of NHEG, we get
\begin{equation}
(\alpha^1\xi_--\alpha^3\xi_+)+(\beta^2\xi_-+\beta^3\xi_0)-(\alpha^1\xi_-+\alpha^2\xi_0+\alpha^3\xi_++\alpha^i \mathrm{m}_i)=0.
\end{equation}
Noting that the above should identically vanish, coefficients of $\xi_a$ and $\mathrm{m}_i$ all should be set to zero. 
\begin{equation}
\alpha^3=\alpha^i=\beta^2=0, \qquad \qquad \alpha^2=\beta^3,
\end{equation}
and hence 
\begin{equation}\label{Sym of pert 4}
[\chi,\xi_{-}]=\alpha^1\xi_-+\alpha^2\xi_0\,,\qquad  [\chi,\xi_{0}]=\beta^1\xi_-+\alpha^2\xi_++\beta^i \mathrm{m}_i.
\end{equation}
Using the redefinition
\begin{equation}\label{Sym of pert 9}
\chi'\equiv\chi+\alpha^1\xi_0+\alpha^2\xi_+-\beta^1\xi_-\,,
\end{equation}
then
\begin{align}
[\chi',\xi_{-}]&=0,\label{Sym of pert 6}\\
[\chi',\xi_{0}]&=\beta^i \mathrm{m}_i\,.\label{Sym of pert 7}
\end{align}
{Therefore,  recalling \eqref{Sym of pert 9}, we have proved that $\chi$ is a linear combination of $\chi'$ with properties \eqref{Sym of pert 6}-\eqref{Sym of pert 7}, and a member of $sl(2,\mathbb{R})$, namely $-\alpha^1\xi_0-\alpha^2\xi_++\beta^1\xi_-$.}
\end{proof}
}
\vspace*{-0.5cm}
\dotfille

One can use Lemma \ref{Lemma pert isometry commut} to find the generic components of $\chi'$ explicitly. 
\eqref{Sym of pert 6} is just $\partial_t \chi'^{\mu}=0$. It means that $\chi'=\chi^\mu\partial_\mu$ where $\chi'^\mu=\chi'^\mu(r,\theta,\varphi^i)$. Inserting it in \eqref{Sym of pert 7}, leads to the following equations.
\begin{equation}\label{Sym of pert 8}
\begin{cases}
(r\partial_r+1)\chi'^t=0\\
(r\partial_r-1)\chi'^r=0\\
r\partial_r\chi'^\theta=0\\
r\partial_r\chi'^{\varphi^i}=\beta^i
\end{cases}
\end{equation}
The above equations fix the $r$ dependence of the $\chi'^\mu$ as follows
\begin{equation}\label{chi-1}
\chi'=\frac{\epsilon^{t}}{r}\partial_t+ r\epsilon^{ r} \partial _r +  \epsilon^\theta \partial _ {\theta}+ (\beta^i\ln r+\epsilon^i)  \partial_{\varphi^i}\, ,
\end{equation}
where $\epsilon^\mu=\epsilon^\mu(\theta,\varphi^i)$. We then fix the constants $\beta^i=0$ since exponentiating such generators would lead to logarithmic terms which would be very irregular at the Poincar\'e horizon. What which remains in \eqref{chi-1} commutes with $\{\xi_-,\xi_0\}$.

\section{}\label{app Z2}
The NHEG background \eqref{NHEG metric} is invariant under the two $Z_2$ transformations; $(r\to -r, \vec{\varphi} \to -\vec{\varphi})$ or $(t\to -t, \vec{\varphi} \to -\vec{\varphi})$.
In Section \ref{sec identify chi}, two families of vector fields were distinguished as generators for the NHEG phase space,
\begin{equation}
\chi_\pm[\epsilon(\vec{\varphi})]=-\vec{k}\cdot \vec{\partial}_\varphi \epsilon (\frac{b }{r}\partial_t+r\partial_r)+\epsilon \vec{k}\cdot\vec{\partial}_\varphi, \qquad b=\pm 1.
\end{equation}
Let us denote the phase spaces generated by $\chi_\pm$ as $\mathcal{G}_\pm [\{F\}]$. Here we show that
\begin{center}
\emph{{The two $Z_2$ transformations} map $\mathcal{G}_+ [\{F\}]$ and $\mathcal{G}_- [\{F\}]$ onto each other.}
\end{center}
\begin{proof}
The background is mapped to itself under { any of the two} $Z_2$ transformations. The $\chi_+[\epsilon]$ is mapped to the $\chi_-[\tilde{\epsilon}]$ in which
\begin{equation}
\tilde{\epsilon}(\vec{\varphi})=-\epsilon(-\vec{\varphi})\,.
\end{equation}
This map provides the bijection relation
\begin{equation}
\mathcal{G}_+[\{F(\vec{\varphi})\}]\leftrightarrow \mathcal{G}_-[\{-F(-\vec{\varphi})\}]\,.
\end{equation}
\end{proof}

\section{}\label{app finite trans}

In this section, the exponentiation of the NHEG symplectic symmetry generators $\chi$ are performed. At the infinitesimal level, one applies the coordinate transformation
\begin{align}
\bar x&\rightarrow x = \bar x-\bar \chi( \bar x).
\end{align}
To find the finite coordinate transformation $ \bar x \rightarrow x(\bar x)$ we use a tricky way composed of two steps. The first step is to constraint the generic shape of coordinate transformations, using some requested conditions. The finite coordinate transformation ought to take the form
\begin{align}\label{finite ansatz}
\bar{\varphi}^i=\varphi^i + k^i F(\vec{\varphi}), \qquad \bar{r} =re^{-{\Psi(\vec{\varphi})}},\qquad \bar{t} =t-\frac{b}{r}(e^{\Psi(\vec{\varphi})}-1),
\end{align}
with functions $F(\vec{\varphi})$ and ${\Psi}(\vec{\varphi})$ periodic in all of their arguments in order to ensure smoothness. 
Indeed, the form of the finite coordinate transformation \eqref{finite ansatz} is constrained by the following facts; (1) $\vec{\epsilon}$ is proportional to $\vec{k}$ and hence $\varphi^i-\bar\varphi^i$ is also proportional to $k^i$; (2) $\bar \chi$ commutes with $\xi_-$ and therefore the time dependence is trivial; (3) there is no $\theta$ dependence; (4) $\bar \chi$ commutes with $\xi_0$ and therefore the radial dependence is uniquely fixed; (5) since $\bar  \chi$ commutes with the vector 
\begin{equation}
\eta_b \equiv \frac{b}{r}\partial_t+r\partial_r, \label{defetab}
\end{equation}
the coordinate
$$
v_b \equiv t+\frac{b}{r},
$$
is invariant.\footnote{In other words, in the coordinates $(v_b,r,\theta,\varphi^i)$ the generator $\bar \chi$ has $\vec{\partial}_\varphi$ and ${\partial}_r$ components. Therefore the coordinate $v_b$ is not affected by the exponentiation of $\bar \chi$.} Note that $v_b$ for $b=\pm 1$ reduces to $v$ and $u$ in the condition (4) in Section \ref{sec identify chi}. This finally fixes the form \eqref{finite ansatz}. We can check that by \eqref{finite ansatz}, always $v_b = \bar t+\frac{b}{\bar r}$. 

The second step is requesting the covariance of the generic shape of $\chi$ over the phase space. Mathematically it is
\begin{equation}\label{chi covariance}
\chi=\chi^\mu[\epsilon(\vec{\varphi})]\partial_\mu=\bar{\chi}^\mu[\bar \epsilon (\vec{\bar\varphi})]\bar{\partial}_{\mu}\,,
\end{equation}
for some periodic functions $\epsilon$ and $\bar\epsilon$. This request relates the functions $F(\vec{\varphi})$ and  ${\Psi}(\vec{\varphi})$ as
\begin{align}\label{Psi-def app}
e^{\Psi}=1+\vec{k}\cdot\vec{\partial}_\varphi F.
\end{align}
In order to prove the claim, beginning from \eqref{ASK}
\begin{equation}\label{ASK-2}
\chi[{\epsilon}(\vec{\varphi})]={\epsilon}(\vec{\varphi})\vec{k}\cdot\vec{\partial}_\varphi-\vec{k}\cdot\vec{\partial}_\varphi{\epsilon}(\dfrac{b}{r}\partial_{t}+r\partial_{r})\,,
\end{equation}
we note that
\begin{equation}\label{chi trans bar chi}
\bar{\chi}^\mu[\bar \epsilon (\vec{\bar\varphi})]=\frac{\partial \bar{x}^\mu}{\partial x^\alpha}\chi^\alpha[\epsilon(\vec{\varphi})]\,.
\end{equation}
Putting \eqref{ASK-2} and the coordinate transformations \eqref{finite ansatz} into \eqref{chi trans bar chi}, RHS is found to be
\begin{equation}\label{finite trans proof 1}
\bar{\chi}^\mu[\bar \epsilon (\vec{\bar\varphi})]={\epsilon}(1+X)\vec{k}\cdot\vec{\partial}_{\bar{\varphi}} -(\dfrac{b}{\bar{r}}\partial_{\,\bar{t}}+\bar{r}\partial_{\,\bar{r}})\Big(\;{\epsilon}\ \vec{k}\cdot\vec{\partial}_\varphi {\Psi}+\vec{k}\cdot\vec{\partial}_\varphi{\epsilon}\Big)\,,
\end{equation}
where $X\equiv \vec{k}\cdot \vec{\partial}_\varphi F(\vec{\varphi})$. Comparing the request \eqref{chi covariance} with \eqref{finite trans proof 1} yields
\begin{align}\label{Psi equation}
\bar\epsilon(\vec{\bar\varphi})\equiv (1+X)\ {\epsilon}\,, \qquad \vec{k}\cdot\vec{\partial}_{\bar{\varphi}} \bar\epsilon={\epsilon}\ \vec{k}\cdot\vec{\partial}_\varphi {\Psi}+\vec{k}\cdot\vec{\partial}_\varphi\ {\epsilon}\,.
\end{align}
To solve \eqref{Psi equation}, we use the fact that
\begin{equation}
\vec{k}\cdot\vec{\partial}_\varphi=-\vec{k}\cdot\vec{\partial}_\varphi {\Psi}(\dfrac{b}{\bar{r}}\partial_{\,\bar{t}}+\bar{r}\partial_{\,\bar{r}})+(1+X)\vec{k}\cdot\vec{\partial}_{\bar{\varphi}}\,,
\end{equation}
which is a result of \eqref{finite ansatz}. Hence, when dealing with functions of  $\vec{\varphi}$ only,
\begin{align}
\vec{k}\cdot\vec{\partial}_{\bar{\varphi}}&=\dfrac{\vec{k}\cdot\vec{\partial}_\varphi}{1+X}\,.
\end{align}
Therefore by the first equation in \eqref{Psi equation}
\begin{align}
\vec{k}\cdot\vec{\partial}_{\bar{\varphi}} \bar\epsilon&=\dfrac{\vec{k}\cdot\vec{\partial}_\varphi}{1+X}\big((1+X)\,\epsilon\big)\cr
&=\dfrac{{\epsilon}\vec{k}\cdot\vec{\partial}_\varphi X}{1+X}+\vec{k}\cdot\vec{\partial}_\varphi{\epsilon}.
\end{align}
Comparison with the second equation of \eqref{Psi equation} then implies ${\Psi}=\ln (1+X)$,  \ie the claimed \eqref{Psi-def app}.
So we have established our ansatz \eqref{finite ansatz} which defines a one-function family of finite coordinate transformations, specified by the function $F(\vec{\varphi})$.

\section{}\label{app eta}

Two special vectors fields are singled out in our construction.
\begin{equation}
\bar \eta_+=\frac{1}{\bar r}\partial_{\bar t}+\bar r\partial_{\bar r},\qquad \bar \eta_-=\frac{1}{\bar r}\partial_{\bar t}-\bar r\partial_{\bar r}. 
\end{equation}
They obey the commutation relation 
\begin{equation}
[\bar \eta_+,\bar \eta_-]=-(\bar \eta_++\bar \eta_-),
\end{equation}
and therefore they form a closed algebra under the Lie bracket. Here are some of their properties:
\begin{enumerate}
\item Although not Killing vectors, $\bar  \eta_\pm$ commute with $\bar \xi_-,\bar \xi_0$ and the $U(1)^{d-3}$ generators $\bar{ \mathrm{m}}_i$ in \eqref{NHEG isometry}.
\item They commute with the respective symmetry generator $\bar \chi_\pm$; i.e.
\begin{equation}\label{commute eta}
\delta_{\bar \eta_+}\bar \chi_+=[\bar \eta_+,\bar \chi_+]=0\,, \qquad \delta_{\bar \eta_-}\bar \chi_-=[\bar \eta_-,\bar \chi_-]=0,
\end{equation}
where $\bar \chi_\pm$ correspond to the choice of $\bar \chi_b$ with $b=\pm1$.
\item As $\bar \eta_+$ commutes with the phase space generating diffeomorphism $\bar \chi_+$, it is invariant in the phase space generated by $\bar \chi_+$  which in turn implies invariance of $\bar\eta_+$ over the phase space, \ie
\begin{equation}
\eta_+ =\frac{1}{r}\partial_t+ r\partial_r.
\end{equation}
The above may be explicitly checked using \eqref{finite transformations}. The same property holds with minuses in the respective phase space. 

\item  $\eta_\pm$, similarly to $\chi_{\pm}$, are mapped to each other by the $Z_2$-transformations discussed in section \ref{sec Z2 isometry}.  
\end{enumerate} 
One can in fact show that $\eta_+$ (or $\eta_-$) are the only vectors with properties 1. and 2. in the above list. Properties 3. and 4. then follow from the first two.

\section{}\label{app regularizing}
On the NHEG phase space, the Lee-Wald symplectic current has singularities at the poles of $\mathcal{H}$. One can calculate it and observe the divergences. Also there can be analytic arguments showing it \cite{CHSS:2015bca}. Here a simple argument is provided to show it indirectly. The argument is as follows. According to the $\{\xi_-,\xi_0\}$ isometry of $g_{\mu\nu}[F]$ and the $\chi$s, Lemma \ref{Lemma NHEG H independence} results that $\delta_{\chi_1}H_{\chi_2}$ should be $\mathcal{H}$ independent. On the other hand, $\boldsymbol{\omega}_{_\text{LW}}$ is also $\{\xi_-,\xi_0\}$ isometric. Hence by the Appendix \ref{app t r dependences}, the $r$ dependency of its components are determined as ${\omega}^t_{_\text{LW}}\propto \frac{1}{r}  $ and ${\omega}^r_{_\text{LW}}\propto r$. These components are non-vanishing on-shell. One can choose a $d\!-\!1$-dim surface $\Sigma$ with boundaries $\mathcal{H}_1$ and $\mathcal{H}_2$. Then, by the Stoke's theorem,
\begin{equation}
\int_\Sigma \boldsymbol{\omega}_{_\text{LW}}(\delta_{\chi_1}\Phi,\delta_{\chi_2}\Phi,\Phi)=\delta_{\chi_1}H_{\chi_2}\Big|_{\mathcal{H}_2}-\delta_{\chi_1}H_{\chi_2}\Big|_{\mathcal{H}_1}\,.
\end{equation}
The LHS does not vanish, because the integration over $r$ would be done simply and factored out, while the integration on the $(\theta,\varphi^i)$ does not vanish. But the RHS was supposed to vanish by the discussion above. So there is seemingly a contradiction. But the point is that $\boldsymbol{\omega}_{_\text{LW}}$ is singular at the poles, undermining the Stoke's theorem.

\section{}\label{app Y fix}
In this section, we show that how requesting $\chi$ to be symplectic symmetry generator, fixes the $\mathbf{Y}$ ambiguity to be the \eqref{NHEG Y theta}. Calculating the LW symplectic current without any $\mathbf{Y}$ contributions, \ie 
\begin{equation}
\boldsymbol{\omega}_{_{\text{LW}}}=\delta_{\chi_1}\mathbf{\Theta}(\delta_{\chi_2}\Phi,\Phi)-\delta_{\chi_2}\mathbf{\Theta}(\delta_{\chi_1}\Phi,\Phi)
\end{equation}
with the standard $\mathbf{\Theta}$ \eqref{EH Theta}, it can be checked that for any $\chi_1,\chi_2\in \{\chi_b\}$
\begin{equation}\label{eta wald}
\mathscr{L}_{\eta_b}\boldsymbol{\omega}_{_\text{LW}}(\delta_{\chi_1}\Phi,\delta_{\chi_2}\Phi,\Phi)=\boldsymbol{\omega}_{_\text{LW}}(\delta_{\chi_1}\Phi,\delta_{\chi_2}\Phi,\Phi)\,.
\end{equation} 
In order to have a symplectic structure such that ${\boldsymbol{\omega}}\approx 0$, one can fix the ambiguity of the symplectic structure as
\begin{equation}
\boldsymbol{\omega}(\delta_{\chi_1}\Phi,\delta_{\chi_2}\Phi,\Phi)\equiv\boldsymbol{\omega}_{_\text{LW}}(\delta_{\chi_1}\Phi,\delta_{\chi_2}\Phi,\Phi)+\mrd\Big(\delta_{\chi_1}{\mathbf{Y}}(\delta_{\chi_2}\Phi,\Phi)-\delta_{\chi_2}{\mathbf{Y}}(\delta_{\chi_1}\Phi,\Phi)\Big)
\end{equation}
such that
\begin{equation}
\mrd\Big(\delta_{\chi_1}{\mathbf{Y}}(\delta_{\chi_2}\Phi,\Phi)-\delta_{\chi_2}{\mathbf{Y}}(\delta_{\chi_1}\Phi,\Phi)\Big)\approx -\mathscr{L}_{\eta_b}\boldsymbol{\omega}_{_\text{LW}}\,.
\end{equation}
If so, then according to the \eqref{eta wald}, it simply follows that
\begin{equation}\label{Y 5}
{\boldsymbol{\omega}}\approx 0\,.
\end{equation}
The argument below identifies the suitable ${\mathbf{Y}}$ term:
\begin{align}
\mathscr{L}_{\eta_b}\boldsymbol{\omega}_{_\text{LW}}(\delta_{\chi_1}\Phi,\delta_{\chi_2}\Phi,\Phi)&=\mathscr{L}_{\eta_b}\Big( \delta_{\chi_1}\mathbf{\Theta}(\delta_{\chi_2}\Phi,\Phi)-\mathscr{L}_{\chi_2}\mathbf{\Theta}(\delta_{\chi_1}\Phi,\Phi)\Big)\label{Y 1}\\
&= \delta_{\chi_1}\mathscr{L}_{\eta_b}\mathbf{\Theta}(\delta_{\chi_2}\Phi,\Phi)-\mathscr{L}_{\chi_2}\mathscr{L}_{\eta_b}\mathbf{\Theta}(\delta_{\chi_1}\Phi,\Phi)\label{Y 2}\\
&\approx\delta_{\chi_1}\mrd\Big(\eta_b\cdot\mathbf{\Theta}(\delta_{\chi_2}\Phi,\Phi)\Big)-\mathscr{L}_{\chi_2}\mrd\Big(\eta_b\cdot\mathbf{\Theta}(\delta_{\chi_1}\Phi,\Phi)\Big)\label{Y 3}\\
&=\mrd \Big(\delta_{\chi_1}(\eta_b\cdot\mathbf{\Theta}(\delta_{\chi_2}\Phi,\Phi))-\mathscr{L}_{\chi_2}(\eta_b\cdot\mathbf{\Theta}(\delta_{\chi_1}\Phi,\Phi))\Big)\\
&= -\mrd\Big( \delta_{\chi_1}{\textbf{Y}}(\delta_{\chi_2}\Phi,\Phi)-\mathscr{L}_{\chi_2}{\textbf{Y}}(\delta_{\chi_1}\Phi,\Phi)\Big)
\end{align} 
in which we have defined 
\begin{equation}\label{Y 4}
{\textbf{Y}}(\delta\Phi,\Phi)\equiv -\eta_b\cdot\mathbf{\Theta}(\delta\Phi,\Phi)\,.
\end{equation}
In deriving \eqref{Y 2} from \eqref{Y 1}, we have used $\mathscr{L}_{\eta_b}\delta_{\chi_b}=\delta_{\chi_b}\mathscr{L}_{\eta_b}$, as a result of $[\eta_b,\chi_b]=0$. In deriving \eqref{Y 3} from \eqref{Y 2} the Cartan magic formula \eqref{Cartan magic} is used, and the $\mrd\mathbf{\Theta}(\delta_\chi\Phi,\Phi)\approx 0$, which is a result of $\delta_\chi \mathbf{L}\approx 0$. For sure, one still can add a $\mathbf{Y}_\text{c}$ term to the \eqref{Y 4} for which 
$$\mrd \big(\delta_{\chi_1}\mathbf{Y}_{\text{c}}(\delta_{\chi_2}\Phi,\Phi)-\mathscr{L}_{\chi_2}\mathbf{Y}_{\text{c}}(\delta_{\chi_1}\Phi,\Phi)\big)\approx 0\,,$$
 without affecting the \eqref{Y 5}.

\section{}\label{app constancy of J}
 
 A direct check for the constancy of $J_i$ and $H_{\xi_+}$ over the $\mathcal{M}$ is presented. For the $\mrm_i$
\begin{equation}
\delta J_i =- \int_{\mathcal H} \boldsymbol k_{{\mathrm{m}}_i}[\delta_\chi \Phi, \Phi]=- \int_{\mathcal \mathscr{S}} \boldsymbol k_{\bar{\mathrm{m}}_i}(\delta_{\bar\chi} \bar \Phi, \bar \Phi)=- \int_{\mathcal H} \boldsymbol k_{\bar{\mathrm{m}}_i}(\delta_{\bar\chi} \bar \Phi, \bar \Phi)=0.
\end{equation}	
The second equality follows from general covariance of all expressions, and noting the $\mathcal{M}$ is built by some coordinate transformations. $\mathscr{S}$ would be the transformed $\mathcal{H}$.  The third equation is a result of Lemma \ref{Lemma independ H}, which enabled us to replace $\mathscr{S}$ with surfaces of constant $(\bar t,\bar r)$. Finally the last equality is a result of the fact that $\bar \Phi$ has axial isometry, and the only $\bar \varphi^i$ dependence coming from $\bar\chi$ makes the integral vanishing. To be more specific, using the Fourier expansion \eqref{Ln expansion} for $\bar \chi$, for the modes with $\vec{n}\neq \vec{0}$, integration over $\vec{\varphi}$ results integration to be zero. For the mode $\vec{n}=\vec{0}$, the isometry condition $\delta_{\vec{0}}\bar\Phi=0$ yields integrand to be zero. 

Hence, $\delta J_i=0$ over all of the points of the phase space. So all points have the same angular momenta as the reference point, which is $J_i$. This argument can also be repeated  for $\xi_a$ charges. Therefore $H_{\xi_a}$ would be constant over the phase space, which by the choice of their reference to be zero on $\bar{g}_{\mu\nu}$, $H_{\xi_a}=0$ over the phase space.

\section{}\label{app central extension} 
In this section, we are intended to find the central extension, appearing in the NHEG algebra
\begin{align}\label{central calc 1}
\{H_{\vec{m}}, H_{\vec{n}}\} = -i\vec{k} \cdot (\vec{m}- \vec{n}) H_{\vec{m}+\vec{n}} + C_{\vec{m},\vec{n}}\,.
\end{align}
Due to the discussion in Section \ref{sec NHEG symplectic symmetry}, central extension is a constant function over $\mathcal{M}$. So we can calculate it on the reference point $g_{\mu\nu}[F\!=\!0]$. The reason for this choice is the  manifest axial isometry of this point, making the calculations easier. Besides, $g_{\mu\nu}[F\!=\!0]$ is the point which we can use to choose the reference point of $H_{\vec{n}}$ in a natural way. 

The LHS of \eqref{central calc 1} can be calculated by
\begin{align}
\{H_{\vec{m}}, H_{\vec{n}}\} &= \oint_\mathcal{H} {\boldsymbol k}_{\chi_{\vec{m}}}(\delta_{\chi_{\vec{n}}}\Phi,\bar\Phi), \\
&=\oint_\mathcal{H} {\boldsymbol k}^{\text{EH}}_{\chi_{\vec{m}}}(\delta_{\chi_{\vec{n}}}\Phi,\bar\Phi)+\oint_\mathcal{H}{\boldsymbol k}^{\mathbf{Y}}_{\chi_{\vec{m}}}(\delta_{\chi_{\vec{n}}}\Phi,\bar\Phi )\,.
\end{align}
The result is
\begin{align}
&\{H_{\vec{m}}, H_{\vec{n}}\}=0\, \qquad \vec{n}\neq -\vec{m}\,,\label{central calc 2}\\
&\{H_{\vec{m}}, H_{-\vec{m}}\}=\frac{-1}{16\pi G}\oint \boldsymbol{\epsilon}_{_\mathcal{H}}\Big(2i(\vec{k}\cdot \vec{m})\big(1+(\vec{k}\cdot \vec{m})^2\big) k^ik^j\gamma_{ij}\Big)\nonumber\\
&\hspace*{2cm}+\frac{-1}{16\pi G}\oint \boldsymbol{\epsilon}_{_\mathcal{H}}\Big(2i(\vec{k}\cdot \vec{m})^3 \big(1-k^ik^j\gamma_{ij})\Big)\,.\label{central calc 3}
\end{align}
\eqref{central calc 2} vanishes because by the axial isometry of the background, the integrand would be a function of $\vec{\varphi}$ only through $e^{-i(\vec{m}+\vec{n})\cdot\vec{\varphi}}$. Hence, the integration over the $\vec{\varphi}$ makes the integral to vanish. In \eqref{central calc 3}, the first integral is from the ${\boldsymbol k}^{\text{EH}}$, and the second is from the ${\boldsymbol k}^{\mathbf{Y}}$. Simplifying it, 
\begin{align}
\{H_{\vec{m}}, H_{-\vec{m}}\}&=\frac{-1}{16\pi G}\oint \boldsymbol{\epsilon}_{_\mathcal{H}}\Big(2i(\vec{k}\cdot \vec{m})k^ik^j\gamma_{ij}\Big)+\frac{-1}{16\pi G}\oint \boldsymbol{\epsilon}_{_\mathcal{H}}2i(\vec{k}\cdot \vec{m})^3 \\
&=-2i\vec{k}\cdot\vec{J}-i(\vec{k}\cdot \vec{m})^3 \frac{S}{2\pi}\,.
\label{central calc 4}
\end{align}
As an overall result, by calculation we have found that
\begin{equation}
\{H_{\vec{m}}, H_{\vec{n}}\}=\Big(-2i(\vec{k}\cdot \vec{m})\vec{k}\cdot\vec{J}-i(\vec{k}\cdot \vec{m})^3 \frac{S}{2\pi}\Big)\delta_{\vec{m}+\vec{n},0}\,.
\end{equation}
Comparing it with \eqref{central calc 1}, the $C_{\vec{m},\vec{n}}$ can be read, if the $H_{\vec{m}+\vec{n}}$ would be determined on the reference point $g_{\mu\nu}[F\!=\!0]$. That is a simple task to do. In the case of $\vec{n}\neq -\vec{m}$, according to the axial isometry of the background, any integrand which would lead to $H_{\vec{m}+\vec{n}}$ has the $\vec{\varphi}$ functionality only through $e^{-i(\vec{m}+\vec{n})\cdot\vec{\varphi}}$. Hence, the integration over the $\vec{\varphi}$ makes the integral to vanish. For the case of $\vec{n}= -\vec{m}$, we have $\chi_{\vec{0}}=-\vec{k}\cdot\vec{\partial}_\varphi$. So according to the reference points of the angular momenta, we have $H_{\vec{0}}=\vec{k}\cdot\vec{J}$. As a result, we end up with
\begin{align}
\{H_{\vec{m}}, H_{\vec{n}}\} = -i\vec{k} \cdot (\vec{m}- \vec{n}) H_{\vec{m}+\vec{n}} -i (\vec{k}\cdot \vec{m})^3 \frac{S}{2\pi}\,\delta_{\vec{m}+\vec{n},0}\,.
\end{align}

\section{}\label{app finite charges}
In this appendix, the calculation of \eqref{H_n}, \ie 
\begin{align}
H_{\vec{n}}&=\frac{i}{\vec{k}\cdot\vec{n}}\{H_{\vec{n}}, H_{\vec{0}}\}\hspace{1cm}\vec{n}\neq \vec{0}\,
\end{align}
is presented. Substituting the RHS by \eqref{Poisson NHEG} we have
\begin{align}
H_{\vec{n}}&= \dfrac{i}{\vec{k} \cdot \vec{n}}\oint_{\mathcal{H}} {\boldsymbol k}_{\chi_{\vec{n}}}(\delta_{\chi_{\vec{0}}}\Phi,\Phi) \\
&=\dfrac{i}{\vec{k}\cdot\vec{n}}\oint_{\mathcal{H}} {\boldsymbol k}^{\text{EH}}_{\chi_{\vec{n}}}(\delta_{\chi_{\vec{0}}}\Phi,\Phi )+\dfrac{i}{\vec{k}\cdot\vec{n}}\oint_{\mathcal{H}}{\boldsymbol k}^{\mathbf{Y}}_{\chi_{\vec{n}}}(\delta_{\chi_{\vec{0}}}\Phi,\Phi )\,.
\end{align}
Using \eqref{EH k} and \eqref{NHEG Y}, calculated on the metrics \eqref{finalphasespace}, results
{	\begin{flalign}\label{Hn charge k term}
 &\Big({\boldsymbol k}^{\text{EH}}_{\chi_{\vec{n}}}(\delta_{\chi_{\vec{0}}}\Phi,\Phi )\Big)_{\theta\varphi^1\dots\varphi^n}=\frac{-\sqrt{-g}\,   e^{-i \vec{n}\cdot\vec{\varphi} }}{16 \pi G \Gamma } \Bigg[2k^ik^j\gamma_{ij}\Big(e^\Psi(i\vec{k}\cdot\vec{n}\Psi'-\Psi'^2-\Psi'')\Big)&\nonumber\\
 &\hspace*{4cm}+\Big(i\vec{k}\cdot\vec{n}\Psi''-\Psi'''\Big)+\Big(e^\Psi(\Psi'^2+\Psi''-i\vec{k}\cdot\vec{n}\Psi')\Big)\nonumber\\
&\hspace*{4cm}+2k^ik^j\gamma_{ij}\Big(\Psi''\Psi'-
  i\vec{k}\cdot\vec{n}\Psi''+e^{2\Psi}\Psi'\Big)\Bigg],
	\end{flalign}}
\vspace*{-0.5cm}	
{\small \begin{flalign}\label{finite charges calc 1}
&\Big({\boldsymbol k}^{\mathbf{Y}}_{\chi_{\vec{n}}}(\delta_{\chi_{\vec{0}}}\Phi,\Phi )\Big)_{\theta\varphi^1\dots\varphi^n}\!\!\!=\!\frac{\sqrt{-g}\,i\vec{k}\cdot\vec{n} (k^ik^j\gamma_{ij}-1) e^{-i\vec{n}\cdot\vec{\varphi}}}{16 \pi G \Gamma} \Bigg[\Psi'^2\!-\!2\Psi''+(\Psi''\!-\!i\vec{k}\cdot\vec{n}\Psi')\Bigg],&
	\end{flalign}}
	
\vspace*{-0.5cm}	
where prime denotes the directional derivative $\vec{k}\cdot \vec{\partial}$. A useful thing to note is that all of the contributions in \eqref{finite charges calc 1} are from the $\mathbf{Y}_c$. The first three parenthesis in ${\boldsymbol k}^{\text{EH}}$ and the last one in ${\boldsymbol k}^{\mathbf{Y}}$ are total derivatives in $\vec{\varphi}$. They are explicitly proportional to  $(\Psi' e^{\Psi-i \vec{n}\cdot\vec{\varphi} })'$, $(\Psi''e^{-i \vec{n}\cdot\vec{\varphi} })'$, $(\Psi' e^{\Psi-i \vec{n}\cdot\vec{\varphi} })'$ and $(\Psi'e^{-i \vec{n}\cdot\vec{\varphi} })'$. Therefore their integration vanishes. Now considering the identity $\int d\theta \sqrt{-g}\frac{k^ik^j\gamma_{ij}}{\Gamma}=2 \int d\theta \frac{\sqrt{-g}}{\Gamma}$, we  have
\begin{align}
H_{\vec{n}}&=\dfrac{i}{\vec{k}\cdot\vec{n}}\oint\boldsymbol{\epsilon}_{\mathcal{H}} \frac{-4e^{-i \vec{n}\cdot\vec{\varphi}}}{16 \pi G } \Big(\Psi''\Psi'- i\vec{k}\cdot\vec{n}\Psi''+e^{2\Psi}\Psi'\Big)-\oint\boldsymbol \epsilon_{\mathcal H}\ \frac{e^{-i \vec{n}\cdot\vec{\varphi}}}{16 \pi G} \left({\Psi'}^2-2\Psi '' \right)\nonumber\\
&=\oint\boldsymbol{\epsilon}_{\mathcal{H}} \frac{e^{-i \vec{n}\cdot\vec{\varphi}}}{16 \pi G } \Big(2\Psi'^2- 4\Psi''+2e^{2\Psi}\Big)-\oint\boldsymbol \epsilon_{\mathcal H}\ \frac{e^{-i \vec{n}\cdot\vec{\varphi}}}{16 \pi G} \left({\Psi'}^2-2\Psi '' \right)\,,
\end{align}
where in the last equation we used integration by parts, and dropped some total derivatives of $\vec{\varphi}$. Finally,
\begin{equation}\label{Hn charge}
H_{\vec{n}}=\oint \boldsymbol \epsilon_{\mathcal H}\ \frac{1}{16 \pi G} \left({\Psi'}^2-2 \Psi '' +2e^{2 \Psi }\right) e^{-i \vec{n}\cdot\vec{\varphi}}.
\end{equation}

\addtocontents{toc}{\vspace{2em}} 

\backmatter


\label{Bibliography}

\lhead{\emph{Bibliography}} 

\bibliographystyle{unsrt-phys} 

\bibliography{References} 

\end{document}